\documentclass[a4paper,11pt]{article}
\usepackage{geometry}
\usepackage{a4wide}
\usepackage[round]{natbib}
\usepackage{color}
\usepackage{enumitem}
\usepackage{tikz}
\usepackage{adjustbox}
\usepackage{verbatim}
\usepackage{longtable}
\usepackage{amsmath, amssymb, graphicx}
\usepackage{mathrsfs}
\usepackage{bm}
\usepackage{setspace}
\usepackage{dsfont}
\usepackage[figuresright]{rotating}
\usepackage{enumitem}
\usepackage{algpseudocode}
\usepackage{algorithm}
\usepackage{algorithmicx}
\usepackage{resizegather}
\usepackage{changepage}
\usepackage[figurewithin = section, tablewithin = section]{caption}
\usepackage{subcaption} 
\usepackage{blkarray}
\usepackage{dcolumn}
\usepackage{hyperref}
\usepackage{lmodern}
\usepackage[auth-sc]{authblk}
\usepackage{dcolumn}

\numberwithin{equation}{section}

\newcommand{\fref}[1]{Figure~\ref{#1}}
\newcommand{\tref}[1]{Table~\ref{#1}}
\newcommand{\secref}[1]{Section~\ref{#1}}

\newcommand{\gb}{\bm{g}}
\newcommand{\ub}{\bm{u}}
\newcommand{\xb}{\bm{x}}

\newcommand{\zb}{\bm{z}}
\newcommand{\Sb}{\bm{s}}
\newcommand{\tb}{\bm{t}}
\newcommand{\db}{\bm{d}}
\newcommand{\parms}{\bm{\phi}}
\newcommand{\Zb}{\bm{Z}}
\newcommand{\mo}{\mathcal{M}}
\newcommand{\Ab}{\bm{A}}
\newcommand{\Sigmab}{\bm{\Sigma}}

\newcommand{\corparms}{\ensuremath{\bm{\phi}}}
\newcommand{\corpar}{\ensuremath{\phi_i}}

\newcommand{\range}{\ensuremath{\lambda}}
\newcommand{\smooth}{\ensuremath{\kappa}}
\newcommand{\rotang}{\ensuremath{\alpha}}
\newcommand{\ratio}{\ensuremath{r}}

\title{ABC model selection for spatial extremes models applied to South Australian maximum temperature data\footnote{\textcopyright\ 2018. This manuscript version is made available under the CC--BY--NC--ND $4.0$ license \newline \url{http://creativecommons.org/licenses/by-nc-nd/4.0/}.}}

\date{ }

\author[1,2,3]{Xing~Ju~Lee}
\author[1,4]{Markus~Hainy \footnote{e-mail: markus.hainy@jku.at}}
\author[1]{James~P.~McKeone}
\author[1,2]{Christopher~C.~Drovandi}
\author[1,2]{Anthony~N.~Pettitt}

\affil[1]{School of Mathematical Sciences, Queensland University of Technology, Brisbane, Australia}
\affil[2]{ARC Centre of Excellence in Mathematical and Statistical Frontiers (ACEMS), Queensland University of Technology, Brisbane, Australia}
\affil[3]{Centre of Research Excellence in Reducing Healthcare Associated Infections (CRE-RHAI), Queensland University of Technology, Brisbane, Australia}
\affil[4]{Department of Applied Statistics, Johannes Kepler University, Linz, Austria}

\begin{document}
 
\maketitle
 
\begin{abstract}
Max-stable processes are a common choice for modelling spatial extreme data as they arise naturally as the infinite-dimensional generalisation of multivariate extreme value theory. Statistical inference for such models is complicated by the intractability of the multivariate density function. Nonparametric, composite likelihood-based, and Bayesian approaches have been proposed to address this difficulty. More recently, a simulation-based approach using approximate Bayesian computation (ABC) has been employed for estimating parameters of max-stable models. ABC algorithms rely on the evaluation of discrepancies between model simulations and the observed data rather than explicit evaluations of computationally expensive or intractable likelihood functions. The use of an ABC method to perform model selection for max-stable models is explored. Three max-stable models are regarded: the extremal-$t$ model with either a Whittle-Mat\'{e}rn or a powered exponential covariance function, and the Brown-Resnick model with power variogram. In addition, the non-extremal Student-$t$ copula model with a Whittle-Mat\'{e}rn or a powered exponential covariance function is also considered. The method is applied to annual maximum temperature data from 25 weather stations dispersed around South Australia.\\[1em]

\end{abstract}

\vspace{3em}

\section{Introduction}
\label{sec:intro}

The statistical analysis of extreme data offers a unique challenge as the interest lies in the tails of the distribution and is of importance in a wide range of application fields, e.g., finance \citep{embrechts1997finance}, hydrology \citep{katz2002hydro} and engineering and meteorology \citep{castillo2004eng}.
Special consideration was given to develop statistical models particular for univariate extreme data, which have now been generalised into a family of distributions known as the generalised extreme value (GEV) distribution with three subfamilies (Fr\'echet, Gumbel and Weibull) following the Fisher-Tippett-Gnedenko theorem. \cite{coles2001} provides an excellent introductory treatment of the topic. 
A particular feature of the GEV distribution is its max-stable property. The max-stable property implies that the maximum of independent copies of a random variable is distributionally invariant up to location and scaling parameters.

Extensions of max-stable models to multivariate extreme data have been limited initially by the intractability of the model likelihood except for some low-dimensional cases. These special cases are of limited use in most applications, where typical multivariate extreme applications arise from spatial data with a large number of spatial locations. 
Even current approaches to parameter estimation for spatial extremes models avoid the computation of the likelihood, relying instead on nonparametric \citep{dehaan2006}, composite likelihood \citep{padoan2010}, or simulated maximum likelihood \citep{koch2014} methods. For threshold exceedance models, full likelihood-based inference has been proposed by \citet{wadsworth2014} and  \citet{engelke2015} for the Brown-Resnick model and \citet{thibaud2015} for the extremal-$t$ model. \citet{stephenson2005} show that the likelihood function simplifies substantially if it is known for which locations the extremal events are occurring at the same time so that the locations can be partitioned accordingly. Regarding these partitions as latent variables, \citet{dombry2017-2} construct a stochastic Expectation-Maximisation (EM) algorithm for exact maximum likelihood estimation of the parameters of multivariate max-stable models. Similarly, \citet{thibaud2016} develop a Gibbs sampler by conditioning on these partitions to conduct Bayesian inference for a Brown-Resnick process. However, these conditional likelihoods are still rather expensive to evaluate as they contain multivariate normal (Brown-Resnick model) or multivariate Student-$t$ (extremal-$t$ model) cumulative distribution functions. \citet{dombry2017-1} extend the idea of \citet{thibaud2016} to several other models and prove the asymptotic normality and efficiency of the posterior median for these models, among them the extremal-$t$ and the Brown-Resnick model. Former attempts to perform Bayesian inference for max-stable models include \citet{Ribatet2012}. They replace the true likelihood with the misspecified composite likelihood, which results in overly precise posterior distributions so that adjustments need to be applied.

To avoid the evaluation of the likelihood for Bayesian inference, approximate Bayesian computation (ABC) \citep{ErhardtSmith2012} methods have been proposed to estimate model parameters.
ABC methods rely on simulations from the max-stable models rather than evaluations of the (approximate) model likelihood in order to perform the required statistical inference. Only model simulations which are `similar' to the observed data are used in the inference, where the measure of similarity is typically a function of informative summary statistics.

There are not many instances where ABC has been applied in the context of extremes. \citet{erhardt2015review} provide an introduction to the use of ABC methods in modelling extremes. The first application of ABC methods in an extremes context is \citet{bortot2007}, where an MCMC ABC algorithm is used for stereological extremes. Subsequent ABC work has focused on spatial extremes \citep{ErhardtSmith2012, barthelme2015, prangle2014-1, ruli2016, hainy2016}. \citet{barthelme2015}, \citet{prangle2014-1} and \citet{ruli2016} use spatial extremes examples to illustrate the performance of their proposed ABC methods for parameter estimation, where the common goal is to alleviate the high computational burden that ABC methods for spatial extremes entail. A different perspective of using ABC methods for spatial extremes applications is provided by \citet{hainy2016}, where the optimal design for estimating the spatial dependence structure of extremes is sought. 

None of the previous work mentioned investigates model selection for spatial extremes applications using ABC. This is the focus of our paper. The aim is to use ABC to determine the posterior model probabilities for models that describe the spatial dependence of the annual maximum temperature data collected from weather stations located around the state of South Australia. These posterior probabilities can be used to select the most preferred model or to perform Bayesian model averaging. An accurate representation of the generating process of the spatial maxima would also allow for a more accurate depiction of the spatial distribution of the maximum temperature across the region. This is useful in estimating the maxima at locations with no observations. For our application, we will consider two extremal-$t$ models \citep{opitz2013} with different correlation functions and the Brown-Resnick model \citep{brown1977, KabluchkoSchlatherdeHaan2009}. By way of contrast, we will also include two non-extremal Student-$t$ copula models \citep{demarta2005} with different correlation functions to our set of possible models. 

To facilitate the task of selecting informative summary statistics, we employ the `semi-automatic' summary selection schemes of \citet{fearnhead2012} for parameter estimation and \citet{prangle2014-2} for model selection, see also \citet{lee2015}. The summary statistics we consider for this purpose are listed in Sections~\ref{subsec:dependence_indicators} and \ref{subsec:composite_score}. Moreover, the method of \citet{prangle2014-2} aims to construct summary statistics for model choice which are close to being Bayes sufficient. This is important because ABC for model choice can lead to inaccurate results if the summary statistics are not sufficient \citep{robert2011}.

For the actual ABC algorithm, we follow \citet{lee2015} and apply the sequential Monte Carlo (SMC) ABC scheme of \citet{drovandi2011}. This algorithm is efficient, straightforward to implement, automatically selects the sequence of target distributions, and has intuitive stopping criteria. For simulating from the extremal-$t$ and Brown-Resnick models, we employ the exact simulation algorithm via extremal functions developed by \citet{dombry2016}. 

The remainder of the paper is structured as follows. The models we considered are introduced in \secref{sec:maxstablemodels}, and the ABC method is elaborated in \secref{sec:abmc}. Implementation details specific to the data set, a description of the data set and the corresponding results as well as a simulation study are contained in Section~\ref{sec:application}. 
A discussion of the results along with limitations and possible extensions is provided in \secref{sec:discussion}.

%
%

\section{Spatial extremes models}
\label{sec:maxstablemodels}

\subsection{Models and parameterisations}

As stated in \secref{sec:intro}, statistical analysis of extremal data uses specialised models developed for such data. In this paper, we consider the use of spatial max-stable models to describe the spatial dependence of the South Australian maximum temperature data. Specifically, we investigate the following three stationary max-stable models: extremal-$t$ model \citep{opitz2013} with either a Whittle-Mat\'ern or a powered exponential correlation function, and the Brown-Resnick model with power variogram \citep{brown1977, KabluchkoSchlatherdeHaan2009}. Furthermore, we will consider to model the data using a non-extremal Student-$t$ copula model \citep{demarta2005} with unit Fr\'echet margins and either a Whittle-Mat\'ern or powered exponential correlation function.

\emph{Max-stable processes} arise as the limiting process of normalised pointwise maxima over infinitely many independent copies of a stochastic processes. If such a limiting process exists, it has to be a max-stable process. That is, if $\forall \: \xb \in \mathcal{X} \subseteq \mathbb{R}^d$ there exist normalising sequences $\{(a_n(\xb), \, b_n(\xb) > 0); \: n \geq 1\}$ such that for the normalised maxima of independent copies of the process $\{X(\xb), \: \xb \in \mathcal{X}\}$ one has
\[ \dfrac{\max_{i=1,\ldots,n} \{X_i(\xb)\} - a_n(\xb)}{b_n(\xb)}  \overset{\mathrm{d}}{\longrightarrow} Z(\xb), \quad n \rightarrow \infty; \quad \xb \in \mathcal{X} \subseteq \mathbb{R}^d,\]
then $\{Z(\xb), \xb \in \mathcal{X}\}$, if not degenerate, is a max-stable process \citep{dehaan2006-2}. The symbol $\overset{\mathrm{d}}{\longrightarrow}$ means convergence in distribution. The stochastic process obtained by taking the appropriately renormalised pointwise maxima over finitely many copies of a max-stable process is equal in distribution to the original max-stable process. The univariate marginal distributions of max-stable processes are members of the \emph{generalised extreme value (GEV)} family. When dealing with max-stable processes, it is often assumed that all univariate margins have a \emph{unit Fr\'echet} distribution ($\Pr(Z(\xb) \leq z) = \exp(-1/z)$), while the focus is on modelling the dependence structure. It is possible to apply simple transformations to each margin to make the univariate margins follow any desired univariate GEV distribution. The converse transformations need to be applied to turn general GEV margins into unit Fr\'echet margins (see Section~\ref{subsec:transformation}).

Their theoretical justification as limiting processes makes max-stable processes a popular choice for the modelling of extreme events in space. In practice, a spatio-temporal data series is usually divided into different blocks (e.g., years) according to the temporal dimension and the pointwise maxima within each block are computed for all locations. Only these pointwise maxima (e.g., annual maxima) are considered for modelling. If each block contains many observations and the different blocks are approximately independent, max-stable process models serve as reasonable approximations to the true spatial process. One does not need to model the unknown underlying process $X$.

Every max-stable process $\{Z(\xb), \xb \in \mathcal{X}\}$ with unit Fr\'echet margins possesses a \emph{spectral representation} of the form
\begin{equation}
Z(\xb) = \max_{i\geq1} \zeta_i Y_i(\xb), \hspace{10pt} \xb \in \mathcal{X} \subseteq \mathbb{R}^d, \label{eq:max_stable_definition}
\end{equation}
where
$\{\zeta_i : i \in \mathbb{N}\}$ are points of a Poisson process on $(0,\infty)$ with intensity $d\Lambda(\zeta) = \zeta^{-2}d\zeta$, and 
$Y_i(\xb)$ are independent realisations of a non-negative stochastic process $Y(\xb)$ with $\mathrm{E}[Y(\xb)] = 1 \; \forall \: \xb \in \mathcal{X}$ (see, e.g., \citet{Ribatet2013}). The functions $\varphi_i(\xb) =  \zeta_i Y_i(\xb)$ are called the spectral functions.
The different max-stable models are determined by the specification of $Y(\xb)$.

The stationary \emph{extremal-$t$ model} is specified as
\[
Y_i(\xb) = c_{\nu} \max\{0, \epsilon_i(\xb)\}^{\nu}, \quad c_{\nu} = \sqrt{\pi} \: 2^{-(\nu-2)/2} \: \Gamma \left( \frac{\nu+1}{2} \right)^{-1}, \quad \nu > 0,
\]
where $\epsilon_i$ are independent copies of a standard stationary Gaussian process with correlation function $\rho(\cdot)$ and $\Gamma(\cdot)$ is the gamma function. We consider two submodels of the extremal-$t$ model which differ in the specification of the correlation function $\rho(\cdot)$. The two correlation function specifications are given in \tref{table:schlather_defns}.
\begin{table}[h]
	\centering
	\caption{\label{table:schlather_defns} The two correlation function $\rho(h)$ specifications considered for the extremal-$t$ and the Student-$t$ copula model, where $h$ is the distance,  $K_\kappa (\cdot)$ is the modified Bessel function of the second kind of order $\kappa$, and $\Gamma(\cdot)$ is the gamma function. 
		The two parameters $\lambda > 0$ and $\kappa > 0$ are generally referred to as the range and the smoothness parameter, respectively.}
	\begin{tabular}{lc}
		\hline
		& $\rho(h)$ \\
		\hline
		\noalign{\vskip 1ex}  
		Whittle-Mat\'ern & $\frac{2^{1-\kappa}}{\Gamma(\kappa)}\left(\frac{h}{\lambda}\right)^\kappa K_\kappa\left(\frac{h}{\lambda} \right)$\\[1ex]
		Powered exponential & $\exp\left[-\left(\frac{h}{\lambda}\right)^\kappa\right]$, \; $0 < \kappa \leq 2$ \\[1ex]
		\hline
	\end{tabular}
\end{table}

The \emph{Brown-Resnick model} has
\begin{equation*}
Y_i(\xb) = \exp \left(\epsilon_i(\xb) - \frac{\sigma^2(\xb)}{2}\right),
\label{eq:Browny}
\end{equation*}
where $\epsilon_i$ are independent copies of a centred Gaussian process and $\sigma^2(\xb) = \mbox{Var}[\epsilon_i(\xb)] \hspace{10pt}\forall \: \xb\in \mathcal{X}$. The Brown-Resnick process is specified by the semi-variogram $\gamma(\xb_1,\xb_2)$ between any two locations $\xb_1$ and $\xb_2$. If the process $\epsilon$ has stationary increments, the semi-variogram is a function of the interpoint distance $h$ only and the resulting Brown-Resnick process is stationary. Most often, the functional form $\gamma(h) = \left( h/\lambda \right)^{\kappa}$ for $\lambda > 0$ and $0 < \kappa \leq 2$ is assumed, in which case the process $\{\epsilon(\xb), \:  \xb\in \mathcal{X}\}$ is a fractional Brownian motion. In our application, we will assume this form of the semi-variogram. The special case $\kappa = 2$ corresponds to the well-known \emph{Smith model} \citep{smith1990}.

In addition to the max-stable extremal-$t$ and Brown-Resnick models, we will also consider a \emph{non-extremal Student-$t$ copula} model with unit Fr\'echet marginal distributions. In that model, only the univariate marginal distributions are modelled by the limiting extremal unit Fr\'echet distribution (after transforming the margins accordingly), whereas the dependence structure is modelled by a standard non-extremal $t$ copula without appealing to the asymptotic max-stable theory. Since the multivariate $t$ distribution is an asymptotically dependent elliptical distribution, the limiting process of rescaled pointwise maxima of a Student-$t$ process is the extremal-$t$ process \citep{opitz2013}. Therefore, the $t$ copula model can be regarded as a tractable approximation to the extremal-$t$ model.

According to Sklar's Theorem \citep{nelsen2006,joe1997}, any joint distribution function $F(z_1,\ldots,z_H)$ can be represented as
\[ F(z_1,\ldots,z_H) = C\{F_1(z_1),\ldots,F_H(z_H)\}, \]
where $C: [0,1]^H \rightarrow [0,1]$ is the copula function defined on the unit hypercube and $F_1,\ldots,F_H$ are the univariate marginal distributions of $F$. The copula function itself is a multivariate distribution function with uniform margins. Therefore, copula models make it possible to model the joint distribution and the marginal distributions separately. If all the margins of $F$ are absolutely continuous, then $C$ is unique.

We will consider the Student-$t$ copula \citep{demarta2005}, which is defined as 
\begin{equation*}
C(u_1,\ldots,u_H) = T_{H;\nu}\{T_{1;\nu}^{-1}(u_1),\ldots,T_{1;\nu}^{-1}(u_H); \bm{\Sigma} \},
\end{equation*}
where $T_{H;\nu}\{\cdots; \bm{\Sigma}\}$ is the cumulative distribution function (CDF) of the $H$-dimensional central Student-$t$ distribution with $\nu$ degrees of freedom and dispersion matrix $\bm{\Sigma}$ and $T_{1;\nu}^{-1}(\cdot)$ is the inverse CDF (quantile function) of the univariate central Student-$t$ distribution with $\nu$ degrees of freedom. The entries of the dispersion matrix $\bm{\Sigma}$ are $\Sigma_{ij} = 1$ for $i = j$ and $\Sigma_{ij} = \rho(h_{ij})$ for $i \neq j$, where $\rho(\cdot)$ is a valid correlation function and $h_{ij}$ is the distance between the points $\xb_i$ and $\xb_j$. We will consider the same correlation functions as for the extremal-$t$ model (Table~\ref{table:schlather_defns}).

The multivariate distributions given by the $H$-dimensional margins of the max-stable extremal-$t$ and Brown-Resnick processes can also be regarded as copula models. However, their copulas are extremal copulas, which must satisfy the max-stable property \citep{joe1997}
\[ C(u_1^n,\ldots,u_H^n) = C^n(u_1,\ldots,u_H), \quad 0 < u_i < 1; \: i = 1,\ldots,H; \: n \in \mathbb{N}. \]
The corresponding copulas for the extremal-$t$ and the Brown-Resnick process are the $t$-EV copula  \citep{demarta2005} and the H\"usler-Reiss copula \citep{husler1989}, respectively.

The correlation functions of the extremal-$t$ and $t$ copula models and the semi-variogram of the Brown-Resnick model contain two parameters: the \emph{range} $(\lambda)$ and \emph{smoothness} $(\kappa)$ parameters. In addition, the extremal-$t$ model and the $t$ copula model have a \emph{degrees of freedom (dof)} parameter, $\nu > 0$. Setting $\nu = 1$ for the extremal-$t$ model yields the widely used \emph{Schlather model} \citep{schlather2002}.

If the distance $h$ is the Euclidean distance between two locations, all the models we consider are isotropic. We allow for \emph{geometric anisotropy} by introducing an anisotropy matrix, which is determined by two additional parameters, see \citet{blanchet2011}. Let $\xb_1 = (x_{1,1}, \: x_{1,2})^T$ and $\xb_2 = (x_{2,1}, \: x_{2,2})^T$ be two locations in $\mathcal{X}$. Then the distance $h$ between these two locations is defined as
\begin{equation*}
h(\xb_1 - \xb_2) = \| \bm{A} \, (\xb_1 - \xb_2) \|,
\end{equation*}
where $\|\cdot\|$ denotes Euclidean distance and the anisotropy matrix $\bm{A}$ is given by
\begin{equation*}
\bm{A} =  \begin{pmatrix}
1 &  0 \\ 
0 & 1/r
\end{pmatrix} \cdot \begin{pmatrix}
\cos \alpha &  \sin \alpha \\ 
- \sin \alpha & \cos \alpha
\end{pmatrix}.
\end{equation*}
Hence, we apply the isotropic models to the transformed space $\bar{\mathcal{X}} = \bm{A} \mathcal{X}$. The two additional parameters governing the extent of geometric anisotropy are the counter-clockwise \emph{rotation angle} $\alpha$ of the correlation contour ellipse ($0 \leq \alpha < \pi/2$) and the \emph{ratio $r > 0$ of the two principal axes} of the correlation contour ellipse.

In our application we assume there is no nugget effect. That is, we assume there is no discontinuity in the correlation function or semi-variogram at $h = 0$. This is a common assumption for environmental processes, see, e.g., \citet{ErhardtSmith2012} and \citet{erhardt2015review}.

\subsection{Marginal transformations to unit Fr\'echet}
\label{subsec:transformation}

All the models we consider have unit Fr\'echet marginal distributions: $\Pr(Z(\xb) \leq z) = \exp\left( - 1/z \right)$, $z > 0$. In general, the limiting distribution for univariate extremal data is from the class of generalised extreme value distributions, which are determined by a location ($\mu \in \mathbb{R}$), scale ($\sigma > 0$), and shape ($\xi \in \mathbb{R}$) parameter \citep{Ribatet2013}: for $Z^*(\xb) \sim GEV(\mu, \sigma, \xi)$, 
\[
\Pr(Z^*(\xb) \leq z) = \exp \left\{  - \left[ 1+ \xi \: \frac{z - \mu}{\sigma} \right]^{-\frac{1}{\xi}} \right\}.
\]

However, it is straightforward to transform $Z^*(\xb) \sim GEV(\mu, \sigma, \xi)$ to a unit Fr\'echet random variable $Z(\xb) \sim GEV(1,1,1)$:
\begin{equation}
Z(\xb) = \left( 1 + \xi \: \frac{Z^*(\xb) - \mu}{\sigma}\right)^{\frac{1}{\xi}}. \label{eq:unit_Frechet_transform}
\end{equation}
For real data, we first estimate the marginal GEV parameters at each location by maximum likelihood and then apply transformation \eqref{eq:unit_Frechet_transform} to the data to (approximately) obtain unit Fr\'echet margins. A nonparametric alternative, which we do not pursue, would be to apply the inverse probability integral transform of the unit Fr{\'e}chet distribution to the observations' rank-based empirical cumulative distribution function values at each location.

\subsection{Indicators of dependence structure}
\label{subsec:dependence_indicators}

The viability of approximate Bayesian computation depends on the availability of informative and easy-to-compute summary statistics of the data. Our summary statistics will incorporate empirical estimates of F-madograms, pairwise and tripletwise extremal coefficients, and Kendall's $\tau$, as well as composite score vectors for all the models considered. Unfortunately, none of these statistics is sufficient for the parameters of any of the models. In order to obtain summary statistics for the ABC procedure that contain as much information about the process as possible, we collect a large set of informative statistics and aggregate them in a sensible way using the Fearnhead-Prangle procedure (see Sections~\ref{subsec:FPstep} and \ref{sec:SAmethod}). Even if various dependence indicators provide partly redundant information, it may prove worthwhile to include all of them to the set of summary statistics. The redundant information will be filtered out by the Fearnhead-Prangle procedure.

The \emph{F-madogram} \citep{cooley2006} is similar to the semi-variogram. The theoretical variogram is not always defined for a max-stable process due to the possibility of a non-finite mean or variance. The F-madogram is always well-defined and is given by
\begin{equation*}
\nu_F(\xb_1, \, \xb_2) = \frac{1}{2}E\left[ |F_{\xb_1}\left\{Z(\xb_1)\right\} - F_{\xb_2}\left\{Z(\xb_2)\right\}| \right],
\end{equation*}
where $F_{\xb_1}$ and $F_{\xb_2}$ are the cumulative distribution functions (CDFs) of the random variable $Z(\xb)$ at locations $\xb_1$ and $\xb_2$, respectively. As we assume that all marginal distributions of our transformed data are unit Fr{\'e}chet, the F-madogram can easily be estimated by
\begin{equation*}
\hat{\nu}_F(\xb_1, \, \xb_2) = \frac{1}{2 n} \sum_{i=1}^{n} |F(Z_i(\xb_1)) - F(Z_i(\xb_2))|,
\end{equation*}
where $n$ is the number of observed extremal process realisations at each location, $Z_i(\xb_j)$ is the observed value at location $j$, and $F(z) = \exp(-1/z)$ is the unit Fr{\'e}chet CDF. In case the marginal distributions are unknown, a nonparametric estimator based on the ranks is commonly used (see \citet{Ribatet2013}).

For max-stable processes, the \emph{extremal coefficient} $\theta(\xb_1, \ldots, \xb_L)$ \citep{SchlatherTawn2003} is another measure of spatial dependence between a subset of $L$ locations. We will consider pairwise ($L = 2$) and tripletwise ($L = 3$) extremal coefficients. Assuming unit Fr\'echet marginal distributions of the process $\{Z(\xb), \: \xb \in \mathcal{X}\}$, the pairwise extremal coefficient between the locations $\xb_1, \xb_2, \ldots, \xb_L$ is defined by the relation
\begin{eqnarray}
\Pr(Z(\xb_1) \leq z, \ldots, Z(\xb_L) \leq z) & = & \Pr(Z(\xb_1) \leq z)^{\theta(\xb_1,\ldots,\xb_L)} \notag \\
& = & \exp\left( -\frac{\theta(\xb_1,\ldots,\xb_L)}{z} \right). \label{eq:def_extremal_coefficient}
\end{eqnarray}

An extremal coefficient value of $1$ indicates complete dependence between the observations, whereas the value of $L$ indicates complete independence.

Note that the extremal coefficient function does not fully characterise the max-stable process (see \citet{Strokorb2015-2}), a notable exception being the Tawn-Molchanov process \citep{Strokorb2015-1}.

\citet{coles1999} propose a simple and fast-to-compute estimator for the pairwise extremal coefficient. \citet{ErhardtSmith2012} generalise the estimator to an arbitrary number of locations. Their proposed extremal coefficient estimator is 
\begin{equation}
\hat{\theta}(\xb_1, \ldots, \xb_L) = \frac{n}{\sum_{i = 1}^{n}{1/\max\{Z_i(\xb_1), \ldots, Z_i(\xb_L)\}}}, \label{eq:estimate_extremal_coef}
\end{equation}
when there are $n$ independent copies of the max-stable process and $Z_i(\xb_j)$ denotes the $i$-th copy of the process observed at location $\xb_j$.

This estimator is easily derived by observing that the random variable \newline $1/\max\{Z(\xb_1),\ldots,Z(\xb_L)\}$ has an exponential distribution with rate parameter $\theta(\xb_1, \ldots, \xb_L)$. It follows that estimator \eqref{eq:estimate_extremal_coef} is the maximum likelihood estimator of the rate parameter.

The extremal coefficient as given by Equation~\eqref{eq:def_extremal_coefficient} is only defined for max-stable models. For other models with unit Fr\'{e}chet margins such as the $t$ copula model, the value of $\theta$ in $$\Pr(Z(\xb_1) \leq z, \ldots, Z(\xb_L) \leq z) = \Pr(Z(\xb_1) \leq z)^{\theta}$$ depends on the level $z$. However, estimator~\eqref{eq:estimate_extremal_coef} is still a useful measure of dependency. If the distribution of the random variable $U = 1/\max\{Z(\xb_1),\ldots,Z(\xb_L)\}  = $ \newline $\min\{1/Z(\xb_1),\ldots,1/Z(\xb_L)\}$ is approximated by an exponential distribution, then estimator~\eqref{eq:estimate_extremal_coef} provides an estimate for its rate parameter. In the limiting cases of complete dependence and complete independence, the distribution of $U$ is exactly exponential with rate parameters $1$ and $L$, respectively, just as for max-stable processes. For simplicity, we will generally refer to the summary statistics $\hat{\theta}$ computed using Equation~\eqref{eq:estimate_extremal_coef} as \emph{extremal coefficient estimates}, even though only max-stable processes possess an extremal coefficient as defined by Equation~\eqref{eq:def_extremal_coefficient}.

There are other estimators for the extremal coefficients with better estimation properties. For max-stable processes, there is a one-to-one relationship between the pairwise extremal coefficient and the F-madogram,
\begin{equation*}
\theta(\xb_1, \, \xb_2) = \frac{1 + 2 \, \nu_F(\xb_1, \, \xb_2)}{1 - 2 \, \nu_F(\xb_1, \, \xb_2)},
\end{equation*}
which can be exploited to construct an estimator. We do not consider this estimator because we already include estimates for the pairwise F-madograms in our set of summary statistics. Two further estimators are proposed by \citet{SchlatherTawn2003}. These estimators are self-consistent and satisfy the boundary conditions, whereas the estimates given by \eqref{eq:estimate_extremal_coef} and the estimates constructed via the F-madogram can fall outside the theoretical bounds.

However, our purpose is to use the extremal coefficients as summary statistics in ABC in order to assess the similarity of data sets. The estimates do not need to adhere to the theoretical properties to be useful for that purpose. Since we will aggregate the estimates across many location pairs and triplets for our summary statistics (see Section~\ref{sec:SAmethod}), we are also not particularly concerned about estimator efficiency or self-consistency. It is more important that the estimates can be computed quickly, which is the case for estimator \eqref{eq:estimate_extremal_coef}.

Another dependence measure we consider is \emph{Kendall's $\tau$} between any pair of locations, which is estimated by 
\begin{equation*}
\hat{\tau}(\xb_1, \, \xb_2) = \frac{2}{n \, (n-1)} \sum_{1 \leq i < j \leq n} \mathrm{sign}[Z_i(\xb_1)-Z_j(\xb_1)] \: \mathrm{sign}[Z_i(\xb_2)-Z_j(\xb_2)],
\end{equation*}
where $\{Z_i(\xb); \: \xb \in \mathcal{X}; \: i=1,\ldots,n \}$ are $n$ observed copies of the process. \citet{dombry2017-3} show that for max-stable processes Kendall's $\tau$ is equal to the \emph{pairwise extremal concurrence probability}. This is the probability that the extremal occurrences at two locations $\xb_1$ and $\xb_2$ are obtained from the same spectral function (cf. Equation~\eqref{eq:max_stable_definition}):
\begin{equation*}
\tau(\xb_1, \, \xb_2) = \Pr\left(\underset{i \geq 1}{\arg \max} \: \zeta_i Y_i(\xb_1) = \underset{i \geq 1}{\arg \max} \: \zeta_i Y_i(\xb_2)\right).
\end{equation*}

\subsection{Composite score vector}
\label{subsec:composite_score}

Our summary statistics will also incorporate the composite score vectors for all the models. \citet{ruli2016} prove that the score vector is a sufficient summary statistic for the parameters of exponential models. Therefore, one can expect that the score vector also provides informative summary statistics for the parameters of many non-exponential models.

In the case of max-stable models, the density functions and hence log-likelihood functions are intractable for moderate to high dimensions for most models. Therefore, classical inference for max-stable models is commonly based on the \emph{composite likelihood} approach \citep{padoan2010}. A composite log-likelihood is a weighted sum of log-likelihoods for marginal or conditional events. For max-stable models, the composite likelihood is usually constructed from pairwise likelihoods. Given $n$ independent copies of the process observed at $H$ locations, the \emph{pairwise log-likelihood} is defined to be
\begin{equation}
p\ell(\parms;\Zb) = \sum_{k=1}^n \sum_{i=1}^{H-1} \sum_{j=i+1}^H w_{ij} \ell_{ij} \{\parms; Z_k(\xb_i), Z_k(\xb_j)\}, \label{pw_loglike_def}
\end{equation}
where $\Zb = \{Z_k(\xb_i); \: k=1,\ldots,n; \: i=1,\ldots,H\}$ are the observations, $\parms$ is the vector of parameters, $\ell_{ij}\{\parms; Z_k(\xb_i), Z_k(\xb_j)\}$ is the bivariate marginal log-likelihood contribution of process copy $k$ observed at locations $i$ and $j$, and the $w_{ij}$ are weights (all set to 1 in our application).
The \emph{composite score vector} is
\begin{equation*}
\Sb(\Zb;\parms) = \nabla_{\parms} \; p\ell(\parms;\Zb). \label{pw_score_vec_def}
\end{equation*}
The \emph{maximum composite likelihood estimator} (MCLE) is $\tilde{\parms} = \arg \max_{\parms} \: p\ell(\parms;\Zb)$, which is obtained by either maximising $p\ell(\parms;\Zb)$ directly or by solving $\Sb(\Zb;\parms) = \bm{0}$. Under the usual regularity conditions on the bivariate marginal likelihoods, $\Sb(\Zb;\parms) = \bm{0}$ is a system of unbiased estimating equations because $\Sb(\Zb;\parms)$ is the sum over the proper score vectors of the bivariate marginal distributions \citep{varin2008}. The MCLE $\tilde{\parms}$ is consistent and asymptotically normal under rather broad assumptions \citep{molenberghs2005}.

\citet{ruli2016} use $\Sb(\Zb; \tilde{\parms}_{\mathrm{obs}})$ as summary statistic for ABC, where \newline $\tilde{\parms}_{\mathrm{obs}} = \arg \max_{\parms} \: p\ell(\parms;\zb_{\mathrm{obs}})$ and $\zb_{\mathrm{obs}}$ is the observed data. This summary statistic can be calculated quickly for any new simulated data set $\Zb$ because $\tilde{\parms}_{\mathrm{obs}}$ has to be computed only once. Note that the value of the summary statistic for the observed data is $\Sb(\zb_{\mathrm{obs}}; \tilde{\parms}_{\mathrm{obs}}) = \mathbf{0}$ by definition of $\tilde{\parms}_{\mathrm{obs}}$ apart from numerical inaccuracies in practice.

The marginal bivariate log-likelihood and score functions are given in Appendices~\ref{sec:bivariate_loglik} and \ref{sec:bivariate_score}.

\section{Approximate Bayesian computation for model selection and parameter estimation}
\label{sec:abmc}

\subsection{Introduction to ABC}

Approximate Bayesian computation (ABC) methods rely on model simulation to conduct approximate Bayesian inference and are particularly useful in circumventing analytically or computationally intractable likelihood evaluations. Spatial extremes analysis is one of many applied disciplines where ABC has been used \citep{erhardt2015review}.

ABC avoids likelihood evaluations by retaining parameter draws that generate simulated data close to the observed data. To avoid the curse of dimensionality, the comparison is typically via a lower-dimensional summary statistic \citep{blum2010}.

Our goal is to obtain posterior inference for the combined objective of model selection and parameter estimation. We consider a model setup with $K$ possible models that could have generated the data. The discrete model indicator random variable $\mo$ can assume the values $k = 0,\ldots,K-1$ with prior probabilities $\Pr(\mo = k)$, which have to satisfy $\sum_{k=0}^{K-1} \Pr(\mo = k) = 1$. For each model, the (intractable) likelihood function is denoted by $f_k(\cdot|\, \parms_k)$, which depends on the model-specific parameter vector $\parms_k$ having prior distribution $\pi(\parms_k |\, \mo = k)$. The observed data $\zb$ and the simulated data $\ub$ are compared via the discrepancy function $d[\cdot,\cdot] $ applied to the summary statistics $\tb(\zb)$ and $\tb(\ub)$. Given a tolerance level $\epsilon$ for accepting simulated draws, the targeted approximate posterior density has the form
\begin{equation}
\pi_{\epsilon}(\parms_k, k | \, \zb) \propto \Pr(\mo = k) \: \pi(\parms_k |\, \mo = k) \: \int_{\ub \in \mathcal{Z}} \, f_k(\ub|\, \parms_k) \: \mathds{1}\{d[\tb(\zb),\tb(\ub)] < \epsilon\} \: \mathrm{d} \ub, \label{eq:ABC_posterior}
\end{equation}
where $\mathds{1}\{d[\tb(\zb),\tb(\ub)] < \epsilon\}$ is the indicator function which is 1 if $d[\tb(\zb),\tb(\ub)] < \epsilon$ and 0 otherwise. If the summary statistics $\tb(\cdot)$ are sufficient for all the model parameters and the model indicator and $\epsilon \rightarrow 0$ when $d[\tb(\zb),\tb(\ub)] \geq 0 \; \forall \: \zb, \ub \in \mathcal{Z}$, one obtains the true joint posterior of parameters and model indicators.

\citet{frazier2018} provide large-sample asymptotic results for ABC concerning consistency, the shape of the posterior distribution, and the asymptotic distribution of the posterior mean under rather general conditions on the convergence of the summary statistics and the rate of $\epsilon \rightarrow 0$. See also \citet{li2018} for similar results for the posterior mean. \citet{marin2014} present necessary and sufficient conditions on the summary statistics for consistent model choice.

The first ABC algorithm that was proposed \citep{pritchard1999, beaumont2002} uses a rejection sampling scheme where the parameters are always drawn from the prior distribution. This scheme is hence referred to as rejection ABC. When extended to account for model selection, the algorithm works as follows: generate $N$ draws $\{k^i, \parms^i\}_{i=1}^N$ by simulating $k^* \sim \Pr(\mo = k)$, $\parms_{k^*}^* \sim \pi(\parms_{k^*}| \, \mo = k^*)$, $\ub^* \sim f_{k^*}(\ub | \, \parms_{k^*}^*)$ and accepting  $(k^*, \parms_{k^*}^*)$ if $d[\tb(\zb),\tb(\ub^*)] < \epsilon$ until $N$ draws are accepted.

Markov chain Monte Carlo ABC (MCMC ABC) \citep{marjoram2003} and later sequential Monte Carlo ABC (SMC ABC) \citep{beaumont2009, toni2009, sisson2007, sisson2009, drovandi2011} were proposed to alleviate the computational burden of rejection ABC by using more efficient parameter proposal distributions to explore the parameter and model space. We use the SMC ABC algorithm of \citet{drovandi2011} for our application.

\subsection{Fearnhead-Prangle step}
\label{subsec:FPstep}

The `semi-automatic' summary statistic selection schemes proposed in \citet{fearnhead2012} for parameter estimation and \citet{prangle2014-2} for model selection provide valuable strategies for selecting low-dimensional yet informative summary statistics.

\citet{prangle2014-2} prove that the posterior model probabilities form Bayes sufficient statistics for model selection, which is important to avoid the ABC model choice inaccuracies reported by \citet{robert2011}. Bayes sufficiency of a statistic $\tb(\zb)$ means that $\parms|\, \zb$ and $\parms|\, \tb(\zb)$ have the same posterior distribution for any prior distribution and almost all $\zb$ \citep[p.~70]{prangle2014-2}. Therefore, \citet{prangle2014-2} propose to construct an estimator of the posterior model probabilities in a preliminary step to the actual ABC algorithm and use the estimated posterior model probabilities as summary statistics in ABC. The posterior model probabilities are estimated based on a set of available statistics $\gb(\zb) = (g_1(\zb),\ldots,g_p(\zb))^T$. \citet{lee2015} call this preliminary step the \emph{Fearnhead-Prangle (FP) step}. 

First, \citet{prangle2014-2} and \citet{lee2015} simulate a large training particle set 
$\{k^i, \parms_{k^i}^i, \ub^i\}_{i=1}^M$ generated from the prior predictive distribution: $k^i \sim \Pr(\mo = k)$, $\parms_{k^i}^i \sim \pi(\parms_{k^i}| \, \mo = k^i)$, $\ub^i \sim f_{k^i}(\ub | \, \parms_{k^i}^i)$. That is, each particle is a random draw from the joint distribution of the model indicator, the parameters, and the simulated data.
Next, they regress the model indicators $k^i$ on the available statistics $\gb(\ub^i)$ of the simulated data to obtain an estimator of the model probabilities. \citet{lee2015} perform a backward stepwise procedure to exclude regressors in $\gb(\ub)$ with little or no explanatory power for the model indicators. In an application with two models to choose from, \citet{prangle2014-2} propose to use logistic regression. \citet{lee2015} generalise the method to $K > 2$ models $\mo = 0,\ldots,K-1$ by fitting a multinomial logistic regression model, 
\begin{equation}
\log\left( \frac{\Pr(\mo = k| \, \ub)}{1 - \sum_{i=1}^{K-1} \Pr(\mo = i| \, \ub)} \right) = \beta_{0,k} + \sum_{i=1}^p \beta_{i,k} \, g_i(\ub), \quad k = 1,\ldots,K-1. \label{eq:logreg_linpredictor}
\end{equation}
In the ABC step, this fit is used to estimate the posterior model probabilities for any sample $\ub^*$ generated during the procedure. This means the data is reduced to a $(K-1)$-dimensional summary statistic $$t_M(\ub^*) = \left( \widehat{\Pr}(\mo = 1 | \, \ub^*), \ldots,  \widehat{\Pr}(\mo = K-1 | \, \ub^*) \right),$$ where
$$\widehat{\Pr}(\mo = k | \, \ub^*) = \frac{\exp\left\{\hat{\beta}_{0,k} + \sum_{i=1}^p \hat{\beta}_{i,k} \, g_i(\ub^*) \right\}}{1 + \sum_{l=1}^{K-1} \exp\left\{\hat{\beta}_{0,l} + \sum_{i=1}^p \hat{\beta}_{i,l} \, g_i(\ub^*)\right\}}, \quad k = 1,\ldots,K-1.$$ The same transformation is applied to the observed data $\zb$. 

If a pilot ABC study has been conducted, it is possible to truncate the parameter priors to accommodate the region of high posterior mass identified by the pilot study. By fitting the regression to a more localised sample, the regression fit is likely to improve. When simulating from truncated priors, a truncation correction has to be performed on the ABC output \citep[p.~73]{prangle2014-2}.

For parameter estimation, \cite{fearnhead2012} propose a similar `semi-automatic' scheme. They show that the choice of the quadratic error loss function yields the posterior mean as the optimal ABC summary statistic for continuous parameters when $\epsilon \rightarrow 0$. Consequently, for each single parameter in $\parms_k = (\phi_{k,1},\ldots,\phi_{k,Q_k})^T$, we follow \cite{fearnhead2012} and \citet{lee2015} and run a backward stepwise linear regression of $\phi_{k,j}^i$ ($j = 1,\ldots,Q_k$) on  $\gb(\ub^i)$ based on the training particle subset $\{k^i, \parms_{k^i}^i, \ub^i: \: k^i = k\}_{i=1}^M$. That is, we use the prior predictive samples from model $k$ to estimate the regression functions for the posterior means of the parameters of model $k$. To reduce the computational expenditure, we re-use the training particle set generated for estimating the posterior model probabilities. In the ABC step, the summary statistic consists of the collection of estimates $\tb_P(\ub) = \{\hat{\phi}_{k,j}(\ub); \: k = 0,\ldots,K-1; \: j = 1,\ldots,Q_k\}$, where 
\begin{equation}
\hat{\phi}_{k,j}(\ub) =  \hat{\beta}_{0,k,j} + \sum_{i=1}^p \hat{\beta}_{i,k,j} \, g_i(\ub). \label{eq:linreg_linpredictor}
\end{equation} Therefore, the dimension of the summary statistic for parameter estimation is $\sum_{k=0}^{K-1} Q_k$.

We use the squared Euclidean distance as a discrepancy function for both model selection and parameter estimation. The model selection discrepancy is 
\[
d_M[\tb_M(\zb), \tb_M(\ub)] = \sum_{k=1}^{K-1} \left[\widehat{\Pr}(\mo = k| \, \zb) - \widehat{\Pr}(\mo = k| \, \ub) \right]^2
\]
and the parameter estimation discrepancy is
\[
d_P[\tb_P(\zb), \tb_P(\ub)] = \sum_{k=0}^{K-1} \sum_{j=1}^{Q_k} \left[ \frac{\hat{\phi}_{k,j}(\zb) - \hat{\phi}_{k,j}(\ub)}{\widehat{\mathrm{sd}}(\hat{\phi}_{k,j})} \right]^2.
\]
The differences between observed and simulated $\hat{\phi}_{k,j}$ are scaled by the prior predictive standard deviation of $\hat{\phi}_{k,j}$. We estimate $\mathrm{sd}(\hat{\phi}_{k,j})$ from simulated observations from the prior predictive distribution.

The overall discrepancy is formed by taking the logarithm of the product of the two discrepancies $d_M$ and $d_P$:
\begin{equation*}
d_T[\tb(\zb), \tb(\ub)] = \log\left\{d_M[\tb_M(\zb), \tb_M(\ub)] \cdot d_P[\tb_P(\zb), \tb_P(\ub)]\right\}.
\end{equation*}
This formulation of $d_T$ ensures that equal proportionate changes in $d_M$ or $d_P$ have the same additive effect on the overall discrepancy. Due to taking the logarithm, this discrepancy can be negative and its minimum is $-\infty$, while the minimum of the strictly monotone transformation $\exp\{d_T[\cdot,\cdot]\}$ is $0$. However, it is not necessary for the discrepancy function to be non-negative for the ABC algorithm to work properly (see Section~\ref{subsec:SMCABCstep}). We only require the discrepancy function to induce an ordering of the simulated data with respect to the observed data.


The FP step establishes the formula for computing the discrepancies that is used in the ABC step. After the FP step, the actual ABC step is performed to obtain the approximate posterior model probabilities and posterior parameter distributions.

\subsection{SMC ABC step}

\label{subsec:SMCABCstep}

The specific ABC algorithm used for our application is the SMC ABC replenishment algorithm \citep{drovandi2011}. This SMC ABC algorithm dynamically determines the non-increasing sequence of tolerances $\epsilon_t$. In each iteration $t = 1,\ldots,T$, the current particle set of size $N$ is ordered according to the discrepancies $d_T[\tb(\zb), \tb(\ub^i)]$ and is given by $\{\psi^i = (k^i, \parms_{k^i}^i, \ub^i)\}_{i=1}^N$. The procedure discards the $N_{\delta} =  \lceil \delta N \rceil$ particles with the highest discrepancies from the ordered particle set, so the particle subset $\Psi = \{\psi^i; \: i = 1,\ldots,N - N_{\delta}\}$ contains the retained particles. The current tolerance level $\epsilon_t$ is lowered to the highest discrepancy value in the retained particle set, $\Psi$.

The other particles are resampled from the retained particles with replacement via multinomial resampling \citep{gordon1993}. For each $j \in \{N - N_{\delta} +1, \ldots, N\}$, we set $\psi^j = \psi^*$, where $\psi^*$ is a random draw from the particle subset $\Psi$. In our case, all particles have equal weights, so $\Pr(\psi^* = \psi^i) = 1/(N - N_{\delta})$ for $i \in \{1,\ldots,N - N_{\delta}\}$.

The resampling step introduces duplicated particles. To diversify the particle set, a reversible jump Markov chain Monte Carlo (RJ-MCMC) kernel is applied $R_t$ times to each resampled particle. The invariant distribution of the RJ-MCMC kernel is the approximate posterior \eqref{eq:ABC_posterior} with the tolerance level set to the current level $\epsilon_t$. The number of RJ-MCMC steps, $R_t$, is set to a value such that the probability that a particle does not move during the $R_t$ steps is approximately equal to a user-defined fixed small number $c$. Given $c$, $R_t$ is estimated by $$\displaystyle R_t = \left\lceil \frac{\log(c)}{\log(1-p_{\mathrm{acc}, t-1})} \right\rceil,$$ where $p_{\mathrm{acc}, t-1}$ is the MCMC acceptance rate of the previous SMC iteration.

The RJ-MCMC proposal distribution for the model indicator is a discrete distribution denoted by $q_{k,k^*}$, which gives the probability of proposing model $k^*$ given the current model $k$. In our application, we assume that all models have the same probability to be proposed irrespective of the current model. Given the proposed model, the proposal distributions for the parameters, denoted by $q_{k}(\parms_k)$, $k=0,\ldots,K-1$, are independent proposals, where the parameters of the proposal distributions are usually estimated from the current particle set. We employ multivariate normal independence proposal distributions, where the mean and variance-covariance matrix are estimated from the retained particle subset.

The algorithm stops when the overall acceptance rate in the RJ-MCMC steps drops below some minimum level or when the desired final tolerance level $\epsilon_{\min}$ is reached. The first condition is reasonable because the algorithm terminates when further moderate reductions of $\epsilon_t$ can only be achieved at excessive computational cost and are therefore not worthwhile given the computing resources. 

The algorithm has two tuning parameters: the proportion of particles to be dropped at each SMC iteration, $\delta$, and the desired probability of moving a particle during the sequence of RJ-MCMC move steps, $1 - c$. 
The choice of these tuning parameters is generally considered to be problem-specific. \citet{drovandi2011} suggest values of $0.5$ and $0.01$ for $\delta$ and $c$, respectively, to be sensible general choices.

The initial particle set for the ABC SMC step is generated by a variant of the ABC rejection algorithm. A large number of particles $\{k^i, \parms_{k^i}^i, \ub^i\}_{i=1}^{N_2}$ is simulated from the prior predictive distribution. From this set, the $N \ll N_2$ particles with the smallest discrepancies form the initial particle set for the ABC SMC algorithm. The initial tolerance level $\epsilon_1$ is the greatest discrepancy among the $N$ remaining particles.

A more detailed discussion of this ABC method is given in \citet{lee2015}. A concise summary of the ABC algorithm for model selection and parameter estimation can be found in Appendix~\ref{sec:abc_summary}.

\section{Application}

\label{sec:application}

\subsection{Implementation details}
\label{sec:SAmethod}

There is an abundance of potential summary statistics available for ABC inference for models of spatial extremes. The advantage of using the FP step described in \secref{subsec:FPstep} is that the summary statistics that best explain a model or its parameters are weighted by the regression coefficients in the linear predictor as given by Equations~\eqref{eq:logreg_linpredictor} and \eqref{eq:linreg_linpredictor}. Furthermore, superfluous summary statistics are eliminated through the backward stepwise procedure using the Akaike information criterion (AIC). 
The summary statistics $\gb(\cdot)$ considered for our application are estimates of the pairwise F-madogram, the pairwise and tripletwise extremal coefficient, the pairwise Kendall's $\tau$, and the composite score vectors for the five possible models, see Sections~\ref{subsec:dependence_indicators} and \ref{subsec:composite_score}. To be consistent across all models, we also use the composite score statistics for the $t$ copula models instead of the score statistics based on the full likelihood.

For the pairwise dependence indicators, the empirical estimates given in \secref{subsec:dependence_indicators} are computed for all pairs of locations and binned into groups from small to large Euclidean distances. The means and standard deviations of the groups are used as regression summary statistics. For our application, we selected four approximately equal-sized groups.

A practical issue in computing the summary statistics for the tripletwise extremal coefficients is determining how similar one location triplet is to another among the ${H \choose 3}$ triplets, where $H$ is the number of locations.
\citet{ErhardtSmith2012} propose clustering the ${H \choose 3}$ triplets into $G$ groups or clusters (with $G\ll {H \choose 3}$) based on the ordered pairwise location distances of the triplet's three locations.

We used the $k$-median clustering algorithm via the function \texttt{kcca} from the R package \texttt{flexclust} \citep{leisch2006}. For our particular application, we found that $G = 100$ groups suggested by \citet{ErhardtSmith2012} is too large. There was great similarity among the $100$ clusters suggested by the algorithm and it was difficult to get reproducible cluster membership with $100$ clusters. Instead, smaller values of $G$ were tested and a $G$ value of $10$ was chosen for our investigation. 
We take the mean and standard deviation for each group and add them to the vector $\gb(\cdot)$ of regression summary statistics.

We selected the following independent priors for all models:
\begin{eqnarray*}
	\log\left(\lambda\right) & \sim & \mathcal{N}(1, \, 4), \\
	\kappa & \sim & \mathcal{U}(0, \, 2), \\
	\alpha & \sim & \mathcal{U}(0, \, \pi/2), \\
	\log(r) & \sim & \mathcal{N}(0, \, 8), \\
	\log(\nu) & \sim & \mathcal{N}(0, \, 1) \text{ truncated on } [-2.5, \, 2.5].
\end{eqnarray*}

These prior distributions put significant prior mass on regions of relevant posterior mass, see \fref{figure:SAdata20150520_Parameters_BR_2} for the Brown-Resnick model. Furthermore, all the observed values of the dependence indicators are in regions of non-negligible prior predictive mass (see Appendix~\ref{sec:prior_predictive_checks}). Since we assume normal prior distributions for $\log(\lambda)$, $\log(r)$, and $\log(\nu)$, we also used the log-transformations of these variables as the dependent variables in the FP step's linear regressions. We did not transform $\kappa$ and $\alpha$. 

For the FP step, $2,\!000$ simulations were generated from the prior predictive distribution of each model to obtain the logistic and linear regression estimates. The SMC ABC replenishment algorithm was run with $N = 2,\!000$ particles and stopped when the particle acceptance probability in the RJ-MCMC steps was less than $p_{\mathrm{acc},\min} = 10^{-2}$. The SMC ABC tuning parameters were set to $\delta = 0.5$ and $c = 0.01$ as suggested by \citet{drovandi2011}. In the initial ABC rejection step, $N_2 = 20,\!000$ particles were simulated from the prior predictive distribution, of which the $10 \%$ with the smallest discrepancies to the observed data were selected to form the initial SMC ABC particle set.

It is straightforward to simulate from the $t$ copula model because it only requires to obtain samples from a multivariate Student-$t$ distribution and to apply the univariate Student-$t$ CDF and the inverse unit Fr\'echet CDF to the margins. Simulating from most max-stable models is more difficult, since one realisation of the max-stable process is the pointwise maximum over infinitely many realisations of the spectral function (see Equation~\eqref{eq:max_stable_definition}). For some max-stable models it is possible to obtain exact simulations by reordering the sampling of the spectral functions appropriately so that the simulation of spectral functions can stop in finite time after meeting a specific stopping rule \citep{schlather2002}. However, for many models like the extremal-$t$ or the Brown-Resnick model only approximate simulations are possible by applying the basic reordering idea. For an overview of simulation methods for max-stable processes, see \citet{oesting2015review}. Recently, exact simulation algorithms have been developed by investigating different representations. For example, \citet{thibaud2015} present an algorithm for exact simulation of extremal-$t$ processes and \citet{Dieker2015} develop a method to obtain exact simulations from the Brown-Resnick process. \citet{dombry2016} introduce two general-purpose algorithms for the exact simulation of max-stable processes, which can be adapted to many max-stable models. For our application, we implemented their Algorithm~1 (simulation via extremal functions) for the extremal-$t$ and the Brown-Resnick model. For our ABC application, which heavily depends on accurate simulations, we prefer to use exact simulation methods. However, exact simulation algorithms are in general computationally more expensive than approximate algorithms. For Algorithm~1 of \citet{dombry2016}, on average $H$ extremal functions need to be sampled to simulate one realisation of the max-stable process at $H$ locations. On dense grids, this simulation method may therefore become too expensive and one has to resort to approximate algorithms. The advantage of Algorithm~1 of \citet{dombry2016} is that the expected number of simulated extremal functions does not depend on any parameters of the model. For example, for most of the other simulation methods the simulation effort of the extremal-$t$ model increases strongly when $\nu$ becomes large.
Since simulation speed is crucial for ABC, we implemented the exact simulation routines in C++ using the \texttt{Rcpp} \citep{eddelbuettel2011} and \texttt{RcppArmadillo} \citep{eddelbuettel2014} interface to R.

\subsection{South Australian annual maximum temperature data}
\label{subsec:data}

South Australia experiences a dry and hot Mediterranean climate. Extreme high temperatures can cause public health and safety concerns, for example heatwaves and bushfires. Understanding the spatial distribution of maximum temperature around the state is a vital component in planning for future adverse events related to extreme high temperatures.

The data set considered in this paper contains annual maximum temperature values for the $18$-year-period spanning from $1979$ to $1996$ at $25$ weather stations around Adelaide, the capital of South Australia. The particular time period $(1979 - 1996)$ is the longest uninterrupted period of temperature recordings for the selected collection of weather stations reasonably close to Adelaide. The publicly available data were obtained from the Australian Bureau of Meteorology website. These maximum temperatures were originally recorded in monthly observation blocks and converted to yearly maximum temperature values. 
The autocorrelation plots of the yearly maxima for each observed location did not show any statistically significant temporal dependence in the data.
 
\begin{figure}[h]
	\centering
	\includegraphics[width=0.6\textwidth]{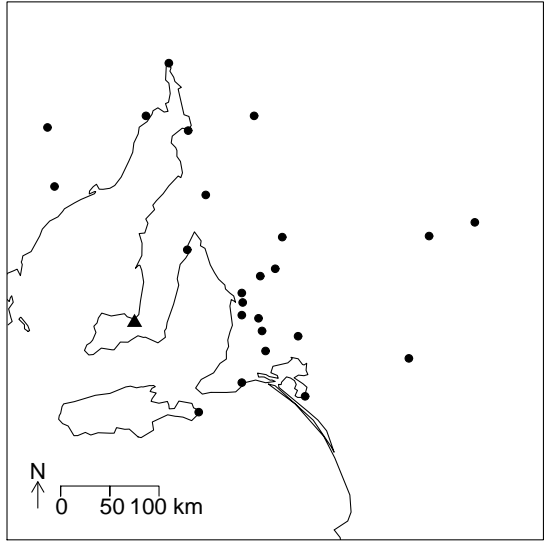}
	\caption{Map of the $25$ land-based weather stations around South Australia where data were collected. The station in Warooka on the peninsula is marked by a triangle.}
	\label{fig:location_map}
\end{figure}

\fref{fig:location_map} depicts the locations of the $25$ weather stations. Appendix~\ref{sec:frechet_transformations} provides detailed information on the marginal transformations of the data to the unit Fr\'echet scale. In addition, the assumption of unit Fr\'echet marginal distributions is checked for the transformed data set via QQ plots and Kolmogorov-Smirnov tests. The unit Fr\'echet assumption is not rejected at any location.

The analysis of the full data set showed that the station in Warooka on the peninsula is an outlier (see Appendix~\ref{sec:analysis_with_Warooka}). There is almost no dependence between Warooka and the surrounding stations, which is not in accordance with any of our models. Therefore we discarded Warooka for the analyses presented in Sections~\ref{subsec:dataset_FPstep} and \ref{subsec:dataset_results}. The results for the full data set can be found in Appendix~\ref{sec:analysis_with_Warooka}. 

\subsection{Simulation study}

We conducted a simulation study to assess the validity and effectiveness of our approach for the South Australian maximum temperature data set when it is assumed that one of the five models we consider is the true model. For each of the five models, we simulated $30$ data sets with the same pattern of locations and the same number of years as the original data set from the prior predictive distribution. Then we employed our ABC procedure (FP step, initial ABC rejection, SMC ABC step) to each simulated data set. To compute the composite score vector summaries, it is necessary to find the MCLEs for all the models for each simulated data set. The procedure for finding the MCLEs is outlined in Appendix~\ref{sec:MCLE_procedure}. In $19$ cases there were numerical problems with the composite score vector summaries when evaluated at the MCLE for at least one of the models. In some cases the computation of the composite scores produced numerical errors, in other cases the composite scores were numerically $0$ for all simulated observations and the standardisation failed. In all of these cases the procedure was aborted automatically. If this would happen for the real data set, one may spend more effort to find a value for the MCLE for which the composite score statistics can be computed. If that is not possible, the composite score statistics for the models causing the problems would have to be removed. In 31 cases, we terminated the SMC ABC algorithm before the acceptance rate fell below its stopping threshold. We incorporate the results of these data sets since the model probabilities typically do not change much over the course of the SMC iterations (see, e.g., Figure~\ref{figure:SAdata20150520_SMCmodelprobsiteration_2}). Table~\ref{tab:failed_attempts} gives the frequencies of unsuccessful attempts and of premature terminations of the SMC ABC step for each of the models:

\begin{center}
	\begin{table}[h!]
		\begin{tabular}{lccccc}
			& E-t WM & E-t PE & B-R & tC WM & tC PE \\
			\hline
			ABC procedure aborted & 1 & 6 & 4 & 3 & 5 \\
			SMC ABC premature stop & 5 & 0 & 1 & 12 & 13
		\end{tabular}
		\caption{Number of simulated data sets for which the ABC algorithm was aborted unsuccessfully (first row) and number of simulated data sets for which the SMC ABC step was stopped prematurely (second row). Abbreviations: E-t: extremal-$t$ model, B-R: Brown-Resnick model, tC: $t$ copula model, WM: Whittle-Mat\'ern correlation function, PE: powered exponential correlation function. \label{tab:failed_attempts}}
	\end{table}
\end{center}

Row $j$ in the following matrix contains the average posterior probabilities for the different models (in the columns) across all data sets generated from model $j$. The column order of the models is the same as the row order. The posterior model probabilities are estimated from the final SMC particle set.

\[
\begin{blockarray}{llccccc}
& & \Pr(1|j) 	&  \Pr(2|j)		& \Pr(3|j) 	& \Pr(4|j) 	& \Pr(5|j)	\\
\begin{block}{ll[ccccc]}
j = 1 & \text{(extremal-$t$ WM)}		& 0.37		& 0.32		& 0.15		& 0.09		& 0.07 \\
j = 2 & \text{(extremal-$t$ PE)}    	& 0.29		& 0.34		& 0.26		& 0.05		& 0.06 	\\
j = 3 & \text{(Brown-Resnick)}  	 & 0.16		& 0.18		& 0.58		& 0.04	& 0.05	\\
j = 4 & \text{($t$ copula WM)} 	 	 & 0.06		& 0.05		& 0.03		& 0.43	& 0.42	\\
j = 5 & \text{($t$ copula PE)} 	 	 & 0.02		& 0.01		& 0.02		& 0.46	& 0.49	\\
\end{block}
\end{blockarray}
\]

\begin{figure}[h]
	\centering
	\begin{tabular}{cc}
		\includegraphics[width=0.4\textwidth]{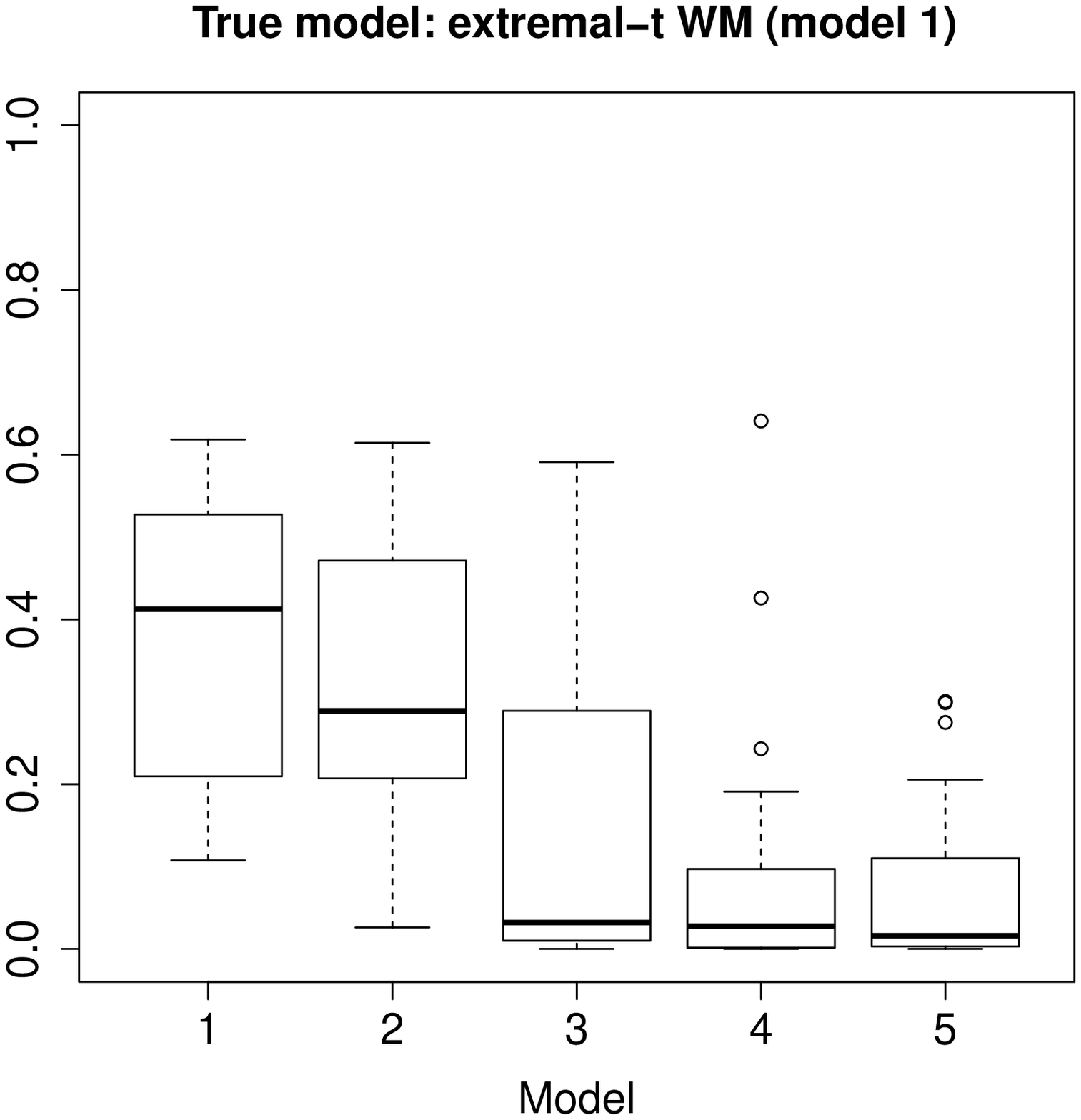} &
		\includegraphics[width=0.4\textwidth]{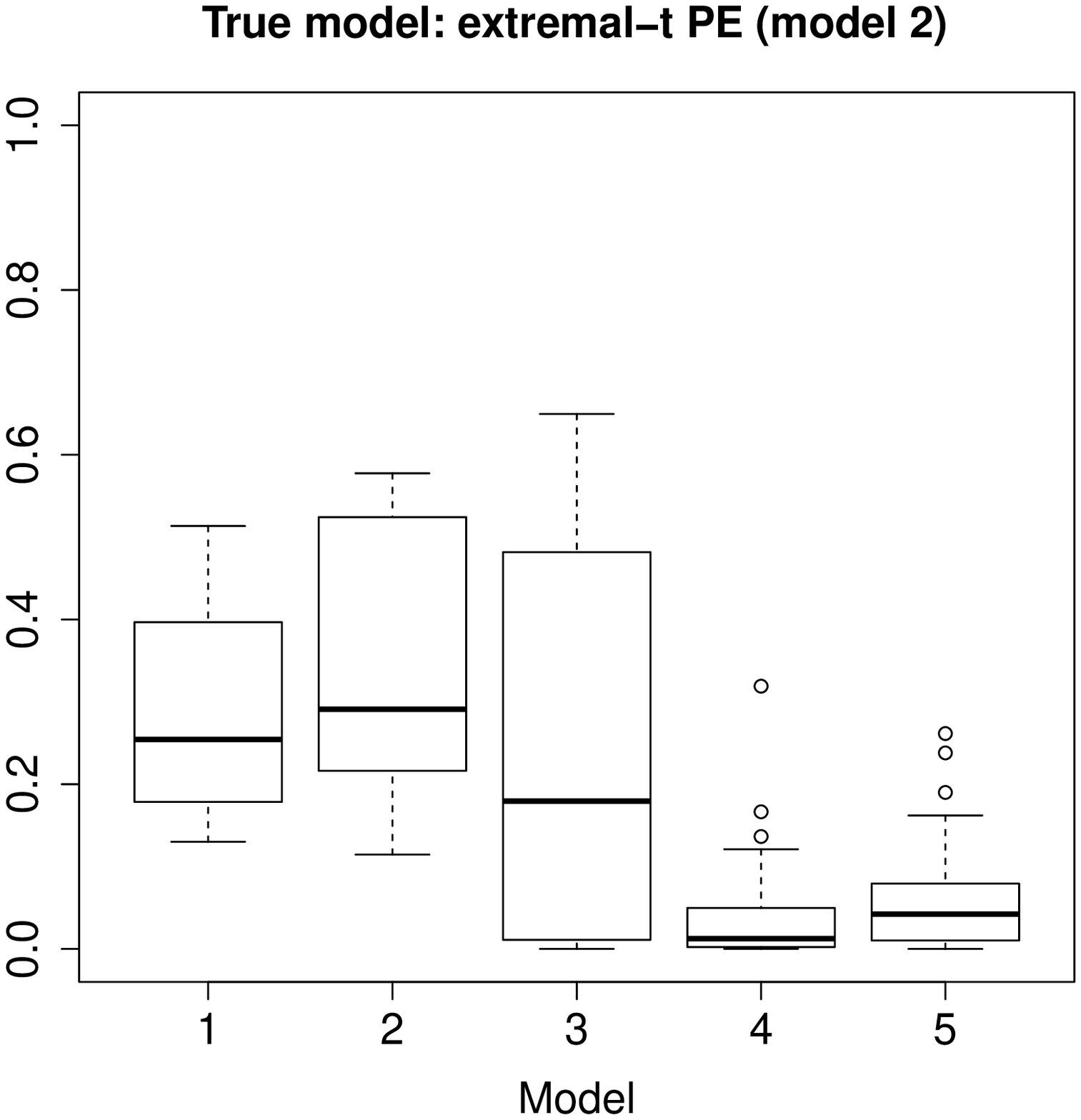} \\
		\includegraphics[width=0.4\textwidth]{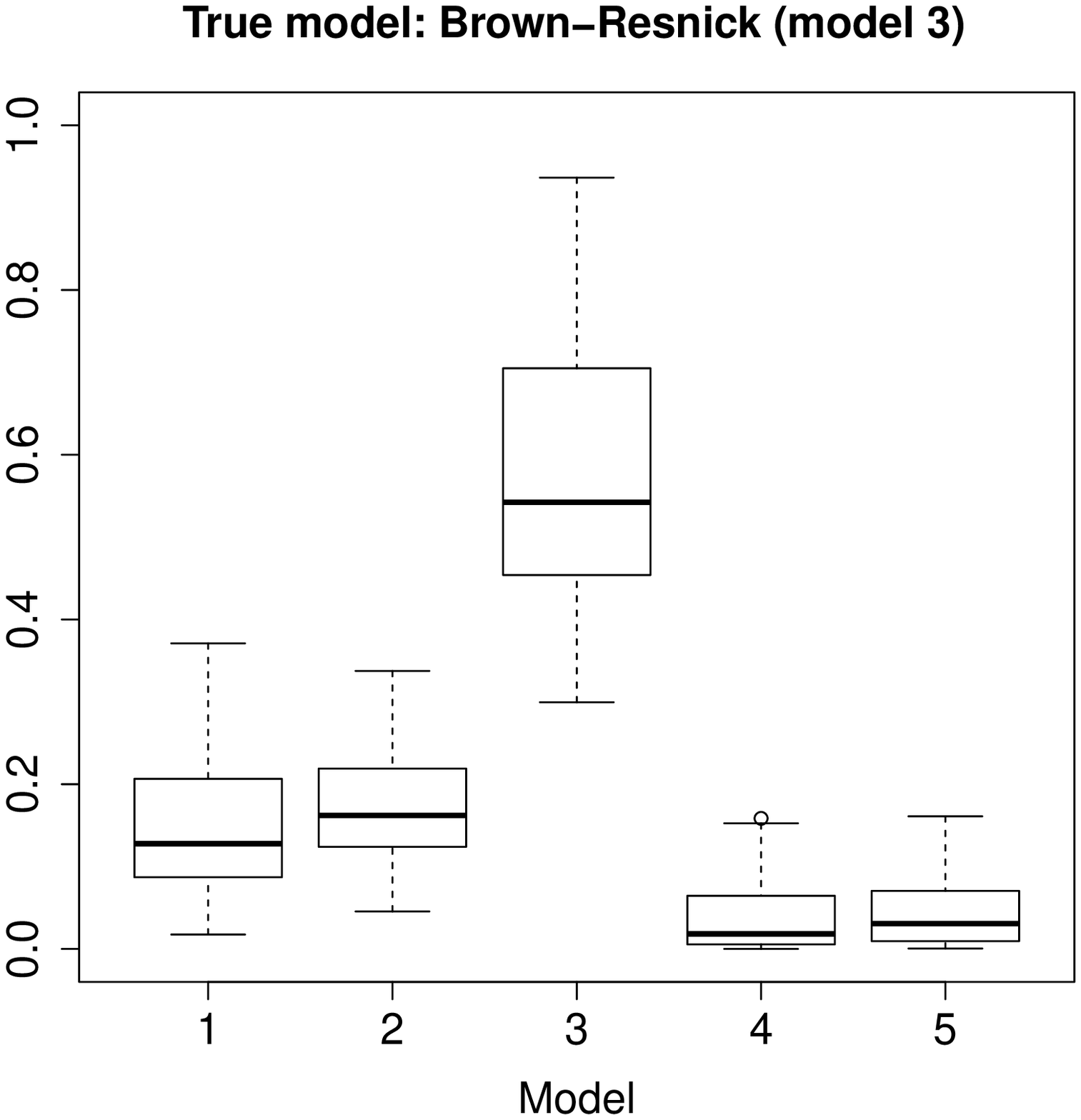} & \\
		\includegraphics[width=0.4\textwidth]{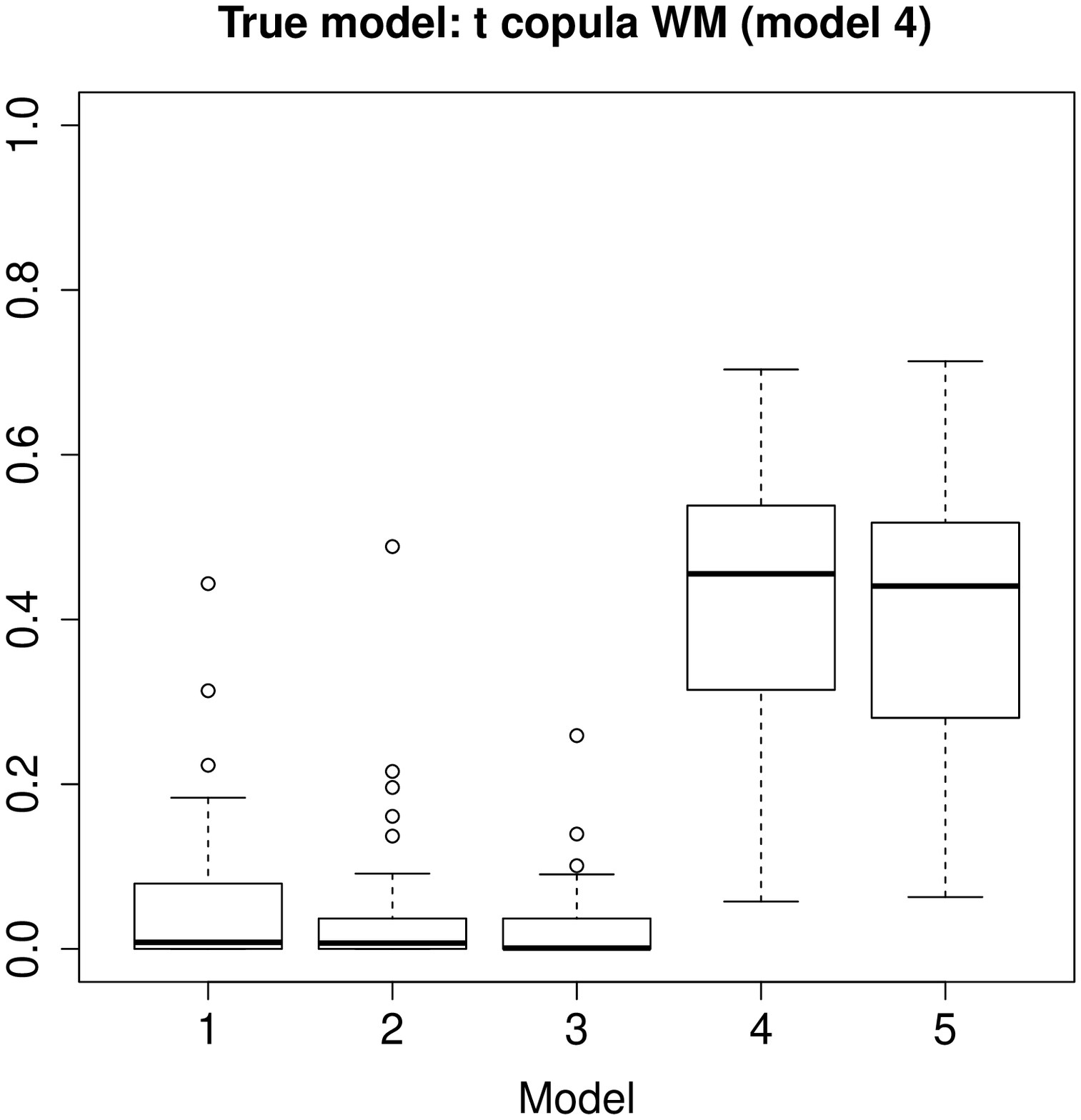} &
		\includegraphics[width=0.4\textwidth]{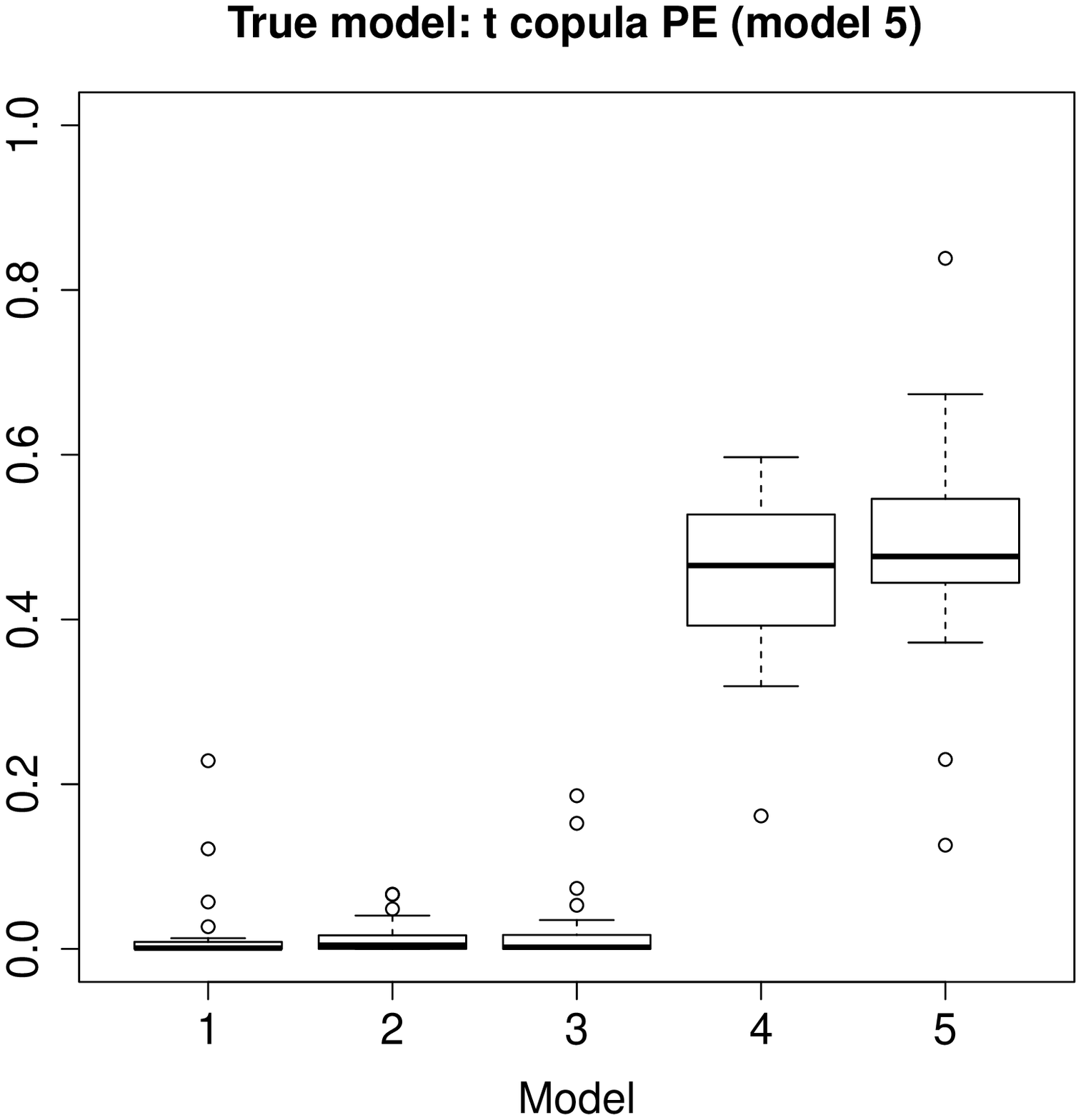}
	\end{tabular}
	\caption{Boxplots of posterior model probabilities for data sets simulated from the prior predictive distribution of the extremal-$t$ model (top) with Whittle-Mat\'ern (top left) and powered exponential (top right) correlation function, the Brown-Resnick model (middle), and the $t$ copula model (bottom) with Whittle-Mat\'ern (bottom left) and powered exponential (bottom right) correlation function.}
	\label{figure:sim_study_boxplots}
\end{figure}

Moreover, \fref{figure:sim_study_boxplots} shows boxplots of the distributions of the posterior model probabilities for simulated data sets from the different models.

The max-stable models and the $t$ copula models are separated very well. The $t$ copula models have low posterior probabilities for the max-stable data sets and vice versa. However, it is very hard to discriminate between the different correlation functions within the model classes. For example, when the true model is the $t$ copula model with Whittle-Mat\'ern correlation function, the distributions of the posterior model probabilities are almost equal between the two $t$ copula models (see Figure~\ref{figure:sim_study_boxplots}). In general, it is possible to discriminate between the Brown-Resnick model and the extremal-$t$ models quite well. One notable exception is when the true model is the extremal-$t$ model with powered exponential correlation function. In that case there is a high chance of a large posterior model probability of the Brown-Resnick model.

The misclassification matrix (see \citet{lee2015}) for the simulation study is provided in Appendix~\ref{sec:misclassification_matrix}. Element $(j,k)$ from the misclassification matrix contains the proportion of data sets from model $j$ that are classified as model $k$, where the classification rule is the highest posterior model probability. It is compared to the misclassification matrix obtained by applying the classical composite likelihood information criterion \citep{padoan2010,davison2012}. 

\clearpage

\subsection{Assessing quality of regression summary statistics obtained by FP step}

\label{subsec:dataset_FPstep}

In this section, we investigate the performance of the summary statistic set obtained by the FP step for the purpose of model selection for the South Australian maximum temperature data. The performance is summarised using the matrix of average posterior model probabilities. 
The ($j$, $k$)-th element of this matrix corresponds to the average posterior probability of model $k$ among all model $j$ simulations in the FP step's training particle set. Given simulated data from model $j$, the posterior model probabilities are estimated according to Equation~\eqref{eq:logreg_linpredictor}. The matrix is
\[
\begin{blockarray}{llccccc}
& & \Pr(1|j) 	&  \Pr(2|j)		& \Pr(3|j) 	& \Pr(4|j) 	& \Pr(5|j)	\\
\begin{block}{ll[ccccc]}
j = 1 & \text{(extremal-$t$ WM)}		& 0.36		& 0.33		& 0.16		& 0.09		& 0.06 \\
j = 2 & \text{(extremal-$t$ PE)}    	& 0.33		& 0.40		& 0.16		& 0.06		& 0.06 	\\
j = 3 & \text{(Brown-Resnick)}  	 & 0.16		& 0.15		& 0.56		& 0.07	& 0.06	\\
j = 4 & \text{($t$ copula WM)} 	 	 & 0.09		& 0.06		& 0.06		& 0.42	& 0.37	\\
j = 5 & \text{($t$ copula PE)} 	 	 & 0.06		& 0.06		& 0.06		& 0.37	& 0.46	\\
\end{block}
\end{blockarray}.
\]

This matrix is similar to the matrix obtained by the simulation study. Max-stable and $t$ copula models are well separated. The estimated posterior model probabilities for models not belonging to the same model class (extremal-$t$, Brown-Resnick, $t$ copula) are generally very low. However, it is difficult to identify the correct correlation function within each class.

In Appendix~\ref{sec:probability_matrix_test_data}, the FP step's matrix of average posterior probabilities is computed on a separately generated test particle set to check for overfitting. This matrix is almost identical to the one given in this section. The regression coefficients of the multinomial logistic regression for the model indicator are provided in Appendix~\ref{sec:regression_coefficients}.

\subsection{Results for South Australian data}

\label{subsec:dataset_results}

\begin{figure}[h]
	\centering
	\begin{adjustbox}{max width = \columnwidth}
		\includegraphics[width=0.8\textwidth]{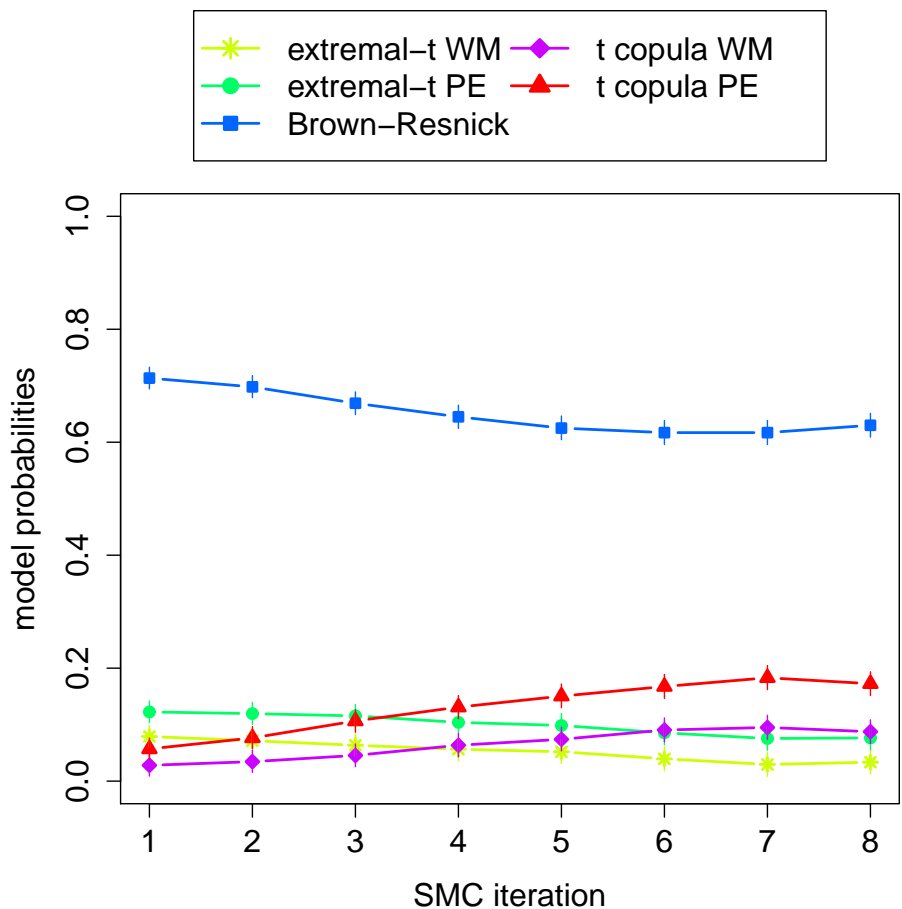}
	\end{adjustbox}	
	\caption{Progression of estimated posterior model probabilities across SMC iterations for the South Australian data set without the station in Warooka.}
	\label{figure:SAdata20150520_SMCmodelprobsiteration_2} 
\end{figure}

For the actual South Australian data without the station in Warooka, \fref{figure:SAdata20150520_SMCmodelprobsiteration_2} shows the progression of the proportions of particles pertaining to the different models across the SMC iterations. 
Simultaneous $95\%$ intervals for these model probability estimates were obtained from the fact that the model indicator is a multinomial variable \citep{sison1995}.

The preferred model is the Brown-Resnick model. In the final particle set after the last SMC iteration, about $63\%$ of the particles are from the Brown-Resnick model. From the remaining particles, $26\%$ belong to one of the two $t$ copula models and $11\%$ belong to one of the two extremal-$t$ models. The model probabilities keep fairly constant over the course of the SMC iterations. The posterior probabilities of the $t$ copula models slightly increase over time at the expense of the other models.

\begin{figure}[h]
	\centering
	\includegraphics[width=0.4\textwidth]{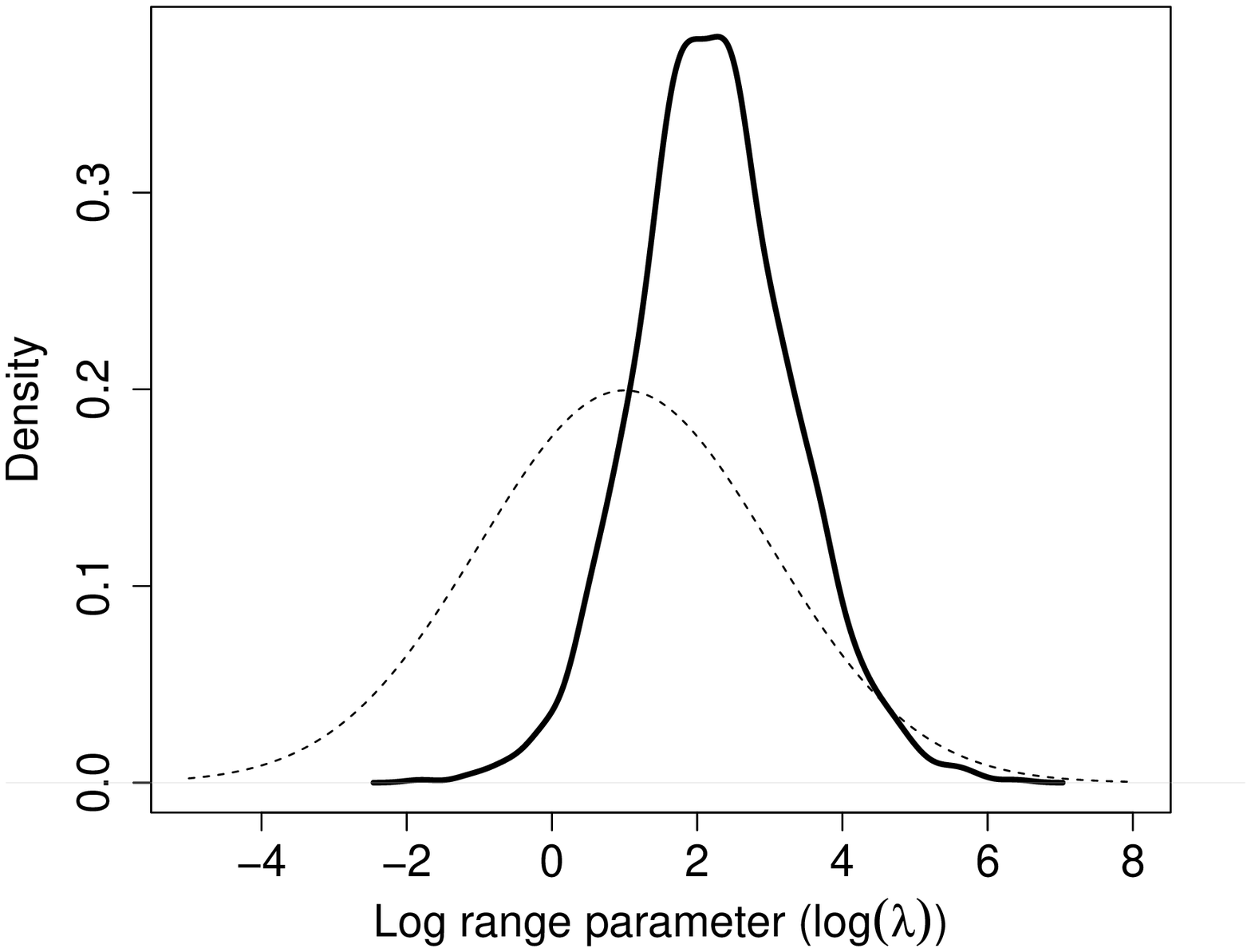} 
	\includegraphics[width=0.4\textwidth]{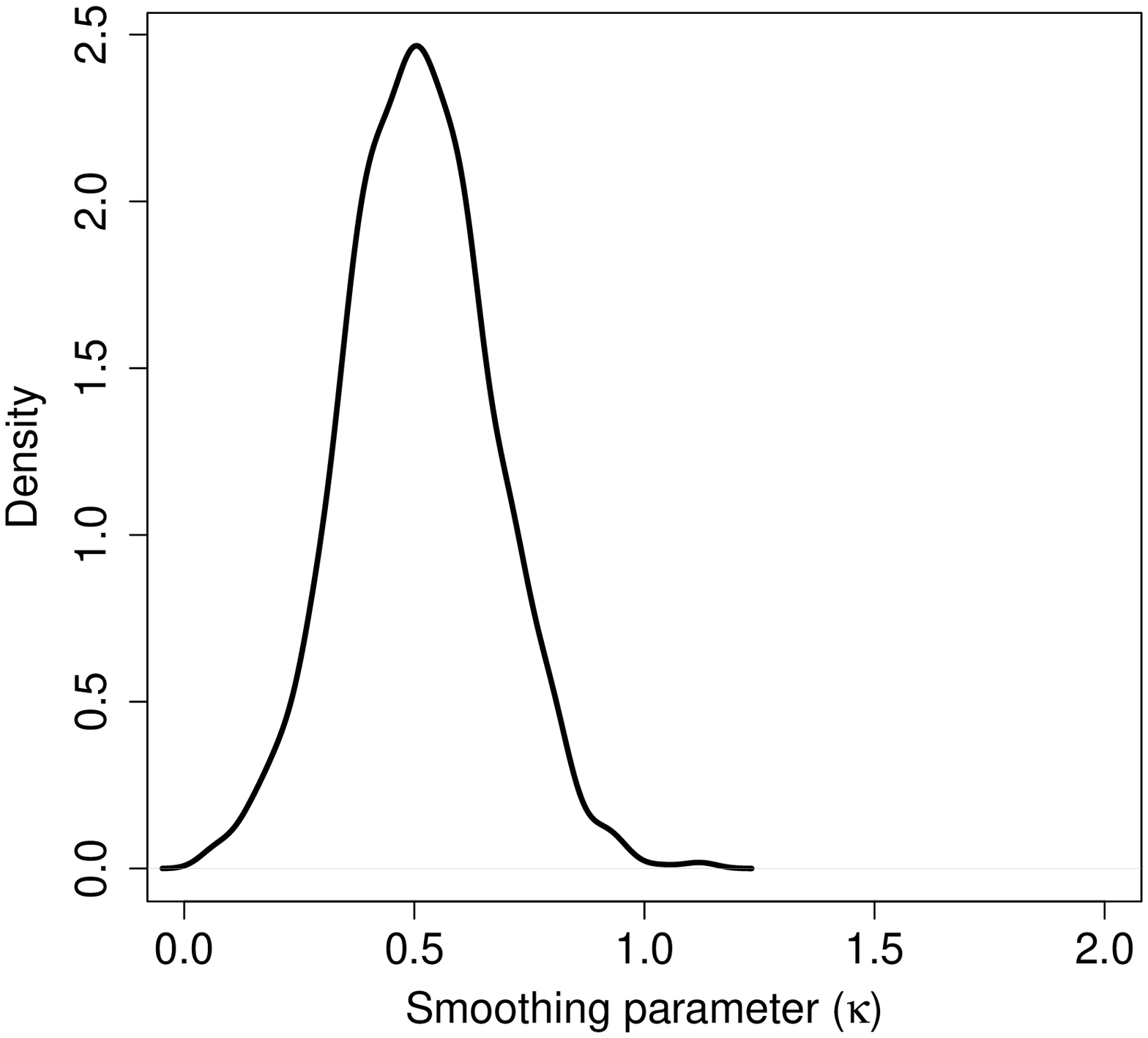} \\
	\includegraphics[width=0.4\textwidth]{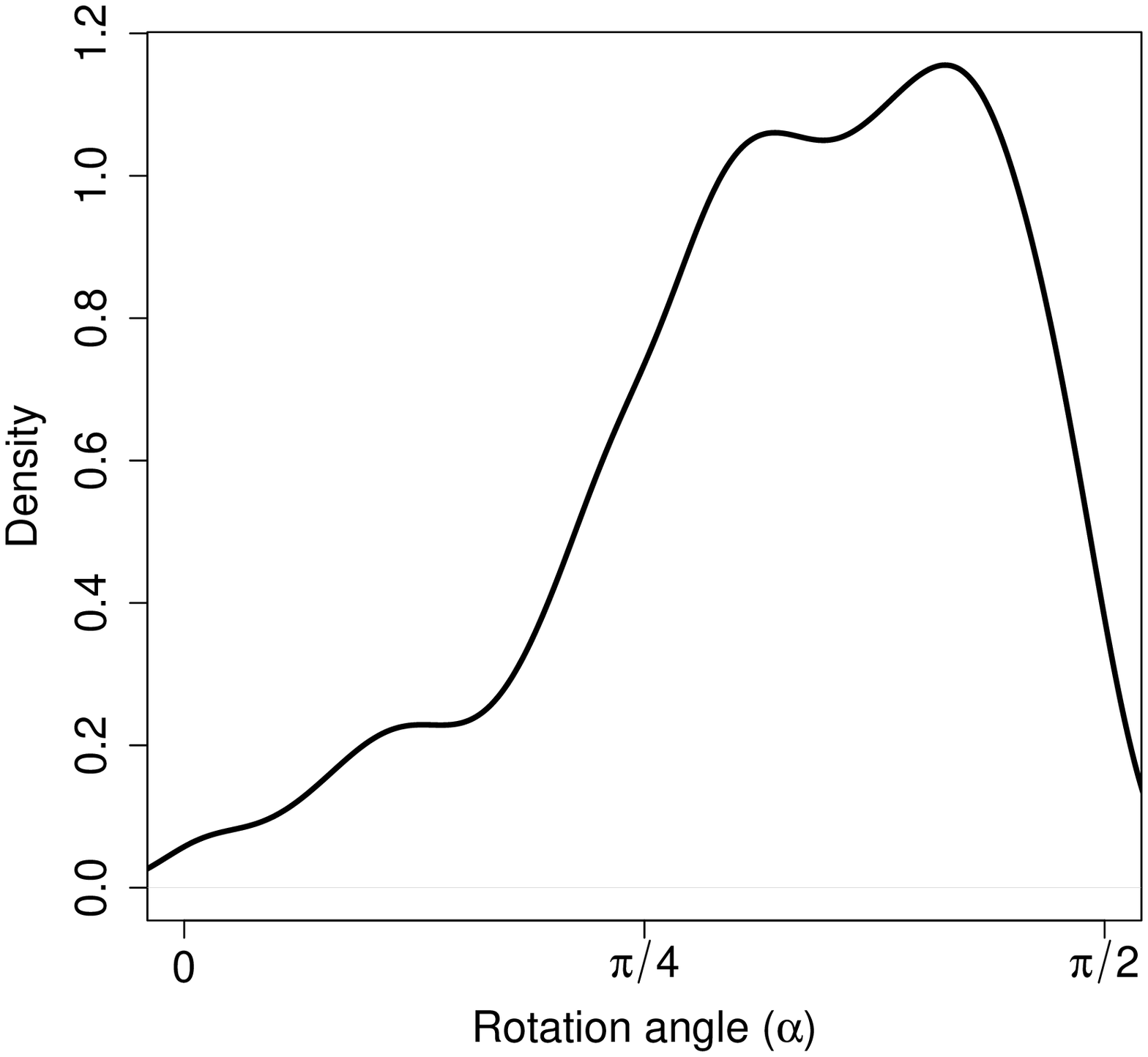}
	\includegraphics[width=0.4\textwidth]{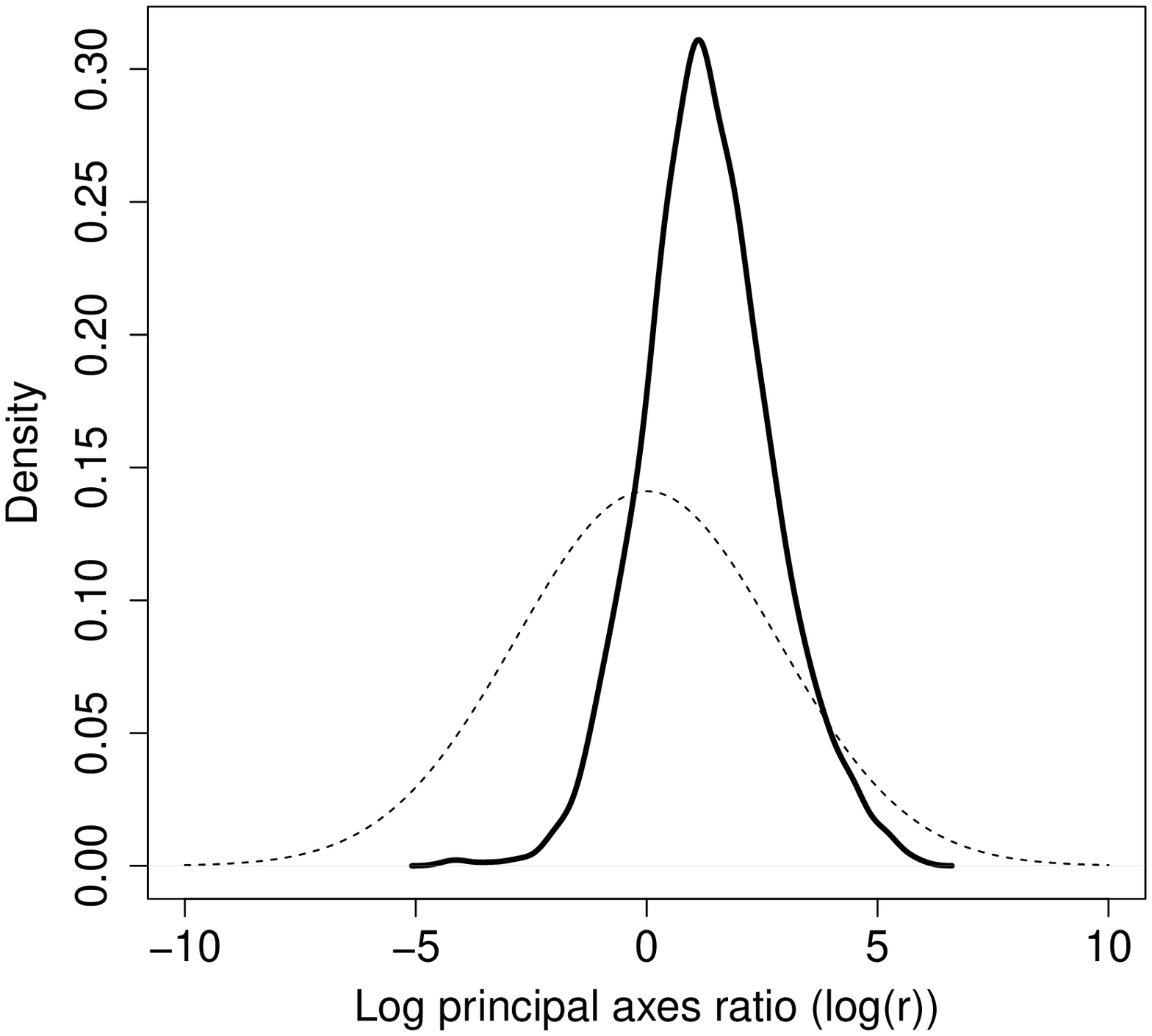}
	\caption{\textbf{Solid lines:} kernel density estimates of marginal posterior distributions for the best-fitting \emph{Brown-Resnick model} when applied to the South Australian data set without the station in Warooka.
		\textbf{Dashed lines:} prior densities. 
		Uniform prior densities for the smoothing and the rotation angle parameter are not displayed. However, for these parameters the abscissa range is equal to the support of the respective uniform prior distribution.}
	\label{figure:SAdata20150520_Parameters_BR_2}
\end{figure}

The estimated approximate marginal posterior distributions for the parameters of the best-fitting Brown-Resnick model are provided in \fref{figure:SAdata20150520_Parameters_BR_2} and \tref{table:SAdata_params}. For all parameters, we get posteriors that are unimodal and more informative than the respective prior distributions.
Parameter estimation results for the other models are provided in Appendix~\ref{sec:posterior_distributions}.

\begin{table}[h]
	\centering
	\caption{\label{table:SAdata_params}Parameter estimates of the best-fitting Brown-Resnick model for the South Australian data set without the station in Warooka. The values of the angle $\alpha$ are given in radiants and are divided by $\pi$.}
	\begin{tabular}{c*{5}{D{.}{.}{2}}}
		\hline
		\multicolumn{1}{c}{Parameter} & \multicolumn{1}{c}{Mean} & \multicolumn{1}{c}{SD} & \multicolumn{1}{c}{$2.5\%$ quantile} & \multicolumn{1}{c}{Median} & \multicolumn{1}{c}{$97.5\%$ quantile}\\
		\hline 
		$\log(\lambda)$ & $2.24$  & $1.08$  & $0.19$  & $2.19$ & $4.48$ \\
		$\lambda$ & $17.74$  & $32.30$  & $1.21$  & $8.98$ & $88.08$ \\
		$\kappa$ & $0.51$ & $0.16$ & $0.20$ & $0.51$ & $0.82$ \\
		$\alpha / \pi$ & $0.33$ & $0.11$ & $0.08$ & $0.34$ & $0.49$ \\
		$\log(r)$ & $1.34$ & $1.39$ & $--1.28$ & $1.27$ & $4.33$ \\
		$r$ & $10.75$ & $24.71$ & $0.28$ & $3.55$ & $76.15$ \\
		\hline
	\end{tabular}
\end{table}

\clearpage

In Appendix~\ref{sec:extremal_plots}, the pairwise extremal coefficient functions evaluated at the median posterior parameter values are plotted for all the models.

To check the overall goodness of fit, we compare the observed pairwise F-madogram and Kendall's $\tau$ estimates with their posterior predictive distributions. To that end, we simulated $10,\!000$ data sets from the posterior predictive distribution. Each data set (18 observations on each of 25 locations) was generated by drawing one particle with replacement from the final particle set and then simulating the data set depending on the selected particle's model indicator and parameter values. Given this sample, we can estimate the posterior predictive distributions of all the dependence indicators for all the location pairs. Note that the final particle set contains particles from all models with an estimated posterior probability greater than $0$, so the posterior predictive distributions are the Bayesian model averages over all models.

\begin{figure}[h]
	\centering
	\begin{tabular}{cc}
		F-madogram & Kendall's $\tau$ \\
		\includegraphics[width=0.4\textwidth]{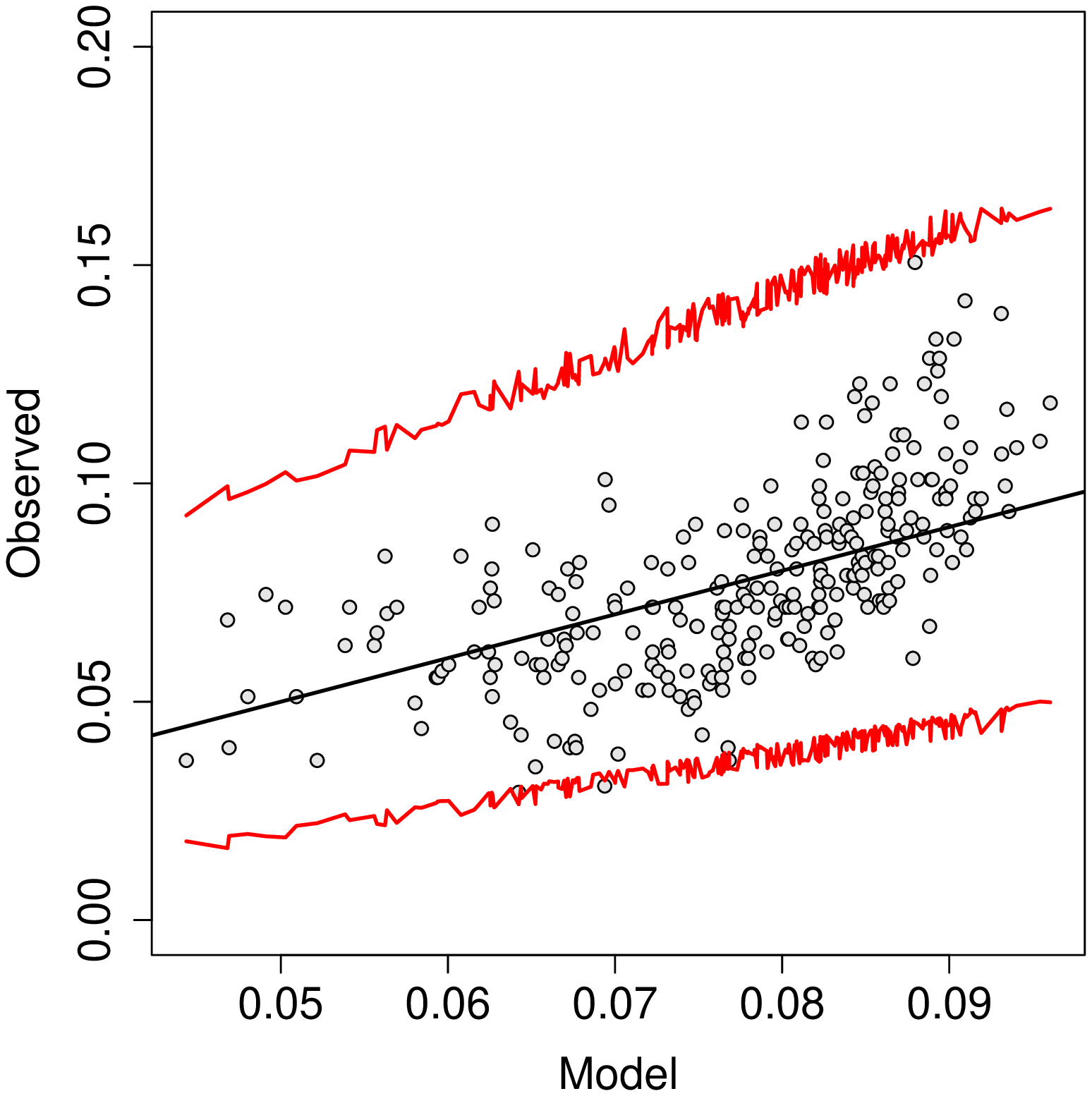} &
		\includegraphics[width=0.4\textwidth]{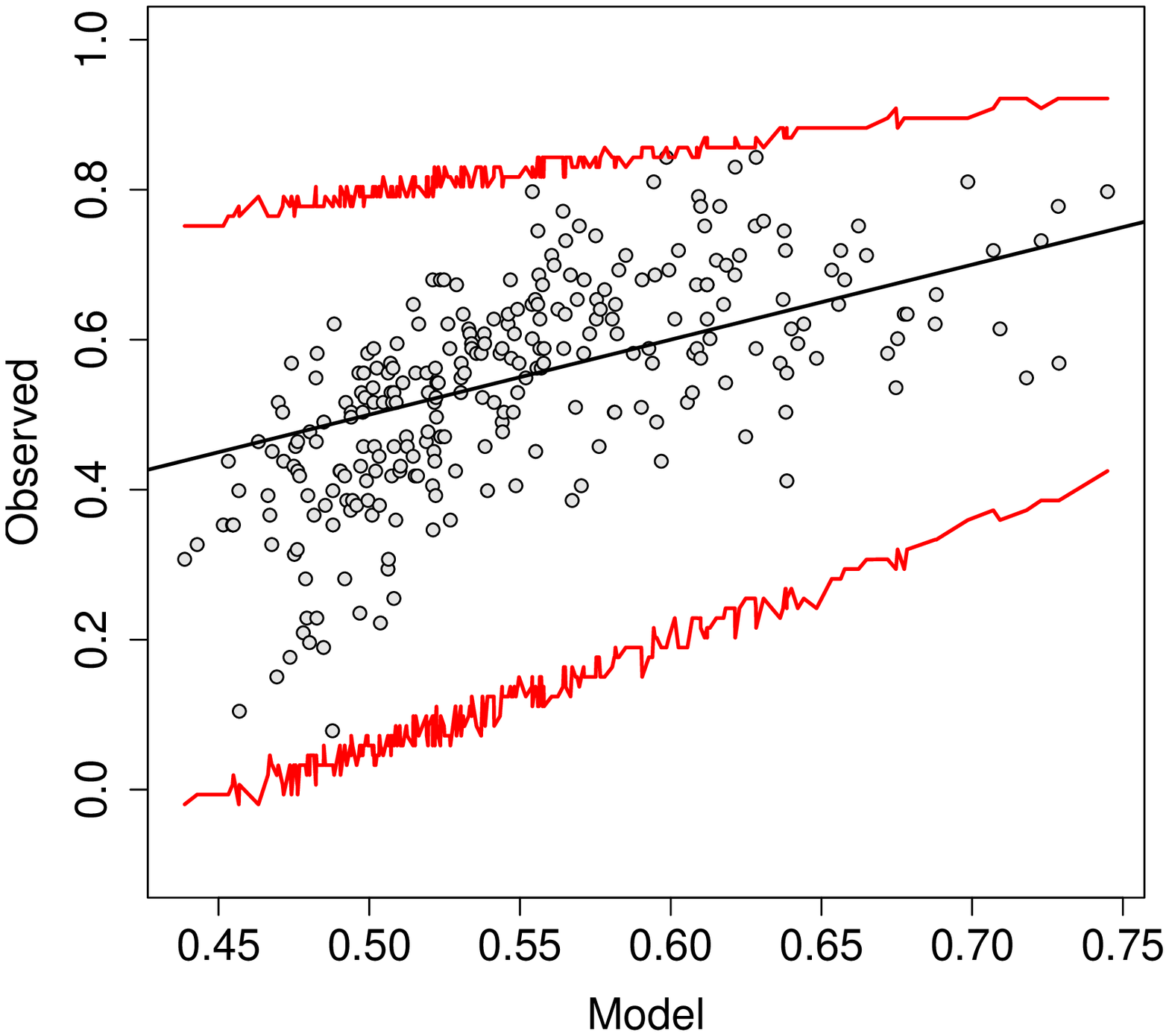} \\
		\includegraphics[width=0.4\textwidth]{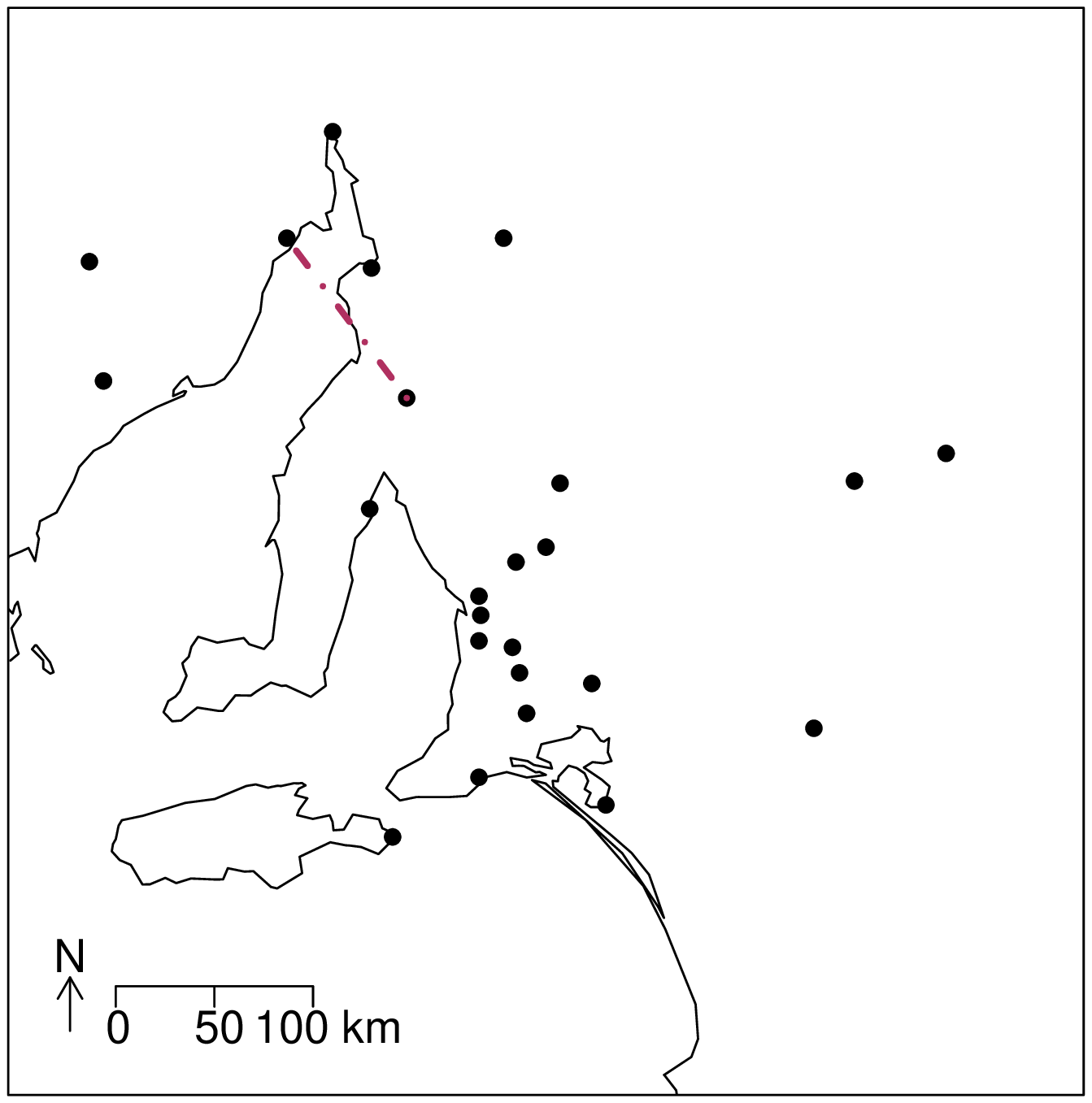} &
		\includegraphics[width=0.4\textwidth]{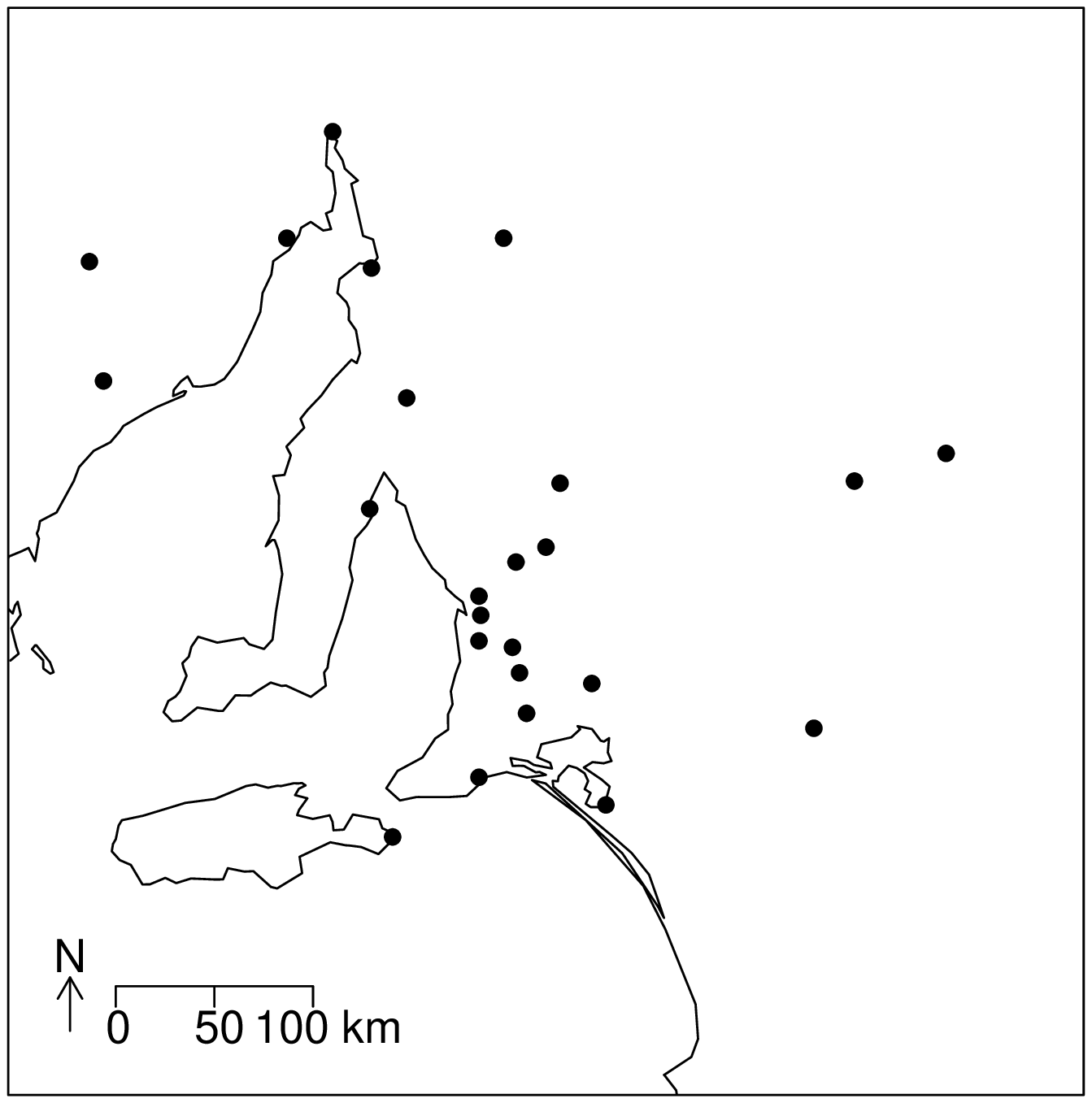}
	\end{tabular}
	\caption{\textbf{Top row:} observed pairwise \emph{F-madogram} (left) and \emph{Kendall's $\tau$} (right) estimates vs.\ posterior predictive means of these estimates for each pair of locations (location 'Warooka' is excluded). The red lines connect the $2.5 \%$ and $97.5 \%$ quantiles of the posterior predictive distributions for the location pairs.
		\textbf{Bottom row:} \emph{maroon dashed lines} connect location pairs with unusually \emph{high observed mutual dependency}. These are location pairs where the observed F-madogram estimate is below the $2.5 \%$ quantile of posterior predictive distribution (left) or the observed Kendall's $\tau$ estimate is above the $97.5 \%$ quantile of posterior predictive distribution (right).
	}
	\label{figure:SAdata_posterior_predictive_2}
\end{figure}

In \fref{figure:SAdata_posterior_predictive_2}, we compare the observed pairwise F-madogram (left plots) and Kendall's $\tau$ (right plots) estimates to their posterior predictive distributions. Only one location pair falls outside the $95 \%$ posterior predictive probability interval for the F-madogram. This location pair is depicted in the left bottom plot of \fref{figure:SAdata_posterior_predictive_2}.

Despite excluding Warooka, our models are not able to perfectly capture the dependence structure. For high distances, the posterior predictive distribution puts too much mass on  summary values that indicate high dependency, for smaller distances it puts too much mass on summary values indicating low dependency, see also the boundaries of the $95 \%$ posterior predictive probability intervals in \fref{figure:SAdata_posterior_predictive_2}. Due to the geography of the locations, especially the vicinity of most stations to the ocean, more complex dependence structures would have to be considered to achieve a better fit.

\section{Discussion}
\label{sec:discussion}

The research presented in this paper expands the use of ABC for spatial extremes applications to include model selection problems. We show that the spatial dependence structure of the maximum temperature data collected around the state of South Australia is best captured by a Brown-Resnick model out of a collection of three max-stable models and two non-extremal Student-$t$ copula models. For this analysis, we exclude the station in Warooka because we consider it to be an outlier.
It should be emphasised that this paper provides a case study of the application of ABC to model selection problems for spatial extremes and is not meant to be directly used to inform decision-making. The models we consider are not able to capture the intricacies of the particular environmental process in their entirety.

We would like to highlight some extensions to the straightforward geometrically anisotropic max-stable models used that would provide a more realistic representation of the spatial structure of the maximum temperature for further research. Our work has ignored any effect that might be attributable to the fact that part of the region in \fref{fig:location_map} is actually a large body of water that would affect the local temperature near the coast. As such, it might be more appropriate to focus on the land mass of the region. One might determine a transformation of the land mass to a regular geometry while keeping the distance measure positive-definite by using complex spatial smoothers \citep{wood2008, sangalli2013}. 
\citet{blanchet2011} discuss several options to model anisotropic behaviour beyond geometric anisotropy.

All the models we consider assume unit Fr\'echet marginal distributions. If that is not the case, one has to transform the data properly prior to model fitting, as we did, or the number of parameters to be estimated has to be increased drastically by three GEV parameters per location. Therefore, the marginal GEV parameters are often modelled themselves by employing spatial smoothing models. This kind of modelling is required if the fitted max-stable model is used to interpolate the measurements at unobserved locations on the original measurement scale. \citet{ErhardtSmith2012} interpolate the GEV parameters on unobserved locations by applying Kriging. The R package \texttt{SpatialExtremes} \citep{Rspatext} facilitates the use of P-splines for GEV response surface modelling. The shape parameter $\xi$ often exhibits little variability and follows no discernible pattern and is therefore set to a constant value over the whole space \citep{Ribatet2013, ErhardtSmith2012}. 
Investigations of the appropriate modelling of the marginal GEV parameters and quantification of the corresponding uncertainty on model results are beyond the scope of this study but would be of interest in further work.
There are also other models for spatial extremes that could be considered, such as those based on latent variable models \citep{davison2012}.

In addition, there are still numerous improvement opportunities for the ABC model selection algorithm and ABC algorithms in general. Use of more sophisticated classification methods such as random forests \citep{pudlo2015} in the FP step could potentially provide improved performance at the expense of increased computation and less straightforward interpretations of the classification method's direct outputs. Optimising the efficiency of the ABC algorithm is also an active research area. 
Such optimisations are particularly vital for applications involving computationally heavy model simulations such as the exact simulation of max-stable models on a dense grid. 
Methods such as Lazy ABC \citep{prangle2014-1} and expectation propagation ABC \citep{barthelme2015} introduce additional approximations to the standard ABC method to decrease the computational time required to obtain the approximate posterior for computationally expensive models. As noted in \citet{barthelme2015}, these methods appear promising for model selection problems as well.

\section*{Acknowledgements}

XJL, CCD and ANP are affiliated with the ARC Centre of Excellence for Mathematical \& Statistical Frontiers (ACEMS), where CCD is an Associate Investigator and ANP is a Chief Investigator. 
XJL received PhD scholarship funding from the Centre of Research Excellence in Reducing Healthcare Associated Infections (NHMRC Grant 1030103).
MH was funded by the Austrian Science Fund (FWF): J3959-N32.
CCD was supported by an Australian Research Council's Discovery Early Career Researcher Award funding scheme DE160100741.
ANP was supported by an Australian Research Council Discovery Project DP 110100159.
Computational resources and services used in this work were provided by the HPC and Research Support Group, Queensland University of Technology, Brisbane, Australia.

\clearpage

\bibliographystyle{abbrvnat}
\bibliography{ABMC_SpatExt}

\clearpage

\appendix

\section{ABC algorithm summary}

\label{sec:abc_summary}

In the following summary of our ABC method, we have omitted the time indices for the particles and the parameter proposal distributions $q_k(\cdot)$ to avoid notational clutter. The two steps of our ABC method for model selection and parameter estimation are:
\begin{enumerate}
	\item FP step
	\newline 
	Inputs: number of model simulations $M$ (assumed to be a multiple of $K$), $K$ candidate models, parameter prior distributions $\pi(\parms_k | \mo = k) \: (k = 0, \ldots, K-1)$, choice of regression summary statistics $\gb(\zb)$.
	\begin{enumerate}
		\item If the model prior is uniform, draw $M/K$ parameter vectors from each candidate model's parameter prior distribution to obtain $P = \{k^i, \parms_{k^i}^i\}_{i=1}^M$, where $\parms_{k^i}^i \sim \pi(\parms_{k^i}| \, \mo = k^i)$. 
		\item Simulate from the respective candidate models for each element in $P$ to obtain the particle set $PP = \{k^i, \parms_{k^i}^i, \ub^i\}_{i=1}^M$, where $\ub^i \sim f_{k^i}(\ub | \, \parms_{k^i}^i)$. Compute the regression summary statistics $\gb(\ub^i)$ for each simulated data set in $PP$. 
		\item Perform a stepwise multinomial logistic regression using all $M$ particles in $PP$, where the model indicator $k^i$ is the outcome variable and the regression summary statistics $\gb(\ub^i)$ are the covariates.
		\item Perform stepwise linear regressions for each model parameter $\phi_{k,j}$ ($k = 0,\ldots,K-1$; $j = 1,\ldots,Q_k$) using the respective $M/K$ parameter draws $\{\phi_{k^i,j}^i: \: k^i = k\}_{i=1}^M$ in $PP$ as the outcome variable and the regression summary statistics $\{\gb(\ub^i): \: k^i = k\}_{i=1}^M$ as covariates.
	\end{enumerate}
	Outputs: regression coefficient estimates $\hat{\bm{\beta}}_k$ ($k=1,\ldots,K-1$) and $\hat{\bm{\beta}}_{k,j}$ ($k=0,\ldots,K-1$; $j = 1,\ldots,Q_k$) from the logistic and linear regressions, respectively.
	\item SMC ABC step
	\newline 
	Inputs: regression coefficient estimates from FP step, simulation size $N_2$ for initial ABC rejection step, number of particles $N$, SMC replenishment tuning parameters $\delta$ and $c$, observed data $\zb$, discrepancy function $d_T[\cdot,\cdot]$, RJ-MCMC model switch proposals $q_{k,k^*}$, RJ-MCMC proposal distribution for model $k$'s parameters $q_k(\cdot)$ $(k = 0, \ldots, K-1)$, stopping criteria (final tolerance $\epsilon_{\min}$ and/or minimum acceptance probability $p_{\mathrm{acc},\min}$).
	\begin{enumerate}
		\item Generate the prior predictive draws $\{k^i, \parms_{k^i}^i, \ub^i\}_{i=1}^{N_2}$ and  perform rejection ABC to obtain the initial particle set $\{k^i, \parms_{k^i}^i, \ub^i\}_{i=1}^N$.
		\item Set the acceptance probability $p_{\mathrm{acc}, 0}$ to a value $>  p_{\mathrm{acc},\min}$ and $R_1$ to some arbitrary value (not too small). Set $t = 1$.
		\item Denote $\epsilon_t$ to be the largest discrepancy value in the current particle set. If $\epsilon_t \leq \epsilon_{\min}$ or $p_{\mathrm{acc},t-1} \leq p_{\mathrm{acc},\min}$, terminate algorithm.
		\item Drop the $N_{\delta} = \lceil \delta N \rceil$ particles with the largest discrepancy values from the particle set. Set $\epsilon_t$ to be the largest discrepancy value of the remaining $N - N_{\delta}$ particles. 
		\item Compute the parameters in $q_k (\cdot)$ using the remaining particles from model $k$. 
		\item Resample $N_{\delta}$ particles with replacement from the remaining particle set until a full set of $N$ particles is recovered. 
		\item To each newly resampled particle $i$ ($i = N-N_{\delta}+1, \ldots, N$), a RJ-MCMC kernel is applied $R_t$ times, where the kernel's Metropolis-Hastings (MH) ratio for acceptance of a move from $(k^i = k, \: \parms_{k^i}^i = \parms_k, \: \ub^i = \ub)$ to the proposed values $(k^i = k^* \sim q_{k,k^*}, \: \parms_{k^i}^i = \parms_{k^*}^* \sim q_{k^*}(\cdot), \: \ub^i = \ub^* \sim f_{k^*}(\cdot| \, \parms_{k^*}^*))$ is 
		\[
		\frac{\Pr(\mo=k^*)}{\Pr(\mo=k)} \frac{\pi(\parms_{k^*}^*|\, \mo = k^*)}{\pi(\parms_k | \, \mo = k)} \mathds{1}\{d_T[\tb(\zb),\tb(\ub^*)] < \epsilon_t\} \frac{q_k (\parms_k)}{q_{k^*} (\parms_{k^*}^*)} \frac{q_{k^*, k}}{q_{k, k^*}}
		\]
		and $\mathds{1}\{\cdot\}$ is the indicator function. 
		
		When $\Pr(\mo=k) = 1/K$ and $q_{k, k^*} = 1/K$ for all $k, k^* \in {1,\ldots,K}$, the model prior and model switch proposal ratios simplify to one and the MH ratio becomes
		\[
		\frac{\pi(\parms_{k^*}^*|\, \mo = k^*)}{\pi(\parms_k | \, \mo = k)} \mathds{1}\{d_T[\tb(\zb),\tb(\ub^*)] < \epsilon_t\} \frac{q_k (\parms_k)}{q_{k^*} (\parms_{k^*}^*)}.
		\]	
		\item Compute the acceptance probability $p_{\mathrm{acc},t} = a_t / (R_t N_{\delta})$, where $a_t$ is the number of accepted proposals in iteration $t$, and set $$\displaystyle R_{t+1} = \left\lceil \frac{\log(c)}{\log(1 - p_{\mathrm{acc},t})} \right\rceil.$$ Increase $t$ by 1. Return to step 2(c).
	\end{enumerate}
	Outputs: final particle set including discrepancies $\left\{ k^i, \parms_{k^i}^i, \ub^i, d_T^i = d_T[\tb(\zb), \tb(\ub^i)]\right\}_{i=1}^N$.
\end{enumerate}

\section{Additional remarks and results for simulation study}

\subsection{Procedure to find maximum composite likelihood estimates (MCLEs)}

\label{sec:MCLE_procedure}

The MCLEs were found through numerical optimisation of the pairwise log-likelihood. For some data sets and models, the result of the optimisation procedure heavily depended on the starting value. Therefore, for each data set and model we ran the optimisation routine repeatedly using random starting values from the prior distribution until we had five runs where the optimisation converged. Then we used the result from the run which led to the highest value of the objective function. For the Student-$t$ copula models, we used the MLEs found for the full log-likelihood as starting values for optimising the pairwise log-likelihood. We employed the box-constrained Broyden, Fletcher, Goldfarb, and Shanno (BFGS) quasi-Newton optimisation method, see \citet{byrd1995}.

\subsection{Misclassification matrices for simulation study}

\label{sec:misclassification_matrix}

\subsubsection{Misclassification matrix obtained by ABC procedure}

Element $(j,k)$ in the misclassification matrix below gives the probability that a data set generated from model $j$ in the simulation study is classified as being from model $k$, where the classification rule is the highest posterior model probability estimated from the final SMC particle set:

\[
\begin{blockarray}{llccccc}
& & \Pr(1|j) 	&  \Pr(2|j)		& \Pr(3|j) 	& \Pr(4|j) 	& \Pr(5|j)	\\
\begin{block}{ll[ccccc]}
j = 1 & \text{(extremal-$t$ WM)}		& 0.38		& 0.28		& 0.21		& 0.07		& 0.07 \\
j = 2 & \text{(extremal-$t$ PE)}    	& 0.17		& 0.38		& 0.42		& 0.04		& 0.04 	\\
j = 3 & \text{(Brown-Resnick)}  	 & 0.04		& 0.00		& 0.96		& 0.00	& 0.00	\\
j = 4 & \text{($t$ copula WM)} 	 	 & 0.04		& 0.04		& 0.04		& 0.44	& 0.44	\\
j = 5 & \text{($t$ copula PE)} 	 	 & 0.00		& 0.00		& 0.00		& 0.52	& 0.48	\\
\end{block}
\end{blockarray}.
\]

\subsubsection{Misclassification matrix obtained by CLIC}

We compare the misclassification matrix for our simulated data sets obtained by our ABC procedure to the misclassification matrix obtained by classifying the simulated data sets according to the \emph{composite likelihood information criterion (CLIC)}, see, e.g., \citet{davison2012} and \citet{padoan2010}. This is the classical criterion traditionally used for model selection of max-stable models, for which only a composite likelihood representation is available. It is a generalisation of the Akaike information criterion and accounts for the model misspecification due to using the composite likelihood.

The CLIC is defined as
\[ 
\mathrm{CLIC} = - 2 \, p\ell(\tilde{\corparms}_{MCLE};\zb) + 2 \, \mathrm{tr}\left(\hat{\bm{J}} \hat{\bm{H}}^{-1}\right),
\]
where $p\ell(\tilde{\corparms}_{MCLE};\zb)$ is the pairwise/composite log-likelihood (Equation~\eqref{pw_loglike_def}) evaluated at the maximum composite likelihood estimate (MCLE), 
$\hat{\bm{J}}$ is an estimate of $\bm{J}(\tilde{\corparms}_{MCLE}) = \mathrm{Var}\left[ \nabla_{\corparms} \: p\ell(\tilde{\corparms}_{MCLE};\zb)\right]$, and $\hat{\bm{H}}$ is an estimate of $\bm{H}(\tilde{\corparms}_{MCLE}) = -\mathrm{E}\left[ \nabla^2_{\corparms} \: p\ell(\tilde{\corparms}_{MCLE};\zb) \right]$. We estimate $\bm{J}$ and $\bm{H}$ as suggested in \citet[p.\ 163]{Ribatet2013}.

In order to be able to compare all models on equal terms, we also computed the CLIC for the Student-$t$ copula models based on their pairwise likelihood representation.

Using the CLIC as classification criterion, the misclassification matrix for the simulated data sets is
\[
\begin{blockarray}{llccccc}
& & \Pr(1|j) 	&  \Pr(2|j)		& \Pr(3|j) 	& \Pr(4|j) 	& \Pr(5|j)	\\
\begin{block}{ll[ccccc]}
j = 1 & \text{(extremal-$t$ WM)}		& 0.57		& 0.33		& 0.00		& 0.07		& 0.03 \\
j = 2 & \text{(extremal-$t$ PE)}    	& 0.37		& 0.40		& 0.03		& 0.13		& 0.07 	\\
j = 3 & \text{(Brown-Resnick)}  	 & 0.10		& 0.50		& 0.37		& 0.00	& 0.03	\\
j = 4 & \text{($t$ copula WM)} 	 	 & 0.03		& 0.03		& 0.10		& 0.38	& 0.45	\\
j = 5 & \text{($t$ copula PE)} 	 	 & 0.00		& 0.00		& 0.00		& 0.40	& 0.60	\\
\end{block}
\end{blockarray}.
\]

The inversion of $\hat{\bm{H}}$ failed sometimes. For a given data set, all models for which the inversion failed were disregarded. To assess the magnitude of the bias and the loss of information incurred by this approach, it is counted for how many models it was not possible to invert $\hat{\bm{H}}$ for each simulated data set. The table of counts across all data sets is given below:

\begin{center}
	
	\begin{tabular}{|l|cccccc|}
		\hline
		\# models where inversion failed & 0 & 1 & 2 & 3 & 4 & 5 \\
		\# simulated data sets & 95 & 29 & 11 & 13 & 1 & 1 \\
		\hline
	\end{tabular}
	
\end{center}

The aim of this section is to provide a rough comparison between the misclassification matrices of the ABC and the classical approach. We consider the quality of our CLIC-based estimate of the misclassification matrix to be sufficiently accurate for our purpose. Otherwise a more elaborate estimation technique for the CLIC would have to be employed.

\section{Marginal transformations to unit Fr\'echet}

\label{sec:frechet_transformations}

\begin{figure}[h]
	\centering
	\begin{subfigure}[t]{0.3\textwidth}
		\includegraphics[width=\textwidth]{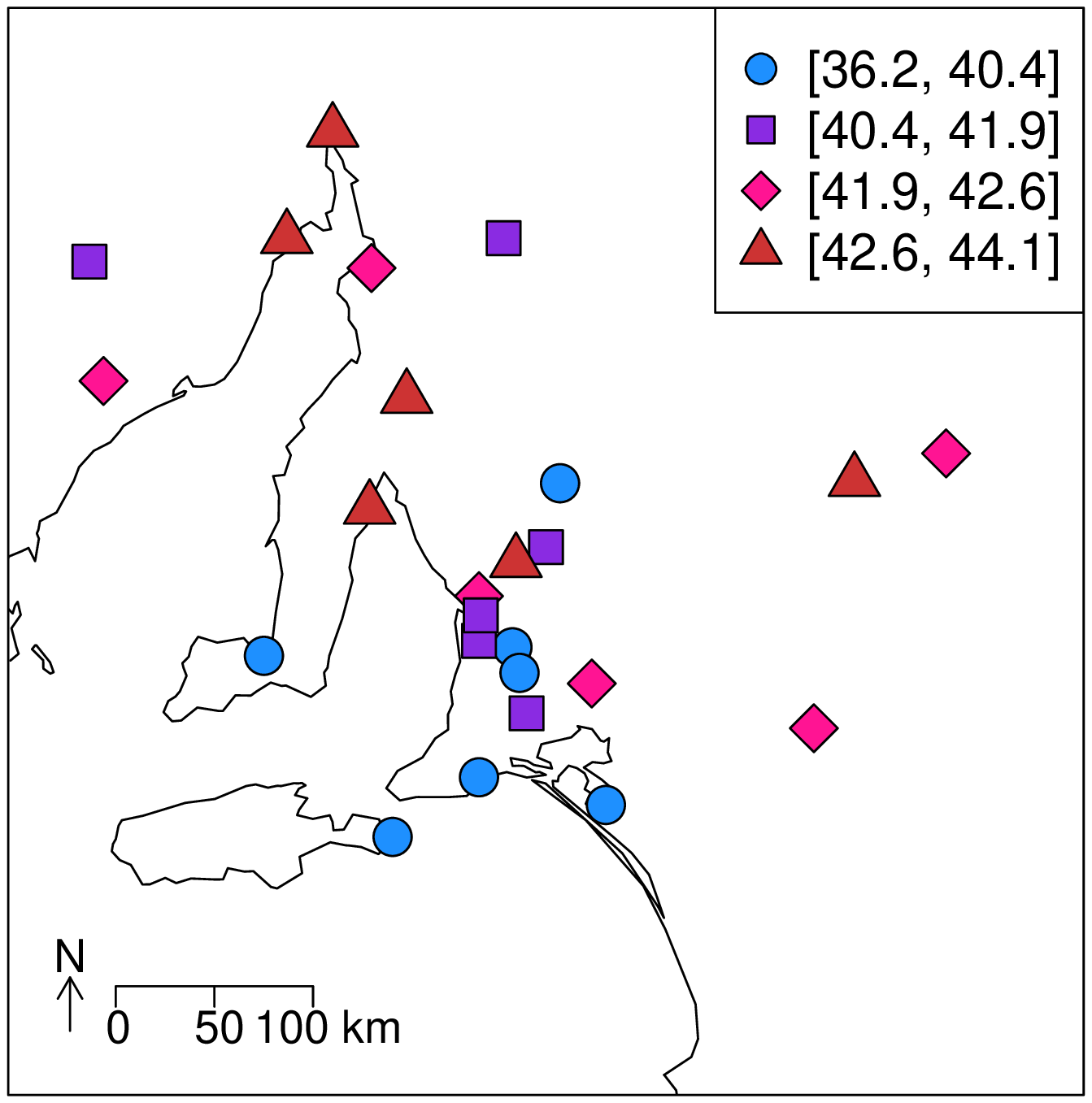}
		\caption{GEV location parameter}
		\label{figure:SAdata_quartiles_notitles_GEV_1}
	\end{subfigure}
	~
	\begin{subfigure}[t]{0.3\textwidth}
		\includegraphics[width=\textwidth]{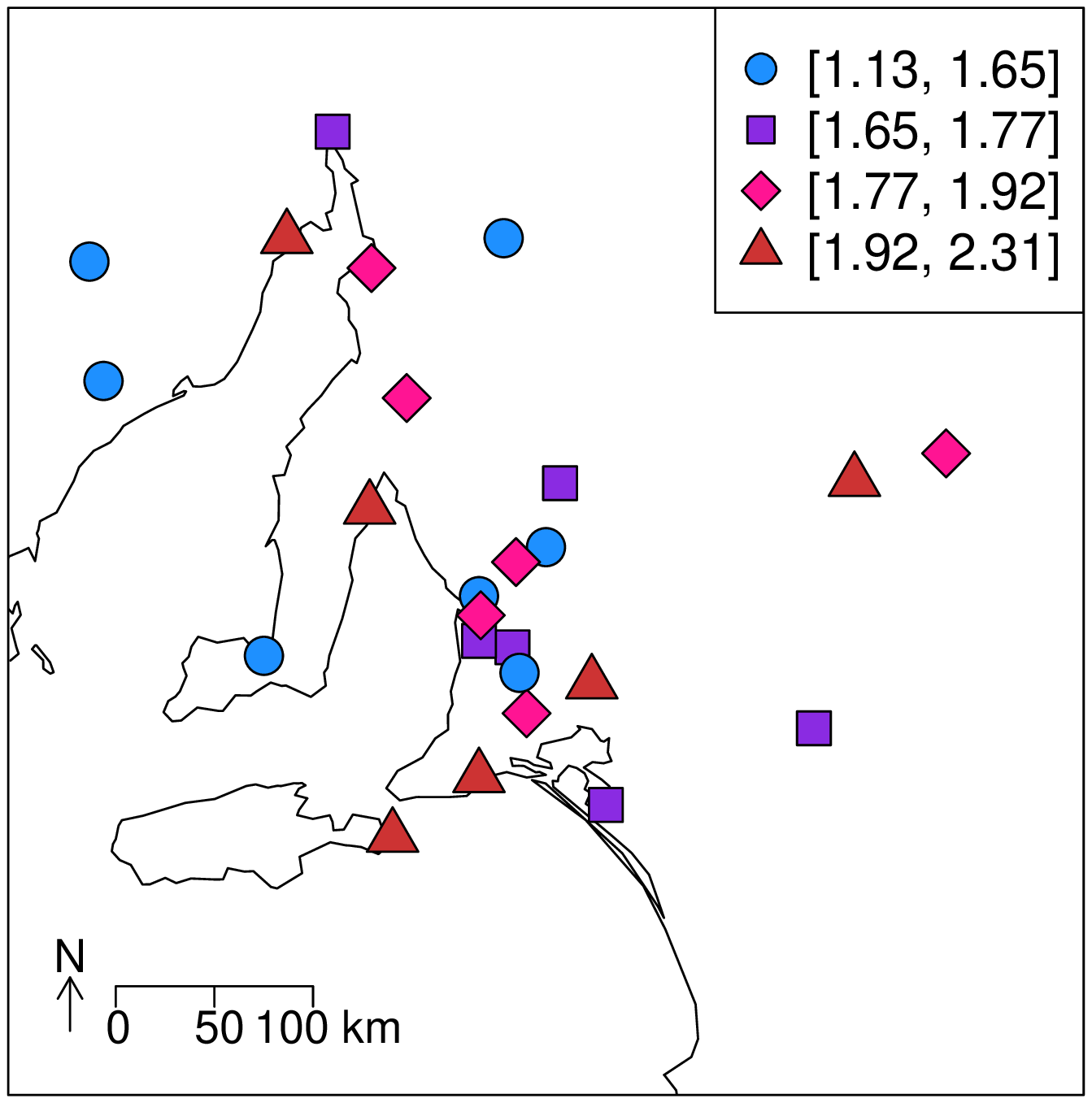}
		\caption{GEV scale parameter}
		\label{figure:SAdata_quartiles_notitles_GEV_2}
	\end{subfigure}
	~
	\centering
	\begin{subfigure}[t]{0.3\textwidth}
		\includegraphics[width=\textwidth]{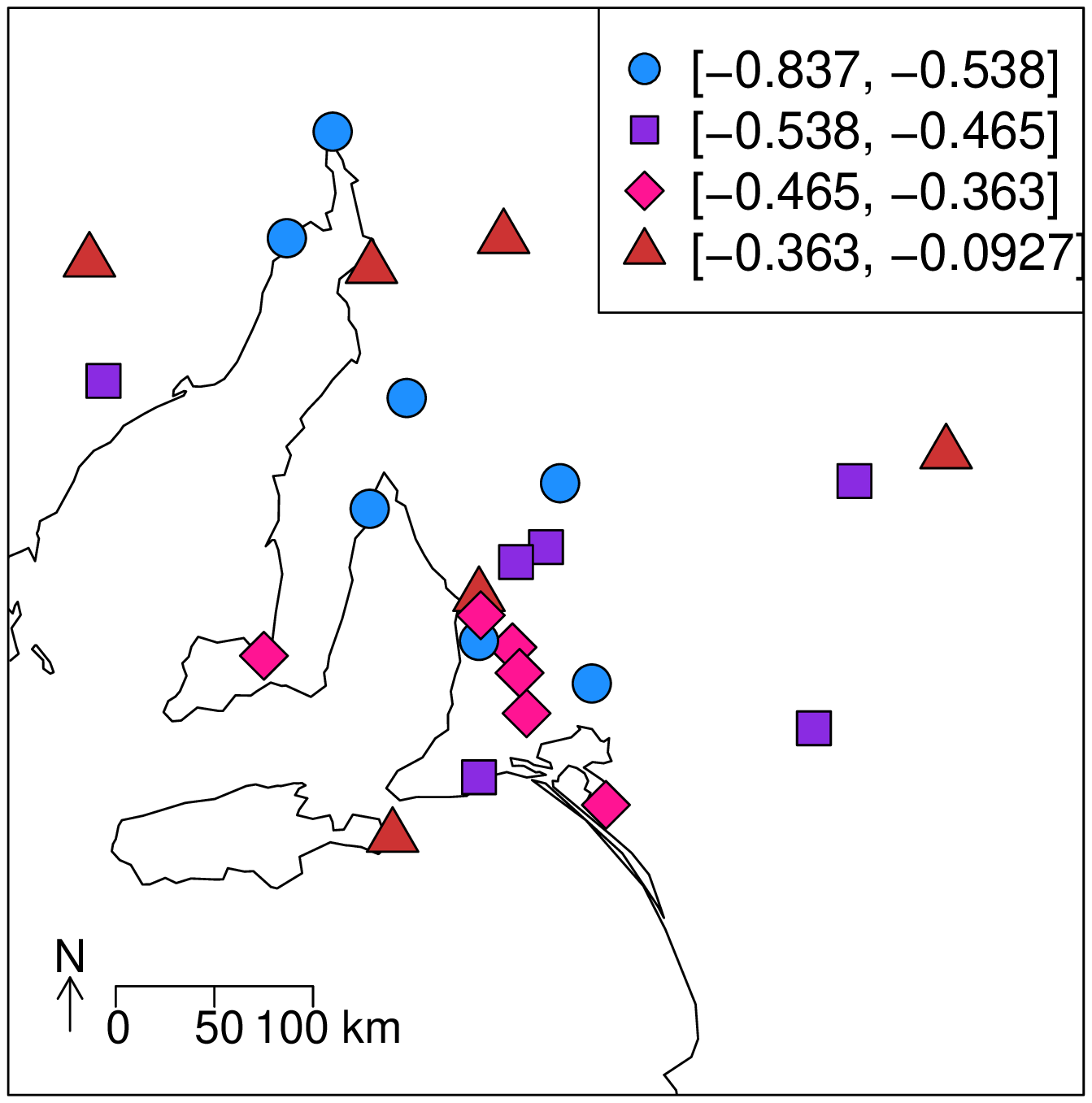}
		\caption{GEV shape parameter}
		\label{figure:SAdata_quartiles_notitles_GEV_3}
	\end{subfigure}
	\caption{Colour-coded quartile sets of the three GEV parameter estimates (location $\hat{\mu}$, scale $\hat{\sigma}$, and shape $\hat{\xi}$) for the $25$ land-based weather stations around South Australia where data were collected. The quartile set with the smallest values is denoted by the blue filled circles. Sets with progressively larger values are denoted by coloured purple squares, pink diamonds and red triangles, respectively. The ranges for the quartiles are provided in the legends.}
	\label{fig:GEV_parameters}
\end{figure}

\fref{fig:GEV_parameters} depicts the $25$ weather stations and the maximum likelihood estimates of the parameters of the univariate marginal generalised extreme value (GEV) distributions (location parameter $\mu$, scale parameter $\sigma$, shape parameter $\xi$). These estimates are used to transform the original data $Z^*_i(\xb_j) \sim GEV\{\mu_j,\sigma_j,\xi_j\}$, $i = 1,\ldots,n$, at each station $\xb_j$, $j = 1,\ldots,H$, to unit Fr\'echet-scaled $Z_i(\xb_j)$ via the transformation

\begin{equation*}
Z_i(\xb_j) = \left( 1 + \hat{\xi}_j \: \frac{Z^*_i(\xb_j) - \hat{\mu}_j}{\hat{\sigma}_j}\right)^{\frac{1}{\hat{\xi}_j}}.
\end{equation*}

\begin{figure}[h]
	\centering
	\includegraphics[width=0.2\textwidth]{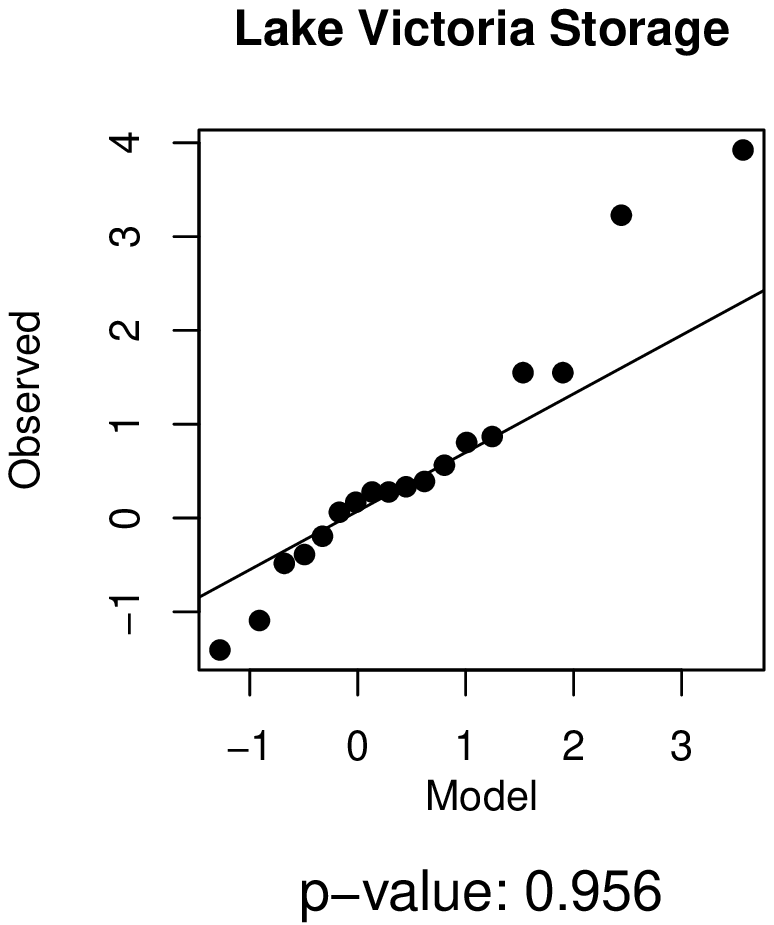}
	\includegraphics[width=0.2\textwidth]{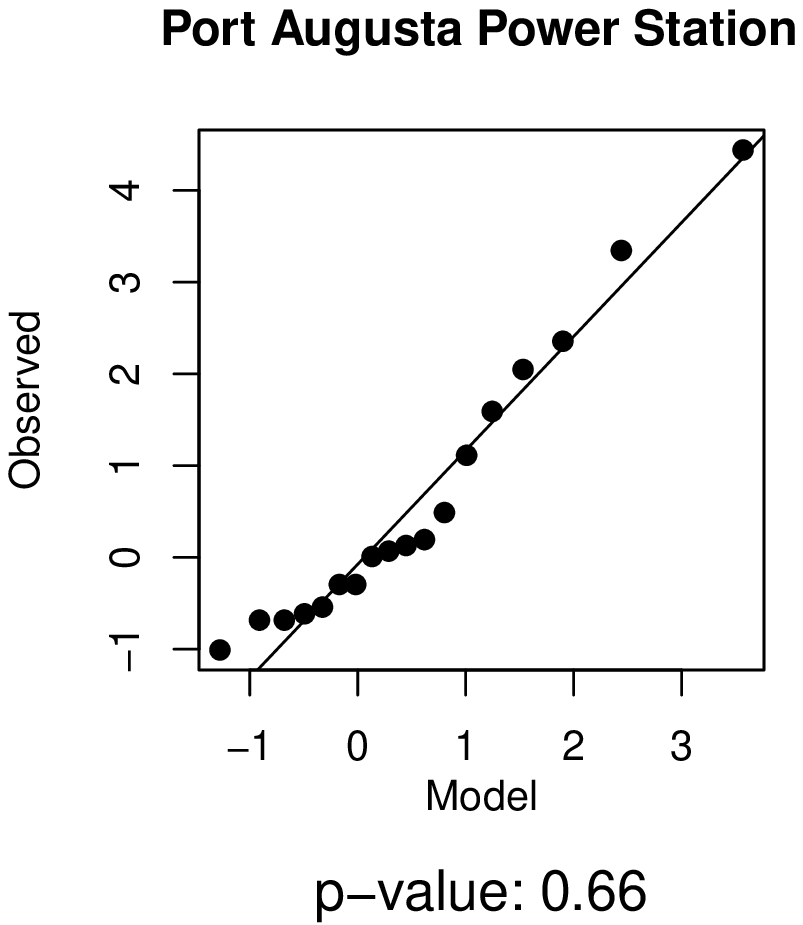} 
	\includegraphics[width=0.2\textwidth]{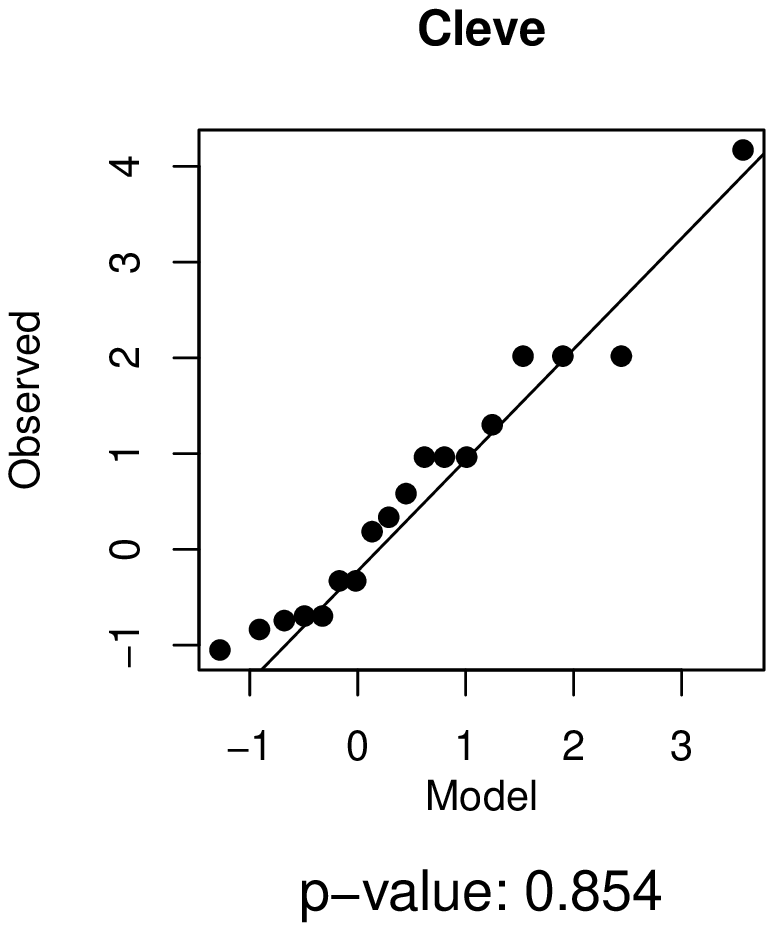} 
	\includegraphics[width=0.2\textwidth]{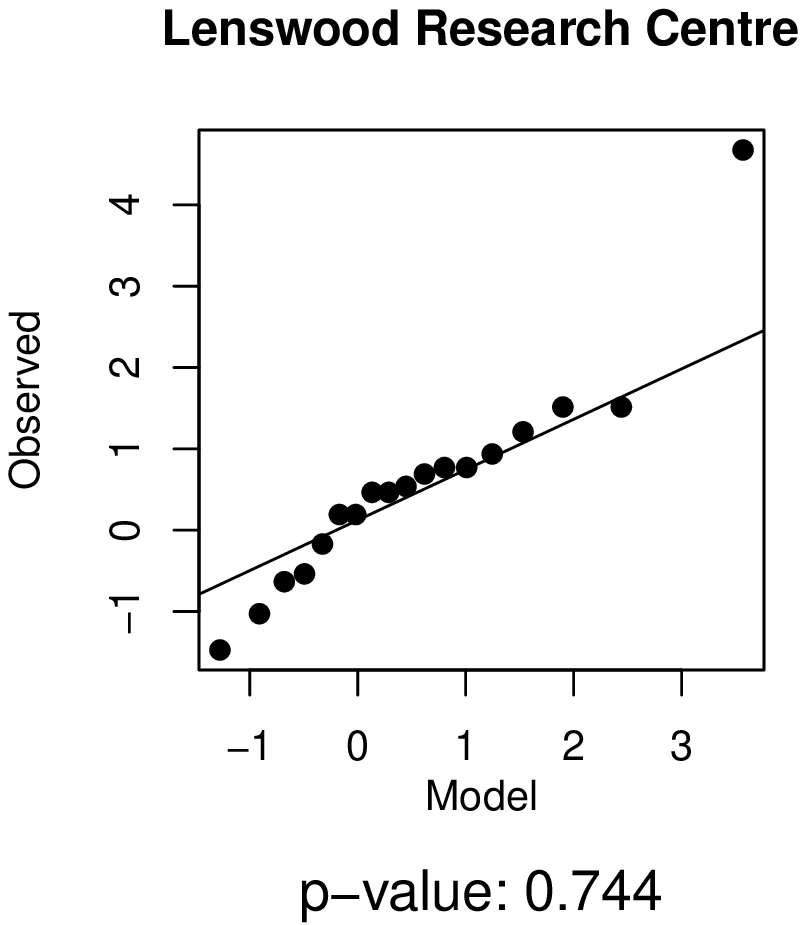} \\
	\includegraphics[width=0.2\textwidth]{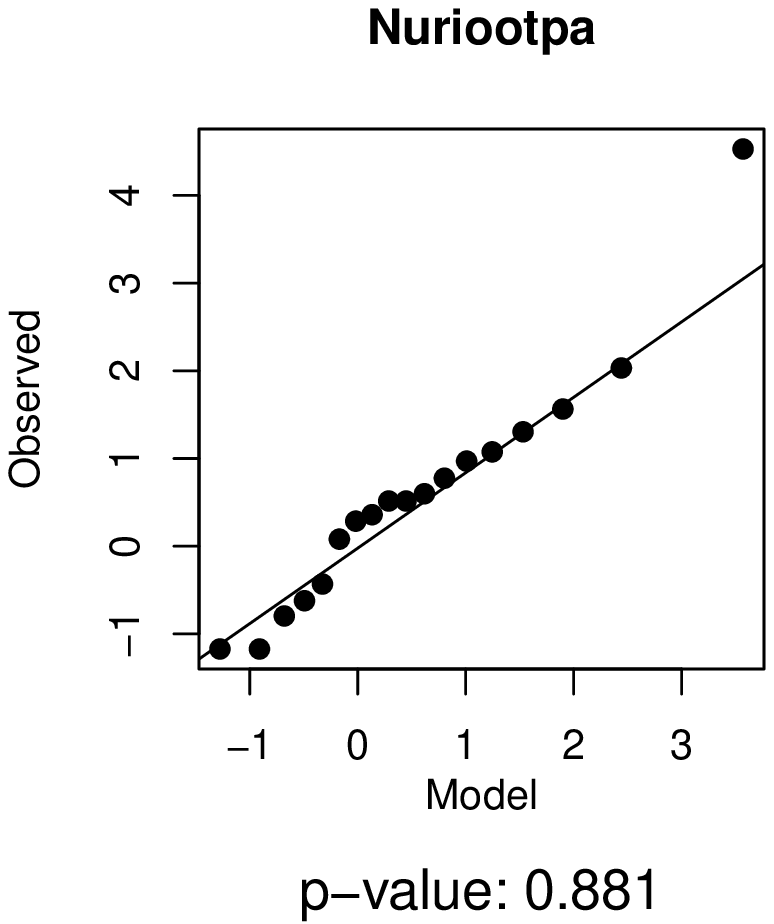} 
	\includegraphics[width=0.2\textwidth]{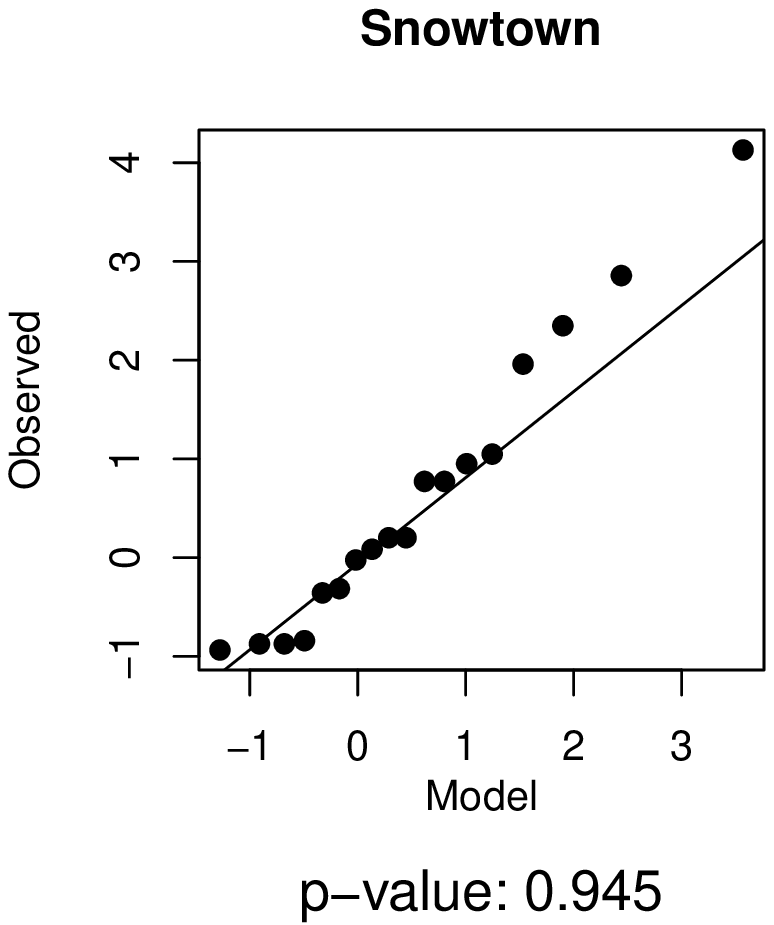} 
	\includegraphics[width=0.2\textwidth]{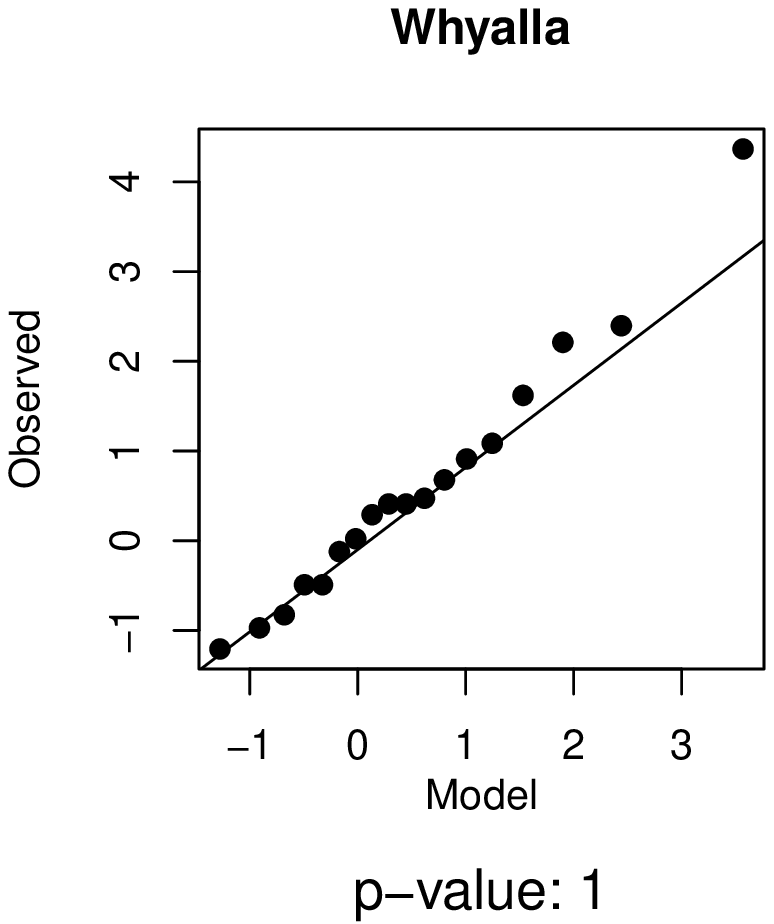}
	\includegraphics[width=0.2\textwidth]{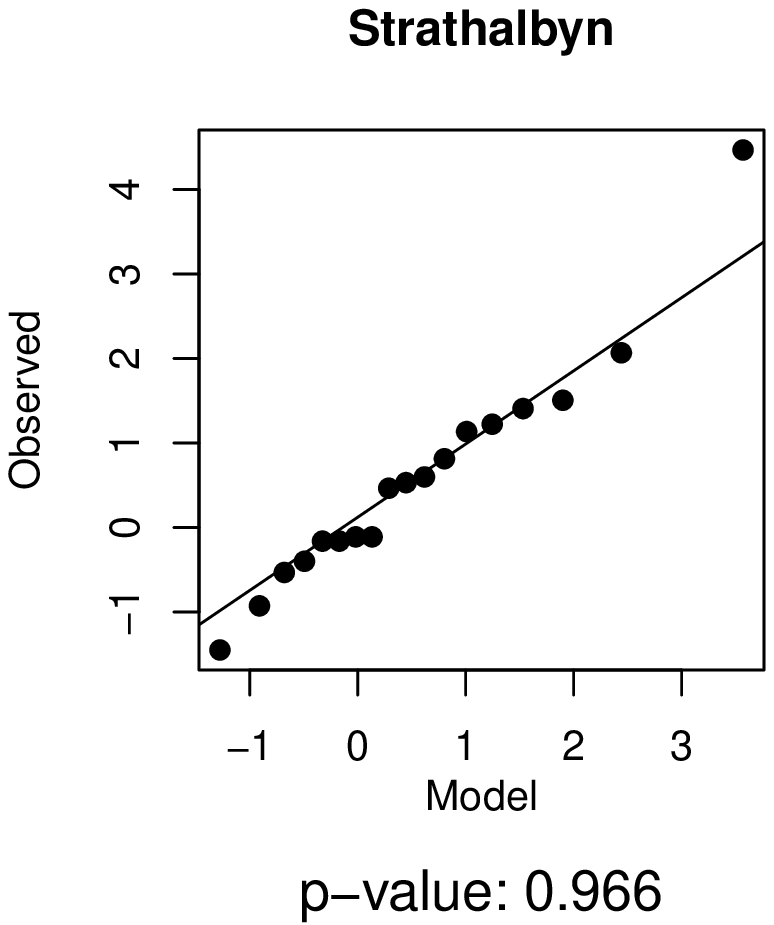} \\
	\includegraphics[width=0.2\textwidth]{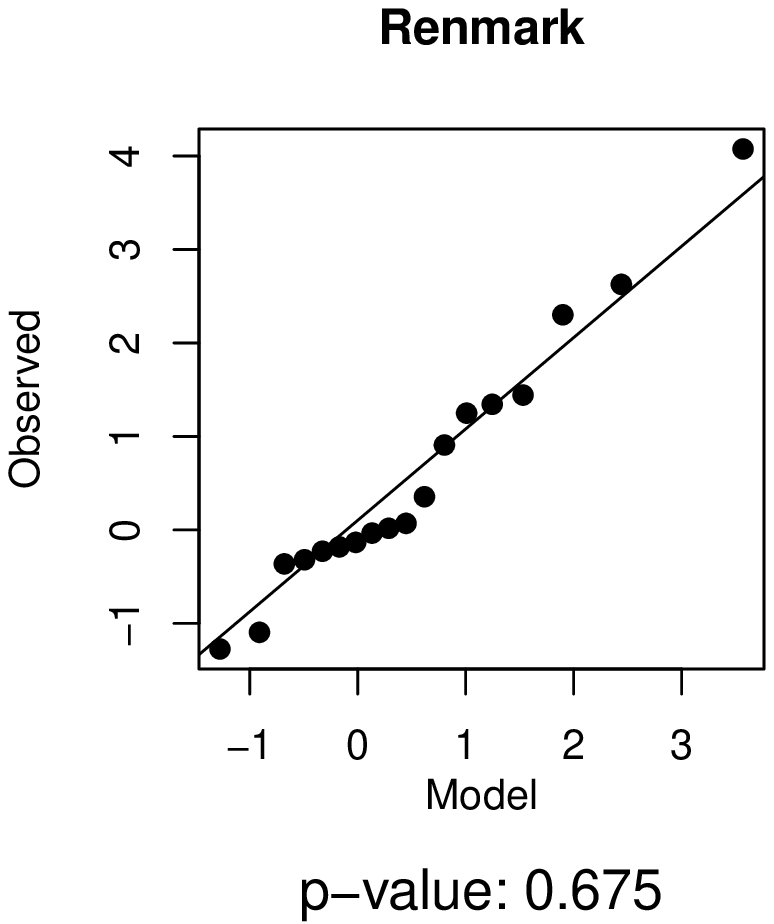} 
	\includegraphics[width=0.2\textwidth]{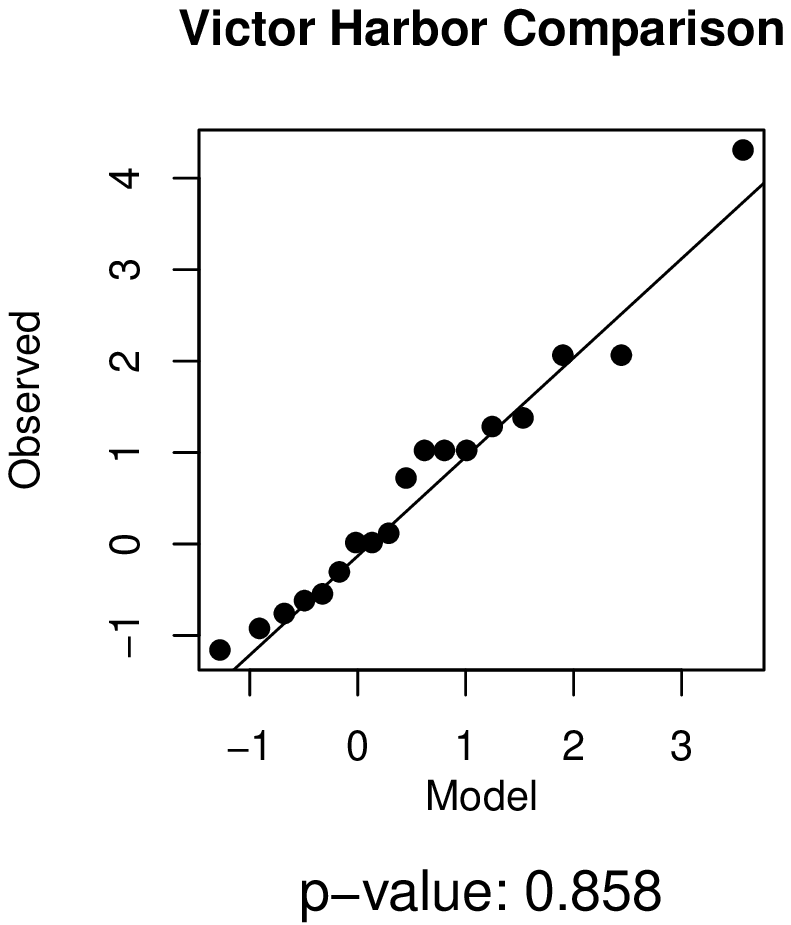}
	\includegraphics[width=0.2\textwidth]{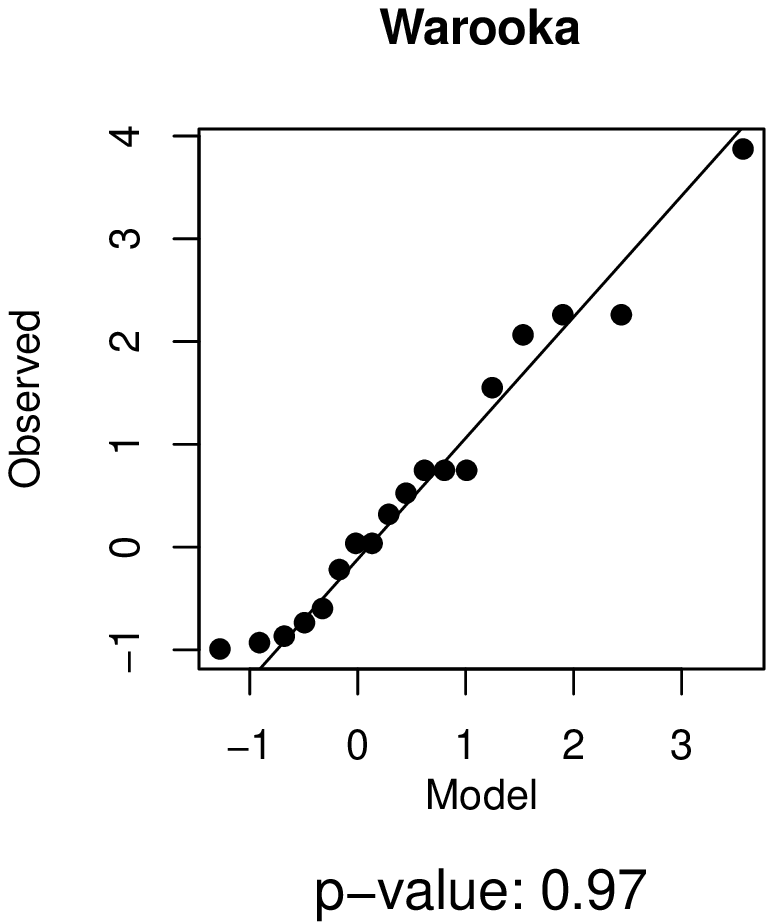} 
	\includegraphics[width=0.2\textwidth]{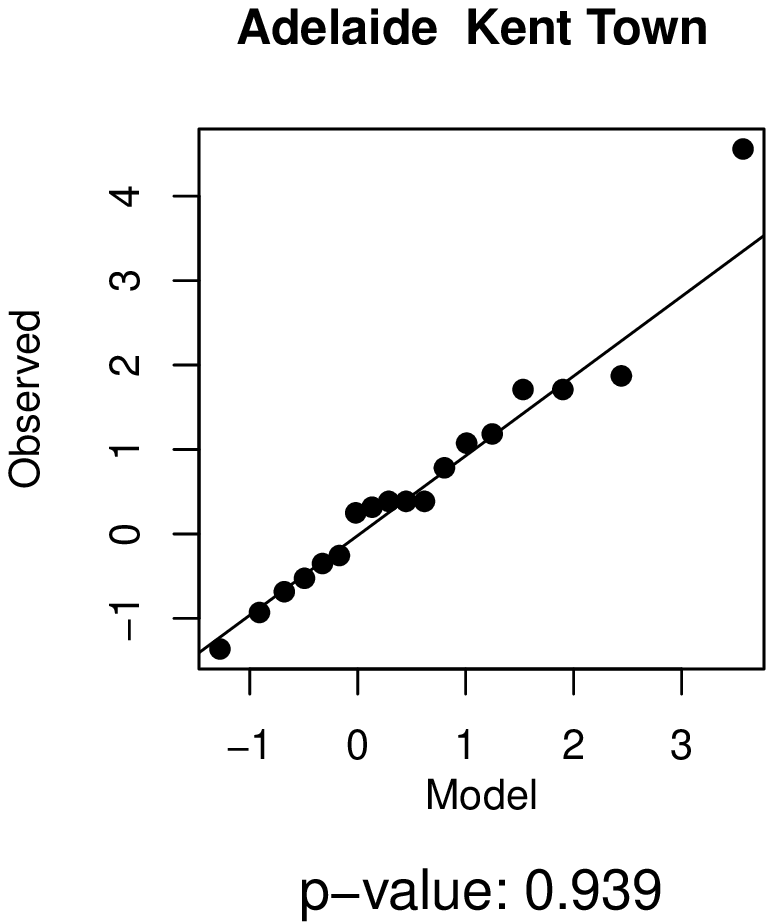} \\
	\caption{QQ plots of the marginal data transformed to Gumbel scale for all stations (part 1). The p-values of a two-sided one-sample Kolmogorov-Smirnov test are also reported.}
	\label{figure:marginalQQ1}
\end{figure}

\begin{figure}[h]
	\centering
	\includegraphics[width=0.2\textwidth]{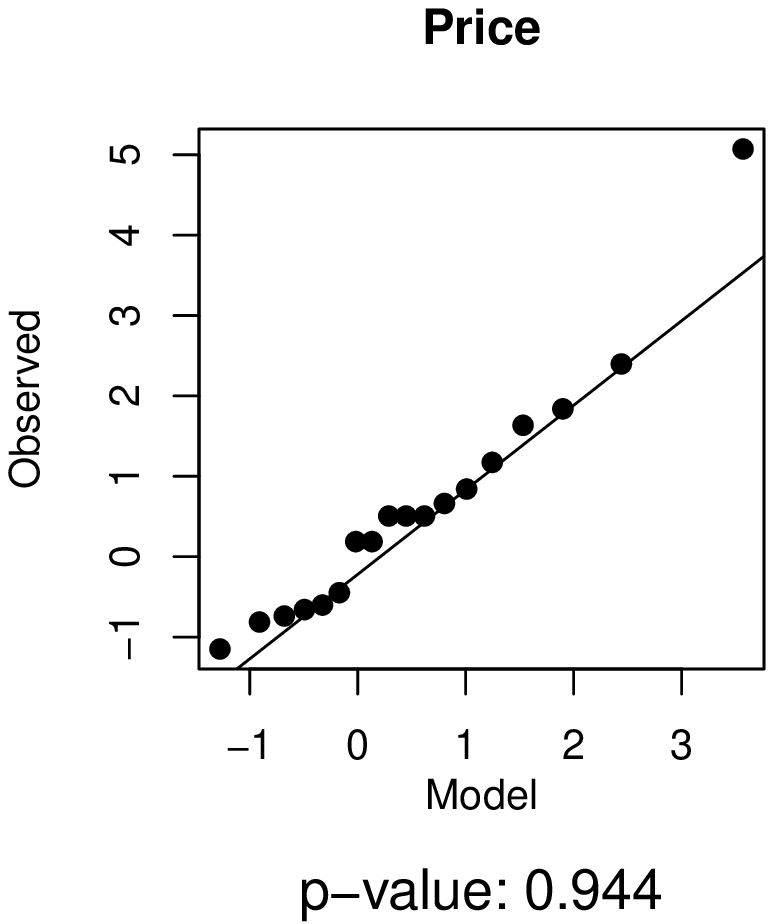} 
	\includegraphics[width=0.2\textwidth]{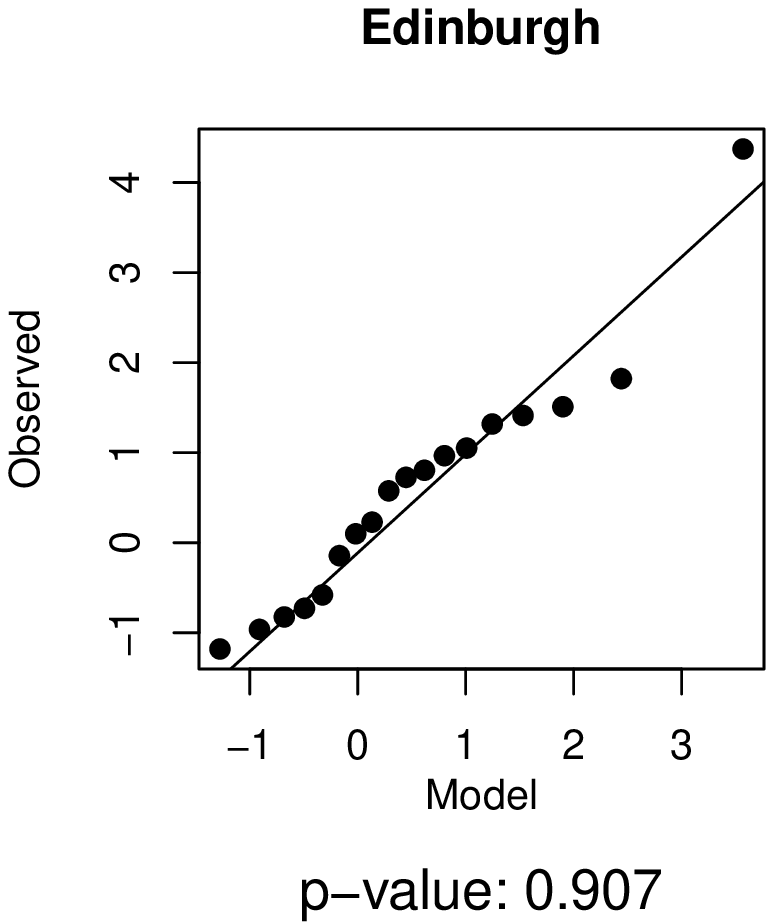}
	\includegraphics[width=0.2\textwidth]{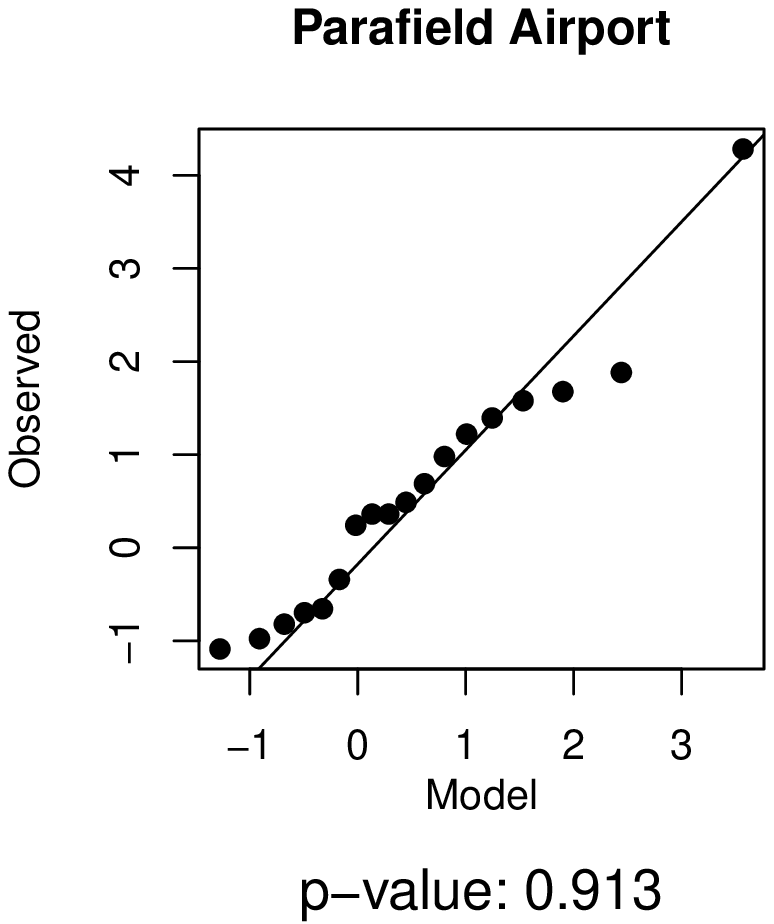} 
	\includegraphics[width=0.2\textwidth]{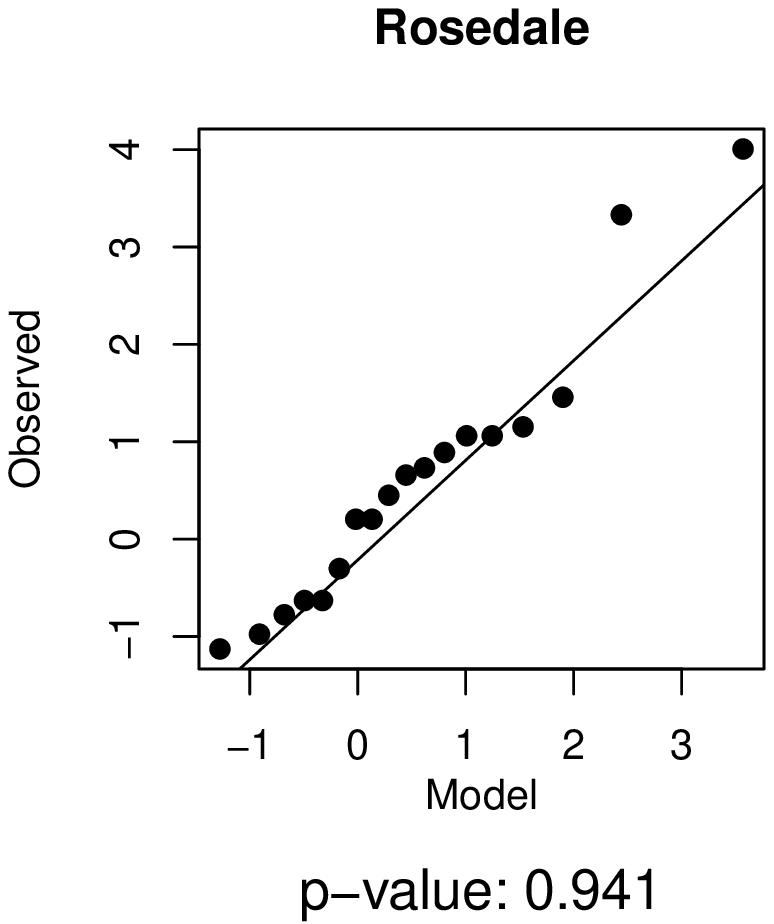} \\
	\includegraphics[width=0.2\textwidth]{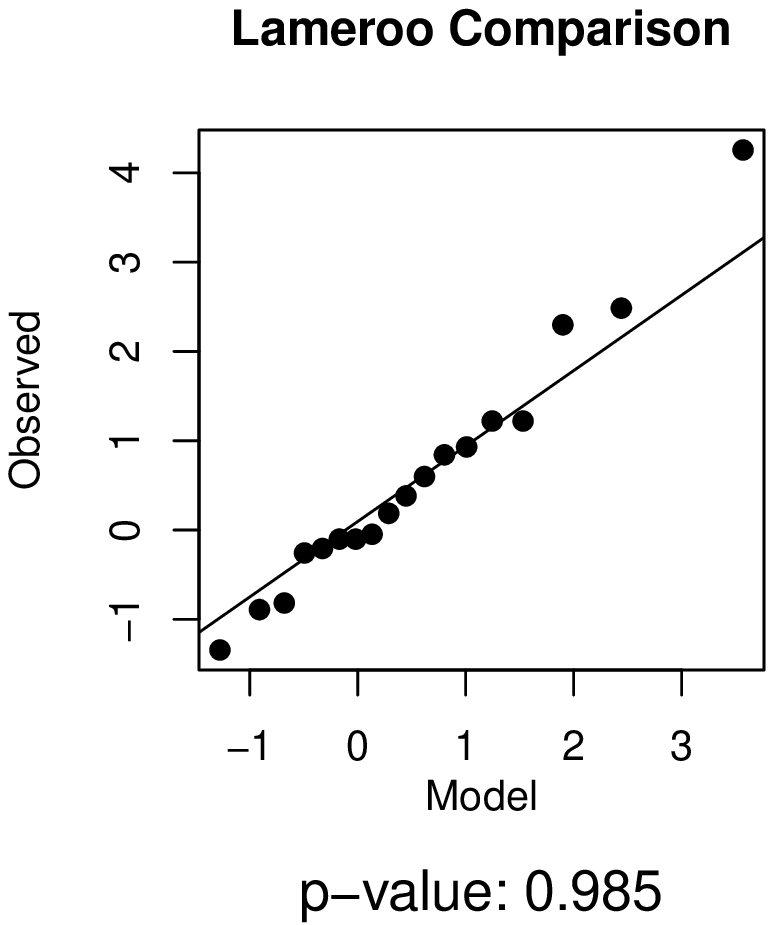}
	\includegraphics[width=0.2\textwidth]{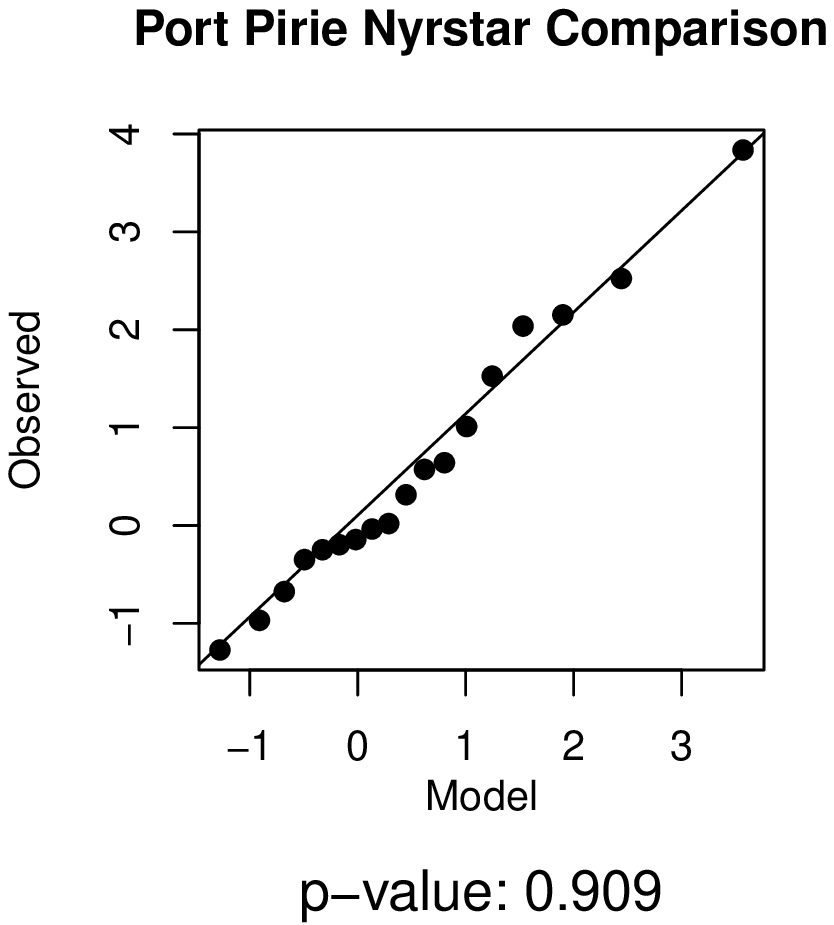} 
	\includegraphics[width=0.2\textwidth]{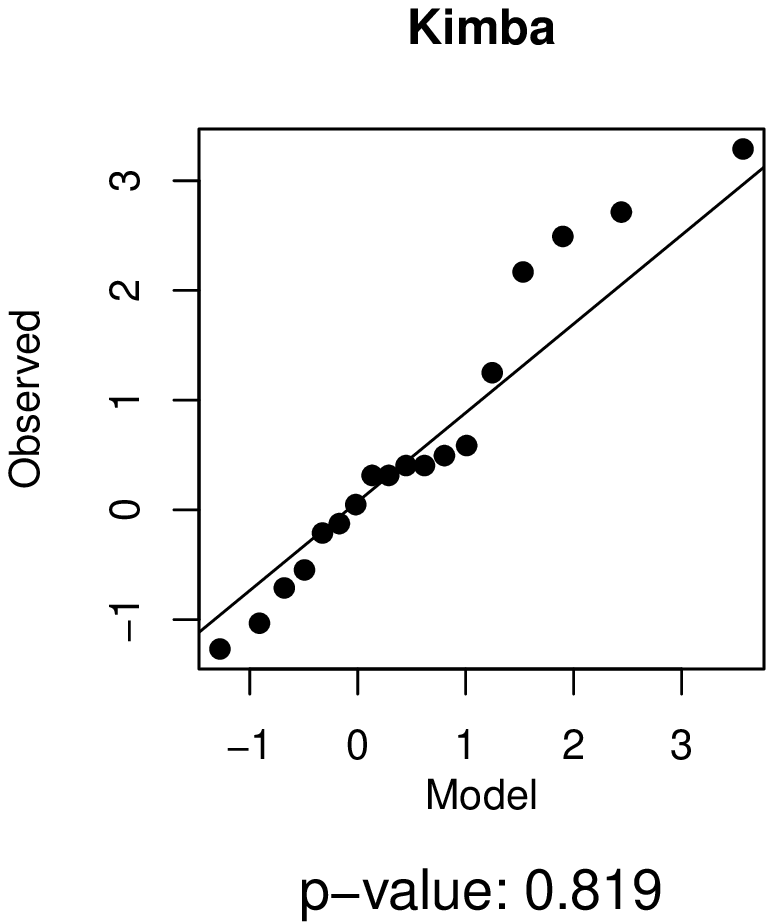} 
	\includegraphics[width=0.2\textwidth]{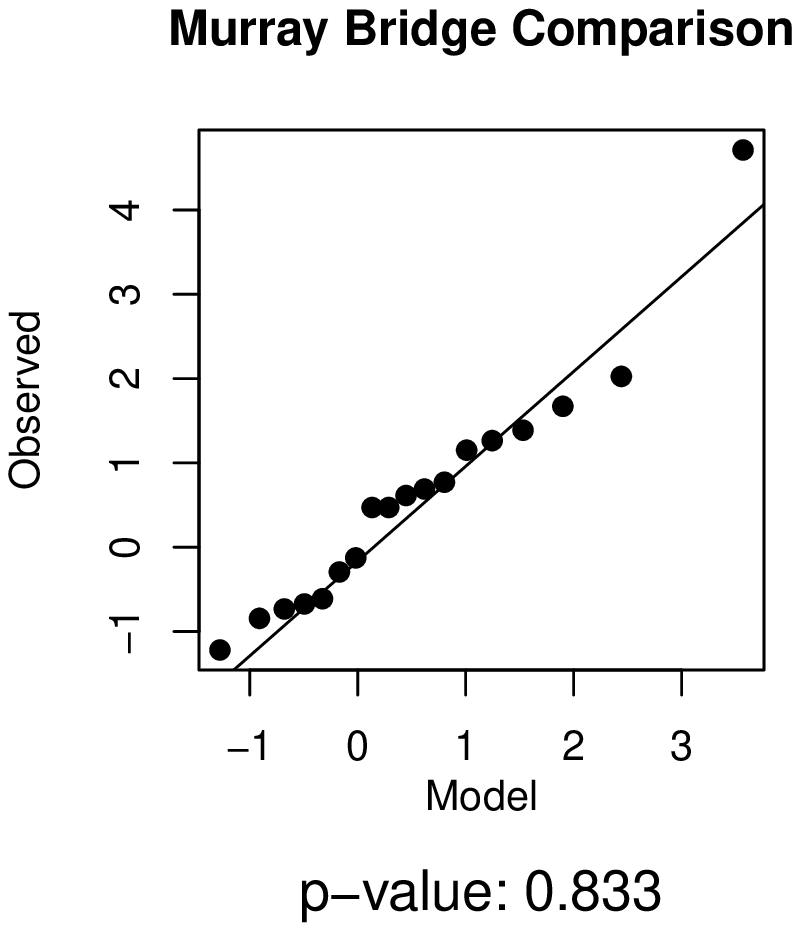} \\
	\includegraphics[width=0.2\textwidth]{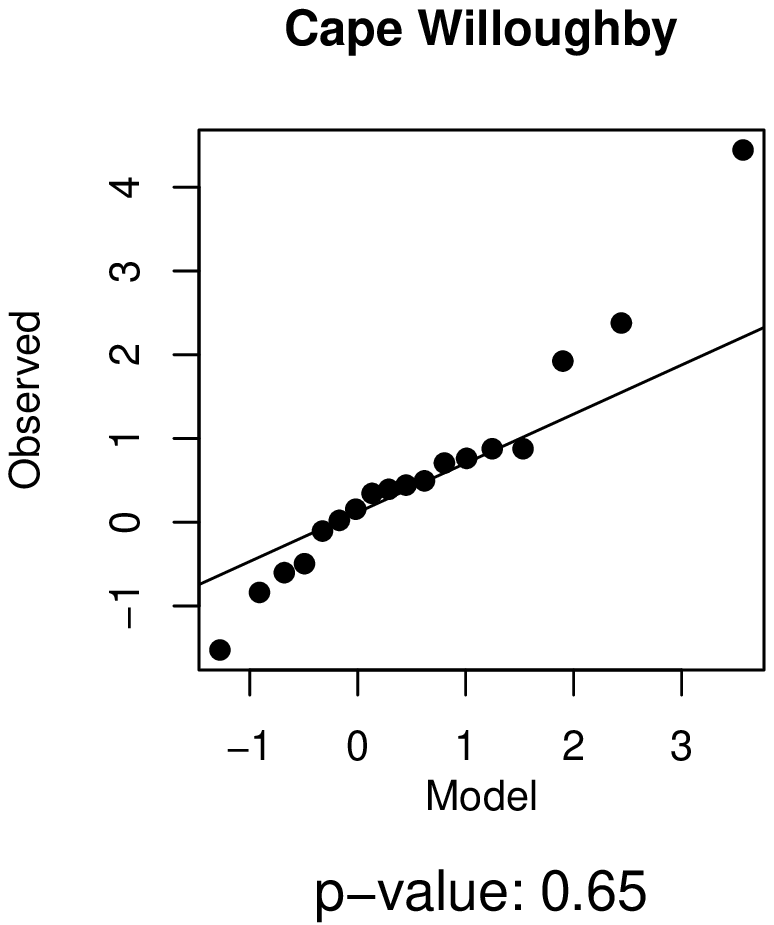} 
	\includegraphics[width=0.2\textwidth]{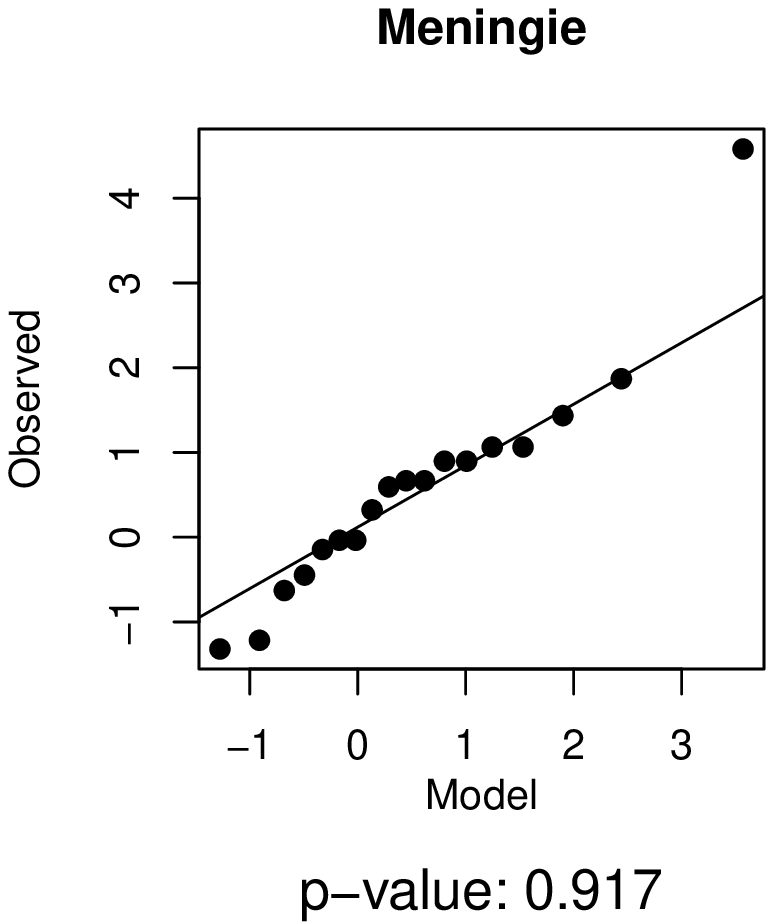} 
	\includegraphics[width=0.2\textwidth]{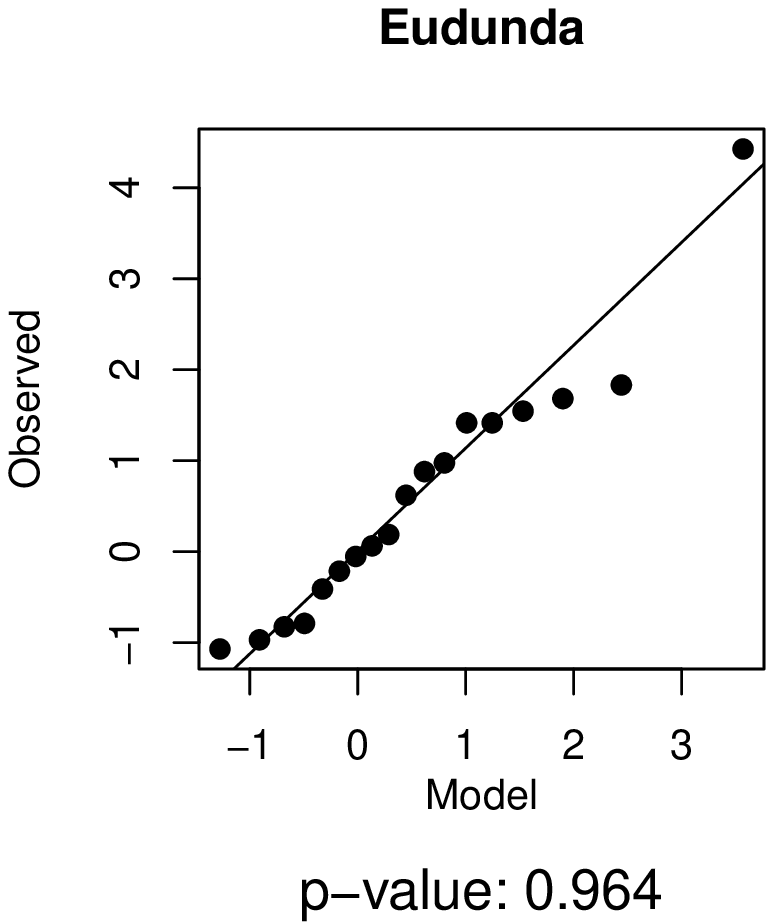}
	\includegraphics[width=0.2\textwidth]{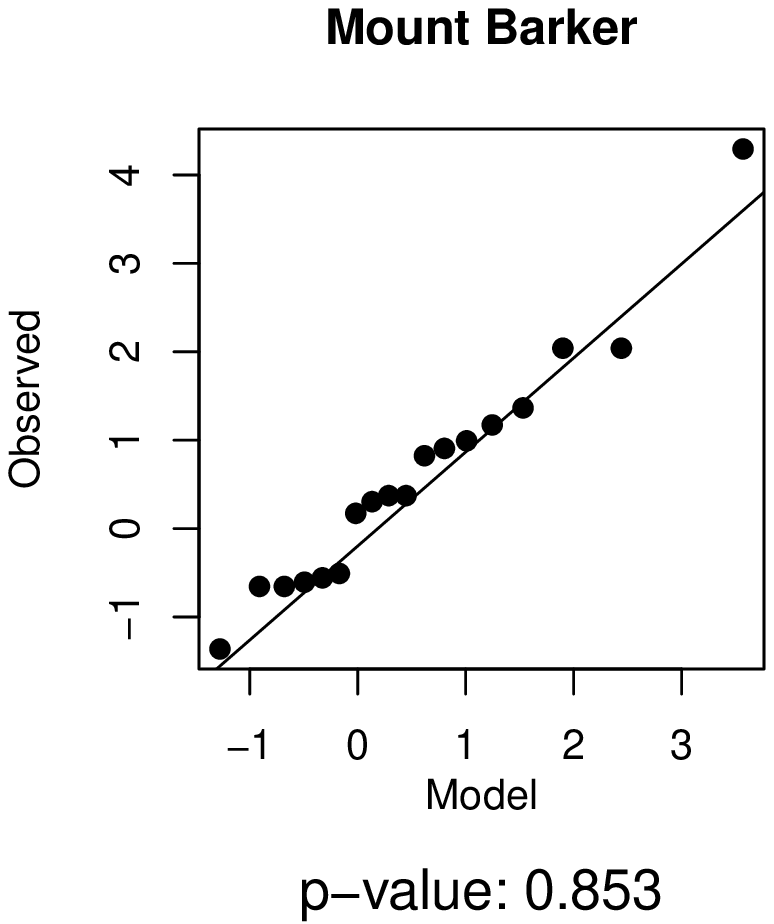} \\
	\includegraphics[width=0.2\textwidth]{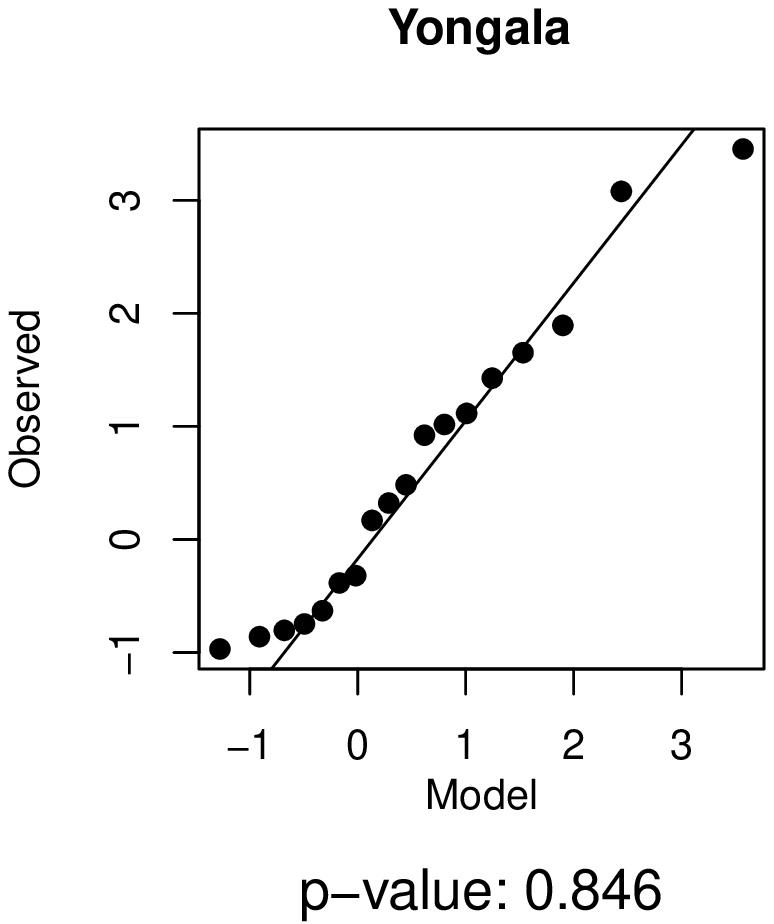} 
	\caption{QQ plots of the marginal data transformed to Gumbel scale for all stations (part 2). The p-values of a two-sided one-sample Kolmogorov-Smirnov test are also reported.}
	\label{figure:marginalQQ2}
\end{figure}

Figures~\ref{figure:marginalQQ1} and \ref{figure:marginalQQ2} show the QQ plots for the transformed marginal data. For the QQ plots, the data have been transformed further to the Gumbel scale by taking the logarithms of the unit Fr\'echet-scaled data in order to obtain more informative plots. The p-values of the two-sided one-sample Kolmogorov-Smirnov tests of the Gumbel-scaled marginal data are also reported. The null hypothesis of Gumbel-distributed data is not rejected at all stations.

\clearpage

\section{Average posterior model probability matrix of FP step computed from test data}

\label{sec:probability_matrix_test_data}

In Section~\ref{subsec:dataset_FPstep}, the FP step's matrix of average posterior model probabilities is created by applying the logistic regression estimates to the training particle set, which is the particle set used to estimate the logistic regression coefficients in the first place. This may lead to overfitting. In this section, we present the average posterior model probability matrix when a separate test particle set of size $M = 10,\!000$ is generated from the prior predictive distribution (with $M/5 = 2,\!000$ draws from each model) and the logistic regression estimates are applied to this test particle set. There are only minimal changes to the matrix reported in Section~\ref{subsec:dataset_FPstep}.

\[
\begin{blockarray}{llccccc}
& & \Pr(1|j) 	&  \Pr(2|j)		& \Pr(3|j) 	& \Pr(4|j) 	& \Pr(5|j)	\\
\begin{block}{ll[ccccc]}
j = 1 & \text{(extremal-$t$ WM)}		& 0.35		& 0.34		& 0.16		& 0.09		& 0.06 \\
j = 2 & \text{(extremal-$t$ PE)}    	& 0.33		& 0.38		& 0.17		& 0.06		& 0.06 	\\
j = 3 & \text{(Brown-Resnick)}  	 & 0.16		& 0.15		& 0.55		& 0.07	& 0.06	\\
j = 4 & \text{($t$ copula WM)} 	 	 & 0.09		& 0.06		& 0.06		& 0.41	& 0.37	\\
j = 5 & \text{($t$ copula PE)} 	 	 & 0.06		& 0.06		& 0.06		& 0.37	& 0.45	\\
\end{block}
\end{blockarray}
\]

\section{Prior predictive checks}

\label{sec:prior_predictive_checks}

\begin{figure}[h]
	\centering
	\begin{tabular}{cc}
		F-madogram & Kendall's $\tau$ \\
		\includegraphics[width=0.4\textwidth]{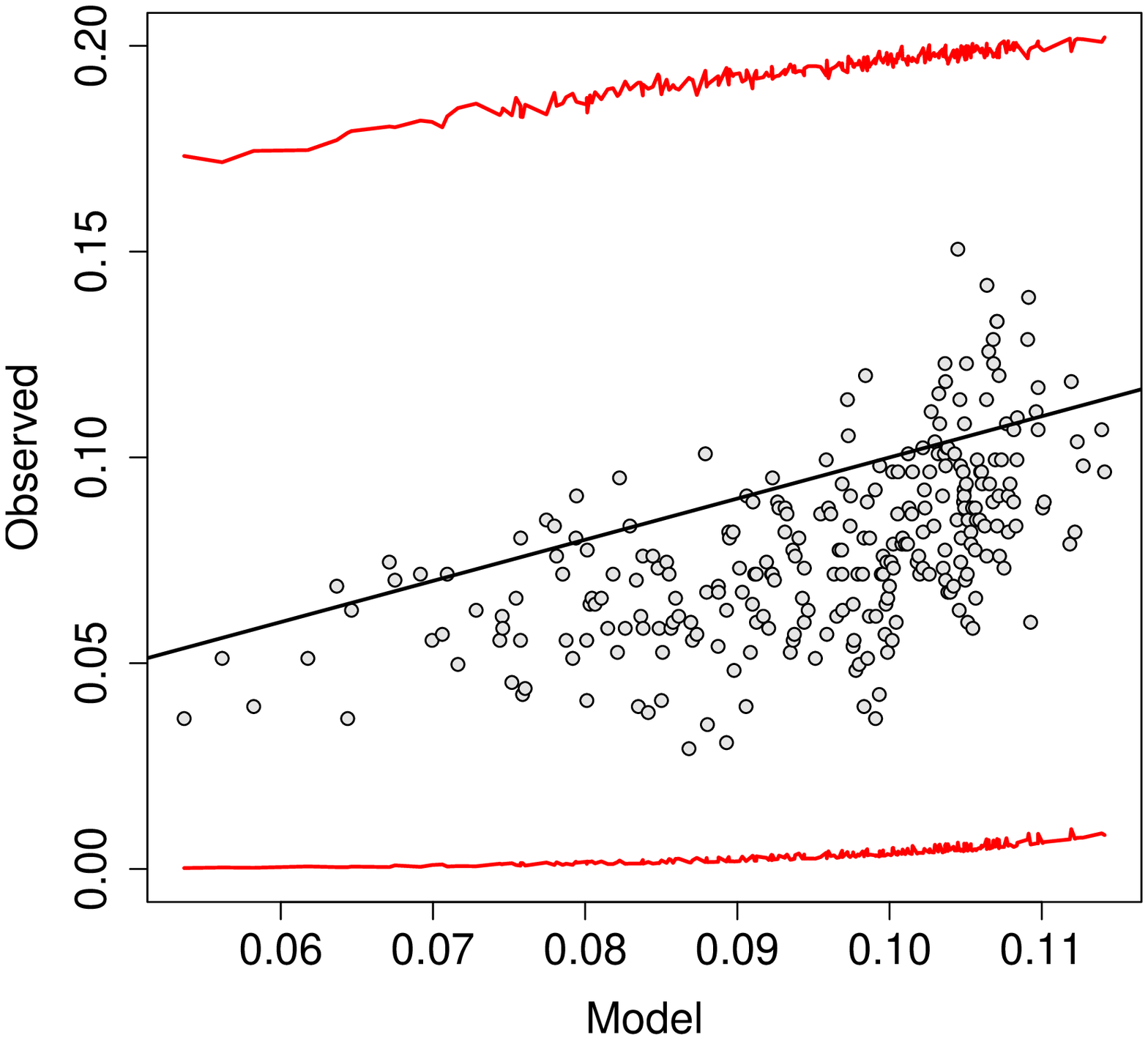} &
		\includegraphics[width=0.4\textwidth]{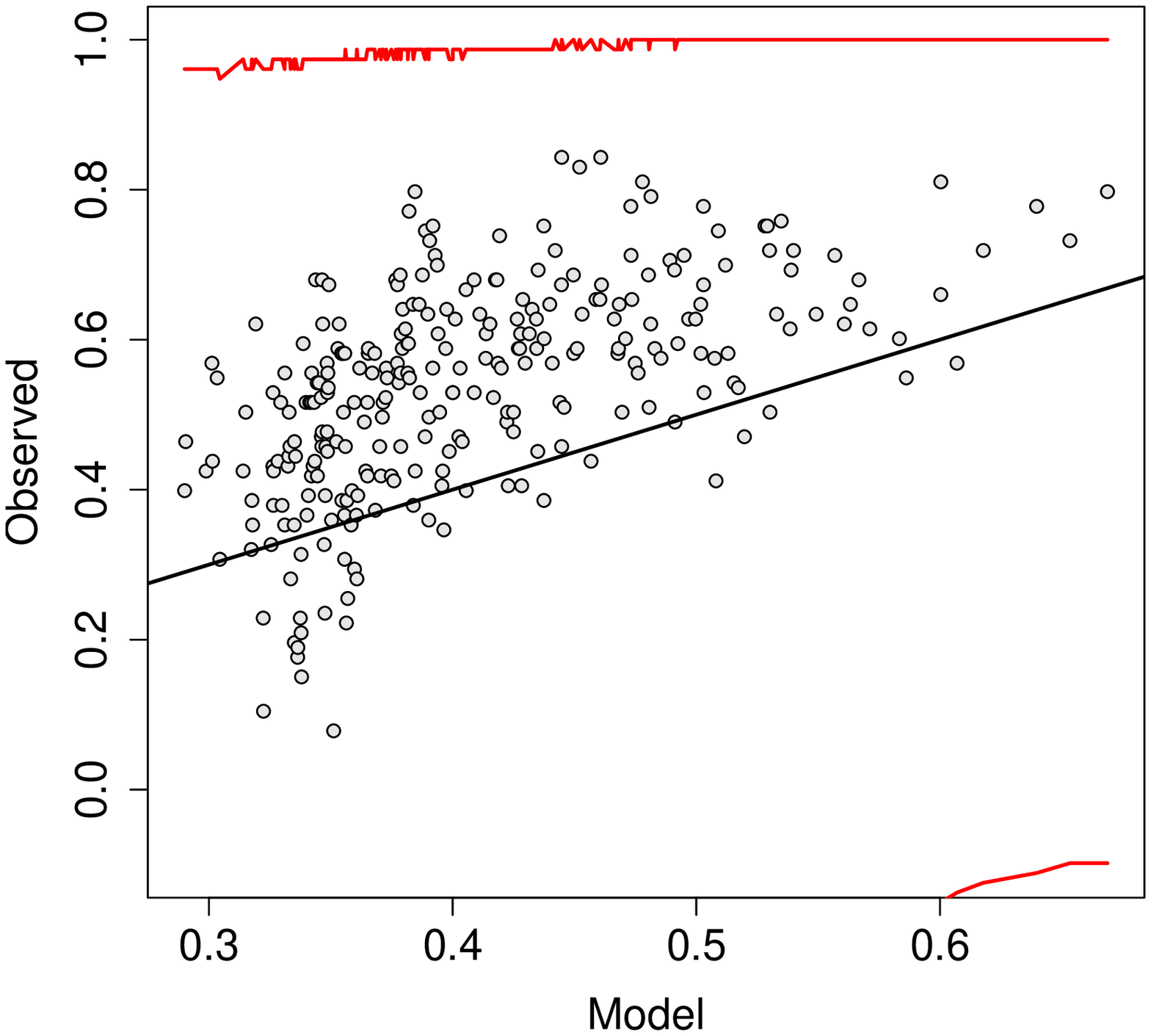} 
	\end{tabular}
	\caption{Observed pairwise \emph{F-madogram} (left) and \emph{Kendall's $\tau$} (right) estimates vs.\ prior predictive means of these estimates for each pair of locations (location 'Warooka' is excluded). The red lines connect the $2.5 \%$ and $97.5 \%$ quantiles of the prior predictive distributions for the location pairs.}
	\label{figure:SAdata_prior_predictive}
\end{figure}

To gauge whether the models and prior choices are appropriate for the South Australian data (when excluding Warooka), we compare the observed F-madogram and Kendall's $\tau$ estimates for all pairs of locations with their prior predictive distributions. \fref{figure:SAdata_prior_predictive} is equivalent to Figure~\ref{figure:SAdata_posterior_predictive_2}, replacing draws from the posterior predictive distribution with draws from the prior predictive distribution. One draw from the prior predictive distribution is generated by choosing one of the five models with equal probability, then simulating one draw from each of the parameter priors of the chosen model (for the prior specifications see Section~\ref{sec:SAmethod}), and then simulating one data set from the chosen model given the parameter draws.

\fref{figure:SAdata_prior_predictive} shows the observed F-madogram (left) and Kendall's $\tau$ (right) estimates for all location pairs vs.~their prior predictive means and the $95 \%$ prior predictive probability intervals. The prior predictive distribution serves its purpose as a preliminary model for the observed data. The $95 \%$ prior probability intervals are very wide and cover all the observed values, so the prior predictive distribution is sufficiently general and not too informative. Compared to the true data, the prior predictive distribution more strongly favours dependence indicators signalling low dependency.

\clearpage

\section{Posterior distributions}

\label{sec:posterior_distributions}

\begin{figure}[h]
	\centering
	\includegraphics[width=0.4\textwidth]{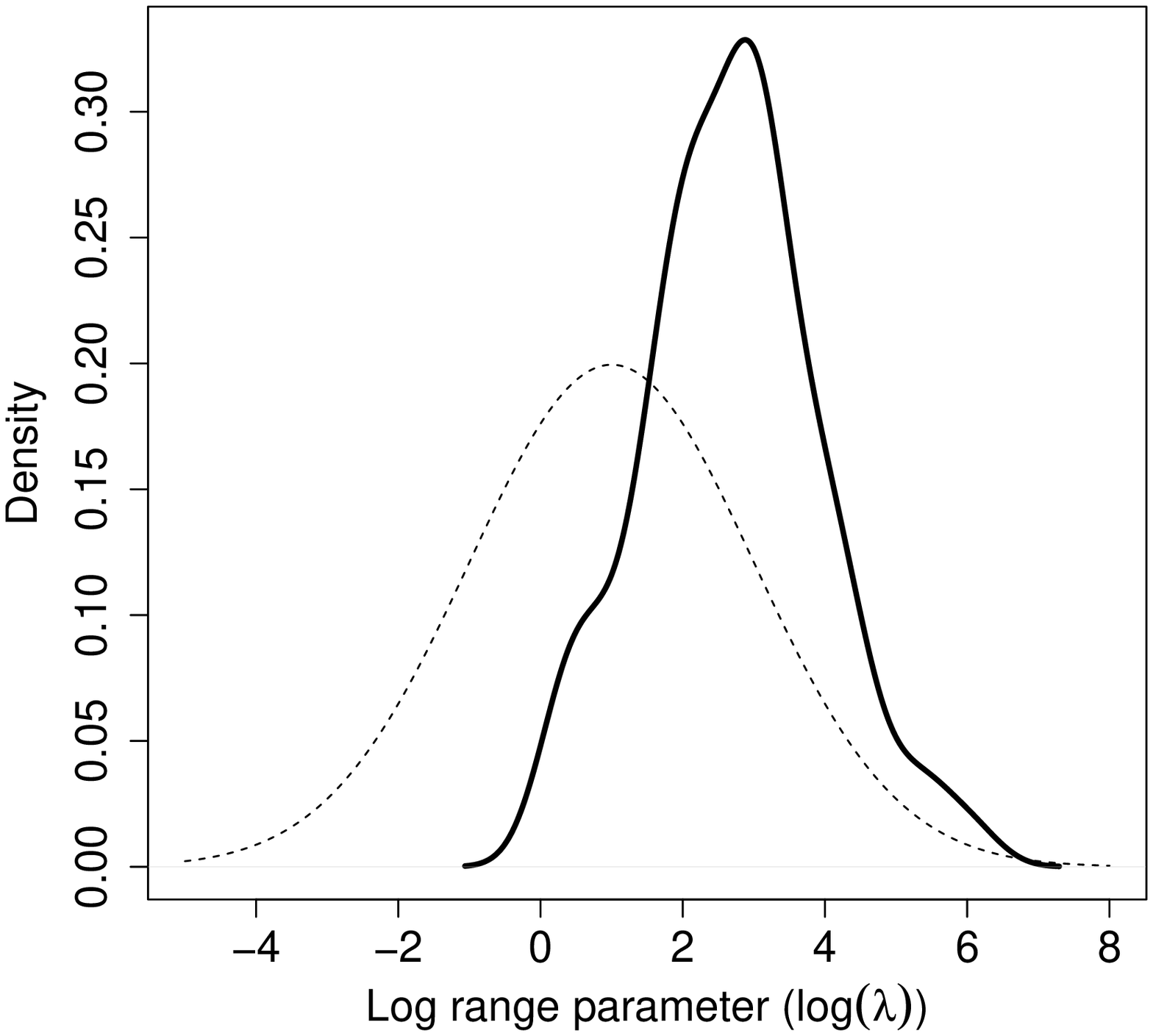} 
	\includegraphics[width=0.4\textwidth]{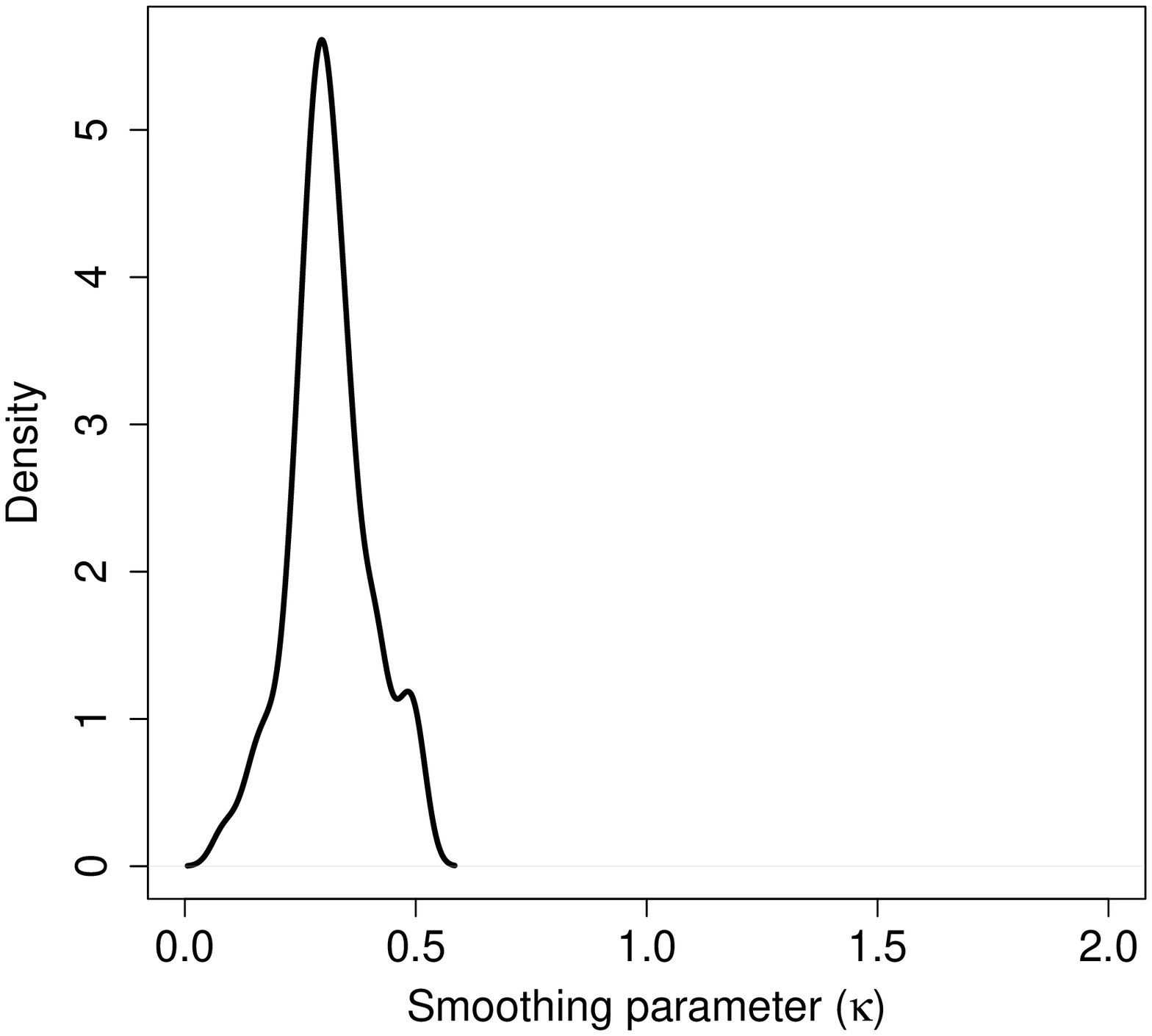} \\
	\includegraphics[width=0.4\textwidth]{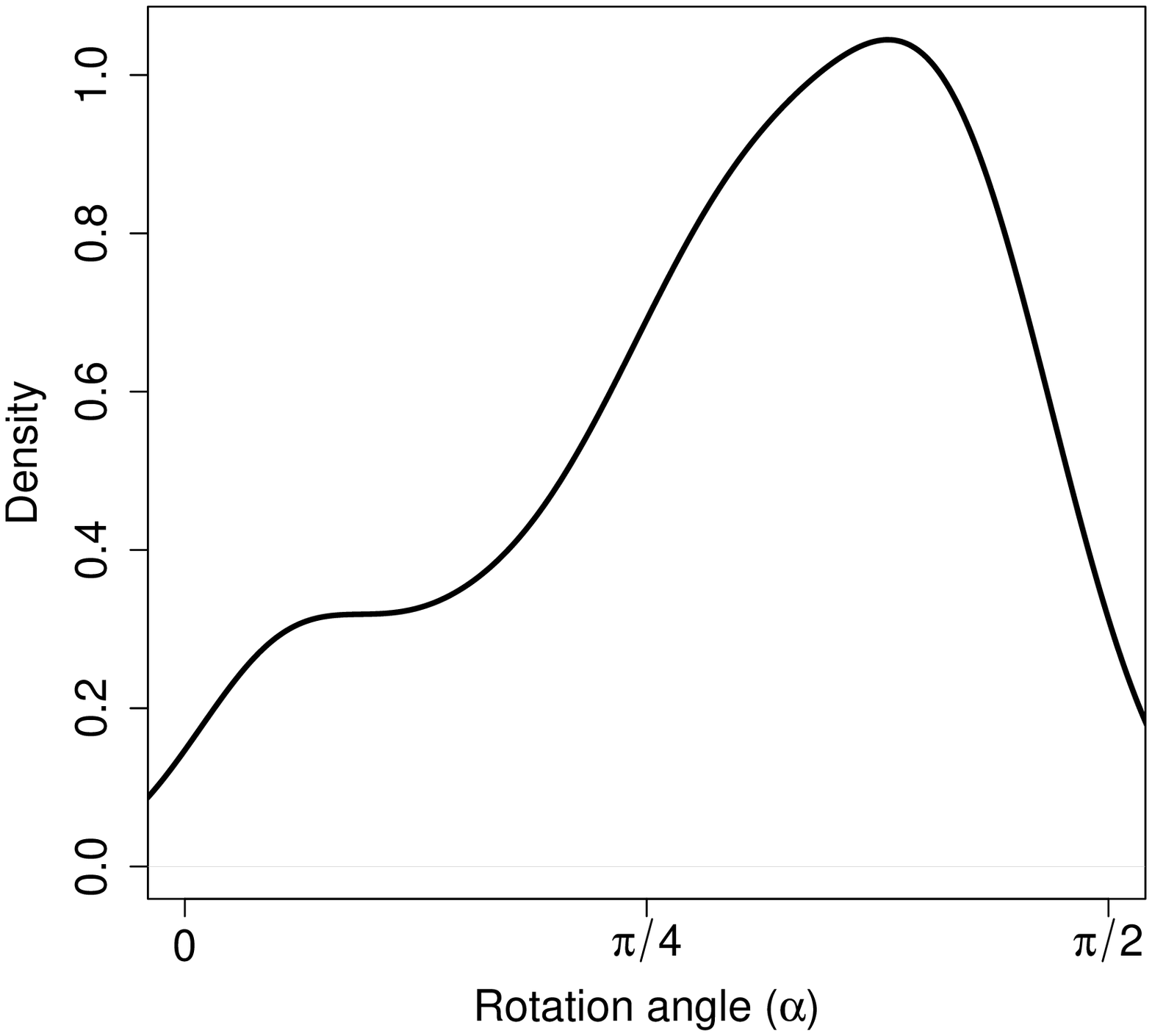}
	\includegraphics[width=0.4\textwidth]{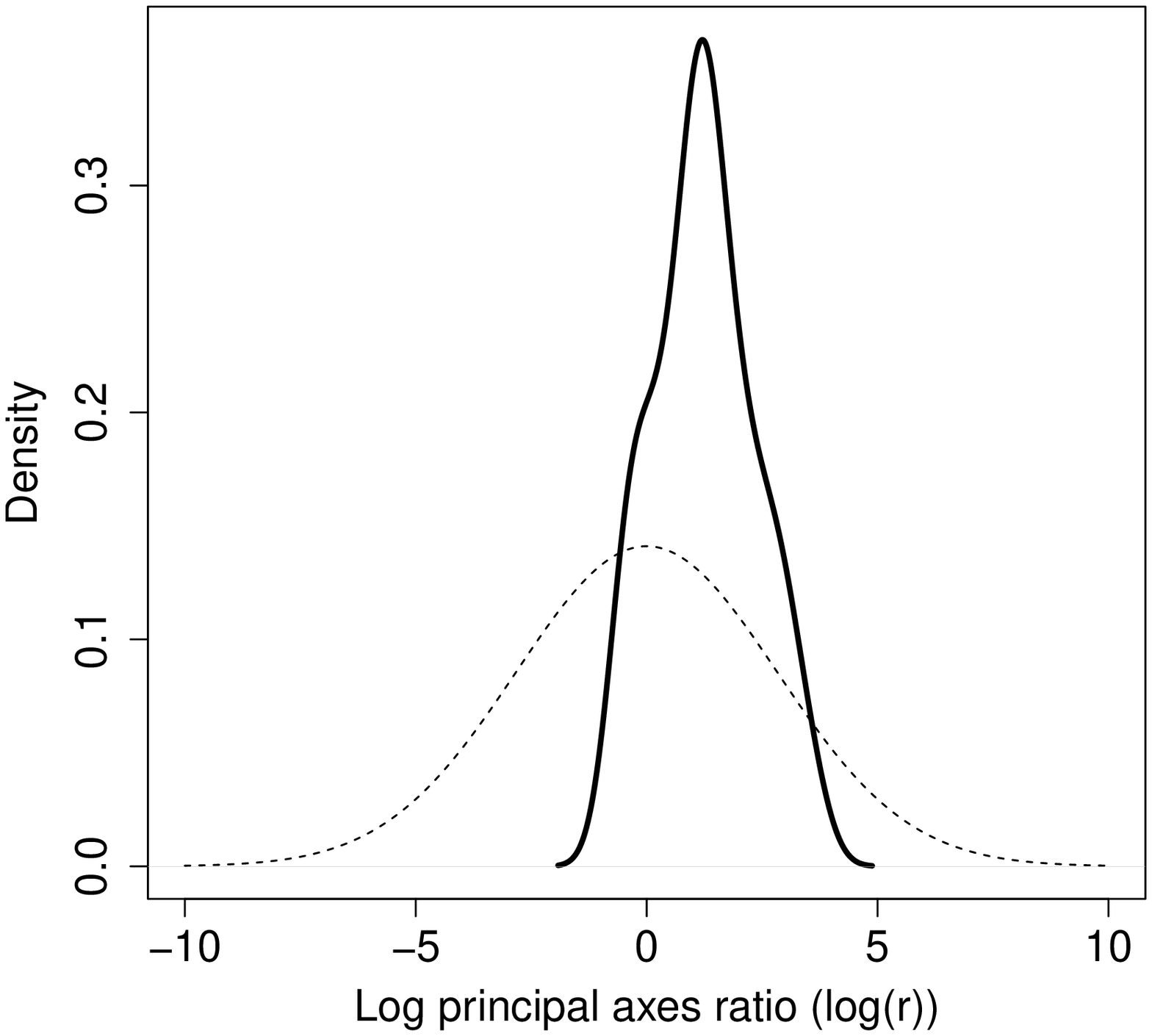} \\
	\includegraphics[width=0.4\textwidth]{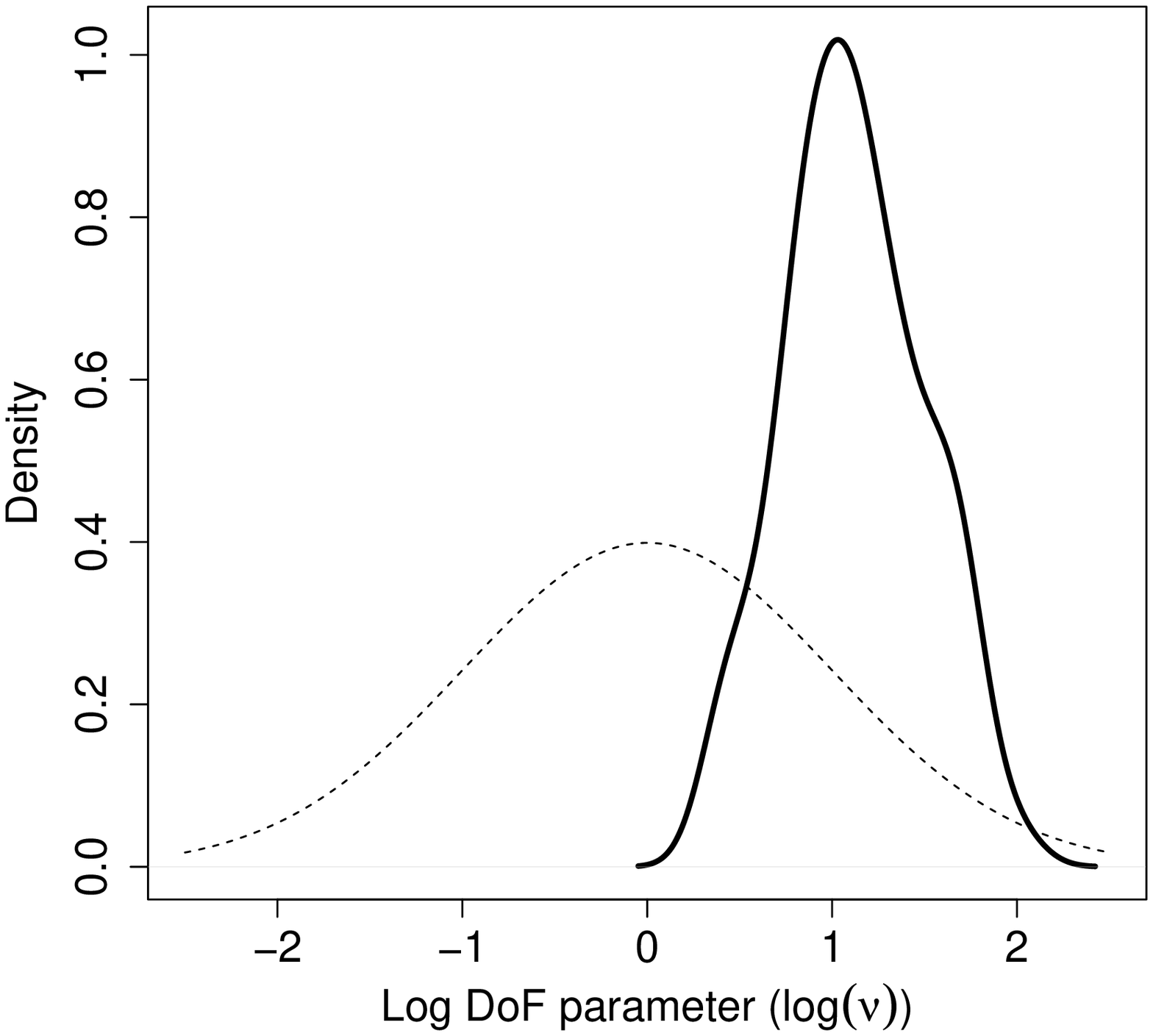}
	\caption{\textbf{Solid lines:} kernel density estimates of marginal posterior distributions for the \emph{extremal-$t$ model with Whittle-Mat\'ern correlation function} when applied to the South Australian data set without the station in Warooka, based on 67 particles.
		\textbf{Dashed lines:} prior densities. 
		Uniform prior densities for the smoothing and the rotation angle parameter are not displayed. However, for these parameters the abscissa range is equal to the support of the respective uniform prior distribution.}
	\label{figure:SAdata20150520_Parameters_WM_2}
\end{figure}

\begin{figure}[h]
	\centering
	\includegraphics[width=0.4\textwidth]{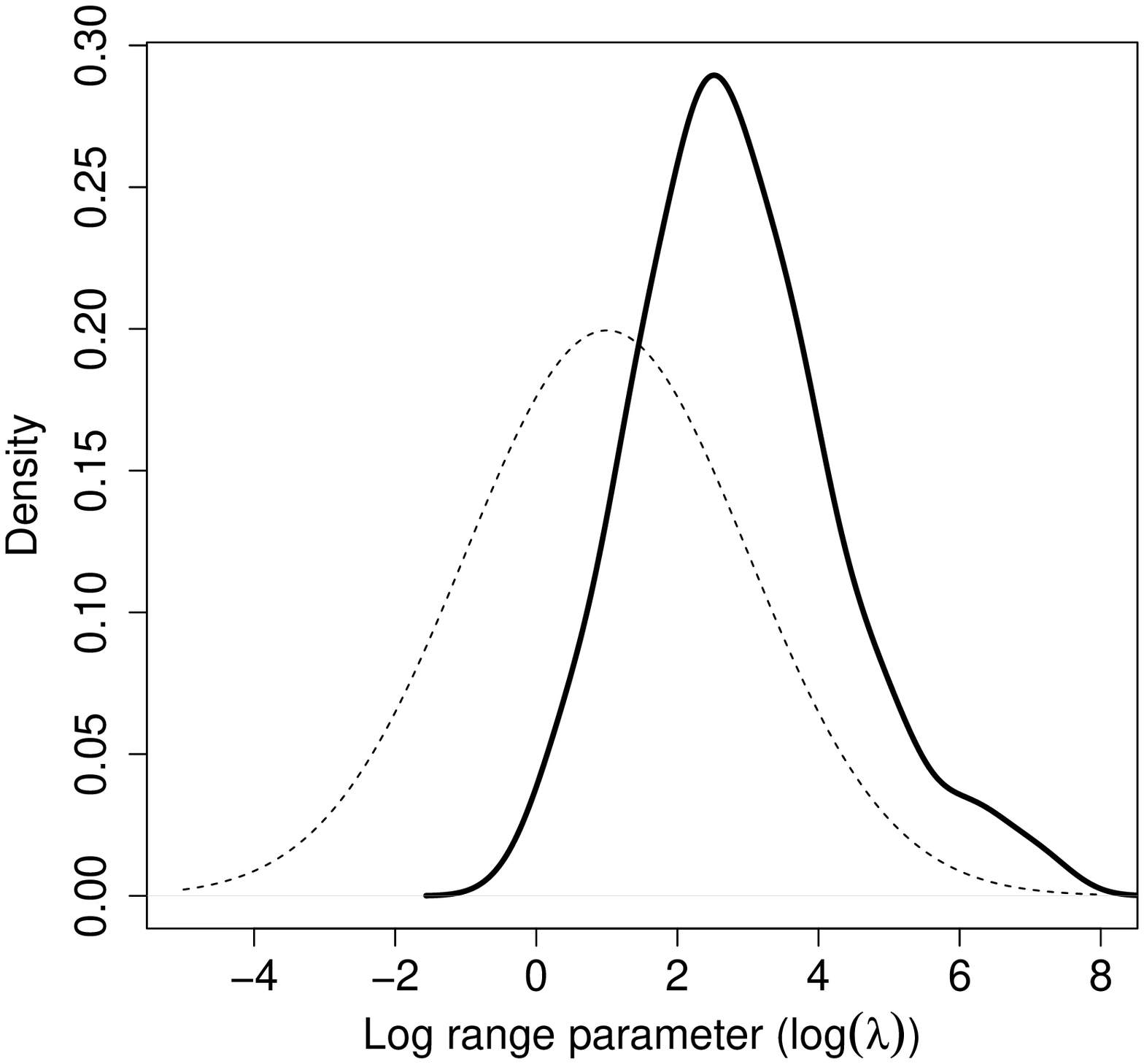} 
	\includegraphics[width=0.4\textwidth]{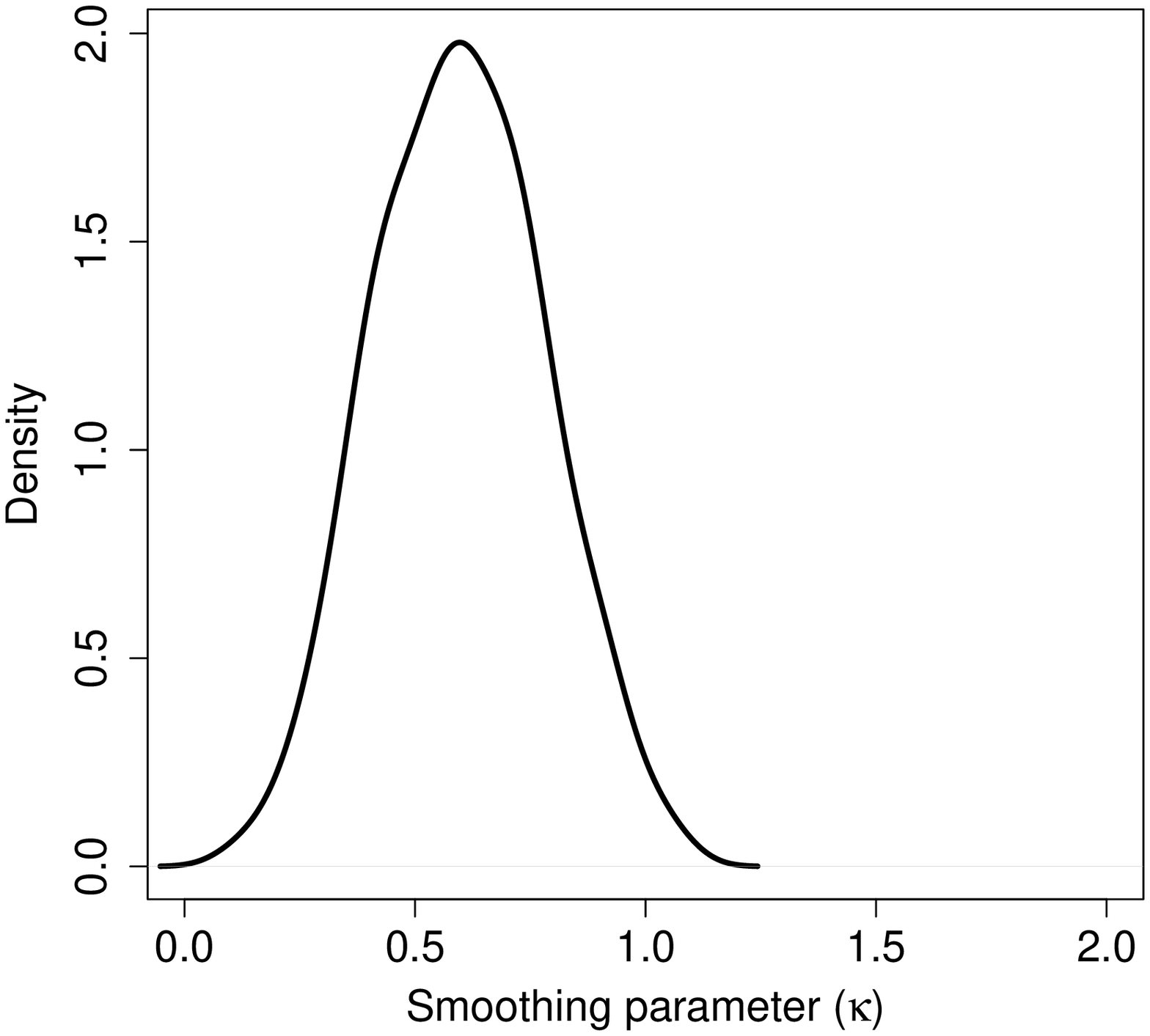} \\
	\includegraphics[width=0.4\textwidth]{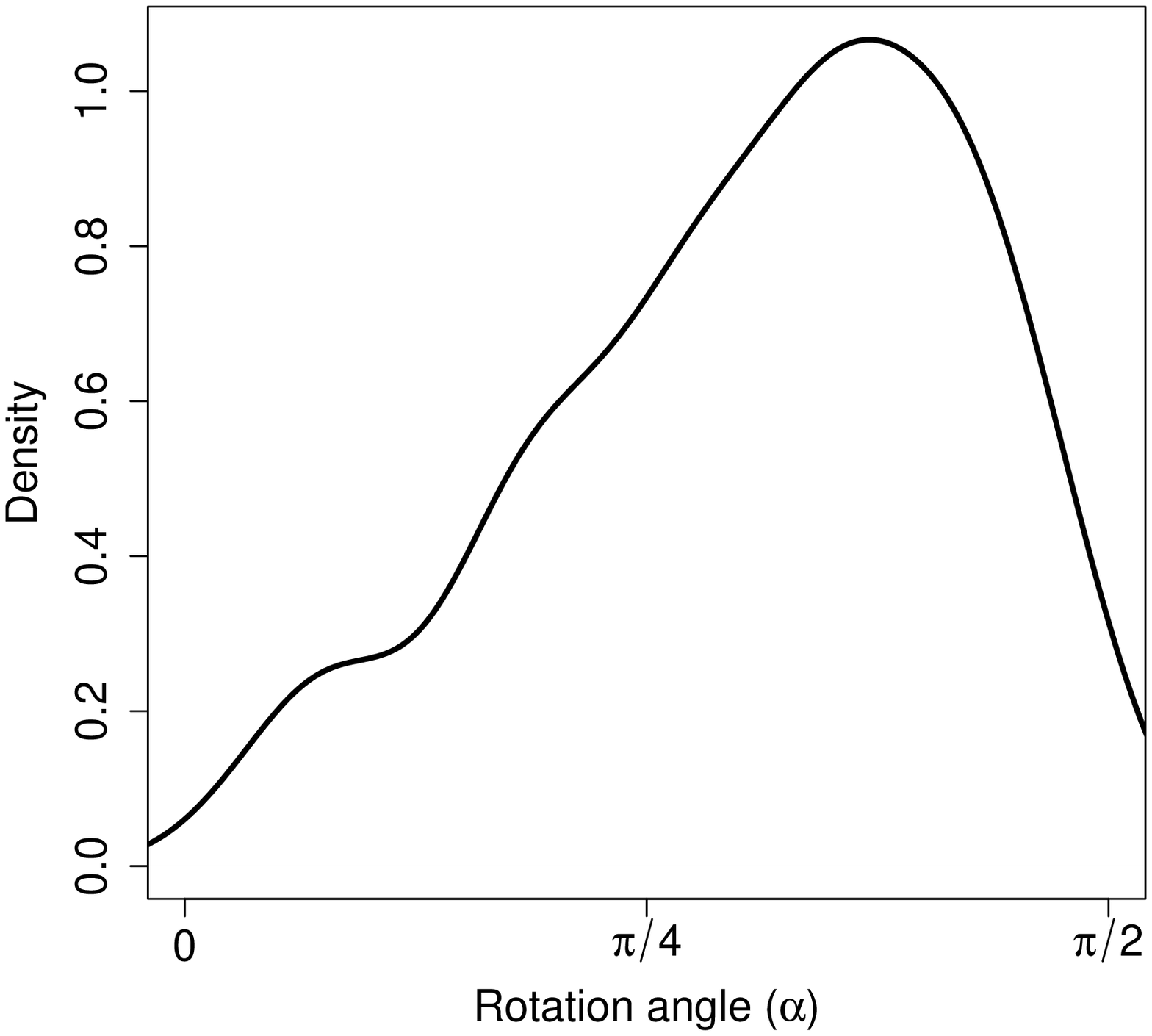}
	\includegraphics[width=0.4\textwidth]{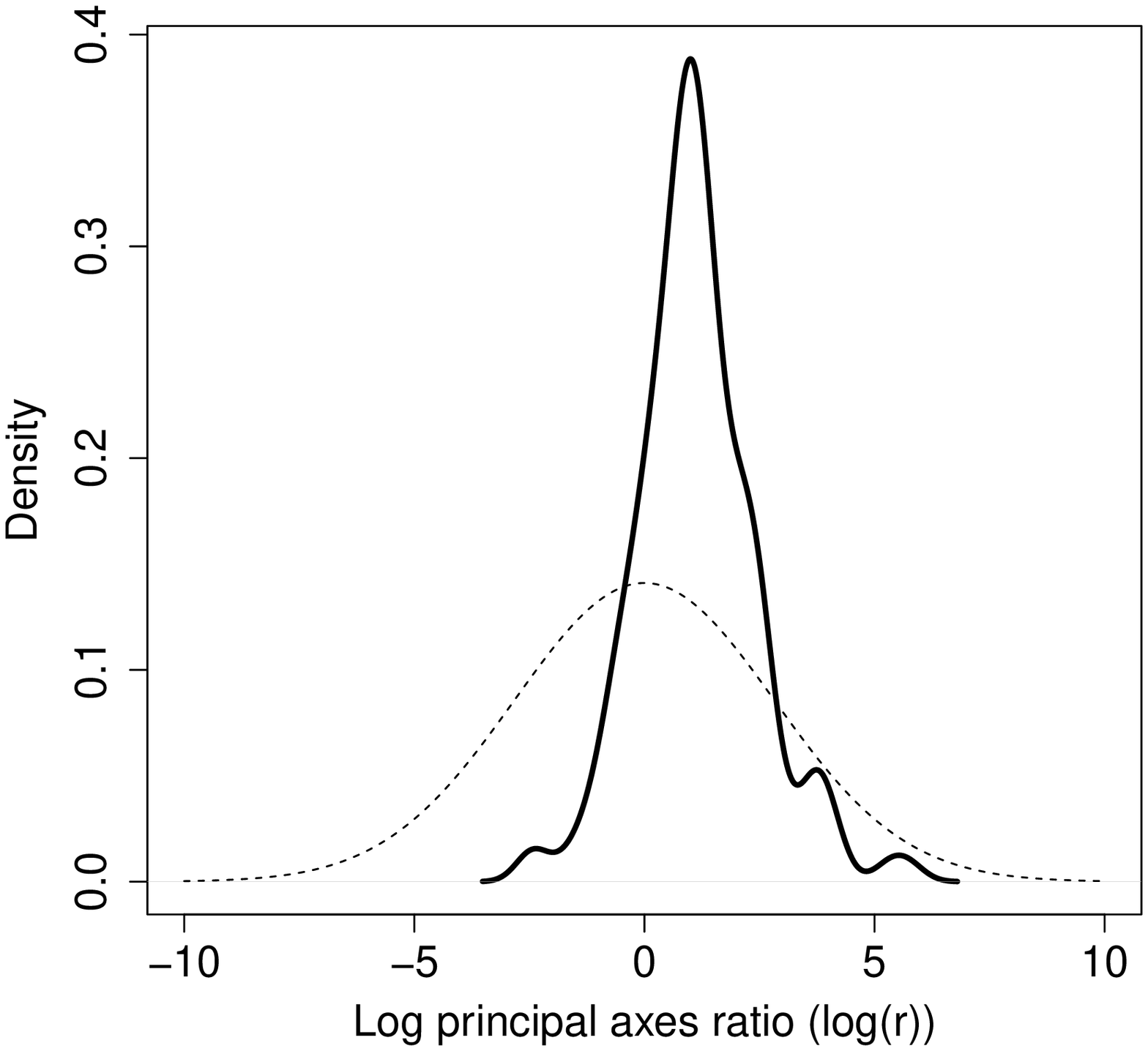} \\
	\includegraphics[width=0.4\textwidth]{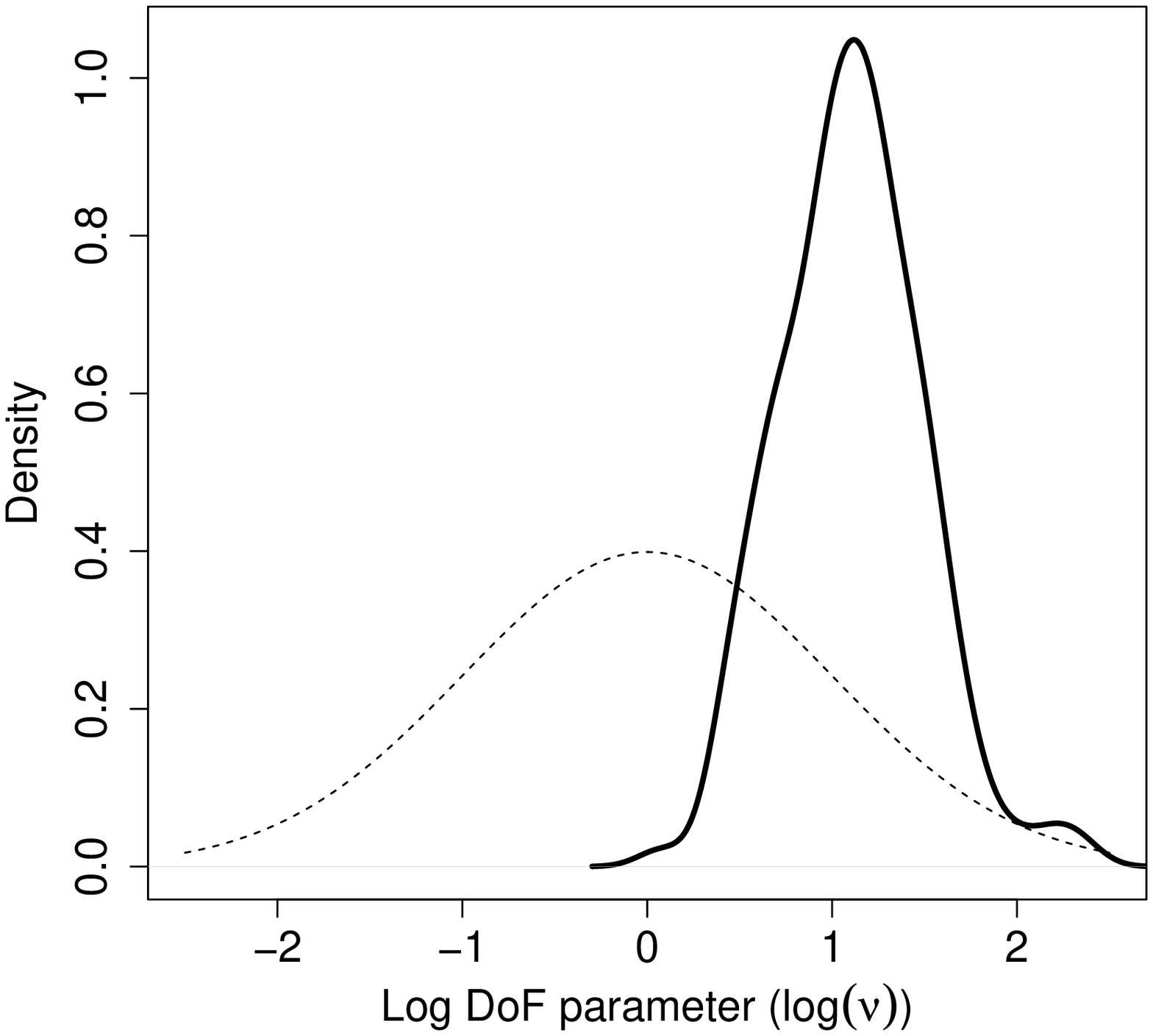} \\
	\caption{\textbf{Solid lines:} kernel density estimates of marginal posterior distributions for the \emph{extremal-$t$ model with powered exponential correlation function} when applied to the South Australian data set without the station in Warooka, based on 153 particles.
		\textbf{Dashed lines:} prior densities. 
		Uniform prior densities for the smoothing and the rotation angle parameter are not displayed. However, for these parameters the abscissa range is equal to the support of the respective uniform prior distribution.}
	\label{figure:SAdata20150520_Parameters_PE_2}
\end{figure}

\begin{figure}[h]
	\centering
	\includegraphics[width=0.4\textwidth]{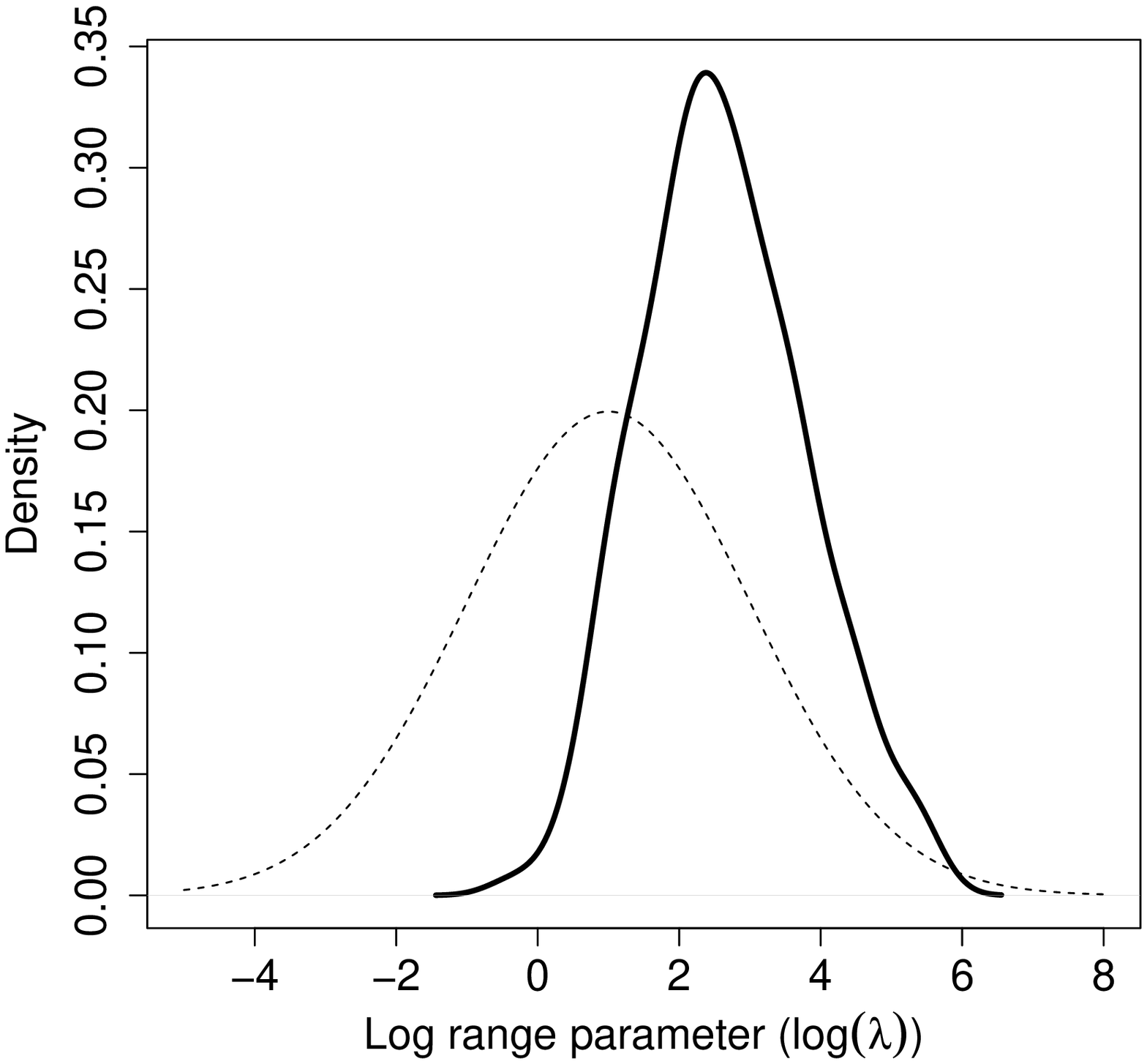} 
	\includegraphics[width=0.4\textwidth]{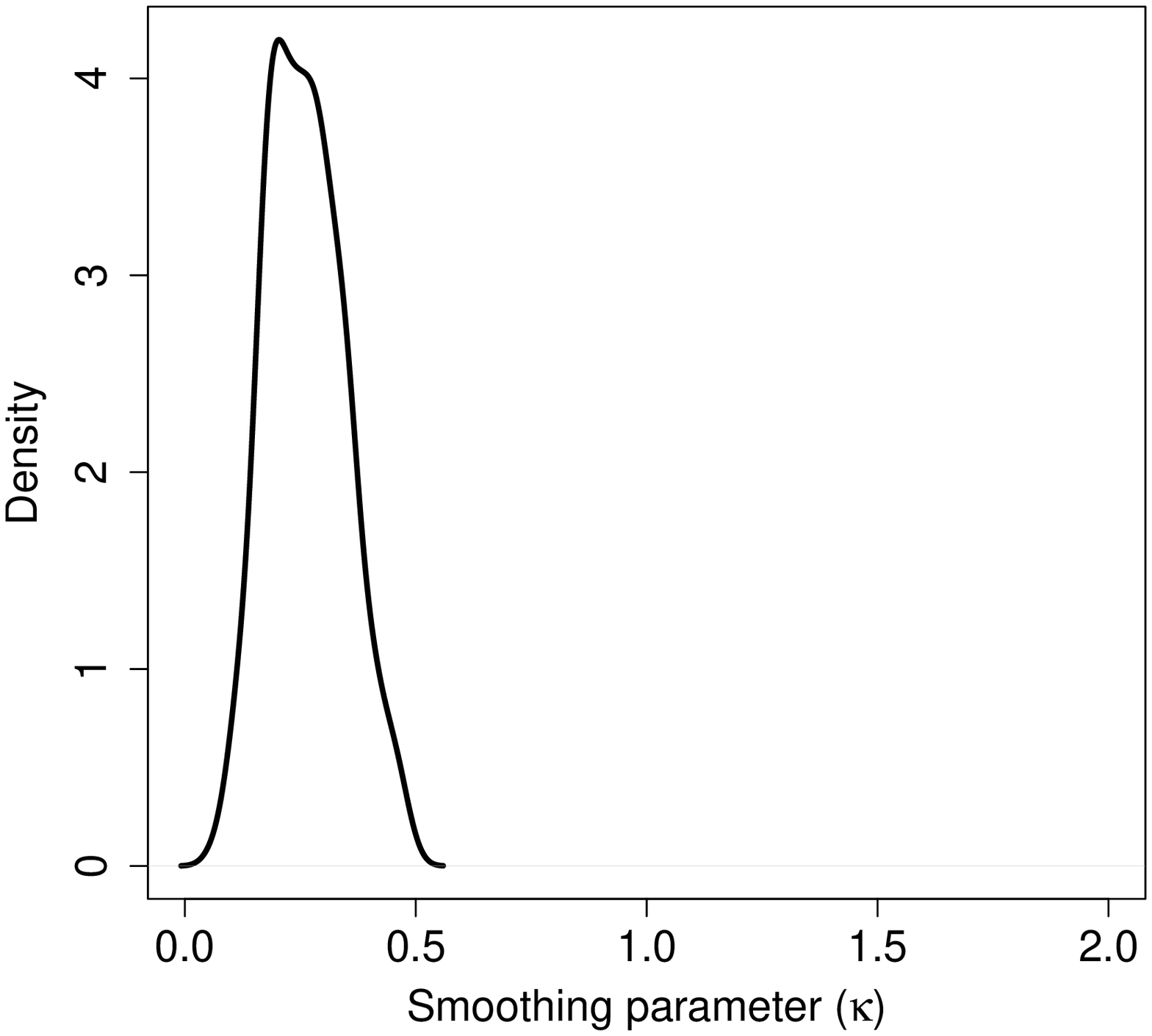} \\
	\includegraphics[width=0.4\textwidth]{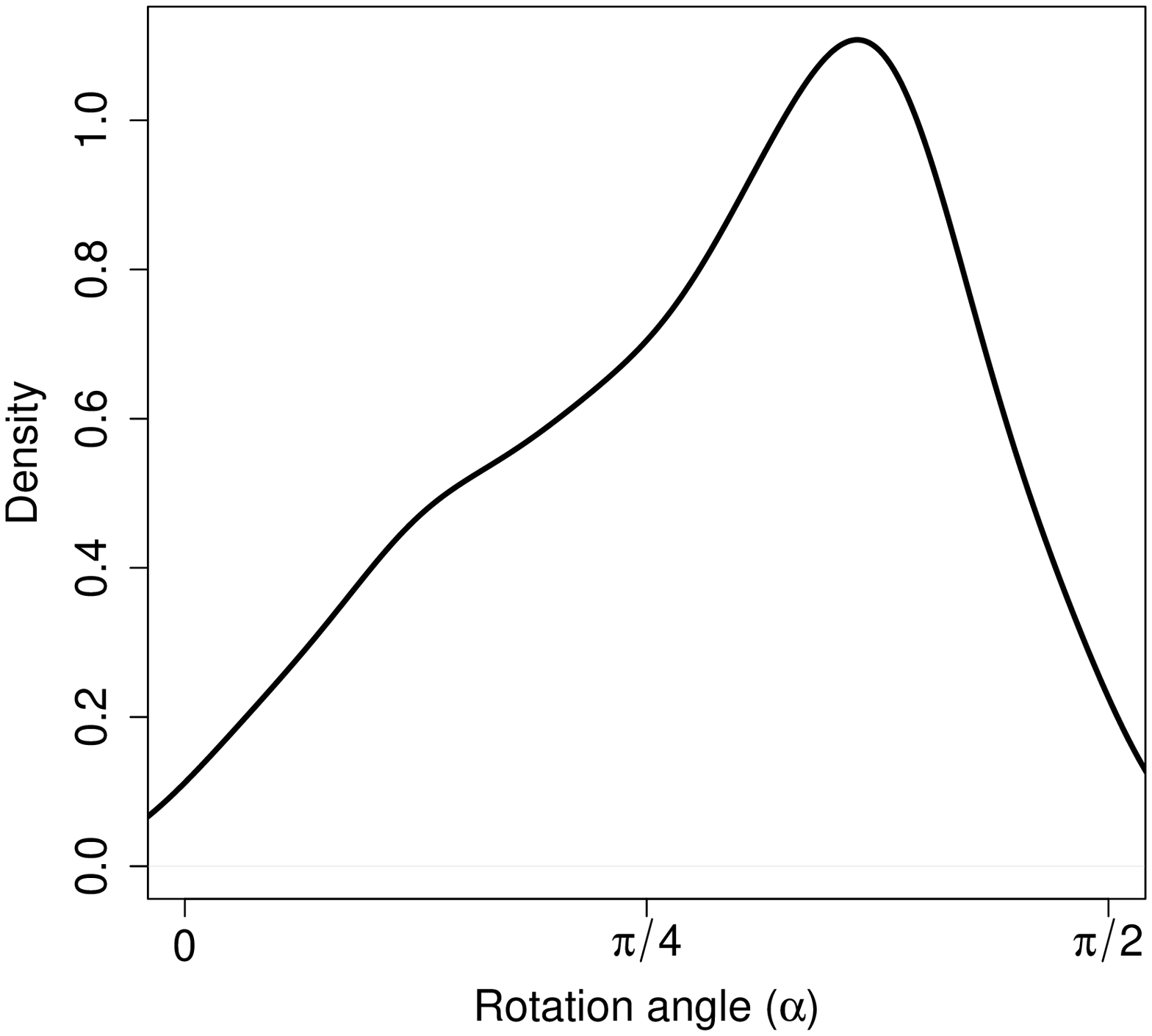}
	\includegraphics[width=0.4\textwidth]{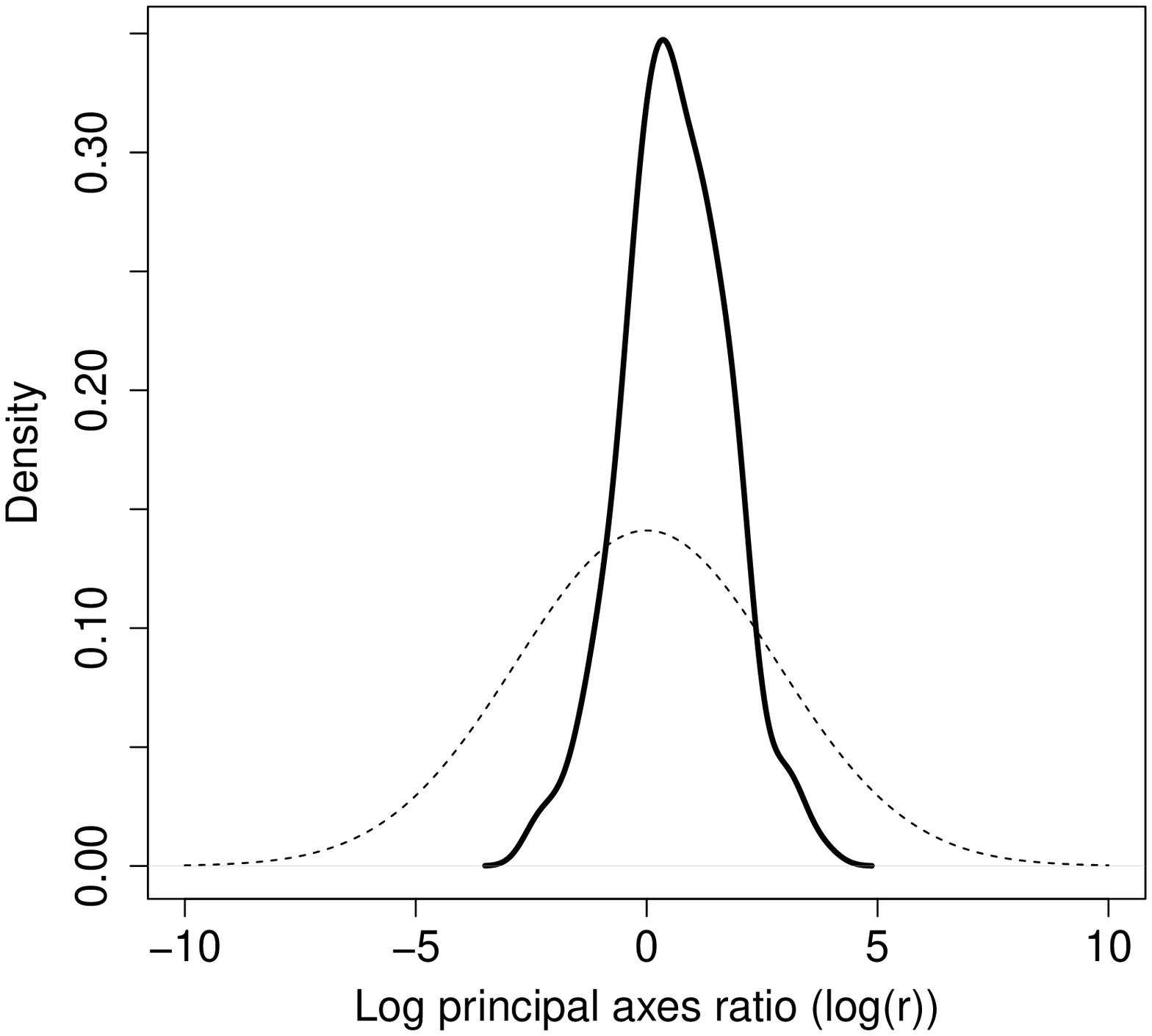} \\
	\includegraphics[width=0.4\textwidth]{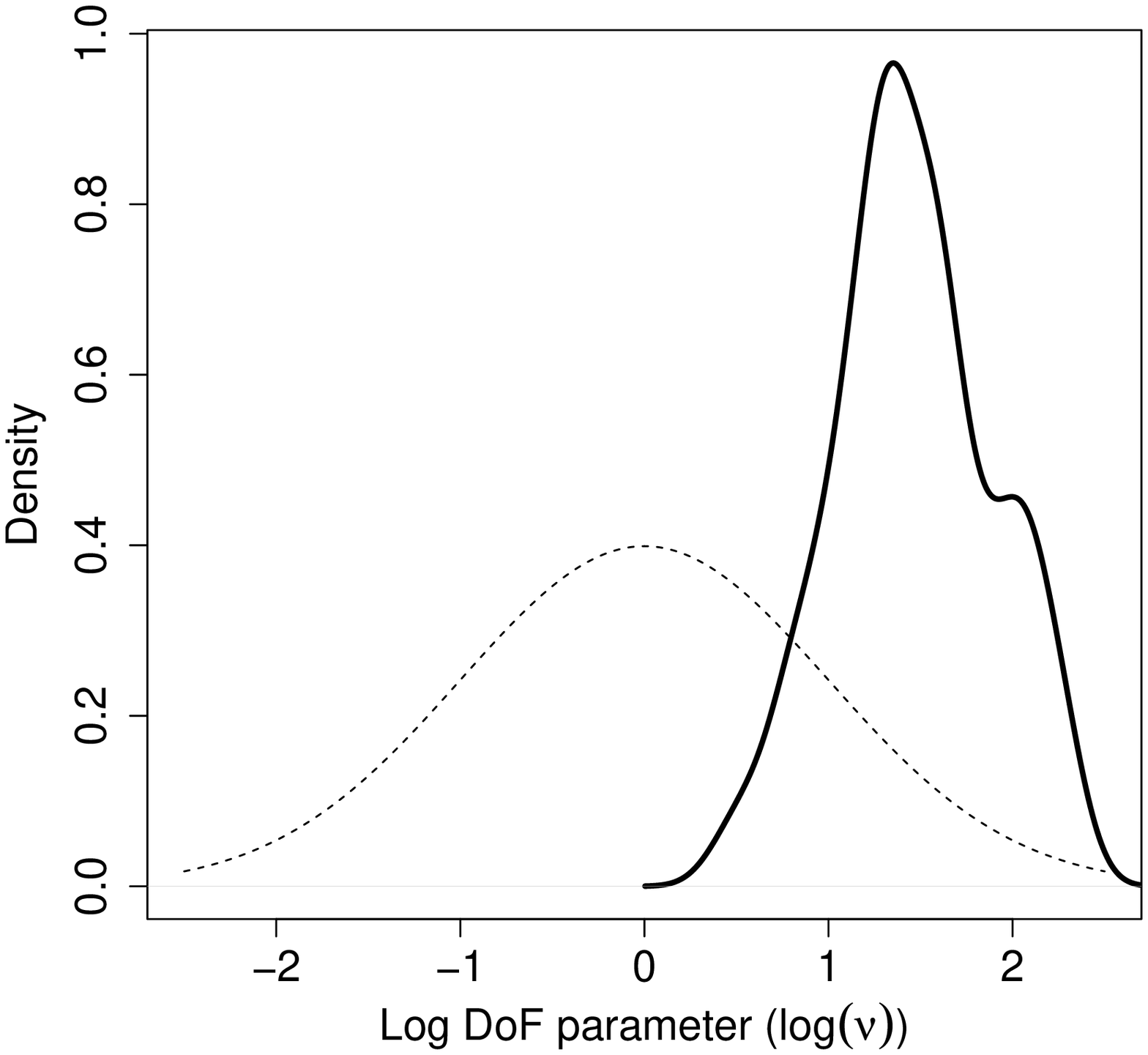}
	\caption{\textbf{Solid lines:} kernel density estimates of marginal posterior distributions for the \emph{Student-$t$ copula model with Whittle-Mat\'ern correlation function} when applied to the South Australian data set without the station in Warooka, based on 175 particles.
		\textbf{Dashed lines:} prior densities. 
		Uniform prior densities for the smoothing and the rotation angle parameter are not displayed. However, for these parameters the abscissa range is equal to the support of the respective uniform prior distribution.}
	\label{figure:SAdata20150520_Parameters_CWM_2}
\end{figure}

\begin{figure}[h]
	\centering
	\includegraphics[width=0.4\textwidth]{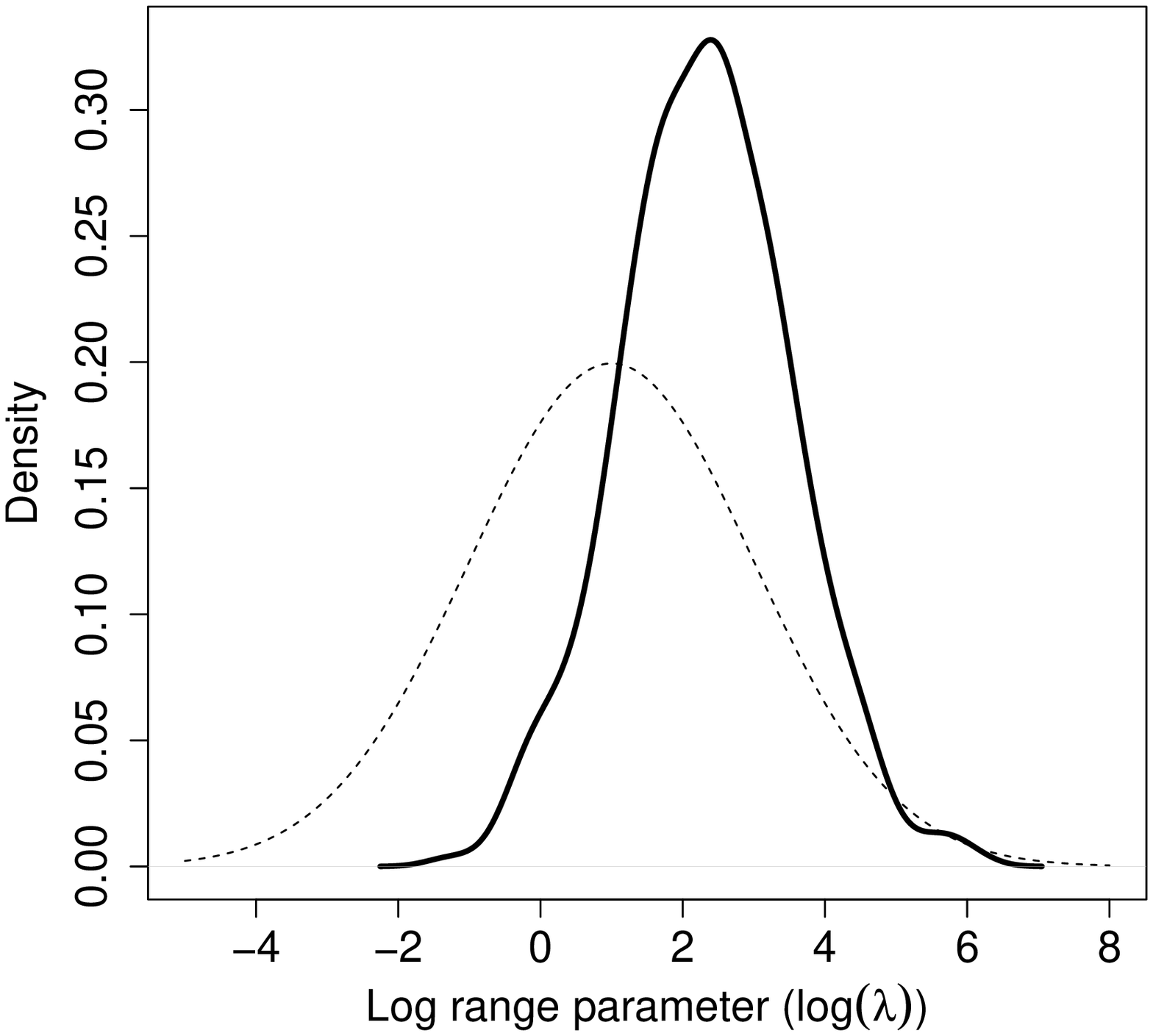} 
	\includegraphics[width=0.4\textwidth]{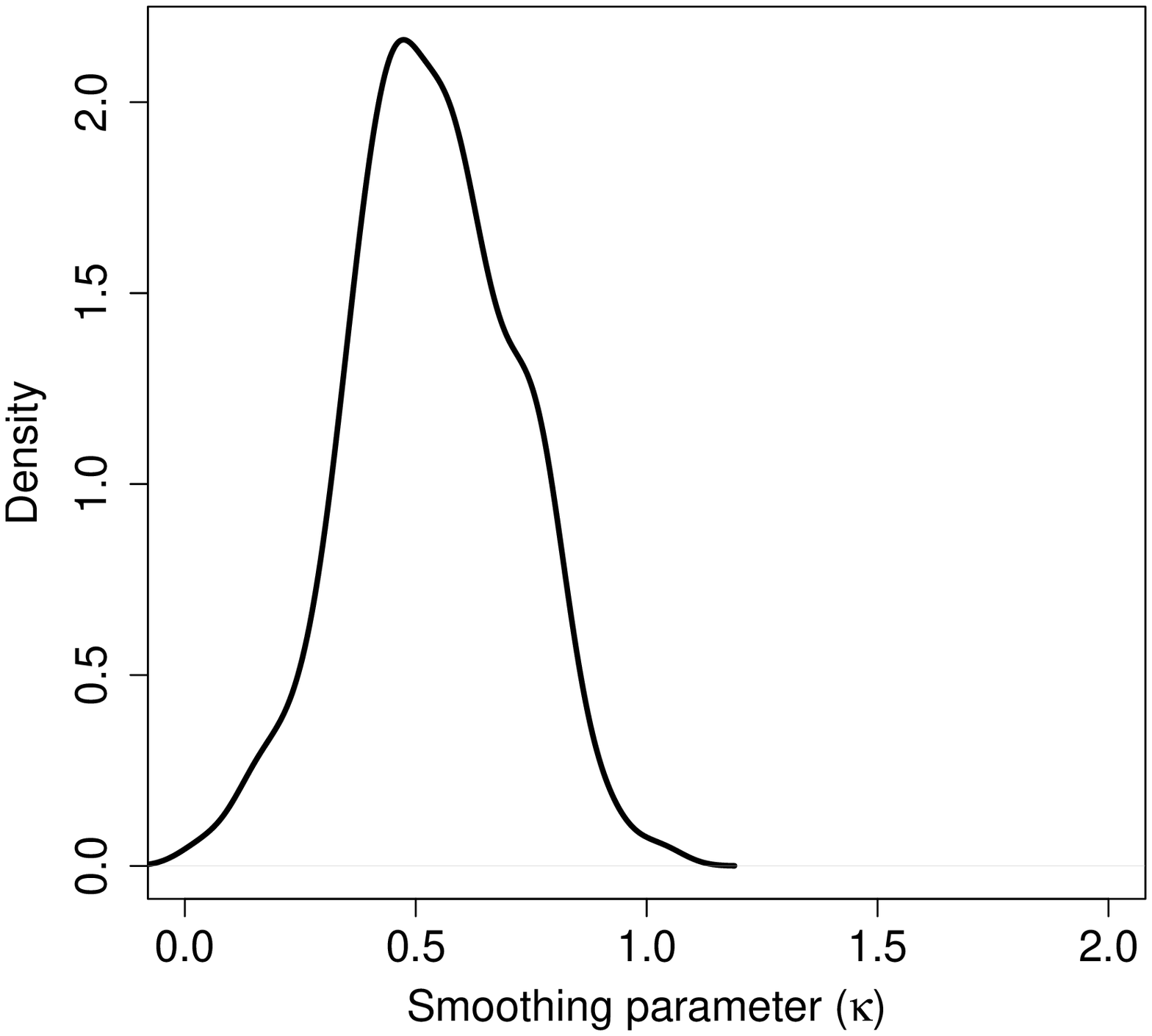} \\
	\includegraphics[width=0.4\textwidth]{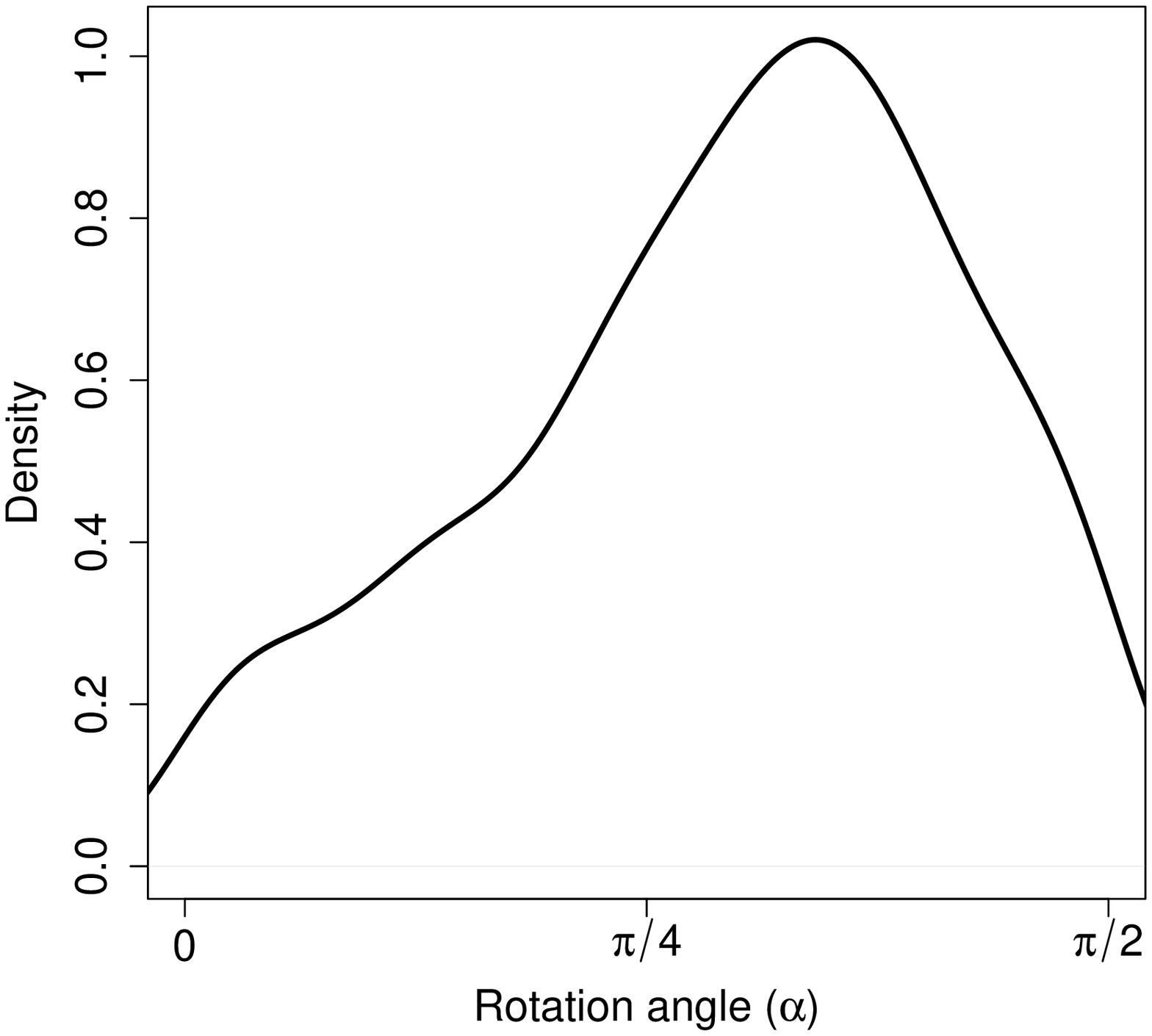}
	\includegraphics[width=0.4\textwidth]{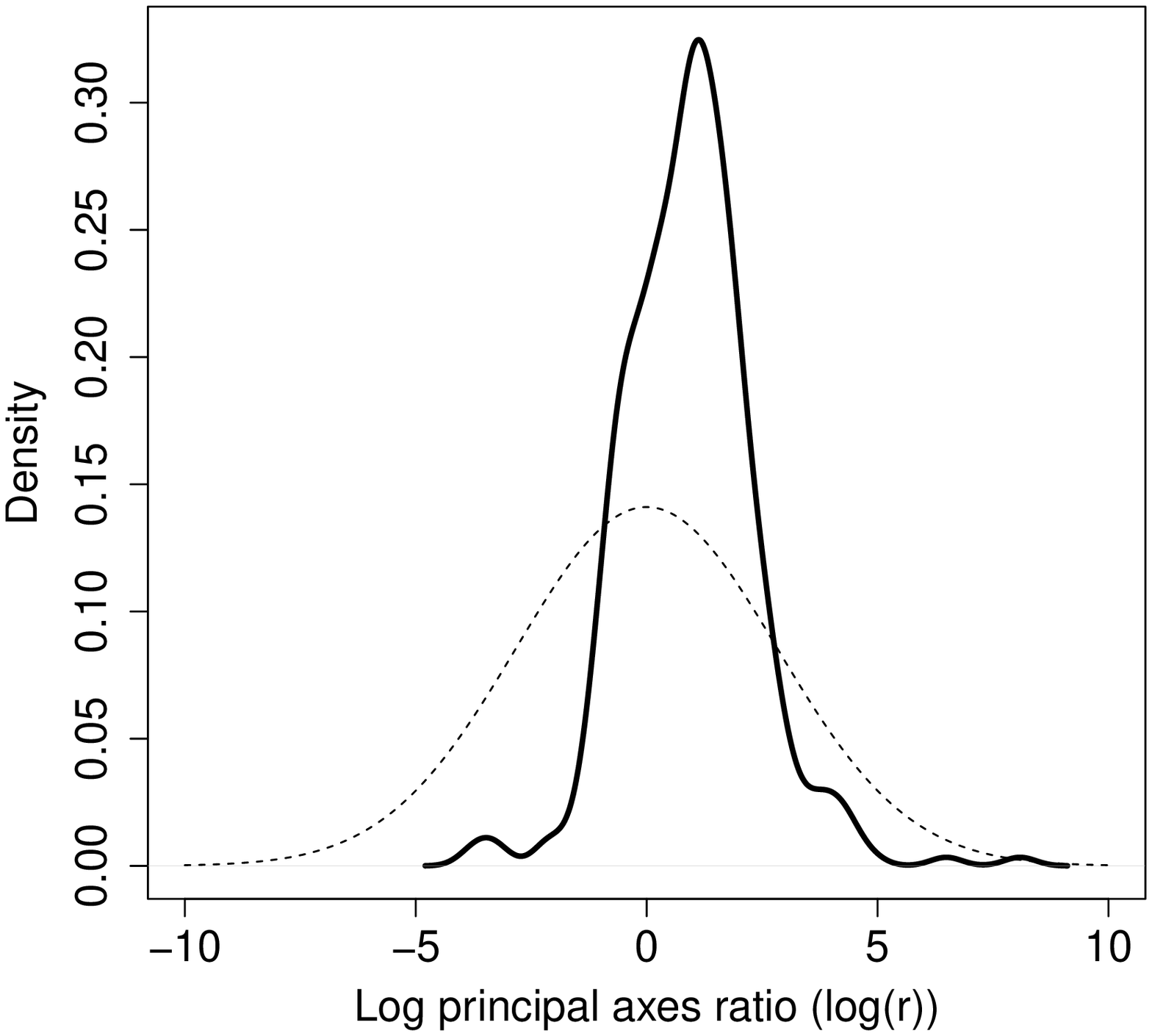} \\
	\includegraphics[width=0.4\textwidth]{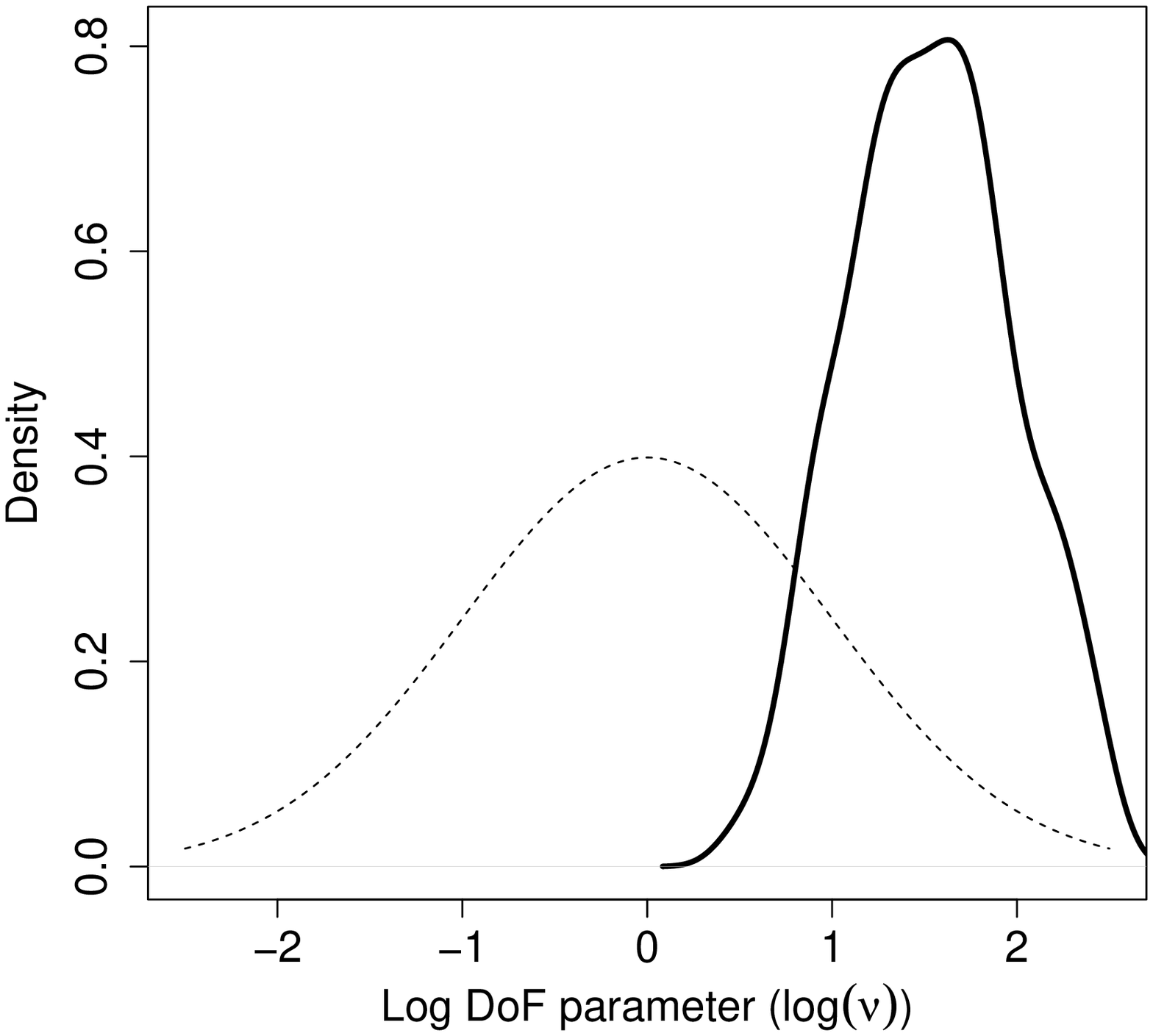}
	\caption{\textbf{Solid lines:} kernel density estimates of marginal posterior distributions for the \emph{Student-$t$ copula model with powered exponential correlation function} when applied to the South Australian data set without the station in Warooka, based on 345 particles.
		\textbf{Dashed lines:} prior densities. 
		Uniform prior densities for the smoothing and the rotation angle parameter are not displayed. However, for these parameters the abscissa range is equal to the support of the respective uniform prior distribution.}
	\label{figure:SAdata20150520_Parameters_CPE_2}
\end{figure}

Figures~\ref{figure:SAdata20150520_Parameters_WM_2}, \ref{figure:SAdata20150520_Parameters_PE_2}, \ref{figure:SAdata20150520_Parameters_CWM_2}, and \ref{figure:SAdata20150520_Parameters_CPE_2} show the posterior distributions of the parameters of the extremal-$t$ models with the Whittle-Mat\'ern and powered exponential correlation function and the Student-$t$ copula models with the Whittle-Mat\'ern and powered exponential correlation function, respectively, when the station in Warooka is excluded. These models do not have very high posterior model probabilities. For example, the estimated posterior model probability of the extremal-$t$ model with Whittle-Mat\'ern correlation function is $3.4 \%$ (67 out of 2000 particles) after the final SMC ABC iteration. Since this is a rather small sample size, the Monte Carlo error of the reported posterior estimates is high. If preciser estimates of the posterior distributions are desired for any of these models, one can run the sequential Monte Carlo ABC algorithm (or any other ABC algorithm) once more exclusively for the particular model of interest with the focus solely on parameter estimation.

\clearpage

\section{Extremal coefficient function plots}

\label{sec:extremal_plots}

A common graphical goodness-of-fit check of spatial max-stable models is to visually assess how well the theoretical pairwise extremal coefficient curve as a function of the distance between the locations fits the cloud of observed pairwise extremal coefficients. Since for our models the distances depend on the parameters governing the geometric anisotropy, we have to assume a specific parameter configuration to compute the interpoint distances for all the points in the cloud. Different anisotropy parameters give rise to different clouds of observed extremal coefficients. For each model, we will set the parameter values to their respective posterior median values.

Specifically, we compute the distance between locations $\xb_1$ and $\xb_2$ as
\begin{equation*}
h(\xb_1, \xb_2) = \frac{\| \tilde{\Ab} (\xb_1 - \xb_2) \|}{\tilde{\range}},
\end{equation*}
where the anisotropy matrix is
\begin{equation*}
\tilde{\bm{A}} =  \begin{pmatrix}
1 &  0 \\ 
0 & 1/\tilde{r}
\end{pmatrix} \cdot \begin{pmatrix}
\cos \tilde{\alpha} &  \sin \tilde{\alpha} \\ 
- \sin \tilde{\alpha} & \cos \tilde{\alpha}
\end{pmatrix},
\end{equation*}
and $\tilde{\range}$, $\tilde{\alpha}$, and $\tilde{r}$ are the posterior medians of the range, rotation angle, and principal axes ratio parameters, respectively. That is, we transform the location space to $\tilde{\mathcal{X}} = \left(\tilde{\bm{A}}/\tilde{\range}\right) \mathcal{X}$ and compute the Euclidean distances in this transformed space.

For these transformed distances $h$, the theoretical extremal coefficient curve of the Brown-Resnick model with power variogram evaluated at the posterior median of the smoothness parameter, $\tilde{\smooth}$, is given by \citep{davison2012}
\[
\theta(h; \tilde{\smooth}) = 2 \, \Phi\left\{a(h; \tilde{\smooth})/2\right\},
\]
where $\Phi\{\cdot\}$ is the CDF of the standard normal distribution and
\[
a(h; \tilde{\smooth}) = \sqrt{2 h^{\tilde{\smooth}}}.
\]

For the extremal-$t$ model, we have \citep{davison2012}

\[
\theta(h; \tilde{\nu},\tilde{\smooth}) = 2 \, T_{1;\tilde{\nu} + 1}\left\{ \sqrt{(\tilde{\nu} + 1) \: \frac{1 - \rho(h; \tilde{\smooth})}{1 + \rho(h; \tilde{\smooth})}} \right\},
\]
where $T_{1;\tilde{\nu} + 1}\{\cdot\}$ denotes the central Student-$t$ CDF with $\tilde{\nu} + 1$ degrees of freedom, $\rho(h; \tilde{\smooth})$ is one of the correlation functions described in Appendix~\ref{sec:deriv_cor_fun}, $\tilde{\nu}$ is the posterior median of the degrees of freedom parameter, and $\tilde{\smooth}$ is the posterior median of the smoothness parameter.

As discussed in Section~\ref{subsec:dependence_indicators}, the extremal coefficient is not defined for the Student-$t$ copula models. We can nevertheless use Equation~\eqref{eq:estimate_extremal_coef} to generate meaningful summary statistics measuring dependence. However, no theoretical extremal coefficient curve is available and the functional form of the underlying coefficient to which the estimates converge is unknown. Therefore, we have to estimate the underlying coefficient for a given set of parameters $\tilde{\smooth}$ and $\tilde{\nu}$ at any distance $h$. For a given distance $h$, we obtain the estimate by computing Equation~\eqref{eq:estimate_extremal_coef} for a sample of size $500,\!000$ generated from a bivariate $t$ copula model with distance $h$. To allow for a concise description of the results, we denote the underlying coefficient to which the estimates converge by extremal coefficient in accordance with the max-stable models.

\begin{figure}[h]
	\centering
	\begin{tabular}{cc}
		\includegraphics[width=0.4\textwidth]{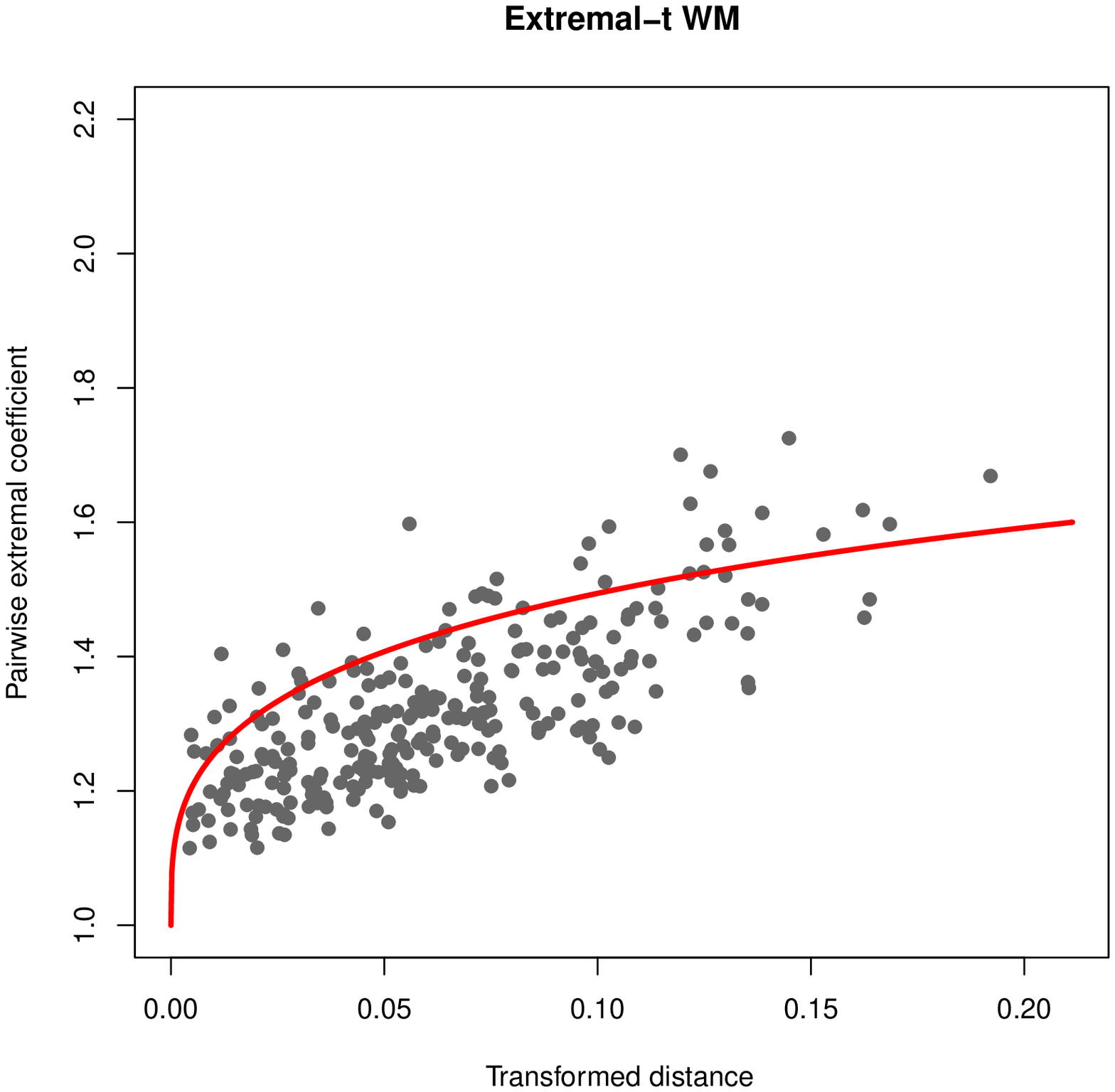} &
		\includegraphics[width=0.4\textwidth]{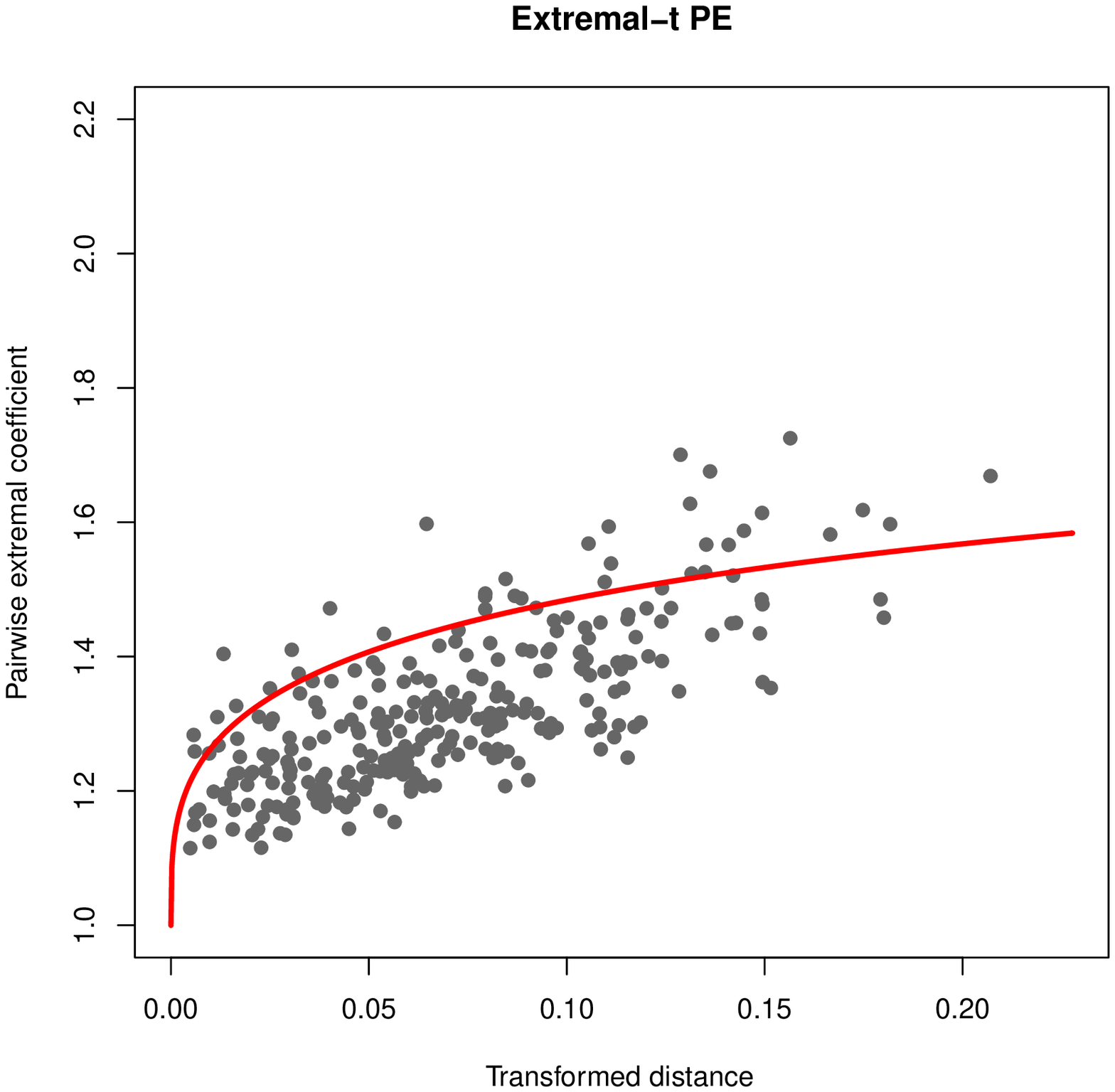} \\
		\includegraphics[width=0.4\textwidth]{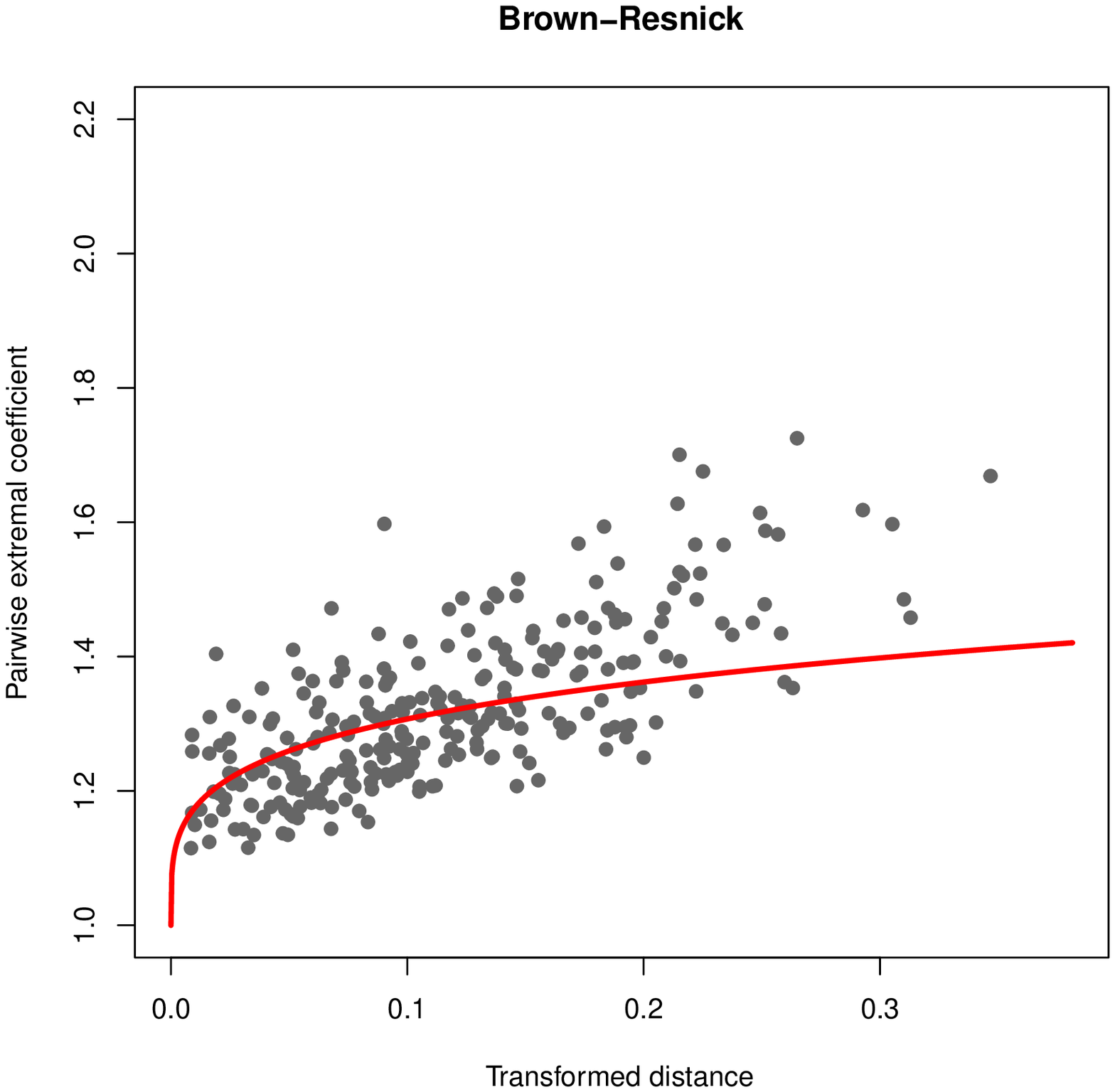} & \\
		\includegraphics[width=0.4\textwidth]{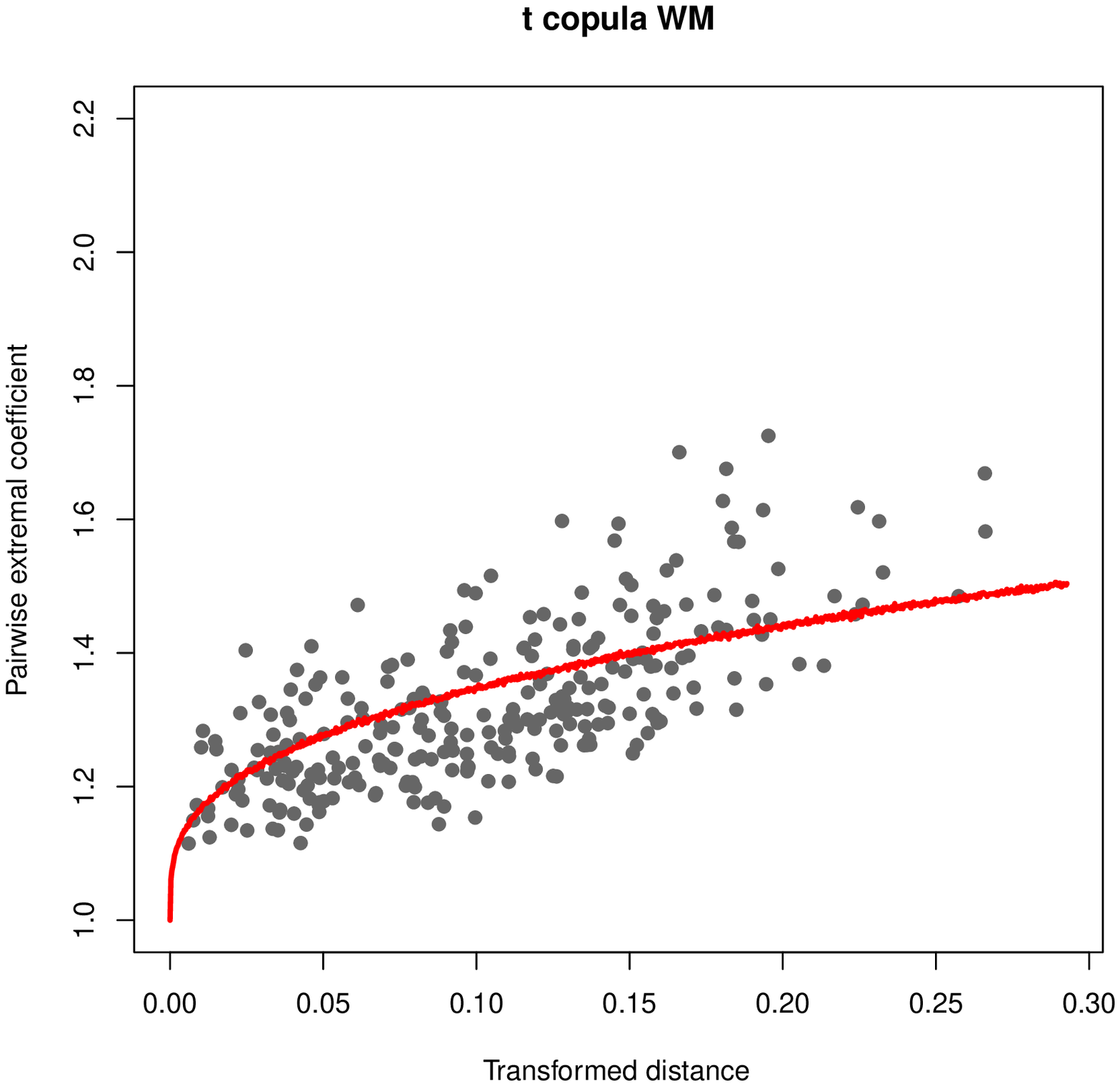} &
		\includegraphics[width=0.4\textwidth]{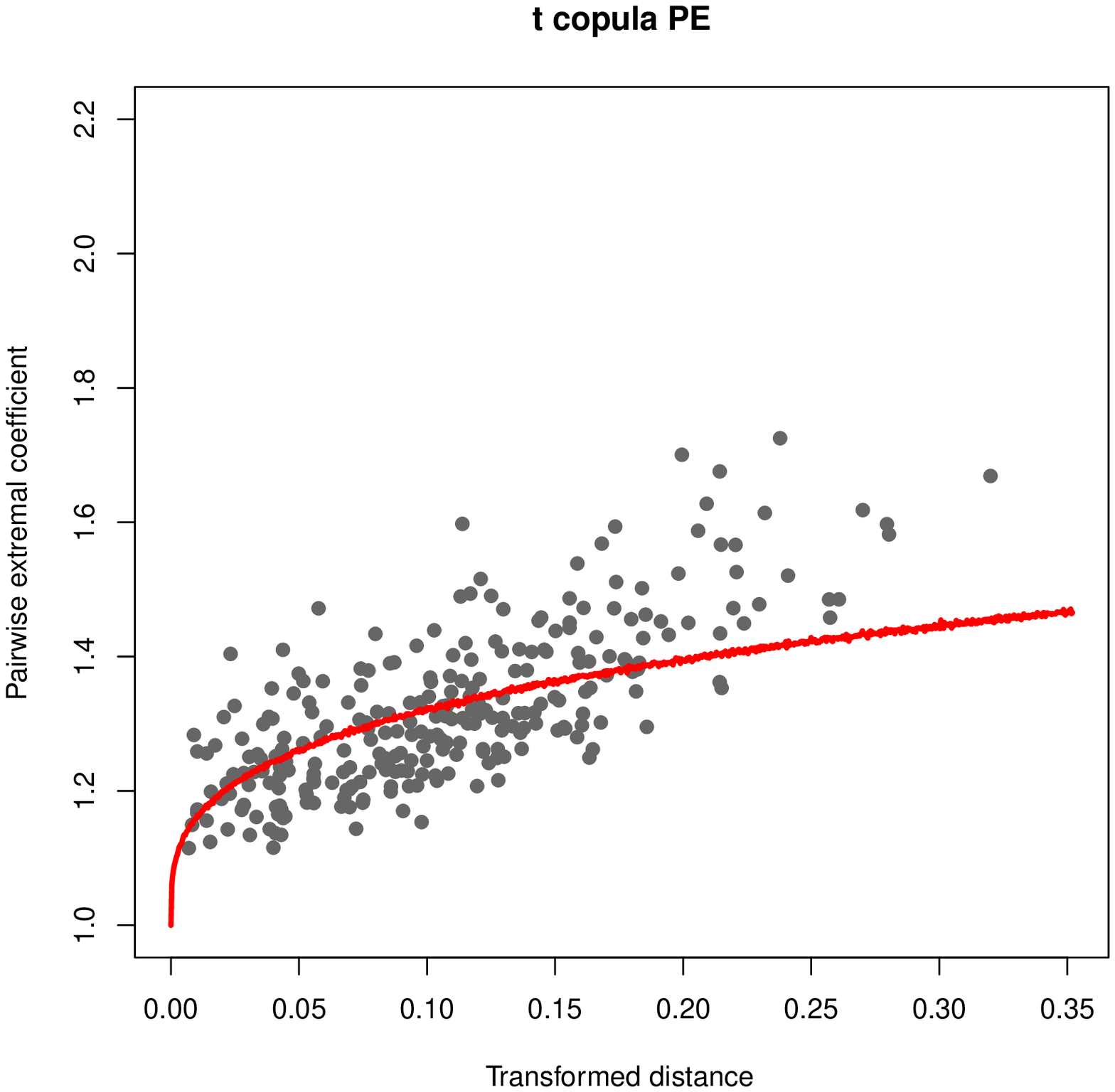}
	\end{tabular}
	\caption{Extremal coefficient function plots when excluding Warooka for the extremal-$t$ model (top) with Whittle-Mat\'ern (top left) and powered exponential (top right) correlation function, the Brown-Resnick model (middle), and the $t$ copula model (bottom) with Whittle-Mat\'ern (bottom left) and powered exponential (bottom right) correlation function. Each plot was generated using the posterior median values of the respective model's parameters.}
	\label{figure:extremal_plots_2}
\end{figure}

Figure~\ref{figure:extremal_plots_2} shows the extremal coefficient function plots for all models when the station in Warooka is excluded, one plot for each model. The extremal-$t$ models fail to capture the extremal coefficients at low and medium distances, where most of the points are concentrated. The Brown-Resnick model, on the other hand, does not capture the extremal coefficients at large distances very well, but it provides the best fit for small and medium distances. The $t$ copula models also provide an adequate fit.

\section{Analysis when including Warooka station}

\label{sec:analysis_with_Warooka}

In this section, we present the results of our ABC analysis when the station in Warooka is included.

\begin{figure}[h]
	\centering
	\begin{adjustbox}{max width = \columnwidth}
		\includegraphics[width=0.8\textwidth]{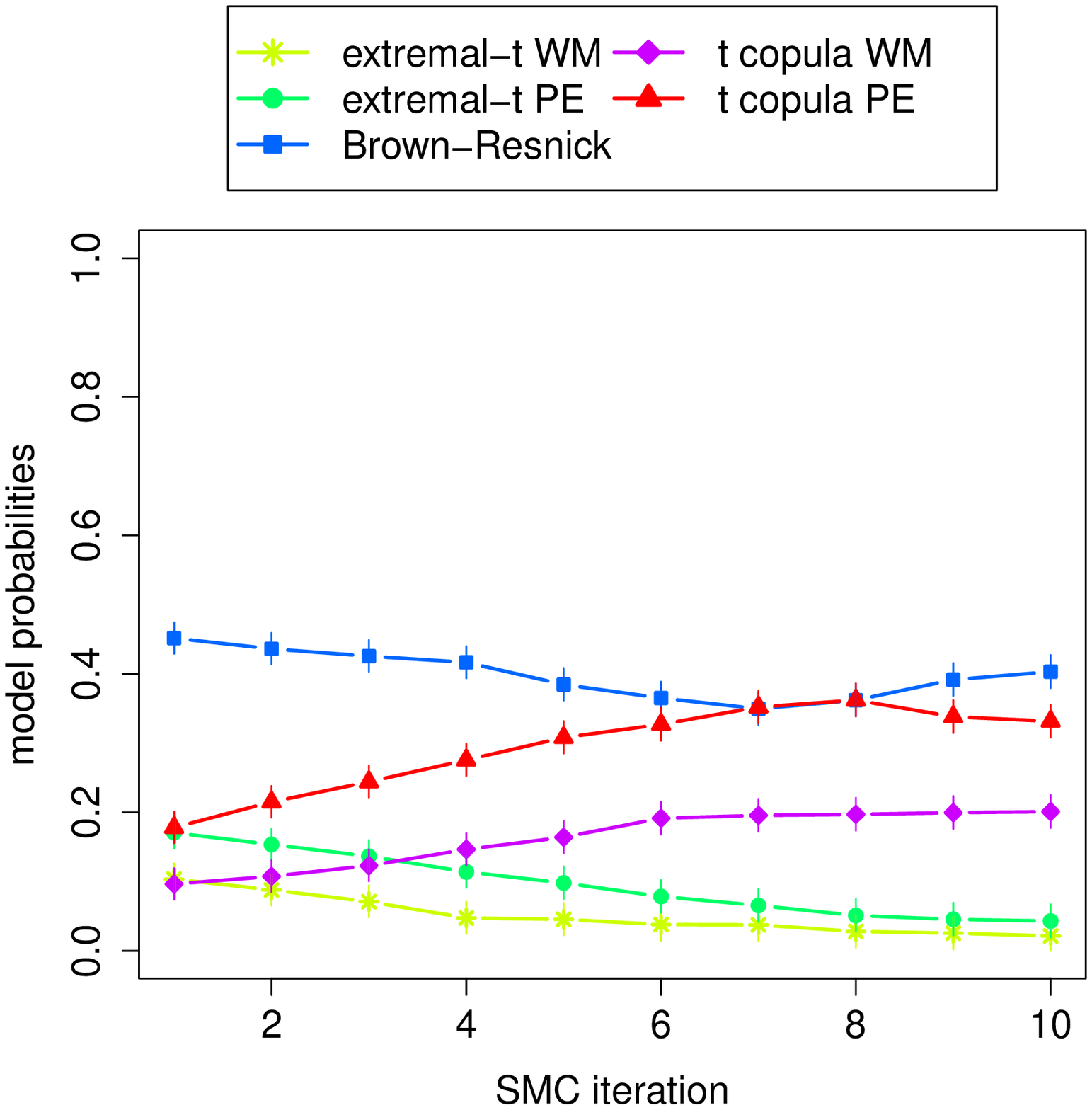}
	\end{adjustbox}	
	\caption{Progression of estimated posterior model probabilities across SMC iterations for the South Australian data set when the station in Warooka is included.}
	\label{figure:SAdata20150520_SMCmodelprobsiteration} 
\end{figure}

The progression of estimated posterior model probabilities is depicted in \fref{figure:SAdata20150520_SMCmodelprobsiteration}. By including Warooka, the estimated posterior model probabilities change significantly. When Warooka is included, the posterior model probability of the Brown-Resnick model is $40 \%$ at the final SMC iteration. When Warooka is excluded, this probability is $63 \%$. On the other hand, the posterior model probabilities of the $t$ copula models rise to $33 \%$ (powered exponential correlation function) and $20 \%$ (Whittle-Mat\'ern correlation function), so the $t$ copula models have a higher combined posterior model probability than the Brown-Resnick model. The posterior model probabilities of the extremal-$t$ models remain low.

\begin{figure}[h]
	\centering
	\includegraphics[width=0.4\textwidth]{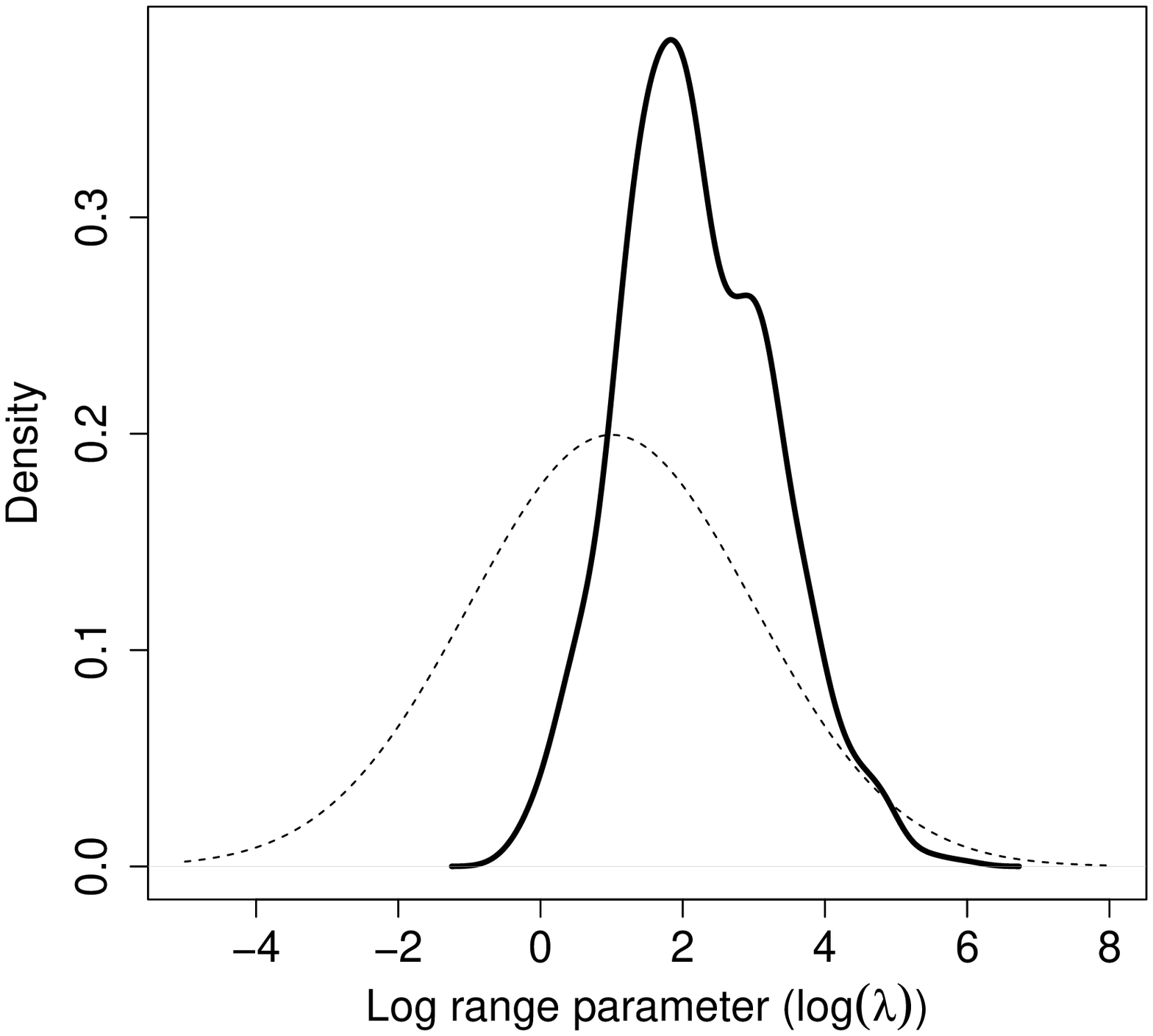} 
	\includegraphics[width=0.4\textwidth]{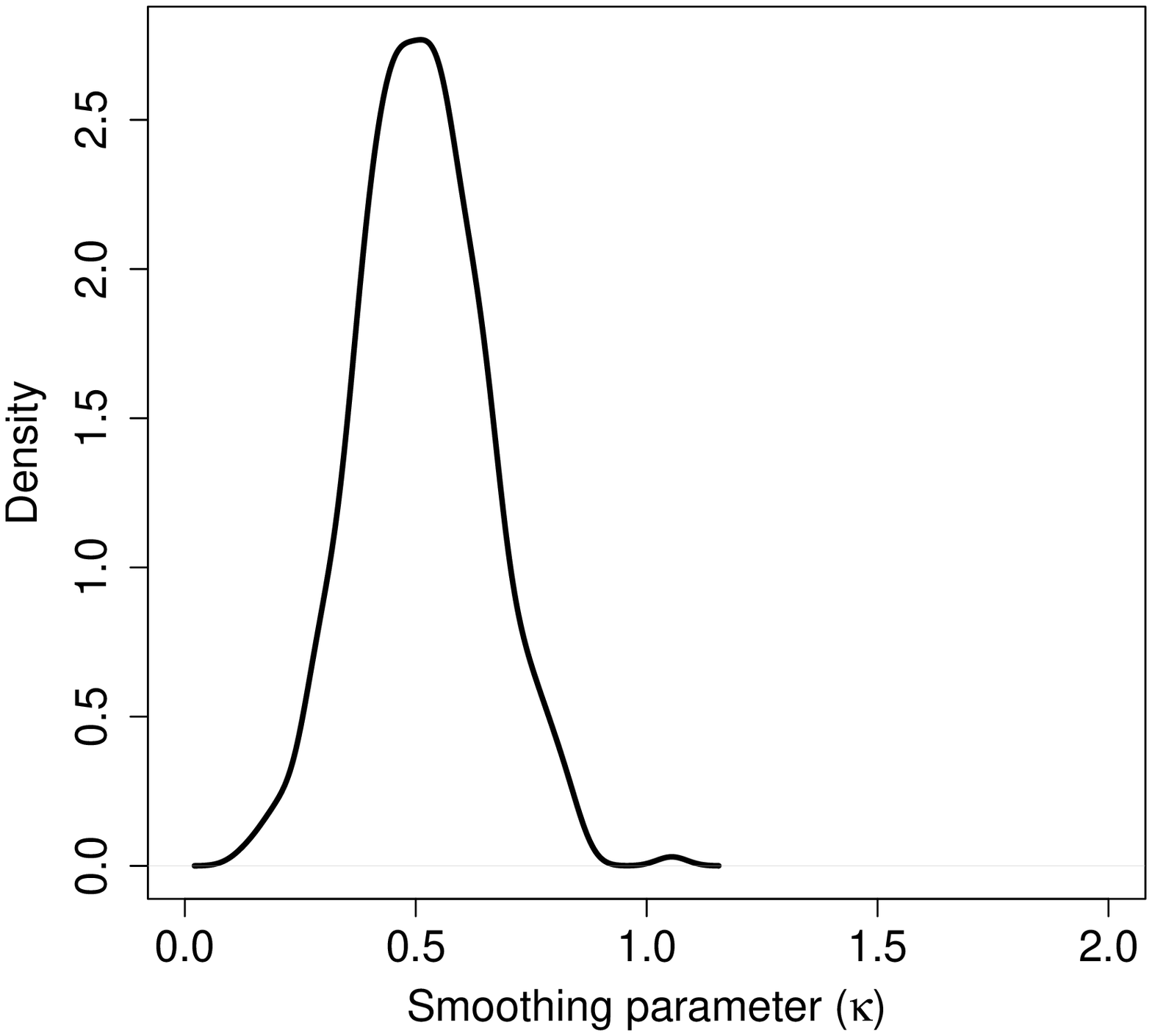} \\
	\includegraphics[width=0.4\textwidth]{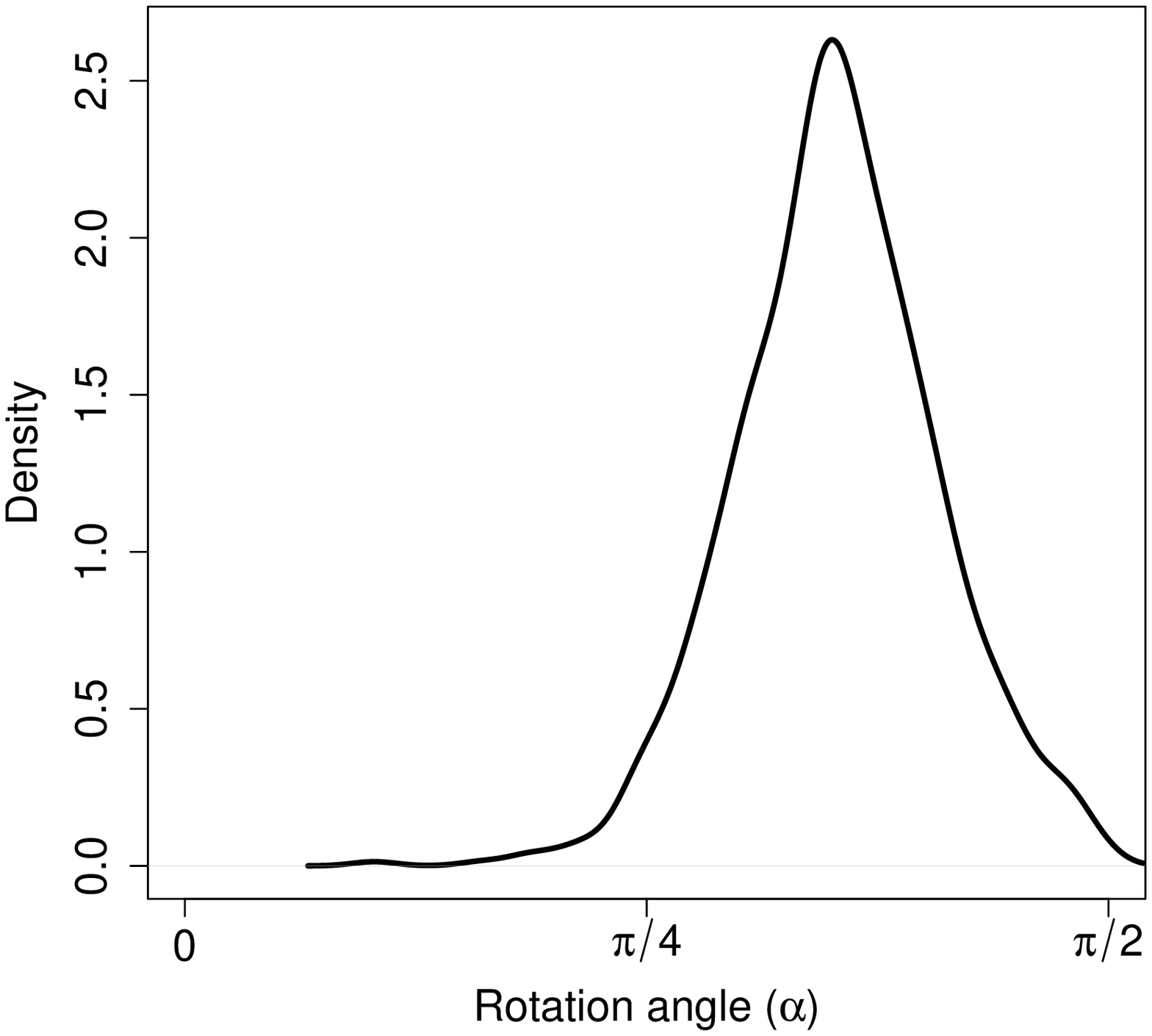}
	\includegraphics[width=0.4\textwidth]{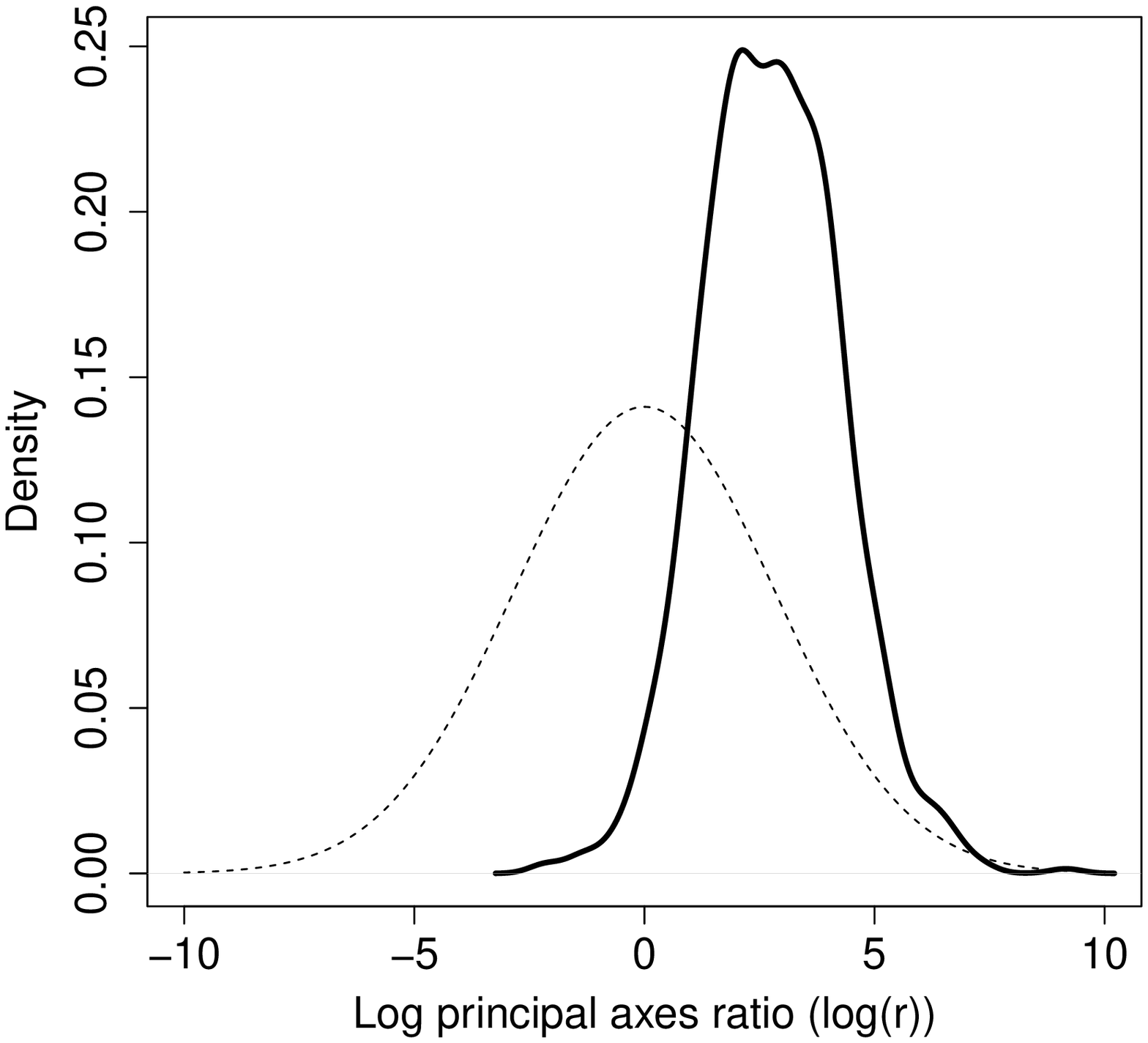} \\
	\caption{\textbf{Solid lines:} kernel density estimates of marginal posterior distributions for the \emph{Brown-Resnick model} when applied to the South Australian data set including the station in Warooka.
		\textbf{Dashed lines:} prior densities.
		Uniform prior densities for the smoothing and the rotation angle parameter are not displayed. However, for these parameters the abscissa range is equal to the support of the respective uniform prior distribution.}
	\label{figure:SAdata20150520_Parameters_BR}
\end{figure}

\begin{figure}[h]
	\centering
	\includegraphics[width=0.4\textwidth]{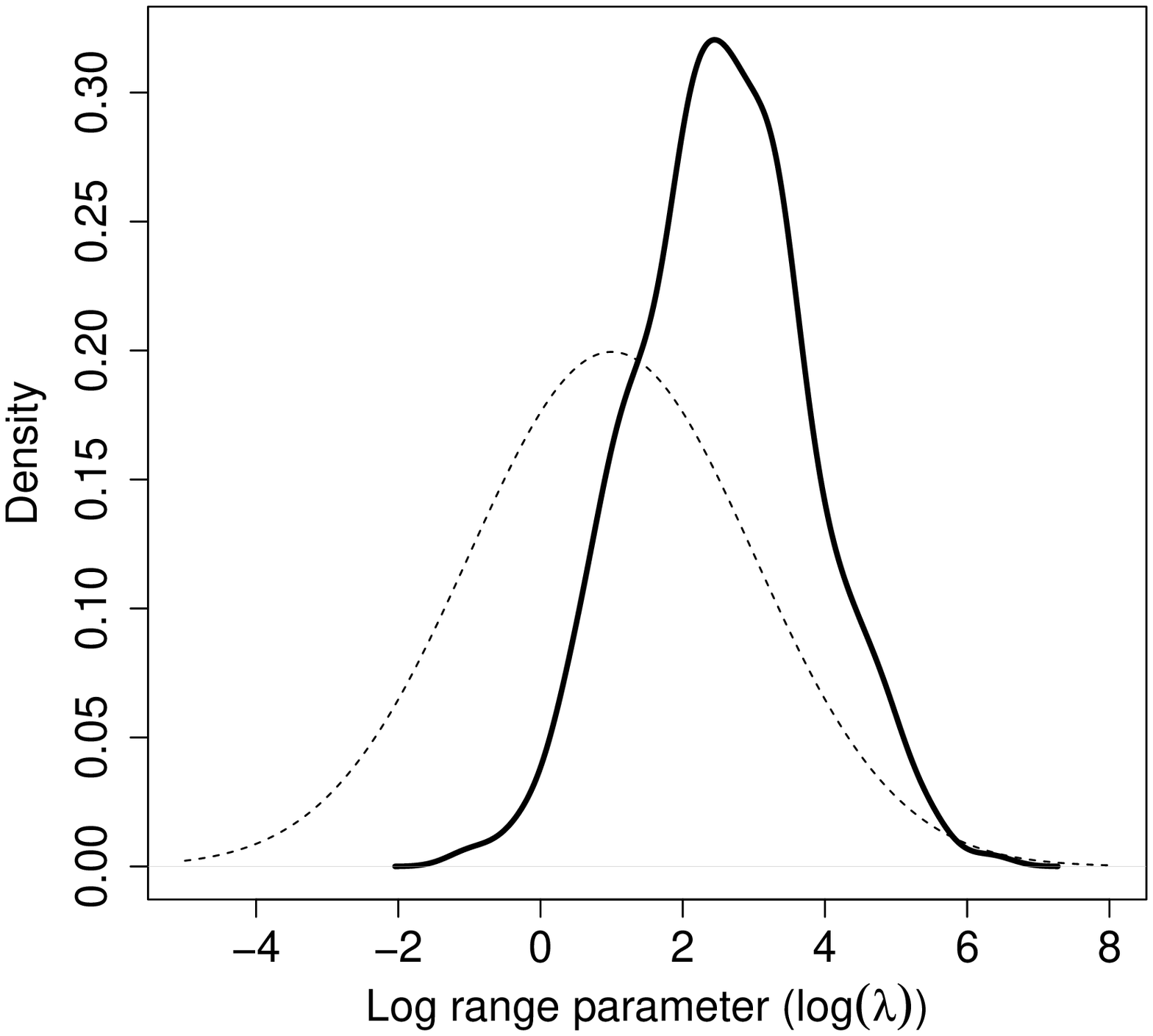} 
	\includegraphics[width=0.4\textwidth]{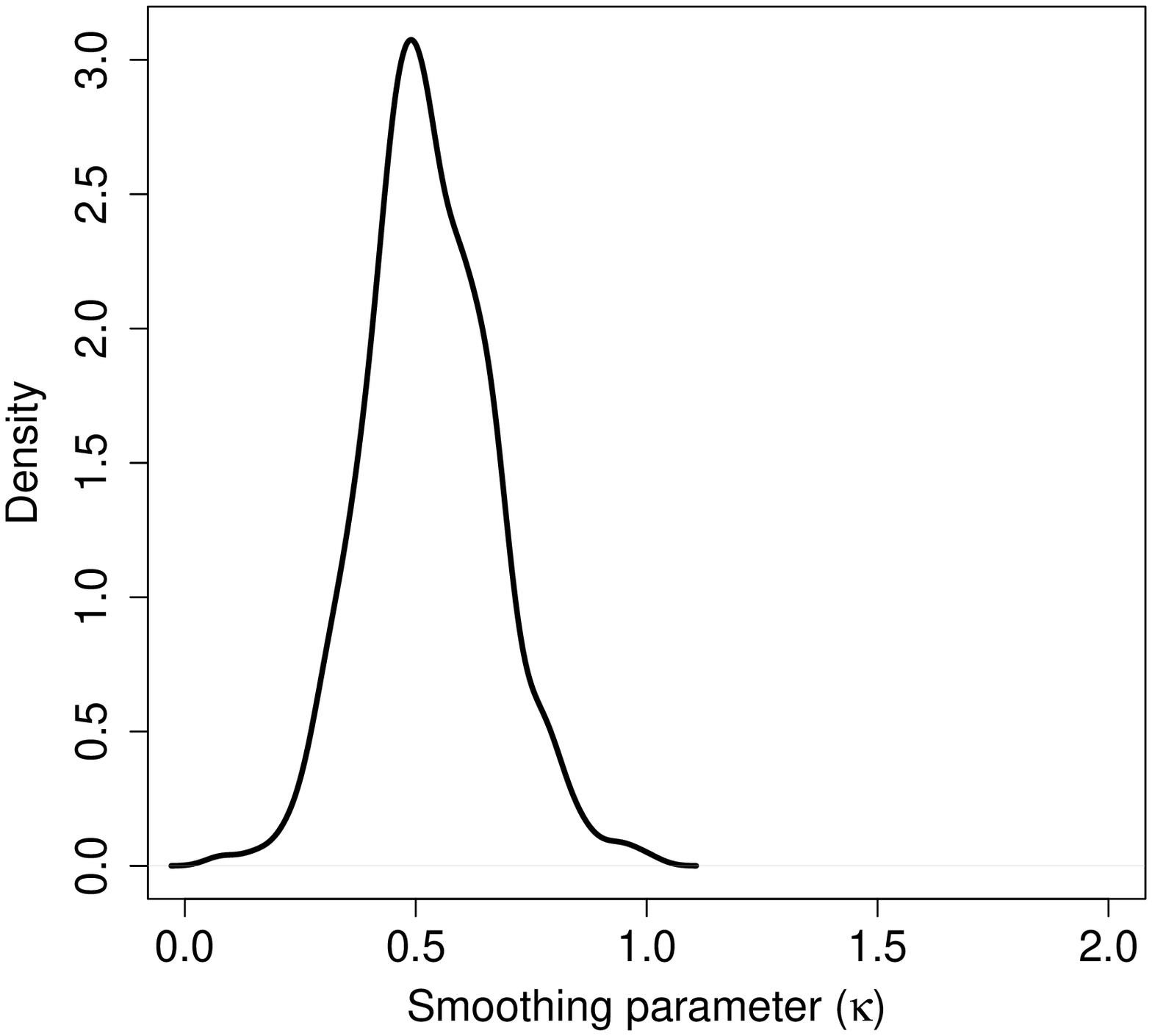} \\
	\includegraphics[width=0.4\textwidth]{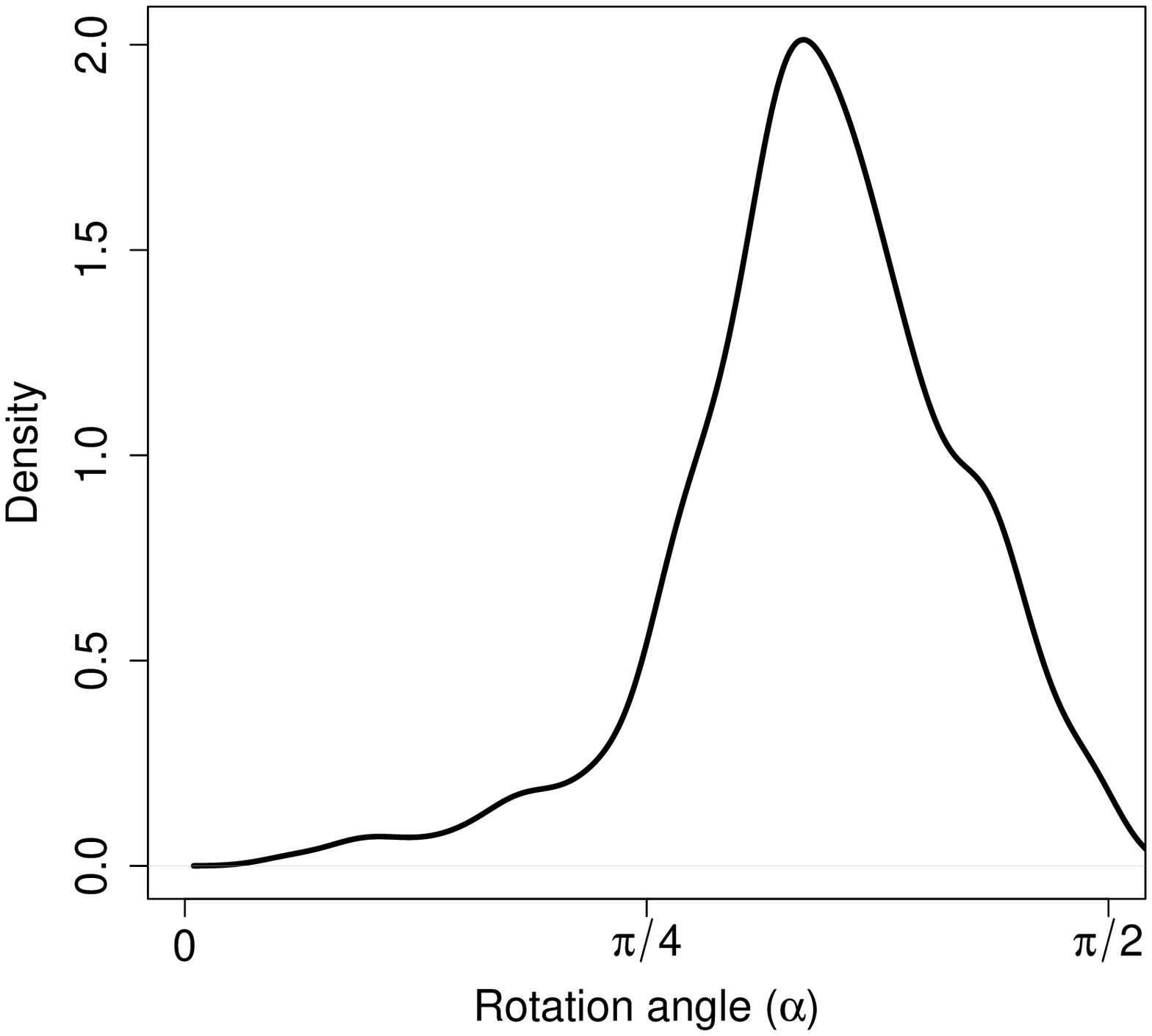}
	\includegraphics[width=0.4\textwidth]{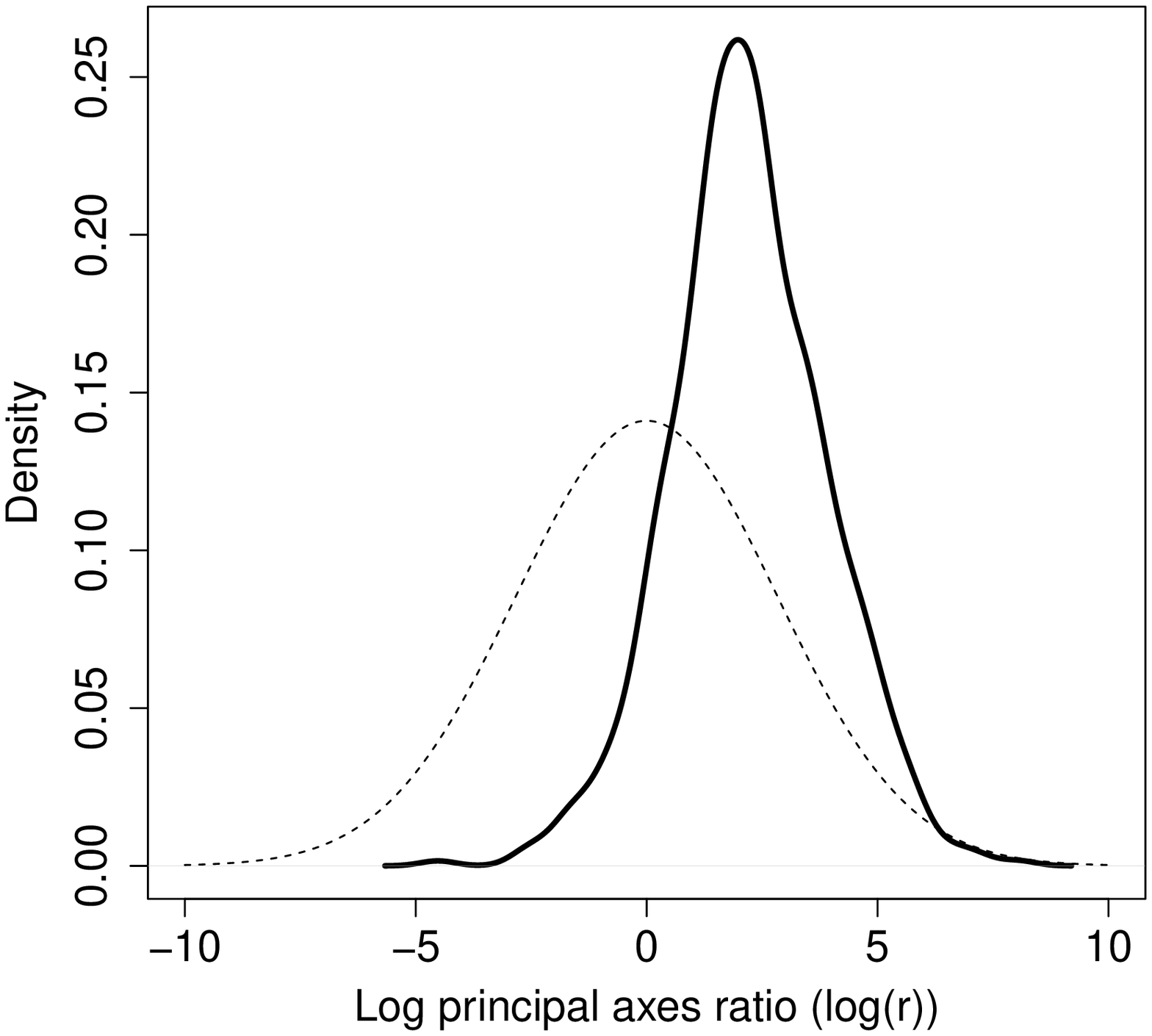} \\
	\includegraphics[width=0.4\textwidth]{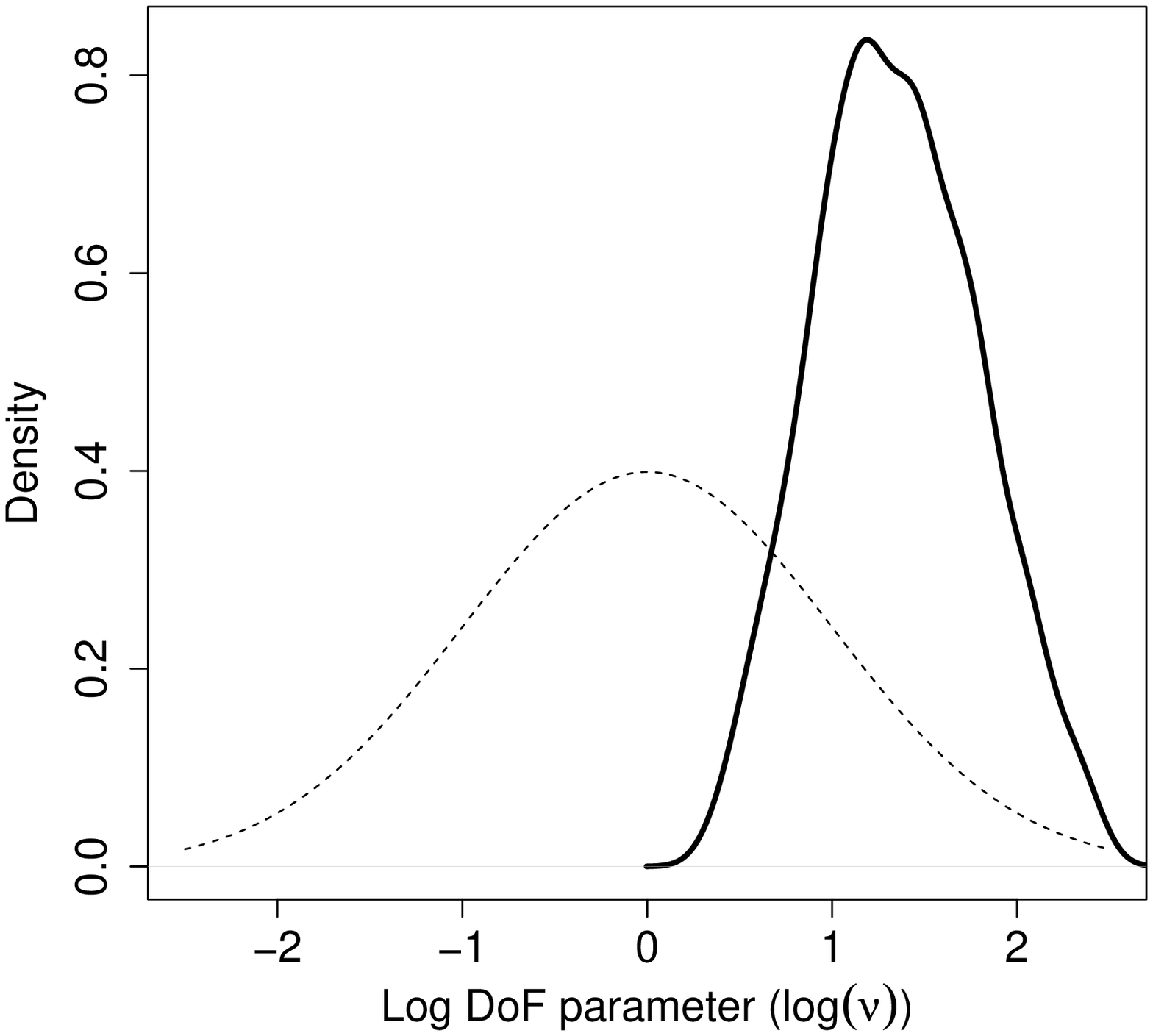}
	\caption{\textbf{Solid lines:} kernel density estimates of marginal posterior distributions for the \emph{$t$ copula model with powered exponential correlation function} when applied to the South Australian data set including the station in Warooka. \textbf{Dashed lines:} prior densities. Uniform prior densities for the smoothing and the rotation angle parameter are not displayed. However, for these parameters the abscissa range is equal to the support of the respective uniform prior distribution.}
	\label{figure:SAdata20150520_Parameters_CPE}
\end{figure}

The marginal posterior distributions for the two most probable models are depicted in \fref{figure:SAdata20150520_Parameters_BR} (Brown-Resnick model) and \fref{figure:SAdata20150520_Parameters_CPE} ($t$ copula model with powered exponential correlation function). They are mostly similar to the posterior distributions when Warooka is excluded (cf.~Figures~\ref{figure:SAdata20150520_Parameters_BR_2} and \ref{figure:SAdata20150520_Parameters_CPE_2}). However, the posterior distributions for the principal axes ratio, $\ratio$, put more mass on very high values. (Note that the densities of the logarithms are plotted.) The posterior distributions for the rotation angle are more concentrated on the upper half. On the other hand, the posterior distribution for the $t$ copula model's degrees of freedom parameter $\nu$ is slightly shifted to the left.

\begin{figure}[h]
	\centering
	\begin{tabular}{cc}
		\includegraphics[width=0.4\textwidth]{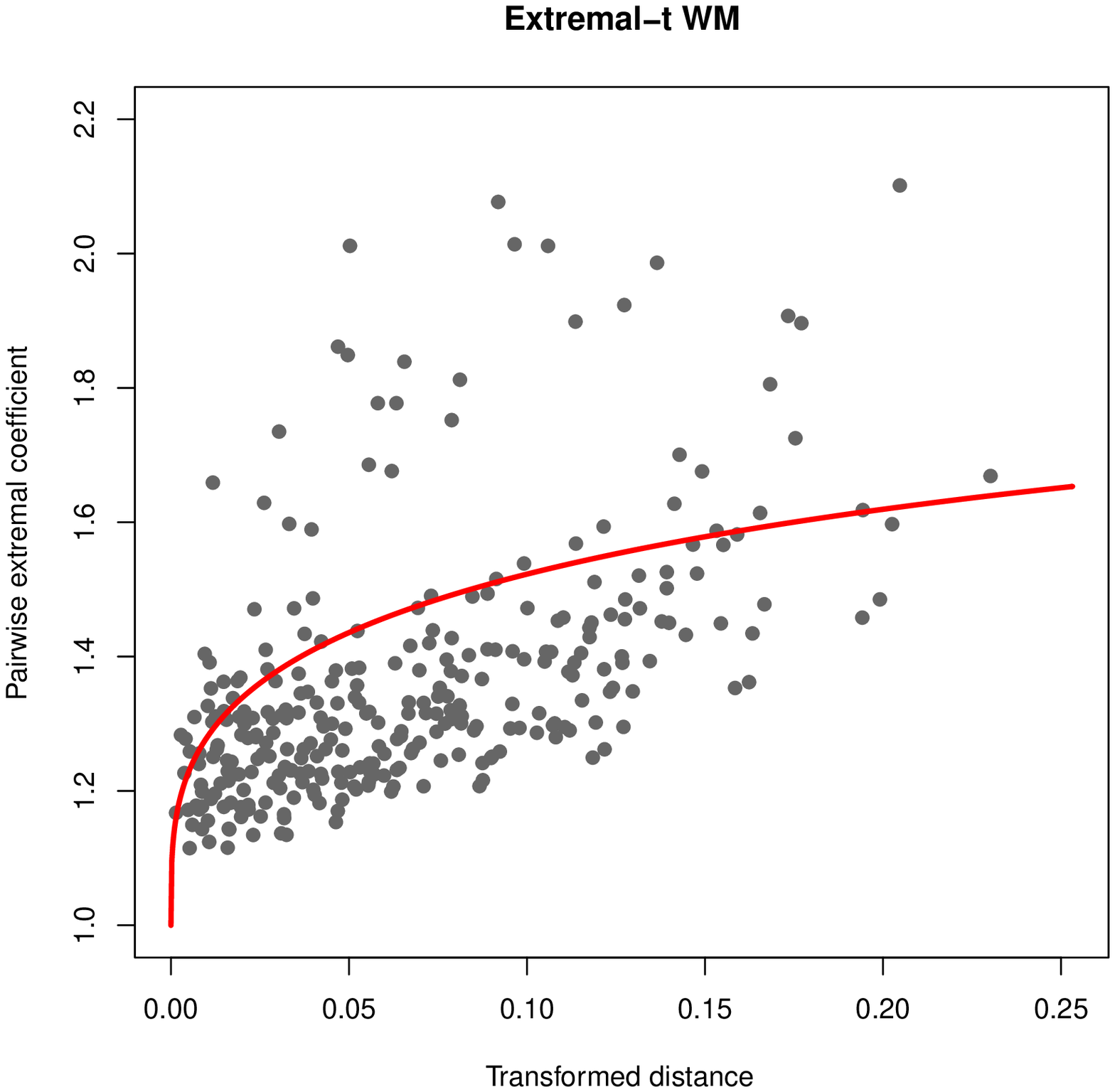} &
		\includegraphics[width=0.4\textwidth]{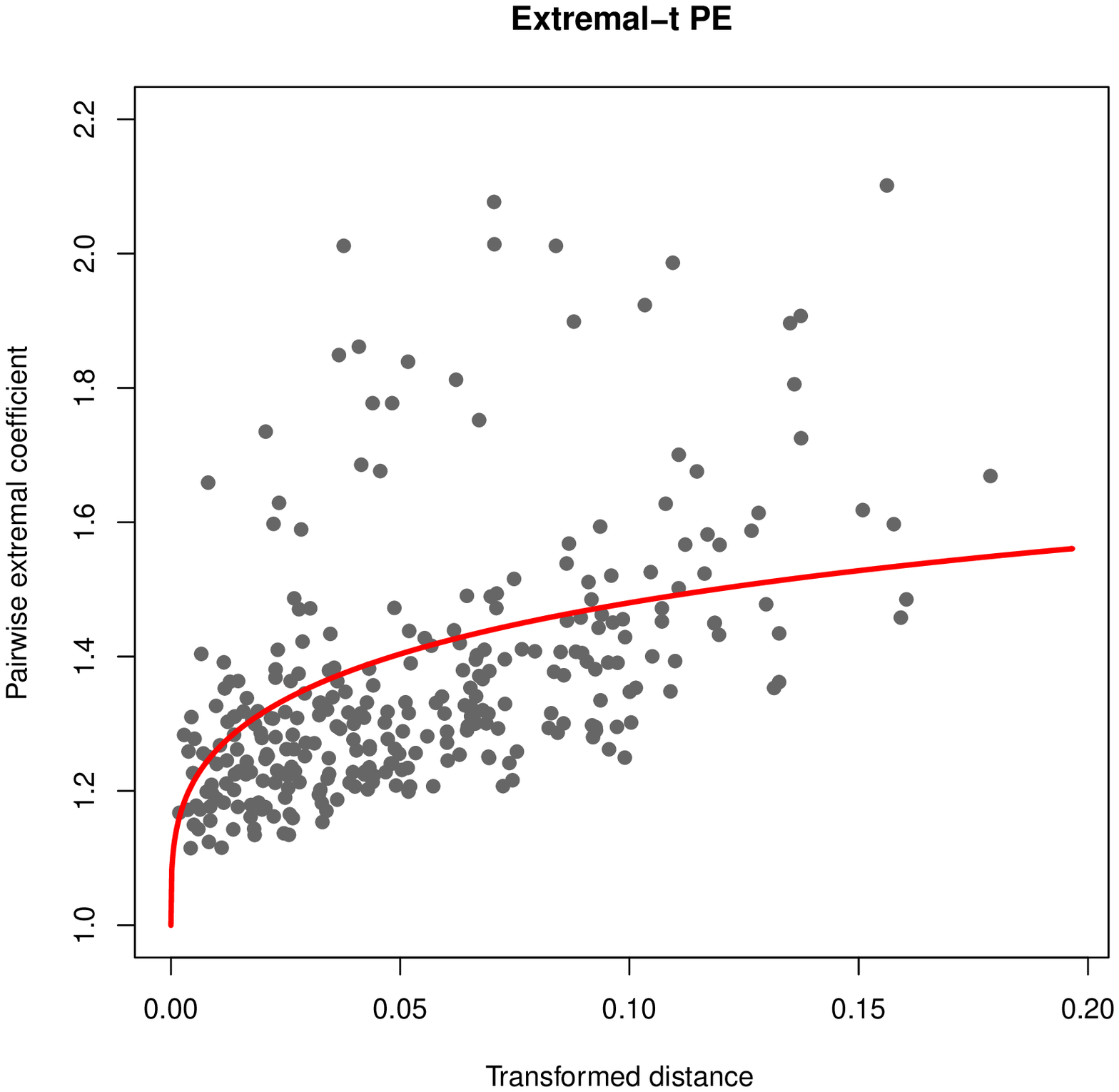} \\
		\includegraphics[width=0.4\textwidth]{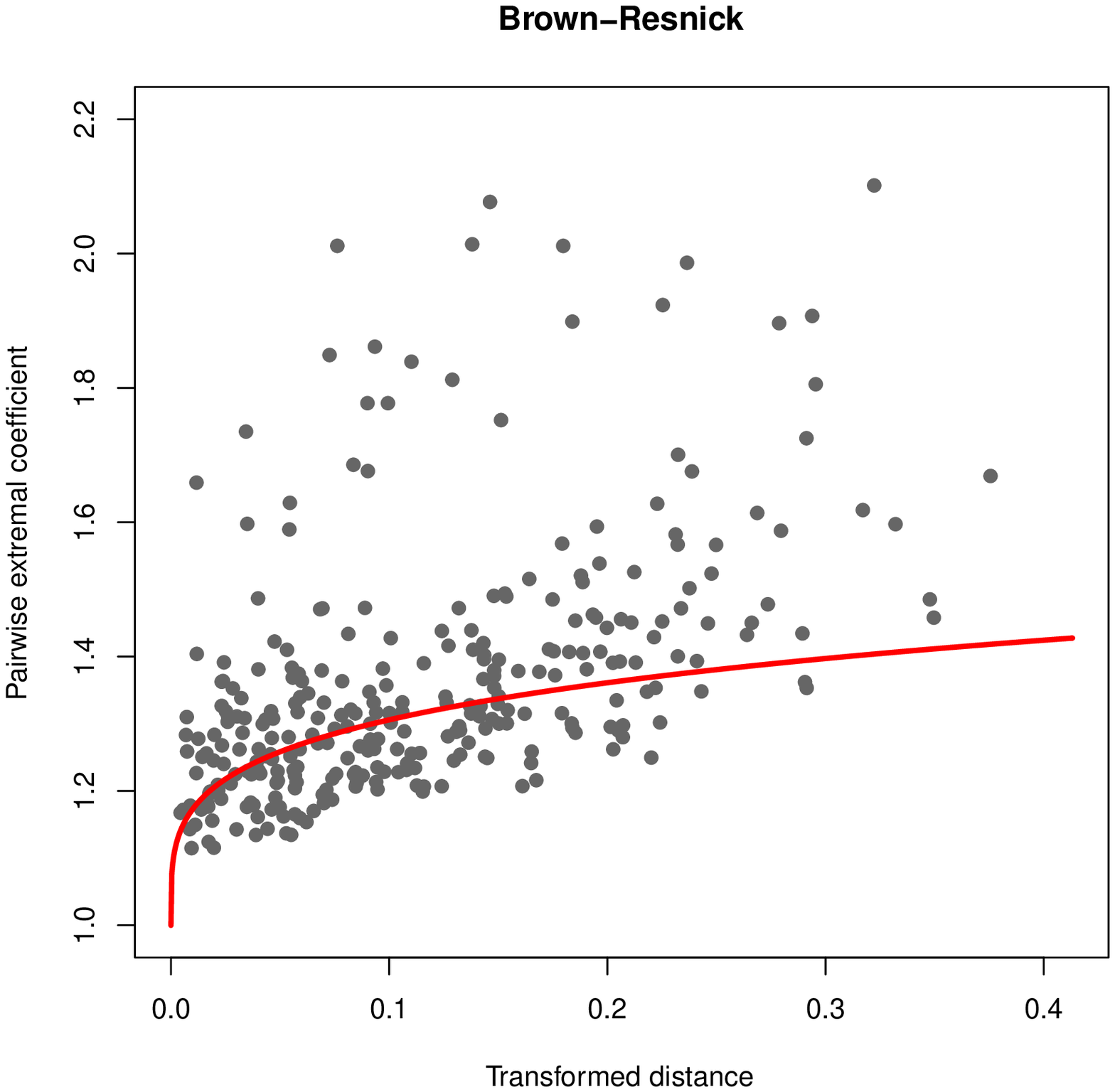} & \\
		\includegraphics[width=0.4\textwidth]{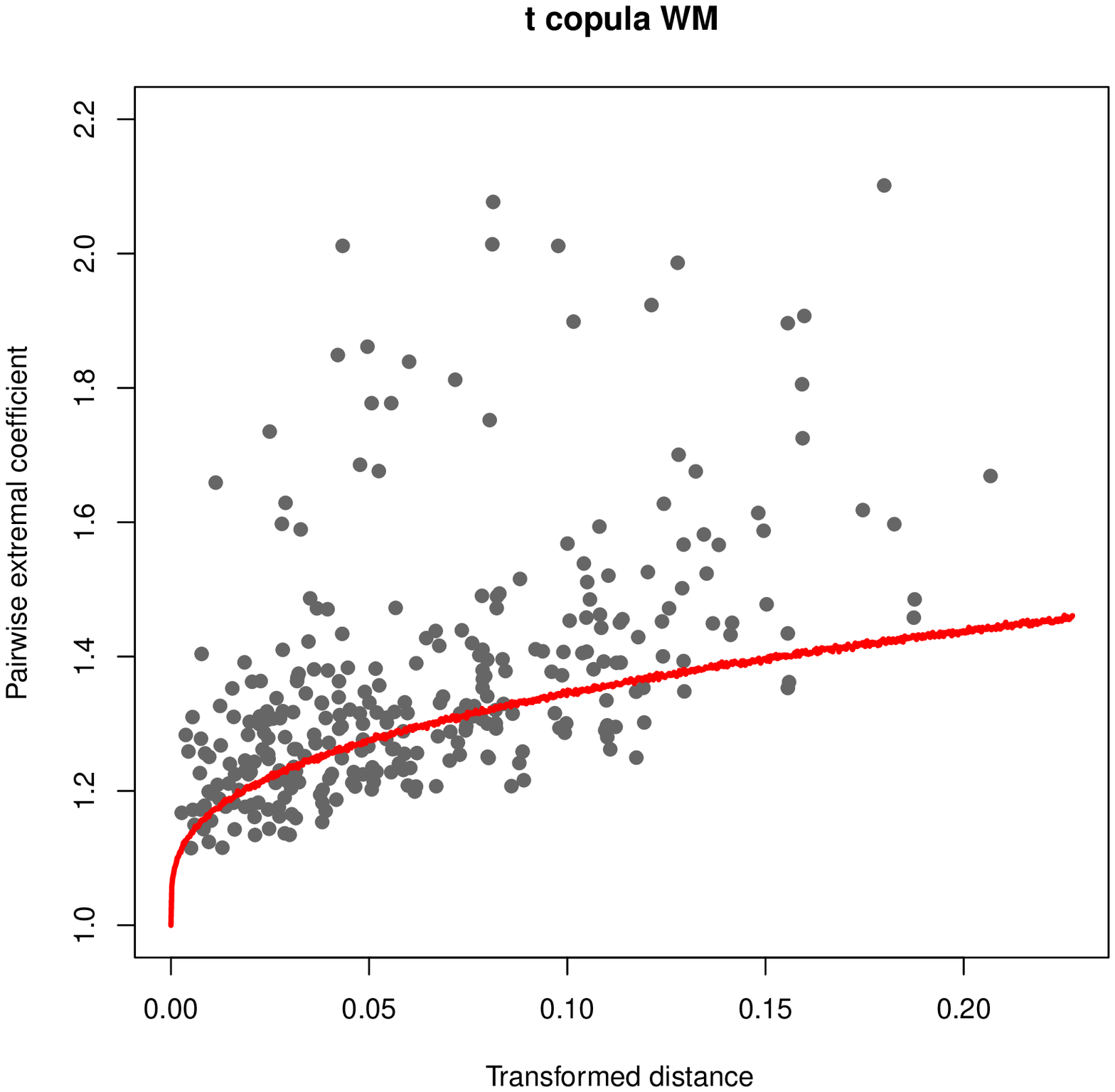} &
		\includegraphics[width=0.4\textwidth]{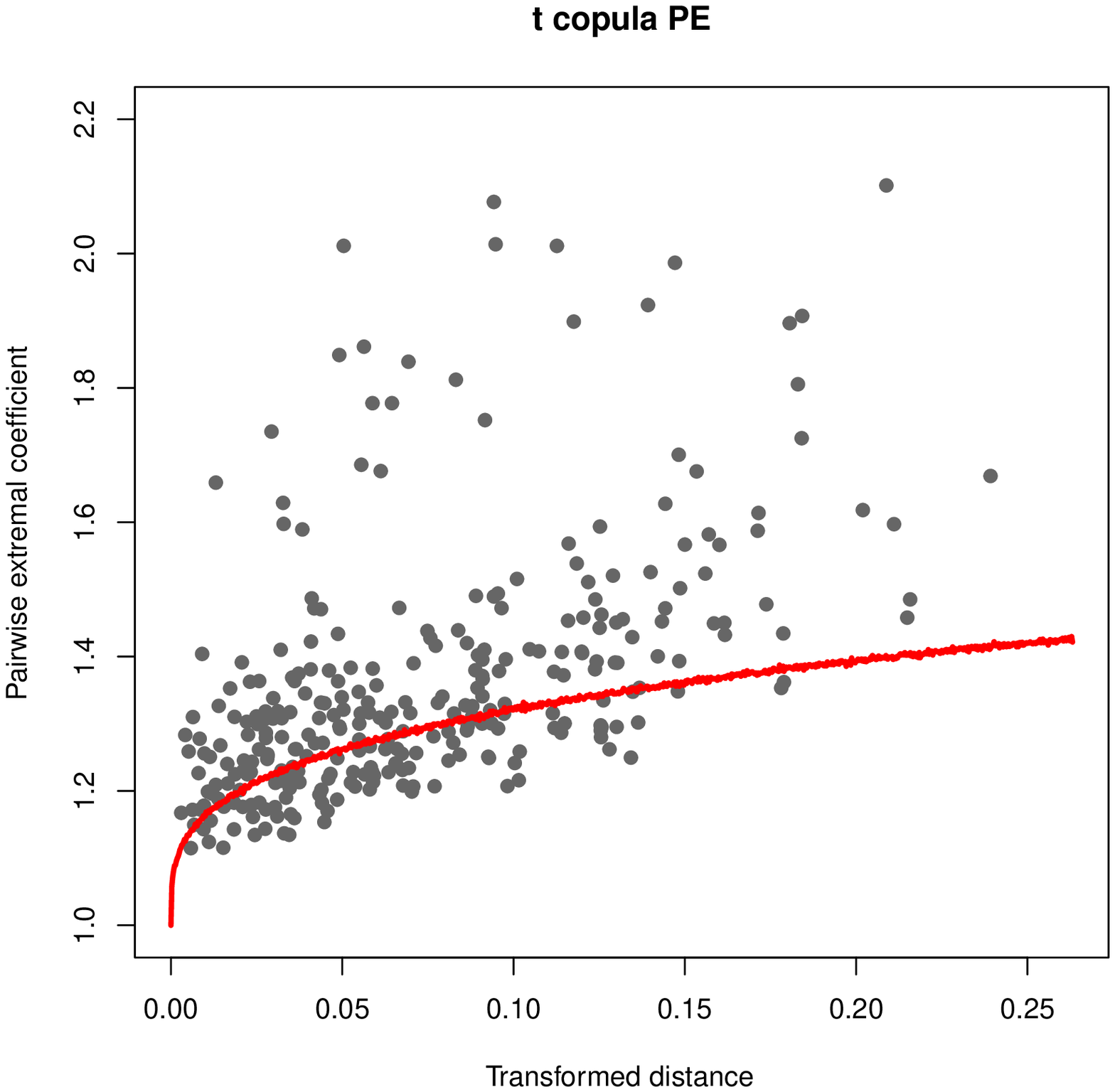}
	\end{tabular}
	\caption{Extremal coefficient function plots when including Warooka for the extremal-$t$ model (top) with Whittle-Mat\'ern (top left) and powered exponential (top right) correlation function, the Brown-Resnick model (middle), and the $t$ copula model (bottom) with Whittle-Mat\'ern (bottom left) and powered exponential (bottom right) correlation function. Each plot was generated using the posterior median values of the respective model's parameters.}
	\label{figure:extremal_plots_1}
\end{figure}

The extremal coefficient function plots are displayed in Figure~\ref{figure:extremal_plots_1}. The pattern is similar to Figure~\ref{figure:extremal_plots_2}.

\begin{figure}[h]
	\centering
	\begin{tabular}{cc}
		F-madogram & Kendall's $\tau$ \\
		\includegraphics[width=0.4\textwidth]{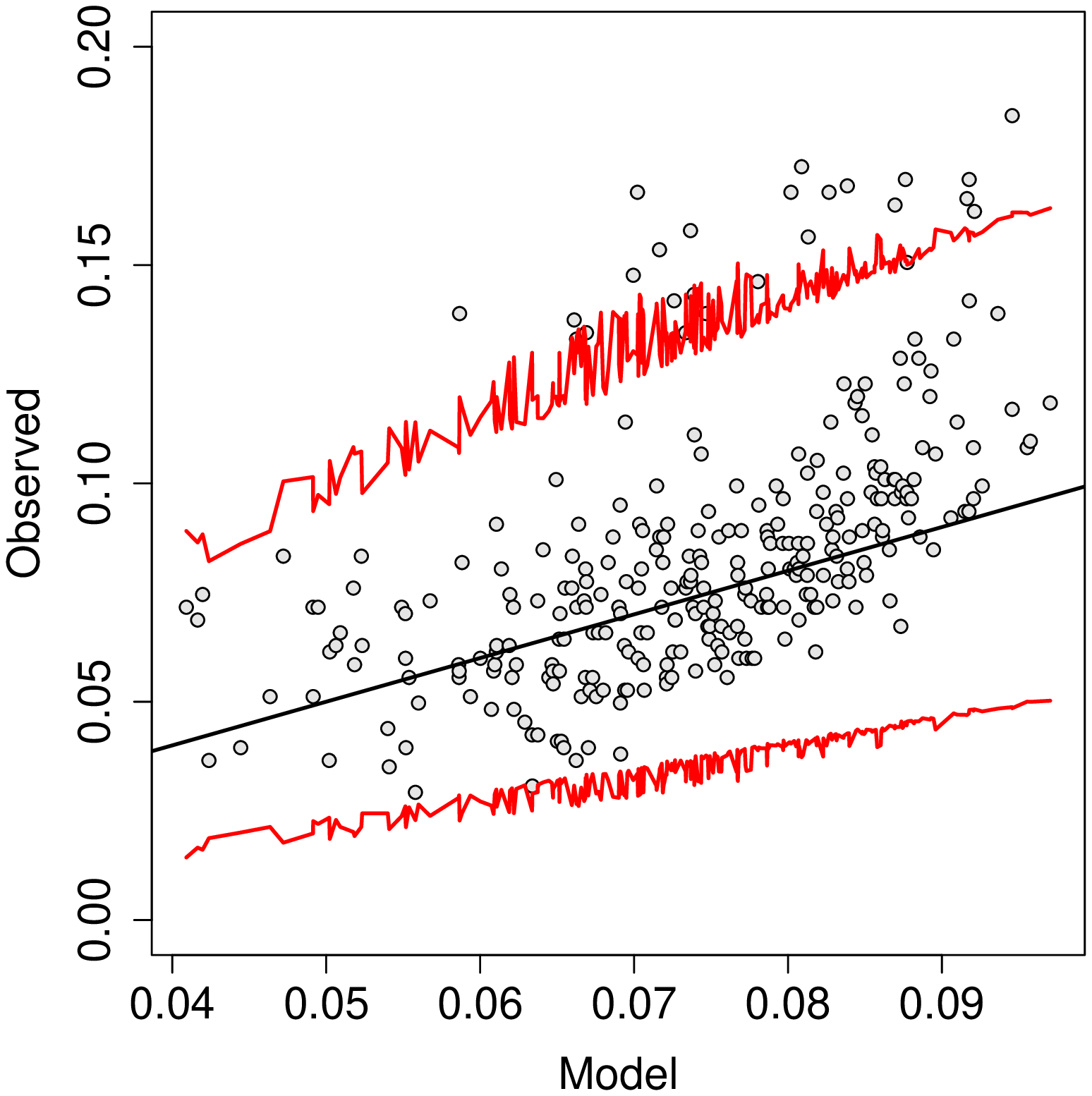} &
		\includegraphics[width=0.4\textwidth]{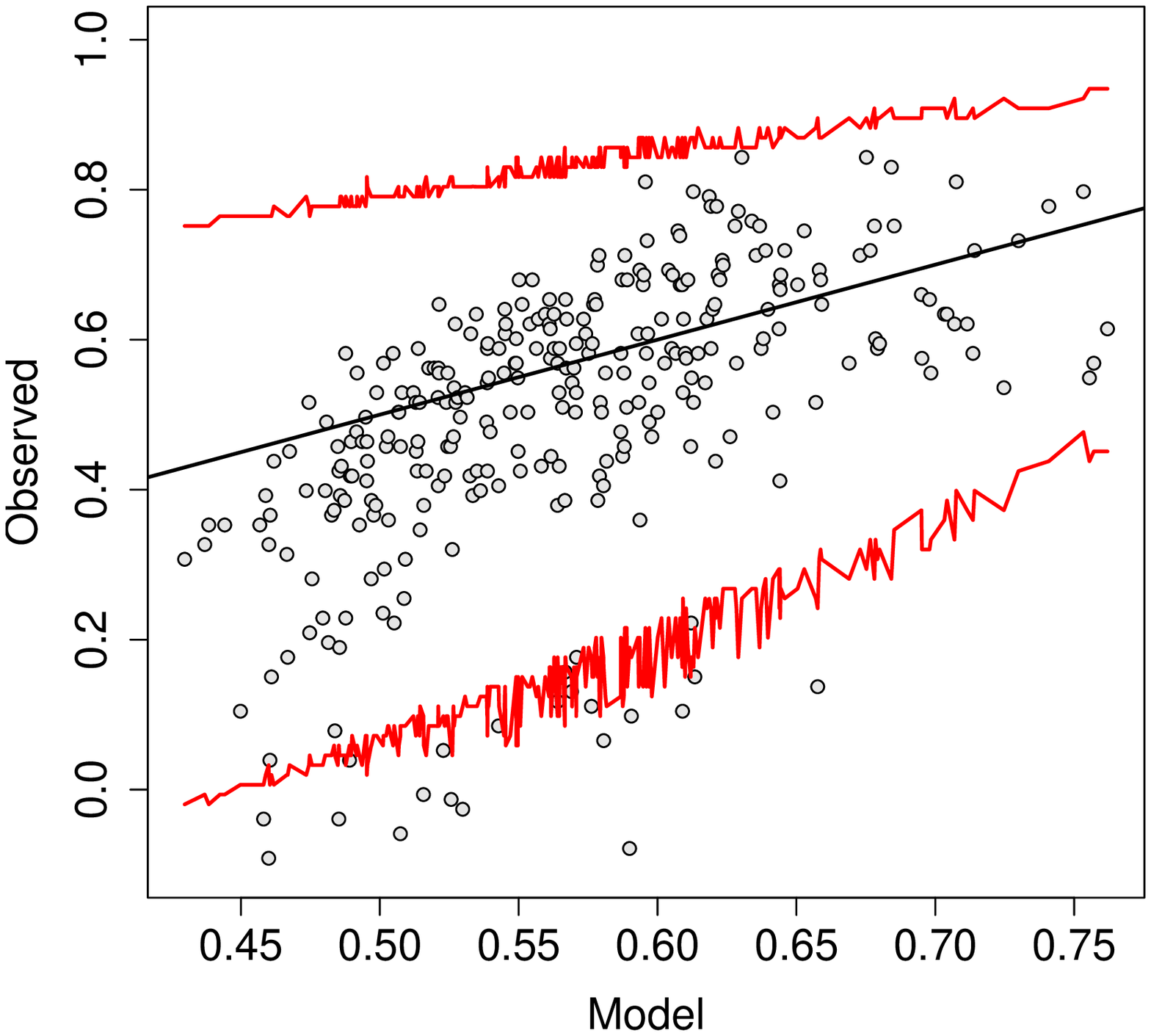} \\
		\includegraphics[width=0.4\textwidth]{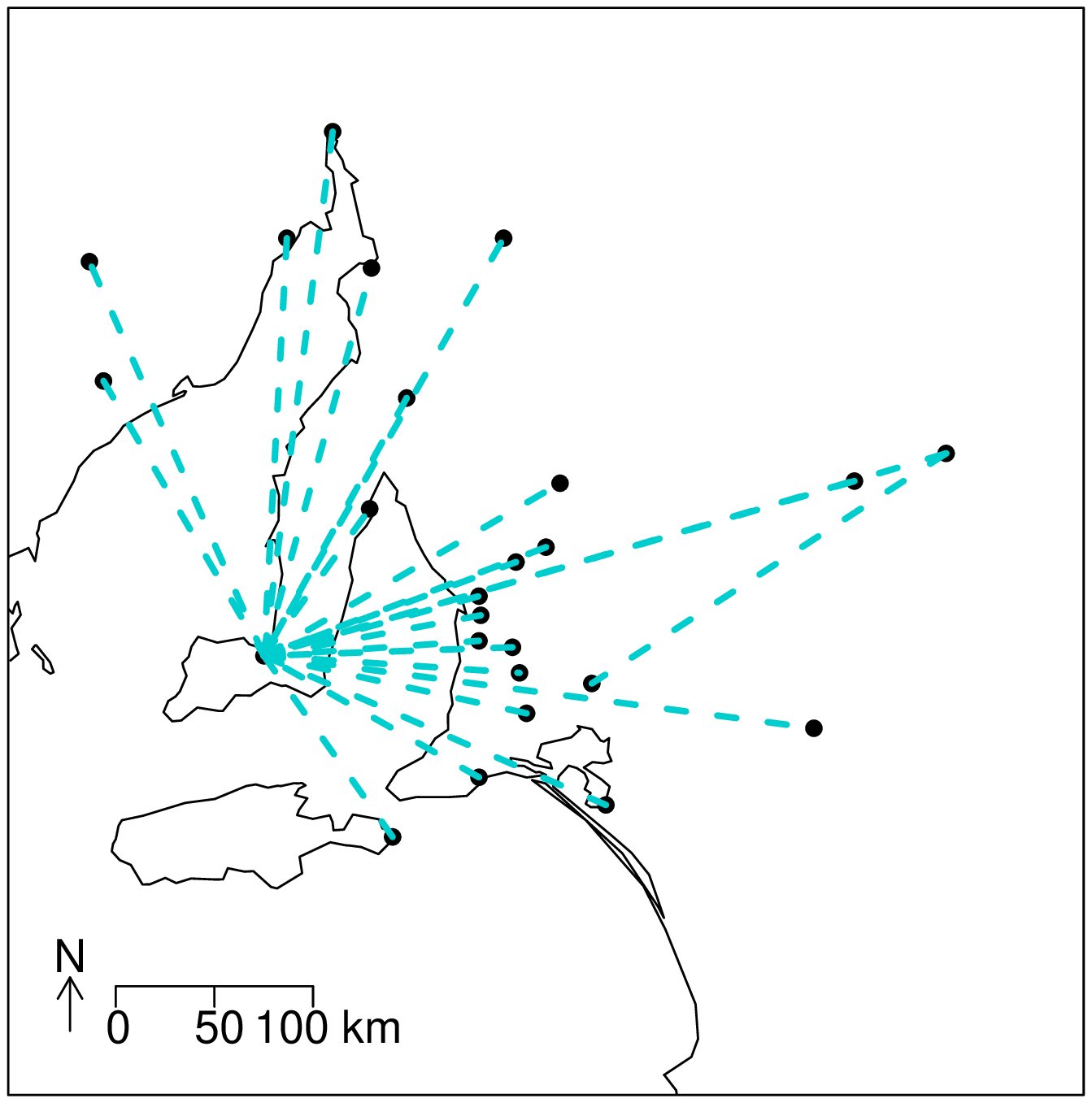} &
		\includegraphics[width=0.4\textwidth]{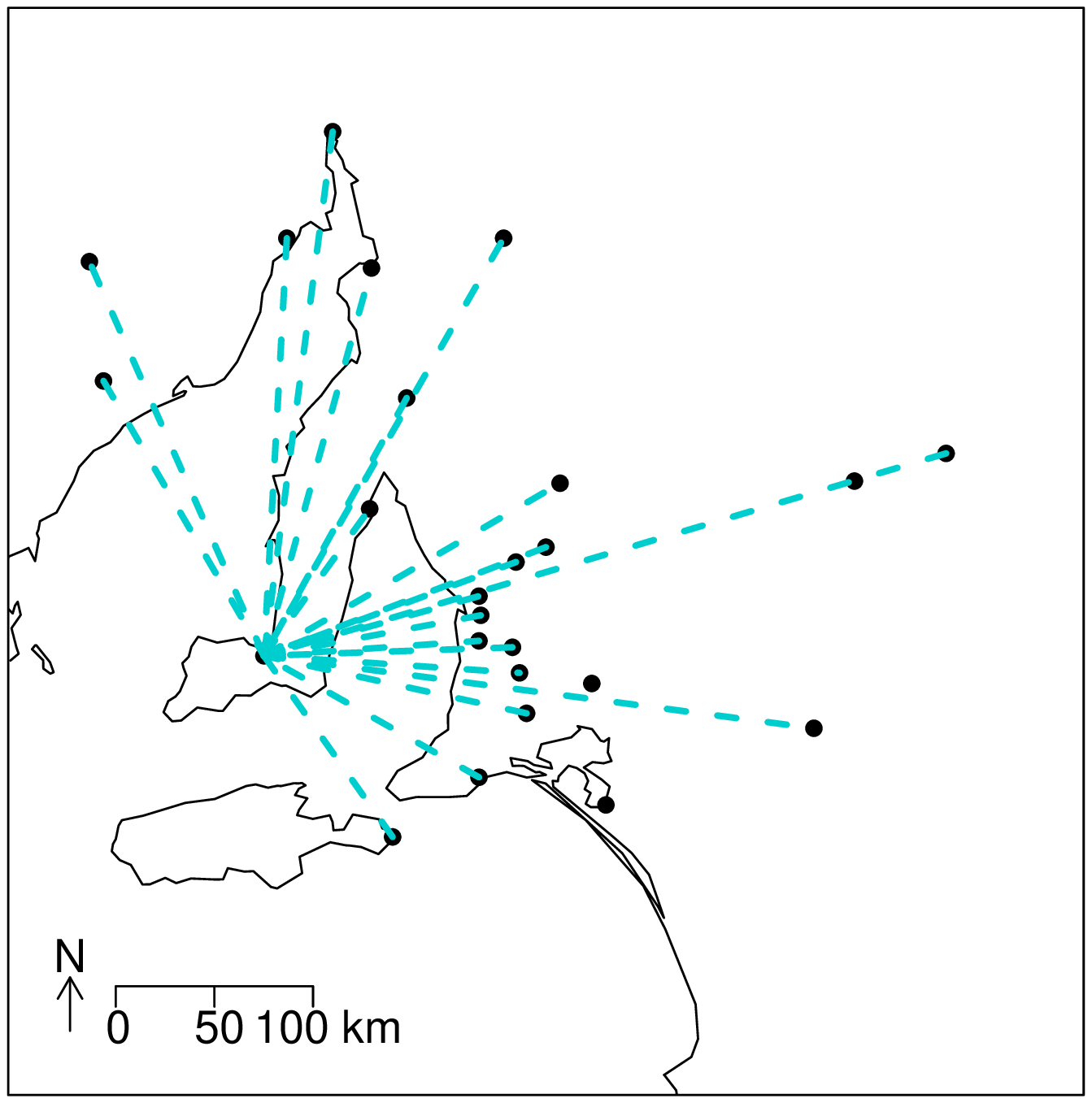}
	\end{tabular}
	\caption{\textbf{Top row:} observed pairwise \emph{F-madogram} (left) and \emph{Kendall's $\tau$} (right) estimates vs.\ posterior predictive means of these estimates for each pair of locations (location 'Warooka' is included). The red lines connect the $2.5 \%$ and $97.5 \%$ quantiles of the posterior predictive distributions for the location pairs.
		\textbf{Bottom row:} \emph{cyan dashed lines} connect location pairs with unusually \emph{low observed mutual dependency}. These are location pairs where the observed F-madogram estimate is above the $97.5 \%$ quantile of posterior predictive distribution (left) or the observed Kendall's $\tau$ estimate is below the $2.5 \%$ quantile of posterior predictive distribution (right).}
	\label{figure:SAdata_posterior_predictive}
\end{figure}

\fref{figure:SAdata_posterior_predictive} shows the posterior predictive analysis for the pairwise F-madogram and Kendall's $\tau$ estimates when Warooka is included. For $8 \%$/$7 \%$ of the location pairs, the observed F-madogram/ Kendall's $\tau$ estimate falls outside the $95 \%$ posterior predictive probability interval. The bottom plots in \fref{figure:SAdata_posterior_predictive} connect the locations in these pairs with lines. All of the outlying location pairs have an unusually low observed mutual dependency and Warooka on the peninsula is included in all of them but one. Therefore, we consider the observations collected at Warooka as being outlying observations according to our models.

\clearpage

\section{Derivations of bivariate log-likelihoods}

\label{sec:bivariate_loglik}

For all max-stable process models, the bivariate cumulative distribution function (CDF) is equal to
\begin{equation*}
F(z_1,z_2) = \exp \left\{ - V(z_1,z_2) \right\},
\end{equation*}
where $V(z_1,z_2)$ is called the \emph{exponent measure} (see \citet{davison2012}). The probability density function (PDF) is obtained by calculating the mixed partial derivative of $F$ w.r.t.~$z_1$ and $z_2$. To shorten notation, we write $\displaystyle V_1(z_1,z_2) = \frac{\partial}{\partial z_1} V(z_1,z_2)$, $\displaystyle V_2(z_1,z_2) = \frac{\partial}{\partial z_2} V(z_1,z_2)$, and $\displaystyle V_{1,2}(z_1,z_2) = \frac{\partial^2}{\partial z_1 \partial z_2} V(z_1,z_2)$. We obtain
\begin{equation*}
f(z_1,z_2) = \exp \left\{ - V(z_1,z_2) \right\} \left[ V_1(z_1,z_2) \cdot V_2(z_1,z_2) - V_{1,2}(z_1,z_2) \right],
\end{equation*}
and so the log-likelihood is
\begin{equation}
\ell(z_1,z_2) = \log f(z_1,z_2) = - V(z_1,z_2) + \log \left[ V_1(z_1,z_2) \cdot V_2(z_1,z_2) - V_{1,2}(z_1,z_2) \right]. \label{general_log_likelihood}
\end{equation}

\subsection{Brown-Resnick model}

\label{subsec_smith_loglike}

The bivariate CDF of the Brown-Resnick model with power variogram has the form \citep{Ribatet2013}
\begin{eqnarray*}
	F(z_1,z_2) & = & \exp \biggl\{ -\frac{1}{z_1} \Phi\left[ \frac{a(\xb_1,\xb_2;\corparms)}{2} + \frac{1}{a(\xb_1,\xb_2;\corparms)} \log \frac{z_2}{z_1} \right] \\
	& & \phantom{\exp \Biggl\{} - \frac{1}{z_2} \Phi\left[ \frac{a(\xb_1,\xb_2;\corparms)}{2} + \frac{1}{a(\xb_1,\xb_2;\corparms)} \log \frac{z_1}{z_2} \right] \biggr\},
\end{eqnarray*}
where $\Phi(\cdot)$ denotes the standard normal CDF,
\begin{equation*}
a(\xb_1,\xb_2; \corparms = (\range, \smooth, \rotang, \ratio)) = \sqrt{2 \, h(\xb_1, \xb_2; \corparms\backslash\{\smooth\})^{\smooth}},
\end{equation*}
and the distance function is
\begin{equation*}
h(\xb_1, \xb_2; \range, \rotang, \ratio) = \frac{\| \Ab (\xb_1 - \xb_2) \|}{\range},
\end{equation*}
with anisotropy matrix
\begin{equation*}
\bm{A} =  \begin{pmatrix}
1 &  0 \\ 
0 & 1/r
\end{pmatrix} \cdot \begin{pmatrix}
\cos \alpha &  \sin \alpha \\ 
- \sin \alpha & \cos \alpha
\end{pmatrix}.
\end{equation*}
There are four parameters in this model, the range parameter, $\range > 0$, the smoothing parameter, $0 < \smooth \leq 2$, the counter-clockwise rotation angle of the confidence ellipse, $0 \leq \rotang < \pi/2$, and the ratio of the principal axes of the confidence ellipse, $\ratio > 0$.

Therefore, the exponent measure is
\begin{equation}
V(z_1,z_2) = \frac{1}{z_1} \Phi\left[ \frac{a(\xb_1,\xb_2;\corparms)}{2} + \frac{1}{a(\xb_1,\xb_2;\corparms)} \log \frac{z_2}{z_1} \right] + \frac{1}{z_2} \Phi\left[ \frac{a(\xb_1,\xb_2;\corparms)}{2} + \frac{1}{a(\xb_1,\xb_2;\corparms)} \log \frac{z_1}{z_2} \right]. \label{smith_v}
\end{equation}

The partial derivatives of the exponent measure are given in \citet[p.~275]{padoan2010}. Setting $\displaystyle w(z_1,z_2;\corparms) = \frac{a(\xb_1,\xb_2;\corparms)}{2} + \frac{1}{a(\xb_1,\xb_2;\corparms)} \log \frac{z_2}{z_1}$ and $v(z_1,z_2;\corparms) = a(\xb_1,\xb_2;\corparms) - w(z_1,z_2;\corparms)$, and writing $v$, $w$, and $a$ without arguments for brevity, the partial derivatives are
\begin{equation}
V_1(z_1,z_2) = \frac{\varphi(v)}{a z_1 z_2} - \frac{\varphi(w)}{a z_1^2} - \frac{\Phi(w)}{z_1^2}, \label{smith_v1}
\end{equation}
\begin{equation}
V_2(z_1,z_2) = \frac{\varphi(w)}{a z_1 z_2} - \frac{\varphi(v)}{a z_2^2} - \frac{\Phi(v)}{z_2^2}, \label{smith_v2}
\end{equation}
and
\begin{equation}
V_{1,2}(z_1,z_2) = - \frac{v\varphi(w)}{a^2 z_1^2 z_2} - \frac{w\varphi(v)}{a^2 z_1 z_2^2}, \label{smith_v12}
\end{equation}
where $\varphi(\cdot)$ denotes the standard normal PDF.

\subsection{Extremal-$t$ model}

\label{subsec_extremalt_loglike}

The bivariate CDF for the extremal-$t$ model is \citep{Ribatet2013}
\begin{eqnarray*}
	F(z_1,z_2) & = & \exp \biggl\{ -\frac{1}{z_1} T_{1;\nu + 1}\left[ a(\xb_1,\xb_2;\corparms,\nu) \left\{ q(z_1,z_2;\nu) - \rho(\xb_1,\xb_2;\corparms) \right\} \right] \\
	& & \phantom{\exp \bigl\{} - \frac{1}{z_2} T_{1;\nu + 1}\left[ a(\xb_1,\xb_2;\corparms,\nu) \left\{ r(z_1,z_2;\nu) - \rho(\xb_1,\xb_2;\corparms) \right\} \right]\biggr\},
\end{eqnarray*}
where $T_{1;\nu + 1}$ denotes the CDF of a univariate Student-$t$ distribution with $\nu + 1$ degrees of freedom, $$a(\xb_1,\xb_2;\corparms,\nu) = \sqrt{\frac{\nu+1}{1 - \rho(\xb_1,\xb_2;\corparms)^2}},$$ $$q(z_1,z_2;\nu) =\left(\frac{z_2}{z_1}\right)^{1/\nu},$$ $$\displaystyle r(z_1,z_2;\nu) = 1/q(z_1,z_2;\nu) = \left(\frac{z_1}{z_2}\right)^{1/\nu},$$
and $\rho(\xb_1,\xb_2;\corparms)$ is one of the correlation functions given in Appendix~\ref{sec:deriv_cor_fun}.

Therefore, the exponent measure is
\begin{eqnarray}
V(z_1,z_2) & = & \phantom{+} \frac{1}{z_1} T_{1;\nu + 1}\left[ a(\xb_1,\xb_2;\corparms,\nu) \left\{ q(z_1,z_2;\nu) - \rho(\xb_1,\xb_2;\corparms) \right\} \right] \notag \\
& & + \frac{1}{z_2} T_{1;\nu + 1}\left[ a(\xb_1,\xb_2;\corparms,\nu) \left\{ r(z_1,z_2;\nu) - \rho(\xb_1,\xb_2;\corparms) \right\} \right]. \label{extremalt_v}
\end{eqnarray}
To save space, let us introduce the functions
\begin{equation*}
w(z_1,z_2;\corparms,\nu) = a(\xb_1,\xb_2;\corparms,\nu) \left\{ q(z_1,z_2;\nu) - \rho(\xb_1,\xb_2;\corparms) \right\}
\end{equation*}
and
\begin{equation*}
v(z_1,z_2;\corparms,\nu) = a(\xb_1,\xb_2;\corparms,\nu) \left\{ r(z_1,z_2;\nu) - \rho(\xb_1,\xb_2;\corparms) \right\}.
\end{equation*}
To shorten notation, we will suppress the arguments in $a(\cdot)$, $q(\cdot)$, $r(\cdot)$, $\rho(\cdot)$, $w(\cdot)$, and $v(\cdot)$. However, one should always be aware of which variables and parameters they depend upon.

Using the facts that $\displaystyle \frac{\partial w}{\partial z_1} = -\frac{a}{\nu z_1} q$, $\displaystyle \frac{\partial w}{\partial z_2} = \frac{a}{\nu z_2} q$, $\displaystyle \frac{\partial v}{\partial z_1} = \frac{a}{\nu z_1} r$ and $\displaystyle \frac{\partial v}{\partial z_2} = -\frac{a}{\nu z_2} r$, it is straightforward to derive the partial derivatives of $V(z_1,z_2)$:
\begin{equation}
V_1(z_1,z_2) = -\frac{1}{z_1^2} T_{1;\nu+1}(w) - \frac{a q}{\nu z_1^2} t_{1;\nu+1}(w) + \frac{a r}{\nu z_1 z_2} t_{1;\nu+1}(v), \label{extremalt_v1}
\end{equation}
\begin{equation}
V_2(z_1,z_2) = -\frac{1}{z_2^2} T_{1;\nu+1}(v) - \frac{a r}{\nu z_2^2} t_{1;\nu+1}(v) + \frac{a q}{\nu z_1 z_2} t_{1;\nu+1}(w), \label{extremalt_v2}
\end{equation}
and
\begin{eqnarray}
V_{1,2}(z_1,z_2) & = & - \frac{(\nu+1) a q}{\nu^2 z_1^2 z_2} t_{1;\nu+1}(w) - \frac{a^2 q^2}{\nu^2 z_1^2 z_2} \frac{\partial t_{1;\nu+1}(w)}{\partial w} \notag\\
&   & - \frac{(\nu+1) a r}{\nu^2 z_1 z_2^2} t_{1;\nu+1}(v) - \frac{a^2 r^2}{\nu^2 z_1 z_2^2} \frac{\partial t_{1;\nu+1}(v)}{\partial v}, \label{extremalt_v12}
\end{eqnarray}
where $t_{1;\nu+1}(x)$ denotes the PDF of the univariate Student-$t$ distribution with $\nu+1$ degrees of freedom and
\begin{equation*}
\frac{\partial t_{1;\nu+1}(x)}{\partial x} = - \frac{x (\nu+2)}{x^2 + \nu + 1} \; t_{1;\nu+1}(x).
\end{equation*}

\subsection{Student-$t$ copula model}

\label{subsec_tcopula_loglike}

The $H$-dimensional marginal CDF for the $t$ copula model with unit Fr\'echet margins is \citep{demarta2005}
\begin{equation*}
F(z_1, \ldots, z_H) = T_{H;\nu}\left\{T_{1;\nu}^{-1}[G(z_1)], \ldots,  T_{1;\nu}^{-1}[G(z_H)]; \: \Sigmab(\xb_1,\ldots,\xb_H;\corparms) \right\},
\end{equation*}
where $G(z) = \exp(-1/z)$ is the CDF of the unit Fr\'echet distribution, $T_{1;\nu}^{-1}[\cdot]$ is the inverse CDF (quantile function) of the univariate central Student-$t$ distribution with $\nu$ degrees of freedom, and $T_{H;\nu}\{\cdots; \:\Sigmab(\xb_1,\ldots,\xb_H;\corparms)\}$ denotes the CDF of the $H$-dimensional central Student-$t$ distribution with $\nu$ degrees of freedom and dispersion matrix 
\[  \Sigmab(\xb_1,\ldots,\xb_H;\corparms) = \begin{pmatrix}
1 & \rho(\xb_1,\xb_2;\corparms) & \cdots & \rho(\xb_1,\xb_H;\corparms) \\ 
\rho(\xb_2,\xb_1;\corparms) & 1 & \cdots & \rho(\xb_2,\xb_H;\corparms) \\ 
\vdots & \vdots & \ddots & \vdots \\
\rho(\xb_H,\xb_1;\corparms) & \rho(\xb_H,\xb_2;\corparms) & \cdots & 1
\end{pmatrix},  \]
where $\rho(\cdot,\cdot;\corparms)$ is one of the correlation functions given in Appendix~\ref{sec:deriv_cor_fun}.

Therefore, the log-likelihood of this model is
\begin{eqnarray}
\ell(z_1, \ldots, z_H) & = & \phantom{- }\log t_{H;\nu}\left\{T_{1;\nu}^{-1}[G(z_1)], \ldots, T_{1;\nu}^{-1}[G(z_H)]; \: \Sigmab(\xb_1,\ldots,\xb_H;\corparms) \right\} \notag  \\
& & - \sum_{i=1}^H \left[ \log t_{1;\nu}\left\{T_{1;\nu}^{-1}[G(z_i)]\right\} - \log g(z_i) \right], \label{eq:loglik_tcopula_1} 
\end{eqnarray}
where
\begin{equation} 
t_{H;\nu}\left\{y_1, \ldots, y_H; \: \Sigmab \right\} = \frac{\Gamma\{(\nu + H)/2\}}{\Gamma\{\nu/2\} (\nu \pi)^{H/2} \, |\Sigmab|^{1/2}} \left\{ 1 + \frac{1}{\nu} \left( (y_1, \ldots, y_H) \, \Sigmab^{-1} \begin{pmatrix} y_1 \\ \vdots \\ y_H \end{pmatrix} \right) \right\}^{-(\nu+H)/2} \label{eq:multiv_t_density}
\end{equation}
is the PDF of the $H$-dimensional central Student-$t$ distribution with $\nu$ degrees of freedom and dispersion matrix $\Sigmab$,
\begin{equation} 
t_{1;\nu}\{y\} = \frac{\Gamma\{(\nu+1)/2\}}{\sqrt{\nu \pi} \, \Gamma\{\nu/2\}} \left\{ 1 + \frac{y^2}{\nu} \right\}^{-(\nu+1)/2} \label{eq:univ_t_density}
\end{equation}
is the PDF of the univariate central Student-$t$ distribution with $\nu$ degrees of freedom, and
\[ g(z) = \frac{\exp(-1/z)}{z^2} \]
is the PDF of the unit Fr\'echet distribution.

In the bivariate case, we can simplify Equation~\eqref{eq:loglik_tcopula_1} further. Plugging the PDFs \eqref{eq:multiv_t_density} and \eqref{eq:univ_t_density} into Equation~\eqref{eq:loglik_tcopula_1} for $H=2$ and simplifying the notation by introducing $\eta_1 = T_{1;\nu}^{-1}[G(z_1)]$ and $\eta_2 = T_{1;\nu}^{-1}[G(z_2)]$ and omitting the arguments of $\rho(\xb_1,\xb_2;\corparms)$ yields

\begin{eqnarray}
\ell(z_1,z_2) & = & \phantom{- }\log (\nu/2) - \frac{1}{2} \log(1 - \rho^2) - \frac{\nu + 2}{2} \log \left[ 1 + \frac{\eta_1^2 - 2 \rho \eta_1 \eta_2 + \eta_2^2}{\nu (1 - \rho^2)} \right]  \notag \\
& & - 2 \log \Gamma\{(\nu+1)/2\} + 2 \log \Gamma\{\nu/2\} + \frac{\nu + 1}{2} \left[ \log\left(1 + \frac{\eta_1^2}{\nu} \right) + \log\left(1 + \frac{\eta_2^2}{\nu} \right) \right] \notag \\
& & - 1/z_1 - 2 \log(z_1) - 1/z_2 - 2 \log(z_2). \label{eq:loglik_tcopula_2}
\end{eqnarray}

\section{Partial derivatives of log-likelihood functions w.r.t.~parameters}

\label{sec:bivariate_score}

Next we compute the partial derivatives of the log-likelihood function w.r.t.\ the parameters of the model (usually correlation parameters). For now, we do not explicitly specify the form of the correlation function $\rho(\cdot,\cdot;\corparms)$ or of the function $a(\cdot,\cdot;\corparms)$ in the Brown-Resnick model. The partial derivatives of specific correlation functions $\rho(\cdot,\cdot;\corparms)$ and of the Brown-Resnick model's function $a(\cdot,\cdot;\corparms)$ with respect to each of their parameters are given in Appendix~\ref{sec:deriv_cor_fun}.

Similar to Appendix~\ref{sec:bivariate_loglik}, when considering the partial derivative of the log-likelihood w.r.t.\ some generic parameter $\corpar \in \corparms$, \footnote{in our case $\corparms = (\range,\smooth,\rotang,\ratio)$} we write $\displaystyle V_{i}(z_1,z_2;\corparms) = \frac{\partial}{\partial \corpar} V(z_1,z_2;\corparms)$, $\displaystyle V_{1,i}(z_1,z_2;\corparms) = \frac{\partial}{\partial \corpar} V_1(z_1,z_2;\corparms)$, $\displaystyle V_{2,i}(z_1,z_2;\corparms) = \frac{\partial}{\partial \corpar} V_2(z_1,z_2;\corparms)$, and $\displaystyle V_{1,2,i}(z_1,z_2;\corparms) = \frac{\partial}{\partial \corpar} V_{1,2}(z_1,z_2;\corparms)$.

Setting $D(z_1,z_2;\corparms) = V_1(z_1,z_2;\corparms) \cdot V_2(z_1,z_2;\corparms) - V_{1,2}(z_1,z_2;\corparms)$, the derivative of the log-likelihood function \eqref{general_log_likelihood} w.r.t.\ the generic parameter $\corpar$ is
\begin{eqnarray*}
	\frac{\partial}{\partial \corpar} \ell(z_1,z_2;\corparms) & = & - V_{i}(z_1,z_2;\corparms) + \frac{1}{D(z_1,z_2;\corparms)} \bigl[ V_{1,i}(z_1,z_2;\corparms) \cdot V_2(z_1,z_2;\corparms) \\
	& & \phantom{- V_{i}(z_1,z_2;\corparms) +} + V_1(z_1,z_2;\corparms) \cdot V_{2,i}(z_1,z_2;\corparms) - V_{1,2,i}(z_1,z_2;\corparms) \bigr].
\end{eqnarray*}

\subsection{Brown-Resnick model}

To obtain $V_{i}(z_1,z_2;\corparms)$, $V_{1,i}(z_1,z_2;\corparms)$, $V_{2,i}(z_1,z_2;\corparms)$, and $V_{1,2,i}(z_1,z_2;\corparms)$ for the Brown-Resnick model, we need to compute the partial derivatives of Equations \eqref{smith_v}, \eqref{smith_v1}, \eqref{smith_v2}, and \eqref{smith_v12}, respectively, w.r.t.~$\corpar$. These formulae are given in \citet[p.~275--276]{padoan2010}. \footnote{Note that there is a mistake in the last line of p.~275 and in the first line of p.~276, where one has to swap $B$ and $C$.} Using the same notation and abbreviations as in Appendix~\ref{subsec_smith_loglike}, they are given by
\begin{equation*}
V_{i}(z_1,z_2;\corparms) = \left[ \frac{v \varphi(w)}{a z_1} + \frac{w \varphi(v)}{a z_2} \right] \frac{\partial a}{\partial \corpar},
\end{equation*}
\begin{equation*}
V_{1,i}(z_1,z_2;\corparms) = \left[ \frac{(1 - v^2) \varphi(w)}{a^2 z_1^2} - \frac{(1 + w v) \varphi(v)}{a^2 z_1 z_2} \right] \frac{\partial a}{\partial \corpar},
\end{equation*}
\begin{equation*}
V_{2,i}(z_1,z_2;\corparms) = \left[ \frac{(1 - w^2) \varphi(v)}{a^2 z_2^2} - \frac{(1 + w v) \varphi(w)}{a^2 z_1 z_2} \right] \frac{\partial a}{\partial \corpar},
\end{equation*}
and
\begin{equation*}
V_{1,2,i}(z_1,z_2;\corparms) = \frac{(w v^2 + 2 v - w) \varphi(w) z_2 + (v w^2 + 2 w - v) \varphi(v) z_1}{a^3 z_1^2 z_2^2} \frac{\partial a}{\partial \corpar}.
\end{equation*}

\subsection{Extremal-$t$ model}

On the one hand, we have to compute $V_{\nu}(z_1,z_2;\corparms,\nu)$, $V_{1,\nu}(z_1,z_2;\corparms,\nu)$, $V_{2,\nu}(z_1,z_2;\corparms,\nu)$, and $V_{1,2,\nu}(z_1,z_2;\corparms,\nu)$, the partial derivatives w.r.t.\ the degrees of freedom parameter $\nu$. On the other hand, we need the partial derivatives w.r.t.\ any of the generic correlation parameters $\corpar$, $V_{i}(z_1,z_2;\corparms,\nu)$, $V_{1,i}(z_1,z_2;\corparms,\nu)$, $V_{2,i}(z_1,z_2;\corparms,\nu)$, and $V_{1,2,i}(z_1,z_2;\corparms,\nu)$. Both set of derivatives are based on the formulae \eqref{extremalt_v}, \eqref{extremalt_v1}, \eqref{extremalt_v2}, and \eqref{extremalt_v12}.

Let us start with the partial derivatives w.r.t.~$\nu$. Before giving these, we calculate some useful quantities that we will need. As in Appendix~\ref{subsec_extremalt_loglike}, we will generally omit any arguments to keep the notation short.

\begin{eqnarray*}
	\frac{\partial a}{\partial \nu} & = & \frac{a}{2 (\nu + 1)}, \\
	\frac{\partial q}{\partial \nu} & = & -\frac{q \log q}{\nu}, \\
	\frac{\partial r}{\partial \nu} & = & -\frac{r \log r}{\nu}, \\
	\frac{\partial w}{\partial \nu} & = & (q - \rho) \frac{\partial a}{\partial \nu} + a \frac{\partial q}{\partial \nu}, \\
	\frac{\partial v}{\partial \nu} & = & (r - \rho) \frac{\partial a}{\partial \nu} + a \frac{\partial r}{\partial \nu}.
\end{eqnarray*}

We will also need the derivatives of the CDF and PDF of a univariate Student-$t$ distribution with $\nu+1$ degrees of freedom w.r.t.~$\nu$ evaluated at $w$ and $v$. Since these functions depend on $\nu$ directly and indirectly via $w$ and $v$, we have to compute the total differential. For the Student-$t$ CDF this means we have to compute
\begin{eqnarray*}
	\frac{\partial T_{1;\nu+1}(w)}{\partial \nu} & = & \frac{\partial T_{1;\nu+1}(w)}{\partial \nu} \biggl|_{w=w(\nu^*)} + \; t_{1;\nu+1}(w) \frac{\partial w}{\partial \nu},\\
	\frac{\partial T_{1;\nu+1}(v)}{\partial \nu} & = & \frac{\partial T_{1;\nu+1}(v)}{\partial \nu} \biggl|_{v=v(\nu^*)} + \; t_{1;\nu+1}(v) \frac{\partial v}{\partial \nu}.
\end{eqnarray*}
To compute $\displaystyle \frac{\partial T_{1;\nu+1}(x)}{\partial \nu} \biggl|_{x=x^*}$, where $x$ is fixed, finite differences are used.

For the PDF, the differentials are
\begin{eqnarray*}
	\frac{\partial t_{1;\nu+1}(w)}{\partial \nu} & = & \frac{\partial t_{1;\nu+1}(w)}{\partial \nu} \biggl|_{w=w(\nu^*)} + \; \frac{\partial t_{1;\nu+1}(w)}{\partial w} \frac{\partial w}{\partial \nu},\\
	\frac{\partial t_{1;\nu+1}(v)}{\partial \nu} & = & \frac{\partial t_{1;\nu+1}(v)}{\partial \nu} \biggl|_{v=v(\nu^*)} + \; \frac{\partial t_{1;\nu+1}(v)}{\partial v} \frac{\partial v}{\partial \nu},
\end{eqnarray*}
so we need an expression for $\displaystyle \frac{\partial t_{1;\nu+1}(x)}{\partial \nu}$ when the argument $x$ is fixed. After some standard calculations, one arrives at
\begin{equation*}
\frac{\partial t_{1;\nu+1}(x)}{\partial \nu} = \frac{t_{1;\nu+1}(x)}{2} \left[ \psi\{(\nu+2)/2\} - \psi\{(\nu+1)/2\} - \frac{1}{\nu+1} - \log\left(1 + \frac{x^2}{\nu+1}\right) + \frac{x^2(\nu+2)}{(\nu+1)(x^2 + \nu + 1)} \right],
\end{equation*}
where $\psi\{\cdot\}$ is the digamma function.

Using the identity $\displaystyle \frac{\partial t_{1;\nu+1}(x)}{\partial x} \equiv - \frac{x (\nu+2)}{x^2 + \nu + 1} \; t_{1;\nu+1}(x)$ from Appendix~\ref{subsec_extremalt_loglike}, one can also show that
\begin{eqnarray*}
	\frac{\partial}{\partial \nu} \left[ \frac{\partial t_{1;\nu+1}(w)}{\partial w} \right] & = & \frac{\partial t_{1;\nu+1}(w)}{\partial w} \biggl[ \frac{1}{w} \frac{\partial w}{\partial \nu}  + \frac{1}{\nu+2}  - \frac{2 w}{w^2 + \nu + 1} \frac{\partial w}{\partial \nu}  \\
	&  & \phantom{\frac{\partial t_{\nu+1}(w)}{\partial w}}- \frac{1}{w^2 + \nu + 1} + \frac{1}{t_{1;\nu+1}(w)} \frac{\partial t_{1;\nu+1}(w)}{\partial \nu} \biggr],\\
	\frac{\partial}{\partial \nu} \left[ \frac{\partial t_{1;\nu+1}(v)}{\partial v} \right] & = & \frac{\partial t_{1;\nu+1}(v)}{\partial v} \biggl[ \frac{1}{v}\frac{\partial v}{\partial \nu} + \frac{1}{\nu+2}  - \frac{2 v}{v^2 + \nu + 1} \frac{\partial v}{\partial \nu} \\
	&  & \phantom{\frac{\partial t_{\nu+1}(v)}{\partial v}} - \frac{1}{v^2 + \nu + 1}  + \frac{1}{t_{1;\nu+1}(v)} \frac{\partial t_{1;\nu+1}(v)}{\partial \nu} \biggr].
\end{eqnarray*}

Now we are ready to state $V_{\nu}(z_1,z_2;\corparms,\nu)$, $V_{1,\nu}(z_1,z_2;\corparms,\nu)$, $V_{2,\nu}(z_1,z_2;\corparms,\nu)$, and $V_{1,2,\nu}(z_1,z_2;\corparms,\nu)$:

\begin{equation*}
V_{\nu}(z_1,z_2;\corparms,\nu) = \frac{1}{z_1} \frac{\partial T_{1;\nu+1}(w)}{\partial \nu} + \frac{1}{z_2} \frac{\partial T_{1;\nu+1}(v)}{\partial \nu},
\end{equation*}

\begin{eqnarray*}
	V_{1,\nu}(z_1,z_2;\corparms,\nu) & = & -\frac{1}{z_1^2} \frac{\partial T_{1;\nu+1}(w)}{\partial \nu} \\
	& & + \frac{a q}{\nu^2 z_1^2} \; t_{1;\nu+1}(w) - \frac{q}{\nu z_1^2} \; t_{1;\nu+1}(w) \frac{\partial a}{\partial \nu} - \frac{a}{\nu z_1^2} \; t_{1;\nu+1}(w) \frac{\partial q}{\partial \nu} - \frac{a q}{\nu z_1^2} \frac{\partial t_{1;\nu+1}(w)}{\partial \nu} \\
	& & - \frac{a r}{\nu^2 z_1 z_2} \; t_{1;\nu+1}(v) + \frac{r}{\nu z_1 z_2} \; t_{1;\nu+1}(v) \frac{\partial a}{\partial \nu} + \frac{a}{\nu z_1 z_2} \; t_{1;\nu+1}(v) \frac{\partial r}{\partial \nu} + \frac{a r}{\nu z_1 z_2} \frac{\partial t_{1;\nu+1}(v)}{\partial \nu},
\end{eqnarray*}

\begin{eqnarray*}
	V_{2,\nu}(z_1,z_2;\corparms,\nu) & = & -\frac{1}{z_2^2} \frac{\partial T_{1;\nu+1}(v)}{\partial \nu} \\
	& & + \frac{a r}{\nu^2 z_2^2} \; t_{1;\nu+1}(v) - \frac{r}{\nu z_2^2} \; t_{1;\nu+1}(v) \frac{\partial a}{\partial \nu} - \frac{a}{\nu z_2^2} \; t_{1;\nu+1}(v) \frac{\partial r}{\partial \nu} - \frac{a r}{\nu z_2^2} \frac{\partial t_{1;\nu+1}(v)}{\partial \nu} \\
	& & - \frac{a q}{\nu^2 z_1 z_2} \; t_{1;\nu+1}(w) + \frac{q}{\nu z_1 z_2} \; t_{1;\nu+1}(w) \frac{\partial a}{\partial \nu} + \frac{a}{\nu z_1 z_2} \; t_{1;\nu+1}(w) \frac{\partial q}{\partial \nu} + \frac{a q}{\nu z_1 z_2} \frac{\partial t_{1;\nu+1}(w)}{\partial \nu},
\end{eqnarray*}
and
\begin{eqnarray*}
	V_{1,2,\nu}(\cdot) & = & \frac{(\nu+2) a q}{\nu^3 z_1^2 z_2} \; t_{1;\nu+1}(w) - \frac{(\nu+1) q}{\nu^2 z_1^2 z_2} \; t_{1;\nu+1}(w) \frac{\partial a}{\partial \nu} - \frac{(\nu+1) a}{\nu^2 z_1^2 z_2} \; t_{1;\nu+1}(w) \frac{\partial q}{\partial \nu} - \frac{(\nu+1) a q}{\nu^2 z_1^2 z_2} \frac{\partial t_{1;\nu+1}(w)}{\partial \nu} \\
	& & + \frac{2 a^2 q^2}{\nu^3 z_1^2 z_2} \frac{\partial t_{1;\nu+1}(w)}{\partial w} - \frac{2 a q^2}{\nu^2 z_1^2 z_2} \frac{\partial t_{1;\nu+1}(w)}{\partial w} \frac{\partial a}{\partial \nu} - \frac{2 a^2 q}{\nu^2 z_1^2 z_2} \frac{\partial t_{1;\nu+1}(w)}{\partial w} \frac{\partial q}{\partial \nu} - \frac{a^2 q^2}{\nu^2 z_1^2 z_2} \frac{\partial}{\partial \nu} \left[\frac{\partial t_{1;\nu+1}(w)}{\partial w}\right] \\
	& & + \frac{(\nu+2) a r}{\nu^3 z_1 z_2^2} \; t_{1;\nu+1}(v) - \frac{(\nu+1) r}{\nu^2 z_1 z_2^2} \; t_{1;\nu+1}(v) \frac{\partial a}{\partial \nu} - \frac{(\nu+1) a}{\nu^2 z_1 z_2^2} \; t_{1;\nu+1}(v) \frac{\partial r}{\partial \nu} - \frac{(\nu+1) a r}{\nu^2 z_1 z_2^2} \frac{\partial t_{1;\nu+1}(v)}{\partial \nu} \\
	& & + \frac{2 a^2 r^2}{\nu^3 z_1 z_2^2} \frac{\partial t_{1;\nu+1}(v)}{\partial v} - \frac{2 a r^2}{\nu^2 z_1 z_2^2} \frac{\partial t_{1;\nu+1}(v)}{\partial v} \frac{\partial a}{\partial \nu} - \frac{2 a^2 r}{\nu^2 z_1 z_2^2} \frac{\partial t_{1;\nu+1}(v)}{\partial v} \frac{\partial r}{\partial \nu} - \frac{a^2 r^2}{\nu^2 z_1 z_2^2} \frac{\partial}{\partial \nu} \left[\frac{\partial t_{1;\nu+1}(v)}{\partial v}\right].
\end{eqnarray*}

Before we can give the partial derivatives w.r.t.\ some generic correlation parameter $\corpar$, we need some further useful quantities.
\begin{eqnarray*}
	\frac{\partial a}{\partial \corpar} & = & \frac{a \rho}{1 - \rho^2} \frac{\partial \rho}{\partial \corpar}, \\
	\frac{\partial w}{\partial \corpar} & = & \left[ - a + \frac{w \rho}{1 - \rho^2} \right] \frac{\partial \rho}{\partial \corpar}, \\
	\frac{\partial v}{\partial \corpar} & = & \left[ - a + \frac{v \rho}{1 - \rho^2} \right] \frac{\partial \rho}{\partial \corpar}.
\end{eqnarray*}

In addition, we need the second derivative of the Student-$t$ density with $\nu+1$ degrees of freedom w.r.t.~$x$ (again making use of $\displaystyle \frac{\partial t_{1;\nu+1}(x)}{\partial x} \equiv - \frac{x (\nu+2)}{x^2 + \nu + 1} \; t_{1;\nu+1}(x)$):
\begin{equation*}
\frac{\partial^2 t_{1;\nu+1}(x)}{\partial x^2} = \frac{1}{x} \frac{\partial t_{1;\nu+1}(x)}{\partial x} + \frac{\nu + 4}{\nu + 2} \; \frac{1}{t_{1;\nu+1}(x)} \left( \frac{\partial t_{1;\nu+1}(x)}{\partial x} \right)^2.
\end{equation*}

Given all these quantities, one obtains
\begin{eqnarray*}
	V_{1,i}(z_1,z_2;\corparms,\nu) & = & - \frac{1}{z_1^2} \; t_{1;\nu+1}(w) \frac{\partial w}{\partial \corpar} \\
	& & - \frac{a q}{\nu z_1^2} \frac{\partial t_{1;\nu+1}(w)}{\partial w} \frac{\partial w}{\partial \corpar} - \frac{q}{\nu z_1^2} \; t_{1;\nu+1}(w) \frac{\partial a}{\partial \corpar} \\
	& & + \frac{a r}{\nu z_1 z_2} \frac{\partial t_{1;\nu+1}(v)}{\partial v} \frac{\partial v}{\partial \corpar} + \frac{r}{\nu z_1 z_2} \; t_{1;\nu+1}(v) \frac{\partial a}{\partial \corpar},
\end{eqnarray*}

\begin{eqnarray*}
	V_{2,i}(z_1,z_2;\corparms,\nu) & = & - \frac{1}{z_2^2} \; t_{1;\nu+1}(v) \frac{\partial v}{\partial \corpar} \\
	& & - \frac{a r}{\nu z_2^2} \frac{\partial t_{1;\nu+1}(v)}{\partial v} \frac{\partial v}{\partial \corpar} - \frac{r}{\nu z_2^2} \; t_{1;\nu+1}(v) \frac{\partial a}{\partial \corpar} \\
	& & + \frac{a q}{\nu z_1 z_2} \frac{\partial t_{1;\nu+1}(w)}{\partial w} \frac{\partial w}{\partial \corpar} + \frac{q}{\nu z_1 z_2} \; t_{1;\nu+1}(w) \frac{\partial a}{\partial \corpar},
\end{eqnarray*}
and
\begin{eqnarray*}
	V_{1,2,i}(z_1,z_2;\corparms,\nu) & = & - \frac{(\nu+1) q}{\nu^2 z_1^2 z_2} \; t_{1;\nu+1}(w) \frac{\partial a}{\partial \corpar} - \frac{(\nu+1) a q}{\nu^2 z_1^2 z_2} \frac{\partial t_{1;\nu+1}(w)}{\partial w} \frac{\partial w}{\partial \corpar} \\
	& & - \frac{2 a q^2}{\nu^2 z_1^2 z_2} \frac{\partial t_{1;\nu+1}(w)}{\partial w} \frac{\partial a}{\partial \corpar} - \frac{a^2 q^2}{\nu^2 z_1^2 z_2} \frac{\partial^2 t_{1;\nu+1}(w)}{\partial w^2} \frac{\partial w}{\partial \corpar} \\
	& & - \frac{(\nu+1) r}{\nu^2 z_1 z_2^2} \; t_{1;\nu+1}(v) \frac{\partial a}{\partial \corpar} - \frac{(\nu+1) a r}{\nu^2 z_1 z_2^2} \frac{\partial t_{1;\nu+1}(v)}{\partial v} \frac{\partial v}{\partial \corpar} \\
	& & - \frac{2 a r^2}{\nu^2 z_1 z_2^2} \frac{\partial t_{1;\nu+1}(v)}{\partial v} \frac{\partial a}{\partial \corpar} - \frac{a^2 r^2}{\nu^2 z_1 z_2^2} \frac{\partial^2 t_{1;\nu+1}(v)}{\partial v^2} \frac{\partial v}{\partial \corpar}.\\
\end{eqnarray*}

\subsection{Student-$t$ copula model}

\label{subsec_tcopula_score}

In the following, we state the derivatives of the $H$-dimensional log-likelihood function \eqref{eq:loglik_tcopula_1} of the Student-$t$ copula model w.r.t.\ $\nu$ and the correlation parameters $\corpar \in \corparms$. As in Appendix~\ref{subsec_tcopula_loglike}, we introduce $\eta_1 = T_{1;\nu}^{-1}[G(z_1)], \ldots, \eta_H = T_{1;\nu}^{-1}[G(z_H)]$ \footnote{Note that all of these quantities are functions of $\nu$.} and omit the arguments of the dispersion matrix $\Sigmab(\xb_1,\ldots,\xb_H;\corparms)$. We denote the vector $(\eta_1,\ldots,\eta_H)^T$ by $\bm{\eta}$ and write 
$Q = \bm{\eta}^T \Sigmab^{-1} \bm{\eta}$.

The derivatives $\displaystyle \frac{\partial \eta_1(\nu)}{\partial \nu}, \ldots, \frac{\partial \eta_H(\nu)}{\partial \nu}$ will be abbreviated as $\eta_1', \ldots, \eta_H'$, respectively, and the corresponding vector of derivatives is denoted by $\bm{\eta}' = (\eta_1',\ldots,\eta_H')^T$. These derivatives are not available analytically, so numerical derivatives are taken.

Plugging Equations~\eqref{eq:multiv_t_density} and \eqref{eq:univ_t_density} into Equation~\eqref{eq:loglik_tcopula_1} and simplifying yields the log-likelihood function 

\begin{eqnarray*}
	\ell(z_1,\ldots,z_H; \corparms, \nu) & = & \phantom{-} \log \Gamma\left(\frac{\nu + H}{2}\right) - H \log \Gamma\left(\frac{\nu + 1}{2}\right) + (H-1) \log \Gamma\left(\frac{\nu}{2}\right) - \frac{1}{2} \log |\Sigmab| \\
	& & - \frac{\nu+H}{2} \log\left( 1 + \frac{Q}{\nu} \right) + \frac{\nu+1}{2} \left[ \sum_{i=1}^H \log \left( 1 + \frac{\eta_i^2}{\nu} \right) \right] + \sum_{i=1}^H \log g(z_i).
\end{eqnarray*}

Using the facts that
$$\frac{\partial \log |\Sigmab|}{\partial \corpar} = \mathrm{tr}\left(\Sigmab^{-1} \frac{\partial \Sigmab}{\partial \corpar}\right),$$
where
$$\frac{\partial \Sigmab}{\partial \corpar} = 
\begin{pmatrix}
0 & \frac{\partial \rho(\xb_1,\xb_2;\corparms)}{\partial \corpar} & \cdots & \frac{\partial \rho(\xb_1,\xb_H;\corparms)}{\partial \corpar} \\ 
\frac{\partial \rho(\xb_2,\xb_1;\corparms)}{\partial \corpar} & 0 & \cdots & \frac{\partial \rho(\xb_2,\xb_H;\corparms)}{\partial \corpar} \\ 
\vdots & \vdots & \ddots & \vdots \\
\frac{\partial \rho(\xb_H,\xb_1;\corparms)}{\partial \corpar} & \frac{\partial \rho(\xb_H,\xb_2;\corparms)}{\partial \corpar} & \cdots & 0
\end{pmatrix},$$ 

as well as 
$$\frac{\partial Q}{\partial \corpar} = \frac{\partial \left( \bm{\eta}^T \Sigmab^{-1} \bm{\eta}\right)}{\partial \corpar} = \bm{\eta}^T \frac{\partial \left(\Sigmab^{-1} \right)}{\partial \corpar}  \bm{\eta} = - \bm{\eta}^T \Sigmab^{-1} \frac{\partial \Sigmab}{\partial \corpar} \Sigmab^{-1} \bm{\eta},$$

and
$$\frac{\partial Q}{\partial \nu} = \frac{\partial \left( \bm{\eta}^T \Sigmab^{-1} \bm{\eta}\right)}{\partial \nu} = \left( \frac{\partial (\bm{\eta}^T \Sigmab^{-1} \bm{\eta})}{\partial \bm{\eta}} \right)^T \bm{\eta}' = \left( 2 \Sigmab^{-1} \bm{\eta} \right)^T \bm{\eta}' = 2 \, \bm{\eta}^T \Sigmab^{-1} \bm{\eta}',$$

we obtain the derivatives

\begin{eqnarray*}
	\frac{\partial}{\partial \nu} \ell(z_1,\ldots, z_H;\corparms,\nu) & = & \frac{1}{2} \Biggl\{ \psi\left(\frac{\nu + H}{2}\right) - H \psi\left(\frac{\nu + 1}{2}\right) + (H-1) \psi\left(\frac{\nu}{2}\right) \\
	& & \phantom{\biggl\{} - \log \left( 1 + \frac{Q}{\nu} \right) - \frac{\nu + H}{\nu (\nu + Q)}\left( 2 \, \nu \, \bm{\eta}^T \Sigmab^{-1} \bm{\eta}' - Q \right) \\
	& & \phantom{\biggl\{} + \sum_{i=1}^H \log\left( 1 + \frac{\eta_i^2}{\nu} \right) + \frac{\nu + 1}{\nu} \left[ \sum_{i=1}^H \frac{2 \, \nu \, \eta_i \eta_i' - \eta_i^2}{\nu + \eta_i^2} \right] \Biggr\},
\end{eqnarray*}
where $\displaystyle \psi(x) = \frac{\frac{\mathrm{d}}{\mathrm{d}x} \Gamma(x)}{\Gamma(x)}$ is the digamma function, and
\begin{equation*}
\frac{\partial}{\partial \corpar} \ell(z_1,\ldots, z_H;\corparms,\nu) = \frac{1}{2} \left\{ - \mathrm{tr}\left( \Sigmab^{-1} \frac{\partial \Sigmab}{\partial \corpar} \right) + \frac{\nu + H}{\nu + Q} \left( \bm{\eta}^T \Sigmab^{-1} \frac{\partial \Sigmab}{\partial \corpar} \Sigmab^{-1} \bm{\eta} \right) \right\}.
\end{equation*}

In the bivariate case, the derivatives are given by
\begin{eqnarray*}
	\frac{\partial}{\partial \nu} \ell(z_1,z_2;\corparms,\nu) & = & \phantom{- }\frac{1}{\nu} - \psi\{(\nu+1)/2\} + \psi\{\nu/2\} - \frac{1}{2} \log \left[ 1 + \frac{Q}{\nu} \right] \\
	& &  - \frac{\nu + 2}{2 \, \nu (\nu + Q)} \: \left[ \frac{2 \, \nu}{1 - \rho^2} \: (\eta_1 \eta_1' - \rho \eta_2 \eta_1' - \rho \eta_1 \eta_2' + \eta_2 \eta_2') - Q \right] \\
	& & + \frac{1}{2} \left[ \log \left( 1 + \frac{\eta_1^2}{\nu} \right) + \log \left( 1 + \frac{\eta_2^2}{\nu} \right) \right] \\
	& & + \frac{\nu+1}{2 \, \nu} \left[ \frac{2 \, \nu \, \eta_1 \eta_1' - \eta_1^2}{\nu + \eta_1^2} + \frac{2 \, \nu \, \eta_2 \eta_2' - \eta_2^2}{\nu + \eta_2^2} \right]
\end{eqnarray*}
and
\begin{equation*}
\frac{\partial}{\partial \corpar} \ell(z_1,z_2;\corparms,\nu) = \left[ \frac{\rho}{1 - \rho^2} + \frac{\nu + 2}{\nu + Q} \: \frac{(1 + \rho^2) \eta_1 \eta_2 - \rho (\eta_1^2 + \eta_2^2)}{(1 - \rho^2)^2} \: \right] \rho',
\end{equation*}
where $\rho = \rho(\xb_1,\xb_2;\corparms)$ and $\displaystyle \rho' = \frac{\partial \rho(\xb_1,\xb_2;\corparms)}{\partial \corpar}$.

\section{Partial derivatives of correlation functions}

\label{sec:deriv_cor_fun}

We will neglect the sill and nugget parameters and assume that their values are $1$ and $0$, respectively (since they have to add up to $1$).

Furthermore, all correlation functions and the function $a(\xb_1,\xb_2; \corparms)$ in the Brown-Resnick model depend on the distance function 
\begin{equation*}
h(\xb_1, \xb_2; \corparms\backslash\{\smooth\} = (\range, \rotang, \ratio)) = \frac{\| \Ab (\xb_1 - \xb_2) \|}{\range}, \label{distance_metric}
\end{equation*}
where
\begin{equation*}
\bm{A} =  \begin{pmatrix}
1 &  0 \\ 
0 & 1/r
\end{pmatrix} \cdot \begin{pmatrix}
\cos \alpha &  \sin \alpha \\ 
- \sin \alpha & \cos \alpha
\end{pmatrix},
\end{equation*}
and $\range > 0$ is the range parameter,  $0 \leq \rotang < \pi/2$ is the counter-clockwise rotation angle of the confidence ellipse, and $\ratio > 0$ is the ratio of the principal axes of the confidence ellipse.

Due to the chain rule, the derivatives of the correlation function with respect to $\corpar = \range$, $\corpar = \rotang$ or $\corpar = \ratio$ are
\begin{equation*}
\frac{\partial \rho}{\partial \corpar} = \frac{\partial \rho}{\partial h} \cdot \frac{\partial h}{\partial \corpar},
\end{equation*}
and so we state $\displaystyle \frac{\partial h}{\partial \corpar}$ for the different parameters before turning to the specific derivatives for the different correlation functions.

Let $\db = (d_1, d_2)^T = \xb_1 - \xb_2$ and 
\begin{eqnarray*}
	Q(\rotang,\ratio) & = & \|\bm{A} \db \|^2 = \db^T \bm{A}^T \bm{A} \db \\
	& = & d_1^2 + d_2^2 + \left[ \frac{1}{\ratio^2} - 1 \right] \left[ d_1^2 \sin^2(\rotang) - d_1 d_2 \sin(2 \rotang) + d_2^2 \cos^2(\rotang) \right].
\end{eqnarray*}

We get
\begin{eqnarray*}
	\frac{\partial h}{\partial \range} & = & - \frac{\|\bm{A} \db \|}{\range^2}, \\
	\frac{\partial h}{\partial \rotang} & = & \frac{ \partial Q(\rotang,\ratio)/\partial \rotang }{2 \|\bm{A} \db \| \range}, \\
	\frac{\partial h}{\partial \ratio} & = & \frac{ \partial Q(\rotang,\ratio)/\partial \ratio }{2 \|\bm{A} \db \| \range},
\end{eqnarray*}
where
\begin{eqnarray*}
	\frac{\partial Q(\rotang,\ratio)}{\partial \rotang} & = & \left[ \frac{1}{\ratio^2} - 1 \right] \left[ (d_1^2 - d_2^2) \sin(2 \rotang) - 2 d_1 d_2 \cos(2 \rotang) \right], \\
	\frac{\partial Q(\rotang,\ratio)}{\partial \ratio} & = &  - \frac{2}{\ratio^3} \left[ d_1^2 \sin^2(\rotang) - d_1 d_2 \sin(2 \rotang) + d_2^2 \cos^2(\rotang) \right].
\end{eqnarray*}

\subsection{Whittle-Mat\'{e}rn}

The Whittle-Mat\'{e}rn correlation function is given by \citep{davison2012}
\begin{equation*}
\rho\{h(\xb_1,\xb_2;\range,\rotang,\ratio),\smooth\} = \frac{2^{1-\smooth}}{\Gamma(\smooth)} h^{\smooth} K_{\smooth}\left( h \right),
\end{equation*}
where $K_{\smooth}(\cdot)$ is the modified Bessel function of the second kind of order $\smooth$, and $\smooth > 0$ is the smoothing parameter. The derivative for the function $K_{\smooth}(x)$ is
\begin{equation*}
\frac{d}{dx} K_{\smooth}(x) = - \frac{1}{2} \left[K_{\smooth - 1}(x) + K_{\smooth + 1}(x)\right].
\end{equation*}
Furthermore, the following recursion formula also holds:
\begin{equation*}
K_{\smooth + 1}(x) = \frac{2 \smooth}{x} K_{\smooth}(x) + K_{\smooth - 1}(x).
\end{equation*}
Using these two relations, one can show that for $\corpar = \range$, $\corpar = \rotang$, or $\corpar = \ratio$,
\begin{eqnarray*}
	\frac{\partial \rho}{\partial \corpar} & = & - \frac{2^{1-\smooth}}{\Gamma(\smooth)} h^{\smooth} K_{\smooth - 1}\left( h \right) \frac{\partial h}{\partial \corpar}\\
	& = & - \frac{2^{1-\smooth}}{\Gamma(\smooth)} h^{\smooth} \left[ K_{\smooth + 1}\left( h \right) - \frac{2 \smooth}{h} K_{\smooth}\left( h \right) \right] \frac{\partial h}{\partial \corpar} \\
	& = &  \rho \left[ - \frac{K_{\smooth + 1} \left(h\right)}{K_{\smooth} \left(h\right)} + \frac{2 \smooth}{h} \right] \frac{\partial h}{\partial \corpar}.
\end{eqnarray*}
For the smoothing parameter $\smooth$, one obtains
\begin{eqnarray*}
	\frac{\partial \rho}{\partial \smooth}  & = & \frac{2^{1-\smooth}}{\Gamma(\smooth)} h^{\smooth} \left[ \left\{ \log\left(\frac{h}{2}\right) - \psi(\smooth) \right\} K_{\smooth} \left(h\right) + \frac{\partial}{\partial \smooth} K_{\smooth} \left(h\right) \right] \\
	& = & \rho \left[ \log\left(\frac{h}{2}\right) - \psi(\smooth)  + \frac{\frac{\partial}{\partial \smooth} K_{\smooth}\left(h\right)}{ K_{\smooth}\left(h\right)} \right],
\end{eqnarray*}
where $\displaystyle \psi(x) = \frac{\frac{\mathrm{d}}{\mathrm{d}x} \Gamma(x)}{\Gamma(x)}$ is the digamma function. The partial derivative $\displaystyle \frac{\partial}{\partial \smooth} K_{\smooth}\left(h\right)$ is not available and is therefore approximated numerically by finite differences.

\subsection{Powered exponential}

The powered exponential (also called ``stable'') correlation function reads \citep{davison2012}
\begin{equation*}
\rho\{h(\xb_1,\xb_2;\range,\rotang,\ratio),\smooth\} = \exp \left(-h^{\smooth} \right),
\end{equation*}
where $\smooth$ is a smoothing parameter with $0 < \smooth \leq 2$.

Its partial derivatives w.r.t.~$\corpar = \range$, $\corpar = \rotang$ or $\corpar = \ratio$ are
\begin{equation*}
\frac{\partial \rho}{\partial \corpar}  =  - \exp \left(- h^{\smooth}\right) \smooth h^{\smooth - 1} \frac{\partial h}{\partial \corpar} 
= - \rho \, \smooth h^{\smooth - 1} \frac{\partial h}{\partial \corpar}
\end{equation*}
and the partial derivative w.r.t.~$\smooth$ is
\begin{equation*}
\frac{\partial \rho}{\partial \smooth} = - \exp \left(- h^{\smooth}\right) h^{\smooth} \log(h) 
= - \rho \, h^{\smooth} \log(h).
\end{equation*}

\subsection{Partial derivatives of function $a$ in Brown-Resnick model}

For the Brown-Resnick model with power variogram, $a(\xb_1,\xb_2;\corparms)$ is given by
\begin{equation*}
a(\xb_1,\xb_2;\corparms) = \sqrt{2 \, h^{\smooth}},
\end{equation*}
where $0 < \smooth \leq 2$.

The partial derivatives w.r.t.~$\corpar = \range$, $\corpar = \rotang$ or $\corpar = \ratio$ are
\begin{equation*}
\frac{\partial a}{\partial \corpar}  = \frac{\smooth h^{\smooth/2 - 1}}{\sqrt{2}} \frac{\partial h}{\partial \corpar} =  a \,\frac{\smooth}{2 h} \frac{\partial h}{\partial \corpar}
\end{equation*}
and the partial derivative w.r.t.~$\smooth$ is
\begin{equation*}
\frac{\partial a}{\partial \smooth} =  \sqrt{\frac{h^{\smooth}}{2}} \log(h) = \frac{a}{2} \log(h).
\end{equation*}


\section{Multinomial logistic regression coefficients obtained by FP step}

\label{sec:regression_coefficients}

Table~\ref{table:log_reg_coefs} contains the estimated multinomial logistic regression coefficients for model choice from the FP step. All covariates were standardised before running the regressions. Empty cells in Table~\ref{table:log_reg_coefs} belong to coefficients which have been excluded by the backward stepwise procedure.

For the pairwise dependence indicators (F-madogram, extremal coefficient, Kendall's $\tau$), the location pairs were grouped according to their Euclidean distance and then the average and the standard deviation of the indicator were computed for each group. The four groups were:

\begin{center}
	\begin{tabular}{cc}
		Group & Distance interval \\
		1 & $(0,1]$ \\
		2 & $(1,2]$ \\
		3 & $(2,3]$ \\
		4 & $(3,6]$
	\end{tabular}
\end{center}

For details on the procedure for clustering the location triplets into ten clusters, see Section~\ref{sec:SAmethod}.

After the backward stepwise procedure most covariates are included in the multinomial logistic regression model. The composite score statistics between the two extremal-$t$ models are clearly substitutes. This exemplifies that it is very hard to distinguish between the different correlation functions within the same class of models.


\begin{longtable}{lrrrr}
	\caption{Multinomial logistic regression coefficients for model choice obtained by FP step. Baseline model: E-t WM.\label{table:log_reg_coefs}} \\
	\hline
	Coefficient/Model & E-t PE & B-R & tC WM & tC PE \\ 
	\hline
	& E-t PE & B-R & tC WM & tC PE \\ 
	\hline
	F-mado avg 1 & 0.27 & 5.27 & -1.58 & 2.44 \\ 
	F-mado avg 2 & 0.49 & -0.59 & 1.40 & 1.96 \\ 
	F-mado avg 3 & 0.32 & 0.98 & 1.65 & 2.03 \\ 
	F-mado avg 4 & 0.84 & 0.47 & -0.66 & -0.80 \\ 
	F-mado sd 1 & -0.53 & -0.71 & -1.12 & -1.98 \\ 
	F-mado sd 2 & 0.24 & 1.15 & -0.71 & -0.42 \\ 
	F-mado sd 3 &  &  &  &  \\ 
	F-mado sd 4 &  &  &  &  \\ 
	Extr 2D avg 1 & 0.03 & 2.58 & 7.09 & 8.17 \\ 
	Extr 2D avg 2 & 0.57 & -1.47 & -0.20 & 1.12 \\ 
	Extr 2D avg 3 & 0.25 & -4.42 & -2.48 & -3.09 \\ 
	Extr 2D avg 4 &  &  &  &  \\ 
	Extr 2D sd 1 & 0.01 & -2.66 & -3.32 & -3.12 \\ 
	Extr 2D sd 2 & 0.20 & -0.88 & -1.14 & -1.32 \\ 
	Extr 2D sd 3 & -0.06 & -2.13 & -1.89 & -1.56 \\ 
	Extr 2D sd 4 & -0.35 & -1.26 & -1.20 & -1.23 \\ 
	Extr 3D avg 1 & -0.19 & 1.57 & 1.33 & 1.24 \\ 
	Extr 3D avg 2 & -0.47 & 2.79 & 2.13 & 1.69 \\ 
	Extr 3D avg 3 & 1.47 & -0.16 & -0.91 & -0.89 \\ 
	Extr 3D avg 4 & 0.99 & 0.93 & -0.73 & -0.58 \\ 
	Extr 3D avg 5 & -0.23 & -1.46 & -2.87 & -3.64 \\ 
	Extr 3D avg 6 & -0.46 & 1.93 & -0.39 & -1.29 \\ 
	Extr 3D avg 7 & -1.04 & -1.94 & -5.30 & -6.50 \\ 
	Extr 3D avg 8 & 0.32 & 1.84 & 0.37 & 0.71 \\ 
	Extr 3D avg 9 & -1.52 & 0.18 & -0.74 & -0.93 \\ 
	Extr 3D avg 10 &  &  &  &  \\ 
	Extr 3D sd 1 & 0.29 & 0.78 & 1.05 & 0.89 \\ 
	Extr 3D sd 2 &  &  &  &  \\ 
	Extr 3D sd 3 & 0.52 & 1.98 & 1.87 & 1.96 \\ 
	Extr 3D sd 4 & 0.16 & 0.48 & 0.84 & 1.20 \\ 
	Extr 3D sd 5 & -0.37 & 0.93 & 1.35 & 1.61 \\ 
	Extr 3D sd 6 & -0.20 & 0.08 & 0.71 & 0.67 \\ 
	Extr 3D sd 7 & 0.15 & 1.19 & 2.06 & 1.81 \\ 
	Extr 3D sd 8 & -0.41 & 1.76 & 1.54 & 1.37 \\ 
	Extr 3D sd 9 &  &  &  &  \\ 
	Extr 3D sd 10 & -0.01 & 0.85 & 0.92 & 0.90 \\ 
	Tau avg 1 & -0.02 & 3.58 & -1.04 & 0.88 \\ 
	Tau avg 2 &  &  &  &  \\ 
	Tau avg 3 &  &  &  &  \\ 
	Tau avg 4 & 0.32 & -0.81 & -2.12 & -2.41 \\ 
	Tau sd 1 & 0.47 & 0.23 & 0.34 & 0.95 \\ 
	Tau sd 2 & -0.03 & -1.24 & 0.39 & 0.22 \\ 
	Tau sd 3 &  &  &  &  \\ 
	Tau sd 4 & -0.05 & -0.49 & 0.21 & 0.28 \\ 
	Score E-t WM dof & 0.25 & 5.23 & 1.09 & 1.78 \\ 
	Score E-t WM range &  &  &  &  \\ 
	Score E-t WM smooth &  &  &  &  \\ 
	Score E-t WM angle &  &  &  &  \\ 
	Score E-t WM ratio & 2.51 & 10.82 & 6.56 & 8.45 \\ 
	Score E-t PE dof &  &  &  &  \\ 
	Score E-t PE range & 9.69 & 2.92 & -13.78 & -1.95 \\ 
	Score E-t PE smooth & -13.41 & -5.05 & -11.13 & -28.13 \\ 
	Score E-t PE angle & -0.17 & -1.17 & -1.02 & -1.83 \\ 
	Score E-t PE ratio &  &  &  &  \\ 
	Score B-R range & -0.02 & 0.01 & -0.15 & -0.15 \\ 
	Score B-R smooth & 0.02 & -0.01 & 0.15 & 0.15 \\ 
	Score B-R angle & 0.02 & -0.01 & 0.15 & 0.15 \\ 
	Score B-R ratio & -0.02 & 0.01 & -0.15 & -0.15 \\ 
	Score tC WM dof & -1.51 & -4.06 & 0.70 & 0.26 \\ 
	Score tC WM range &  &  &  &  \\ 
	Score tC WM smooth & -1.93 & 24.47 & 21.78 & 12.44 \\ 
	Score tC WM angle & -0.11 & 0.58 & -0.83 & -1.27 \\ 
	Score tC WM ratio & -1.53 & -13.14 & -3.15 & -5.21 \\ 
	Score tC PE dof & 1.47 & 3.98 & -8.18 & -6.91 \\ 
	Score tC PE range &  &  &  &  \\ 
	Score tC PE smooth & 6.12 & -8.39 & 5.13 & 22.11 \\ 
	Score tC PE angle & -0.02 & -0.02 & 2.58 & 3.66 \\ 
	Score tC PE ratio &  &  &  &  \\ 
	\hline	
\end{longtable}

\end{document}